%% file: main.tex
\documentclass[
reprint,
amsmath,amssymb,
aps,
pra,
]{revtex4-2}
\usepackage{nopageno}
\usepackage{url}
\usepackage[usenames, dvipsnames]{color}
\definecolor{darkgreen}{rgb}{0.0, 0.42, 0.24}
\definecolor{lightgray}{rgb}{0.75, 0.75, 0.75}
\definecolor{darkgray}{rgb}{0.33, 0.33, 0.33}

\usepackage{hyperref}
\usepackage{amsmath, amsthm, amssymb}
\usepackage{mathtools}
\usepackage{cases} 
\usepackage{braket}

\usepackage{thmtools}
\usepackage{thm-restate}

\usepackage{blindtext}
\usepackage{cleveref}
\usepackage{graphicx,subcaption} 
\usepackage{dcolumn}
\usepackage{bm}

\newtheorem{lemma}{Lemma}

\usepackage{tikz}
\usetikzlibrary{decorations.text}
\usetikzlibrary{decorations.pathreplacing}
\usepackage{pgfplots}
\pgfplotsset{compat=1.18}
\usepgfplotslibrary{groupplots} 

\newcommand{\dom}[1]{\mathcal{D}_{1}}

\newcommand{\linfnorm}[1]{\left|\left|#1\right|\right|_{\infty}}
\newcommand{\opnorm}[1]{\left|\left|#1\right|\right|_{op}}

\newcommand{\trace}[1]{\text{Tr}\left(#1\right)}
\usepackage{soul}

\definecolor{c1}{RGB}{44, 62, 80}
\definecolor{darkgray176}{RGB}{176,176,176}
\definecolor{c3}{RGB}{58, 125, 68}
\definecolor{c2}{RGB}{125, 128, 127}
\definecolor{c4}{RGB}{164, 166, 46}
\definecolor{c5}{RGB}{212, 180, 76}
\definecolor{c6}{RGB}{44, 62, 80}
\definecolor{c7}{RGB}{58, 125, 68}

\begin{document}
\title{Prospects of Quantum Error Mitigation for Quantum Signal Processing}

\author{Ugn\.e Liaubait\.e}\email{ugne.liaubaite@itp.uni-hannover.de}
\author{S. E. Skelton}%
\email{shawn.skelton@itp.uni-hannover.de}
\affiliation{%
 Institute for Theoretical Physics\\
 University of Leibniz Hannover
}%
\date{\today}
\begin{abstract}
Quantum error mitigation (QEM) protocols have provably exponential bounds on the cost scaling; however, exploring which regimes QEM can recover usable results is still of sizable interest. The expected absence of complete error correction for near-term and intermediate-term quantum devices means that QEM protocols will remain relevant for devices with low enough error rates to attempt small examples of fault-tolerant algorithms.

Herein, we are interested in the performance of QEM with a template for quantum algorithms, quantum signal processing (QSP). QSP-based algorithms are designed with an especially simple relation between the circuit depths and the algorithm's parameters, including the required precision. Hence, they may be a useful playground for exploring QEM's practical performance and costs for a range of controlled parameters. As a preliminary step in this direction, this work explores the performance of zero-noise-extrapolation (ZNE) on a Hamiltonian simulation algorithm designed within QSP under local depolarizing noise. 

We design a QSP-based Hamiltonian simulation of a modified Ising model under depolarizing noise for low precision and varying simulation times. We quantify for which noise and depth regimes our ZNE protocol can recover an approximation of the noiseless expectation value.
We discuss existing bounds on the sample budget, eventually using a fixed number of shots. While this does not guarantee the success of QEM, it gives us usable results in relevant cases. Finally, we briefly discuss and present a numerical study on the region where ZNE is unusable, even given an unlimited sample budget. 
\end{abstract}

\maketitle

\section{Introduction}
\input{intro-draft}

\section{Preliminaries}
\input{preliminaries}

\section{Results}\label{sec:results}
\input{results}
\section{Discussion}
\input{discussion}

\begin{acknowledgments}
Discussions with Tobias Osborne and Sabhyata Gupta are gratefully acknowledged. This work was directly or indirectly supported by Quantum Valley Lower Saxony, the Deutsche Forschungsgemeinschaft  project SFB
1227 (DQ-mat),  Germany’s Excellence Strategy EXC-2123 QuantumFrontiers 390837967, the BMWK project ProvideQ, and by the BMBF projects QuBRA, ATIQ. S.E.S.S acknowledges the support of the Natural Sciences and Engineering Research Council of Canada (NSERC), PGS D - 587455 - 2024.

Cette recherche a été financée par le Conseil de recherches en sciences naturelles et en génie du Canada (CRSNG), PGS D - 587455 - 2024. 
\end{acknowledgments}
\section{Data Availability Statement}
 Our simulations are conducted within python based quantum simulation package Pennylane. Code is available at \url{https://github.com/Skeltonse/prospects-qem-qsp}
\bibliography{QEMbiblio}

\appendix
\section{QSP Hamiltonian Simulation}\label{sec:qspapp}
\input{qsp-methods}
\section{Quantum Error Mitigation}
\input{noiseappendix}
\input{results-math}

\end{document}

%% file: intro-draft.tex
Quantum error mitigation (QEM) is a common name referring to all techniques aimed at reducing the effects of noise in quantum computation. Most error mitigation techniques aim to recover noise-free measurement statistics. In contrast to quantum error correction, QEM techniques do not usually require a qubit overhead but rather a sampling overhead. It is known that quantum error mitigation requires exponential resources \cite{FI_pap, stat_pap, Eisert}. However, QEM protocols remain popular in hopes that QEM may be useful for extracting results from shallow circuits on NISQ devices. Alternately, QEM may be used to enhance performance in fault-tolerant scenarios, as was recently discussed in \cite{newpaper}. Within QEM techniques, zero noise extrapolation (ZNE) is a common choice because it requires no knowledge of the noise profile of the device.

Quantum signal processing (QSP) is a template for developing quantum algorithms introduced by \cite{lowoptimal2017} and extended to quantum singular value transformation (QSVT) in \cite{gilyen_quantum_2019}. Since 2017, it has received increasing attention, including modifications, new applications, and some simple implementations. QSP relies on identifying a scalar function which, when applied to the eigenvalues of a matrix, produces some desired unitary. For example, $e^{i\tau x}$ is a function that, applied to the eigenvalues of Hamiltonian $H$, builds Hamiltonian simulation $e^{i\tau H}$.  As has been often noted, see, for example, \cite{magano2022}, QSP and QSVT circuits allow for an easy relation between circuit depth and precision.

 It is natural to ask how well the QSP can be integrated into other very general protocols, notably protocols for error mitigation or correction. Although QSP algorithms are usually complex routines outside the scope of noisy intermediate scale quantum (NISQ) architectures, there are a few bench-marking protocols that utilize QSP \cite{Dong_2022, Dong2020RandomCB}. Some error mitigation strategies to cope with QSP errors have been introduced within very specialized noise models \cite{tan2023errorcorrectionquantumalgorithms, Kikuchi_2023}. However, the performance of QSP with standard error mitigation protocols has not been studied explicitly.

 The central question of this work is how well error mitigation can improve the performance of short quantum signal processing (QSP) circuits. As a preliminary step in this direction, we study the performance of ZNE on a Hamiltonian Simulation algorithm designed within QSP. Using a simple spin-lattice model and local Pauli measurements, we show that ZNE is sometimes able to succeed at small simulation times for reasonable near-term noise levels.

Hamiltonian simulation of local lattice models with Trotter-Lie-Suzuki product formulas \cite{trotter1959, suzuki1976, suzuki1985} are known to perform exceptionally well for short times. The expectation is that QSP and other fault tolerance techniques for Hamiltonian simulation will only become worth the overhead expense for strongly correlated or long-time regimes where Trotter struggles. 

Here, we explore QSP-based Hamiltonian simulation in low-precision regimes $\mathcal{O}(10^{-2})$ and $\mathcal{O}(10^{-4})$. We include Trotter simulations to precision  $\mathcal{O}(10^{-2})$ under the same noise and QEM conditions. This allows us to identify noise and simulation times for which the Trotter simulation already struggles, and a short-depth QSP circuit might admit a performance advantage. However, our test cases use QSP circuits of less than expected length for the Hamiltonian considered, because our classical simulation skirts decomposing the QSP-oracle into hardware-specific gates. This can be understood as assuming access to an oracle with no internal noise, which might apply to a scenario where QEM protocols are used to complement error correction protocols on the oracle, such as the ones explored in \cite{ZNE+QEC} and \cite{QEM+encoded}. Thus, our results are a preliminary study highlighting regions of interest for future work.

We also consider general bounds on QEM resources applied to our QSP-QEM in \cref{sec:appendixbounds}, which allows us to discuss the practicality of such bounds in our given examples in \cref{sec:results}. Finally,  we discuss limits where QEM should not be expected to succeed with depolarizing noise, even given infinite resources.

%% file: preliminaries.tex
\subsection{Quantum Signal Processing}
Many conventions for QSP have been developed, each with particular conditions on $\mathcal{P}(z)$, oracle structures, and circuit lengths. We use the original QSP sequence introduced in \cite{lowoptimal2017, low_method_2016} with notation following \cite{Haah2019product}. In this presentation, QSP is a technique for applying a polynomial $\mathcal{P}(z)$ to the eigenstates of a unitary $U$ with eigenvalues $e^{i\theta}$. Typically, we want to apply a function $f$ on eigenvalues $\lambda$ of Hamiltonian $H$, and we relate $U, H$ through $\arccos(\lambda)=\theta$. We do so by designing a polynomial $\mathcal{P}(z)$ which acts on variable  $z=e^{i\theta}$, where $\theta=\arccos(\lambda)$ and $\linfnorm{\mathcal{P}(z)-f(\lambda)}\leq \epsilon_{approx}$ for some accuracy $\epsilon_{approx}\in[0, 1)$. 
$\mathcal{P}(U)$ denotes the unitary
 \begin{equation}
     \mathcal{P}(U)=\sum_{j=0}^{\text{rank}(H)}\mathcal{P}(z_j)\ket{\lambda_j}\bra{\lambda_j},
 \end{equation}
 where $\ket{\lambda_j}, z_j$ are corresponding eigenstates and eigenvalues of $U$.

The QSP operator corresponding to unitary $U$ and $2n$ $SU(2)$ bases $\{\ket{p}, \ket{q}\}$ is
\begin{align}\label{eq:QET_operator}
U_{QSP}&=E_0\otimes I_{\text{dim}(U)}\prod_{k=1}^{n} C_{p_k}UC_{p_{k+1}}U^{\dag},\\
C_{p}U&=\ket{p}\bra{p}\otimes U+\ket{q}\bra{q}\otimes I=VC_0UV^{\dag}\label{eq:generalizedcontrolgate}
\end{align}
Such a QSP sequence can build a circuit general enough to approximate any Fourier or Chebyshev expansion. 

\begin{lemma}[Laurent QSP, paraphrased from \cite{Haah2019product}]
\label{lemma:QSP_Haah}
    Consider $2\pi$-periodic function $\mathcal{P}(e^{i\theta})=\mathcal{A}(e^{i\theta})+i\mathcal{B}(e^{i\theta})$, where $\mathcal{A}(e^{i\theta})^2+\mathcal{B}(e^{i\theta})^2\leq1$ and $\mathcal{A, B}$ are real-on-circle, and reciprocal or anti-reciprocal. Then, there exists a unique decomposition into $E_p$'s and a unitary 
    $F(e^{i\theta/2})=E_0E_{P_1}(e^{i\theta/2})...E_{P_{2n}}(e^{i\theta/2})$ such that 
    \begin{equation}
        \opnorm{\mathcal{P}(U)-\bra{+}U_{QSP}\ket{+}}\leq C(n)\epsilon_{approx}
    \end{equation}
    for some function of the polynomial degree $C(n)$ and $\epsilon_{approx}\in[0, 1)$
\end{lemma}
This circuit uses the minimal number of ancilla qubits, and the minimal circuit depth possible without incorporating the factor of two speed-up from \cite{berrydoubling} for G-QSP.

 The QSP rotation bases for the Jacobi-Anger expansion are calculated using the Fejer-Wilson method from \cite{skelton2025hitchhikersguideqsppreprocessing}; this provides an analytic bound on $C(n)$, but in practice, we find much better agreement than the worst-case error propagation through the QSP-processing. See \cref{sec:qspapp} for details on the Jacobi-Anger expansion and the degree truncation used herein. Within the main text, the accuracy of the QSP circuit is $\epsilon_{qsp}=C(n)\epsilon_{coeff}$.

 Assuming we prepare the initial state $\ket{+}\bra{+}\otimes \rho$ and post-select on the $+$ outcome on the QSP ancilla qubit measured in the computational basis, the circuit has prepared $\rho_{qsp}$ which is $\epsilon_{qsp}$ close to $\mathcal{P}(U)^{\dag}\rho \mathcal{P}(U)$. In our case, $\mathcal{P}(U)$ is a truncated Jacobi-Anger expansion expressed in variable $z=e^{i\theta}$. This fulfills the criteria of \cref{lemma:QSP_Haah}, so the QSP routine encodes an approximation of $\sum e^{i\tau \lambda}\ket{\lambda}\bra{\lambda}$.
 After post-selection, a noiseless measurement of operator ${O}$ is performed with
 \begin{equation}
     \braket{{O}}=\trace{\left(H\otimes {O}\right)\rho_{qsp}}.
 \end{equation}

\subsection{Error Mitigation}
One of the most popular QEM strategies is zero-noise extrapolation (ZNE), introduced by Temme et al. in \cite{temmeerrormitigation2017}; ZNE is attractive because it is agnostic to device noise. The protocol constructs an estimator for a noiseless expectation value $E_0$ using samples from noisy circuit runs. Usually, one assumes the desired expectation value can be realized as a function of a noise parameter $\lambda$, such that the noiseless expectation value is the value of the function at $\lambda = 0$.

ZNE is performed in two stages: first, one obtains expectation values at various noise levels $\lambda_m$. In the second stage, one uses extrapolation to construct an estimator for the noiseless expectation value, that is, at $\lambda = 0$.

The scaling stage can pose a challenge when implementing ZNE, as the noise parameter is not directly controllable during computation. There are several techniques available to increase the noise parameter and obtain samples of expectation value, such as unitary folding and pulse control \cite{DQEM}. The results in this paper are gathered using a classical simulator; therefore, the noise is controlled directly. This gives an advantage in performance, as scaling factors are precisely known. However, for real quantum hardware, the inaccuracy arising from scaling can be mitigated by averaging several expectation values obtained for the same scaling.

The choice of the extrapolation method is crucial for the success of ZNE. While there have been some works on choosing an extrapolation function for the ZNE protocol \cite{ZNEbestpractices}, there is no definitive way to determine the most suitable fit without testing various fits and choosing one giving the least error. Herein, we consider linear, Richardson (up to 3rd degree), and exponential fits. The fit giving the least error depends on the length of the circuit as well as the noise strength of the initial circuit. More information on extrapolation methods can be found in \cref{sec:appendixextrapolation}.

The success of the mitigation procedure can be quantified by the mean squared error (MSE). MSE encompasses both statistical error in the form of variance of the estimator, $\text{Var}[E_O]$, and the  absolute error via bias:
\begin{equation}
    \text{MSE}[E_O] = \text{Var}[E_O]+b_{E_O}^2.
\end{equation}
Bias $b_{E_O}$ is defined as the difference between the predicted result and the ideal result, $ E_O-\langle O\rangle_{ideal}$.

The cost of QEM lies predominantly in the cost of sampling. Bounds on the sampling costs of general error mitigation protocols were discussed in  \cite{stat_pap, Eisert, FI_pap}; all of these works quantify the sample number required for successful QEM differently, but prove that the lower bound sampling cost for mitigation grows exponentially with at least the depth of the circuit. A detailed discussion of the bounds on sampling cost can be found in \cref{sec:appendixbounds}.

\subsection{Model}
We consider the transverse field Ising model (TFIM), an extremely simple spin chain model with $2$-qubit interactions and site-local magnetic fields.
 We assume a $1D$ lattice and use open boundary conditions. For $N$ sites, the Hamiltonian is
\begin{align}
    H=-\alpha\left(\sum_{i=1}^{N-1}\left(J_Z\sigma_{i}^Z\sigma_{i+1}^Z+J_X\sigma_{i}^X\sigma_{i+1}^X\right)+h_x\sum_{i=0}^{N-1}\sigma_i^{x}\right),
\end{align}
where $J_Z=1, J_X=0.1$ are the nearest neighbor coupling strengths, and $h_x=0.1$ is the transverse field strength. We chose $\alpha=\frac{7}{50}$ to ensure that $H$ is always normalized, a requirement of the QSP-simulation. This Hamiltonian is extremely similar to a model used in \cite{javanmard2022quantumsimulationdynamicalphase}; as $h_x\rightarrow 0$,  it reduces to the exactly solvable transverse field Ising model. The model has $P=2(N-1)+N$ Pauli terms.

We consider local depolarizing noise with parameter $p$,
\begin{equation}
    \mathcal{D}(\rho) = (1-p)I + \frac{p}{3}(\sigma^X+\sigma^Y+\sigma^Z).
\end{equation}
$D$ acts on every qubit in every layer of the circuit. 

Generic QSP oracles are expensive circuits, usually prepared with a linear combination of unitaries. With the LCU strategy, the QSP oracle is a block-encoding of the Hamiltonian and uses a generalization of QSP, the quantum eigenvalue transformation (QET)\cite{low_hamiltonian_2019}\footnote{Preparing the exponentiation of $\arccos(H)$ into an oracle with the same dimension as $H$ is not a standard routine; one generally embeds it within a larger unitary known as a block-encoding. For example, one can use quantum walks to generate a block-encoding. However, the block-encoding is prepared, it will have eigenvalues $ie^{\pm i\arccos\lambda}$ and then one can act on the appropriate eigensubspaces of the Hilbert space of the block-encoding.}. For this Hamiltonian, a state-of-the-art method with $n_{anc}$ ancilla qubits would require $\mathcal{O}\left(PN\log(1/\epsilon)\right)$ depth (suppressing the depth scaling factors from ancilla qubits)\cite{Zhang_2024}, leading to oracles $\mathcal{O}(10^1, 10^2)$.

To skirt the extra qubit and block encoding complexity for QET, we classically compute the QSP oracle and apply it as a single gate. This is a major simplification to avoid the extra qubit and complexity costs of building a block encoding.

We then simulate QSP circuits at different levels of depolarizing noise. Finally, we measure Pauli observables and perform ZNE on the noisy expectation values. In \cref{sec:results}, we compare the resulting simulations to the desired expectation value $\braket{O}_{ideal}$ computed classically at different noise parameters.

\subsection{Trotter Decomposition}
Trotter-Lie product formulas are the standard way to perform Hamiltonian simulation on early digital quantum computers.

Consider time-independent $k$-local Hamiltonian $H = \sum_{i=1}^N H_i$ where each $H_i$ acts non trivially on $k$ qubits. If one takes time $\tau$ to be small, then the time evolution operator $e^{-i\tau H}$ is approximated by the Trotter-Lie formula $e^{-i\tau H_N}...e^{-i\tau H_1}$. For larger times, one can divide the evolution into a number of steps $r$ each of size $d\tau$, resulting in $(e^{-id\tau H_N}...e^{-id\tau H_1})^r$, with $d\tau = \tau/r$. This is also known as first-order Trotterisation \cite{Trotter}.

Higher-order approximations exist with better algorithmic error scaling. Here we are interested in low-precision simulation, therefore we use only the first-order approximation to set the smallest precision and ensure the short-circuit depths for Trotter to match low-precision QSP cases.

%% file: results.tex
Initially, we generate QSP and Trotter circuits within $\tau\in[0.1, 20]$ for $4$-qubit instances and $\epsilon_{QSP}=\mathcal{O}(10^{-2})$. Our initial state is always spin state, $\rho=\ket{0^{\otimes N}}\bra{0^{\otimes N}}$ and we measure 
 \begin{equation}
     {O}={I_0\otimes \sigma_{Z_1}\otimes \sigma_{Z_2}\otimes I_{N-3}}
 \end{equation}
after each evolution. Our simulations are conducted within the Python-based quantum simulation package Pennylane.

\begin{figure}[h]
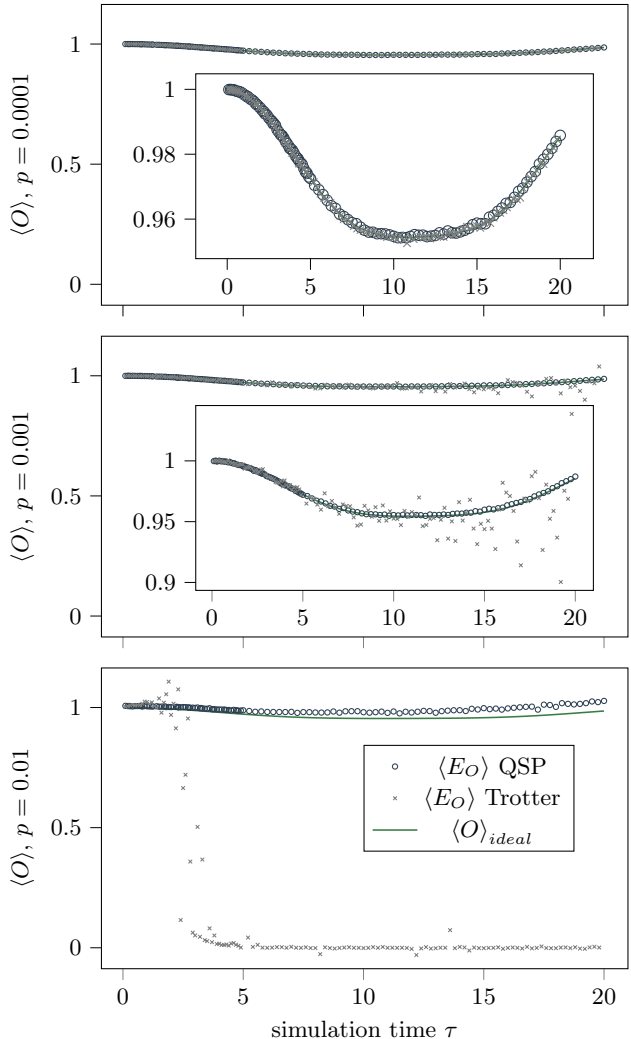

     \centering
     \begin{subfigure}[b]{\linewidth}
         \centering
         \resizebox{\linewidth}{!}{
        \include{mainexpvstau0.1}
        }
     \end{subfigure}
     \hfill
     \begin{subfigure}[b]{\linewidth}
         \centering
         \resizebox{\linewidth}{!}{
         \include{mainexpvstau1.0}
         }
     \end{subfigure}
     \hfill
     \begin{subfigure}[b]{\linewidth}
         \centering
         \resizebox{\linewidth}{!}{
         \include{mainexpvstau10.0}
         }
     \end{subfigure}
        \caption{Best performing QEM estimates of expectation values for $4$-qubit Trotter and QSP TFIM simulations at noise levels $p=1\cdot 10^{-4}$ (above), $p=1\cdot 10^{-3}$ (center), and $p=1\cdot 10^{-2}$ (below).}
        \label{fig:mainexpvstau}
\end{figure}

\begin{figure}[h]
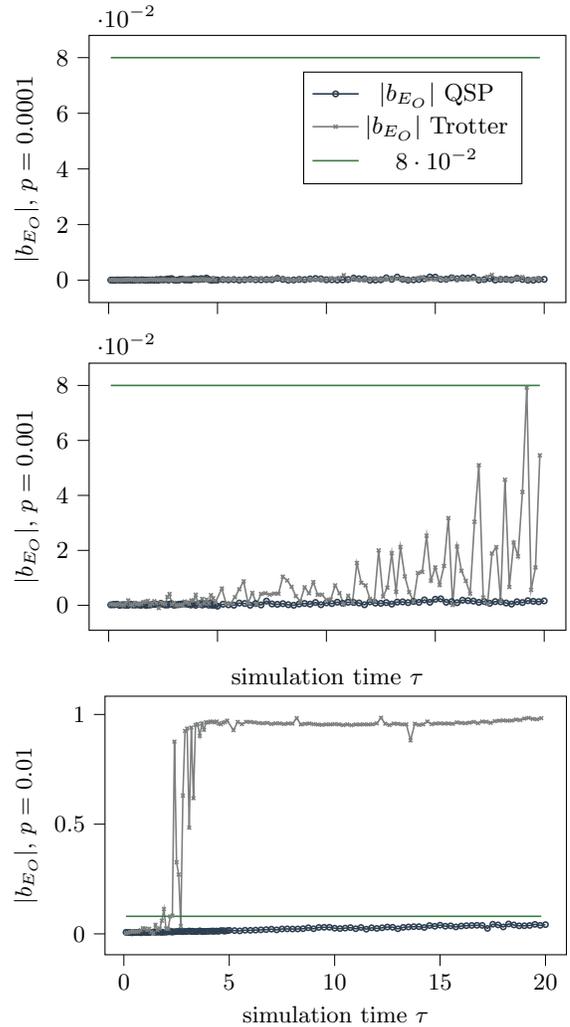

     \centering
     \begin{subfigure}[b]{0.9\linewidth}
         \centering
         \resizebox{\linewidth}{!}{
         \include{biasvstau_p0.1}
         }
     \end{subfigure}
     \hfill
     \begin{subfigure}[b]{0.9\linewidth}
         \centering
         \resizebox{\linewidth}{!}{
       \include{biasvstau_p1.0}
       }
     \end{subfigure}
     \hfill
     \begin{subfigure}[b]{0.9\linewidth}
         \centering
         \resizebox{\linewidth}{!}{
         \include{biasvstau_p10.0}
         }
     \end{subfigure}
        \caption{Biases of the best performing QEM  for $4$-qubit Trotter and QSP TFIM simulations at noise levels $p=1\cdot 10^{-4}$ (above), $p=1\cdot 10^{-3}$ (center), and $p=1\cdot 10^{-2}$ (below).}
        \label{fig:mainbiasvstau}
\end{figure}

Because the number of qubits is much less influential on the performance of the error mitigation protocol than the noise level (see \cref{fig:allqexpvstau} and the surrounding discussion in \cref{sec:numericextrapolationresults}), we reproduce results for only the $4$-qubit cases in this section. The mean-squared-errors of $6,8$-qubit QSP and Trotter simulations can be found in \cref{fig:scalingplots} in \cref{sec:numericextrapolationresults}. The final results of the extrapolated expectation value of the two-site correlation for noise parameters $[0.0001, 0.001, 0.01]$  are presented in \cref{fig:mainexpvstau} and the corresponding statistical biases are in \cref{fig:mainbiasvstau}.

We consider three noise levels: noise within a range reported on current devices, such as $2$-qubit average error rates $p=1\cdot 10^{-3}, 1\cdot 10^{-2}$ \cite{LOTSTEDT2024140975, ibmQuantum, quantiniummh1datasheet}, and noise which may be achievable in the near-term $p=1\cdot 10^{-4}$. For the QEM protocol to be successful, the estimator has to lie within the chosen error threshold of an actual value, that is, have bias below the chosen threshold. We set this threshold $\epsilon_{QEM}$ to be $\mathcal{O}(\epsilon_{QSP})$, stipulating $\epsilon_{QEM}=1\cdot 10^{-2}$. This is a natural choice because if the QEM protocol can not recover the same order precision of the QSP circuit, then the higher-precision QSP circuit wastes resources.

The mitigation is performed by scaling noise in increments of $[1, 2, 3]$ or $[1, 1.25, 1.5]$. We find that the larger scaling factors provide better performance for shorter circuits and smaller initial noise parameters. As the circuit length grows or if $p$ is increased, then the smaller scaling factors perform better. The scaling choice for each data set is discussed in \cref{sec:numericextrapolationresults}. 

The exponential extrapolation fit almost always outperforms Richardson and linear fits. However, for larger noise and scaling, an exponential fit could not be found. When no exponential fit is found, the Richardson fit is used to extrapolate zero noise value. Note that for very low depths and  $p=0.0001$, the linear fit performed comparably to the exponential fit. Tests for various extrapolation techniques can be found in Figure \ref{sec:numericextrapolationresults} in the appendix. 

 \begin{figure}[h]
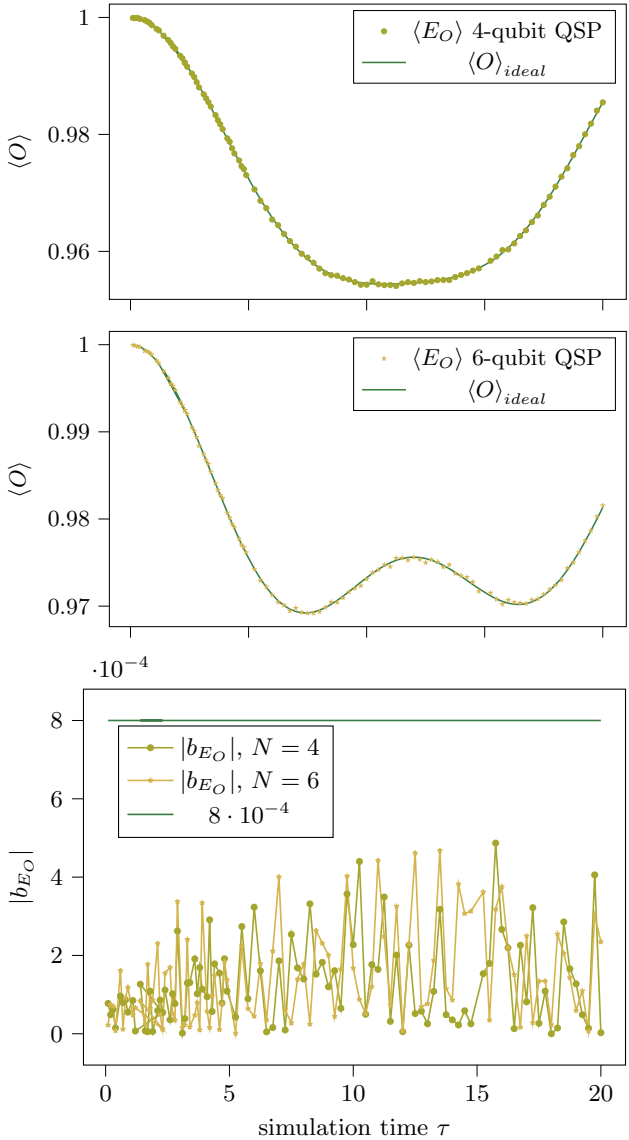

     \centering
     \begin{subfigure}[b]{\linewidth}
         \centering
         \resizebox{\linewidth}{!}{
         \include{mainexpvstauprecise4q}
         }
     \end{subfigure}
     \hfill
     \begin{subfigure}[b]{\linewidth}
         \centering
         \resizebox{\linewidth}{!}{
       \include{mainexpvstauprecise6q}
       }
     \end{subfigure}
     \hfill
     \begin{subfigure}[b]{\linewidth}
         \centering
         \resizebox{\linewidth}{!}{
       \include{biasvstau_precise}
       }
     \end{subfigure}
        \caption{Higher-precision QSP. Biases of the best performing QEM  for $4$-qubit (above) and $6$ qubit (middle) QSP TFIM simulations at $p=1\cdot 10^{-4}1$. The biases for each are plotted together (below)}
        \label{fig:highprecisionexpvstau}
\end{figure}
We find that the $p=1\cdot 10^{-4}$ and the $p=1\cdot 10^{-3}$ instances are successfully mitigated. While all of our data points were collected with fewer samples than the predicted requirement in the exponential bound, it is predominantly the higher noise instances that showed a departure of the estimator from an ideal value on a short time scale. Besides undersampling, we can expect the $p=1\cdot 10^{-2}$ instances to struggle because of the evolution of the circuit to a steady state under depolarizing noise, as discussed in \cref{sec:qemfeasibleregimes}.  Note that the relative success of QSP over Trotter is expected for two reasons: first, the QSP circuits are artificially short. Second, the QSP algorithm requires post-selection, which constitutes a form of error mitigation.

We can push the limits of the QSP circuits in two ways:  we can increase the time of evolution for QSP, or we can increase the required precision. First, we set $\epsilon_{QEM}=1\cdot 10^{-4}$ and $p=1\cdot 10^{-4}$ and consider the $4, 6$-qubit cases in \cref{fig:highprecisionexpvstau}.

The QEM protocol used is successful even with the same number of samples as the less precise cases. Next, we consider QSP circuits for $\tau\in[25, 50,....375, 390.0]$, simulated with circuit depth $811$; we find the mitigation still succeeds in this regime.

\begin{figure}[h]
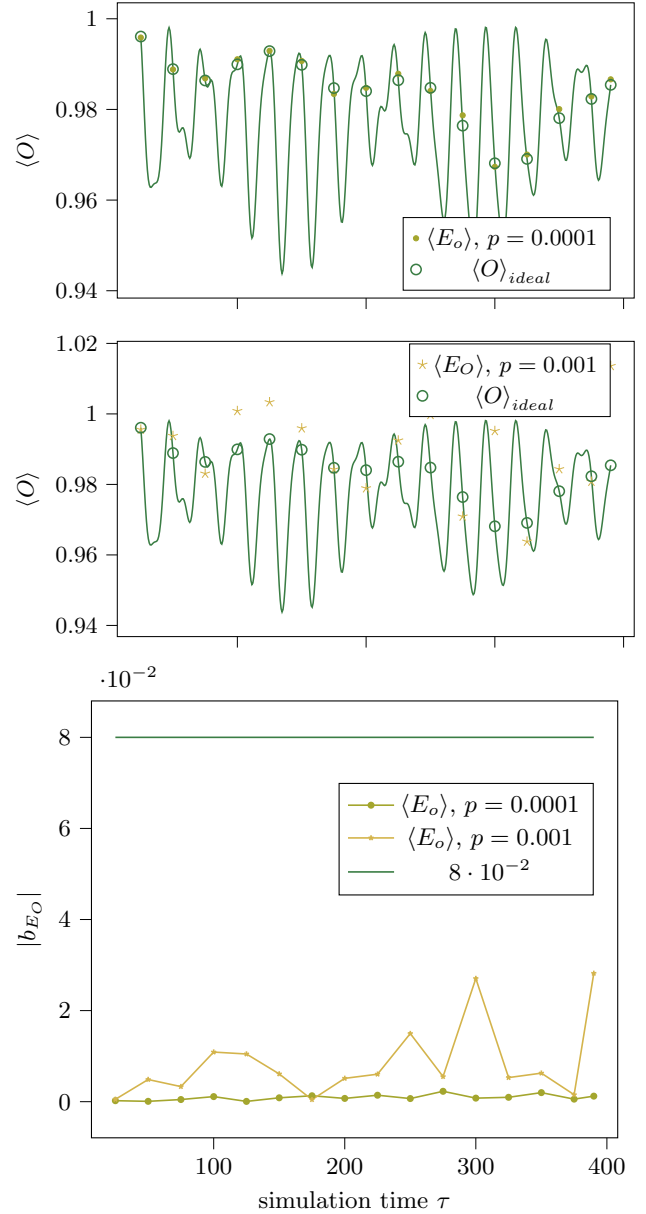

    \centering
     \begin{subfigure}[b]{\linewidth}
         \centering
         \resizebox{\linewidth}{!}{
          \include{mainexpvstaulong0.1}
          }
     \end{subfigure}
     \hfill
     \begin{subfigure}[b]{\linewidth}
         \centering
         \resizebox{\linewidth}{!}{
        \include{mainexpvstaulong1.0}
        }
     \end{subfigure}
     \hfill
     \begin{subfigure}[b]{\linewidth}
         \centering
         \resizebox{\linewidth}{!}{
        \include{biasvstau_long}
        }
     \end{subfigure}
    \caption{Long time evolution QSP. Best performing mitigation for $p=1\cdot 10^{-4}$ (above) and $p=1\cdot 10^{-3}$ (middle) with $4$-qubit QSP circuits in $\tau\in\{25,50...375, 390\}$. Biases are shown together (below)}
    \label{fig:longtimeexpvstau}
\end{figure}

\begin{figure}[h]
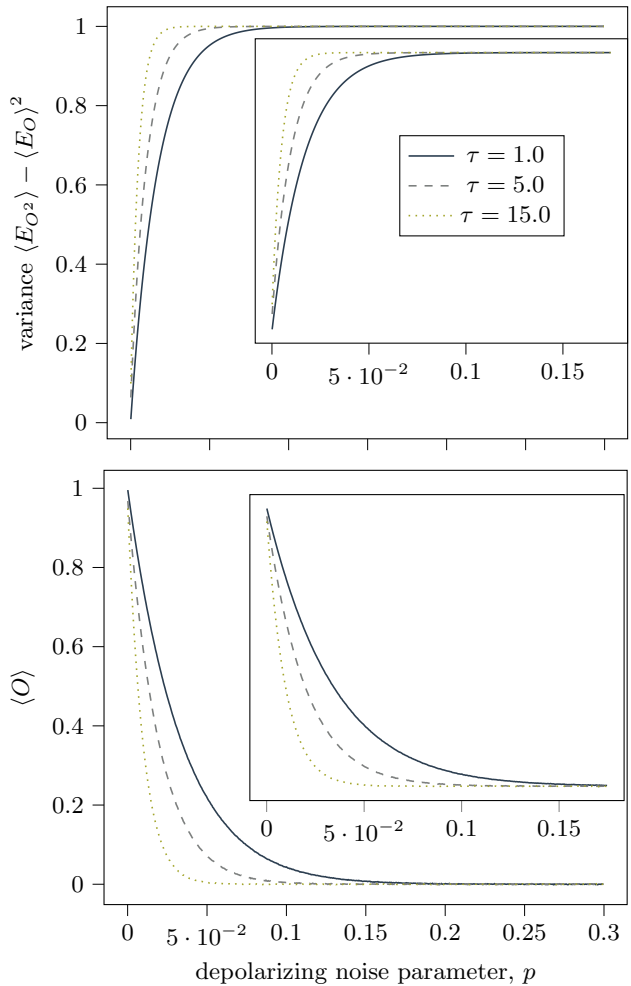

    \centering
    \begin{subfigure}[b]{\linewidth}
    \resizebox{\linewidth}{!}{
      \include{varvsp}
      }
    \end{subfigure}
    \begin{subfigure}[b]{\linewidth}
    \resizebox{\linewidth}{!}{
       \include{expvsp}
       }
    \end{subfigure}
    \caption{Limiting scenarios for our $4$ qubit QEM-QSP protocol. Above: variance vs increasing noise. Below: noisy expectation values vs increasing noise. }
    \label{fig:varvsnoise}
\end{figure}

Simulating local depolarizing noise with larger noise strength, such as $p=0.01$, or longer times, we observed the evolution approaching a steady state. In these cases, increasing the noise factor no longer alters the expectation value. This creates a no-go regime in ZNE, which has nothing to do with the exponential increase in sampling cost\textemdash the variance is stable at the point the steady state is reached. In our case, that happens to be when the system reaches the fully mixed state, which happens to be the steady state of our noise model. Note that the variance approaches $1$ more quickly. We observe this phenomenon clearly in \cref{fig:varvsnoise} for increasing noise. Similar results for increasing circuit depth are given in  \cref{sec:numericextrapolationresults}.

%% file: mainexpvstau0.1.tex
\begin{tikzpicture}

\begin{axis}[
tick align=outside,
tick pos=left,
x grid style={darkgray176},
xticklabel=\empty,
xmin=-0.895, xmax=20.995,
xtick style={color=black},
y grid style={darkgray176},
ylabel={$\braket{O}$, $p=0.0001$},
ymin=-0.0871216912978824, ymax=1.16469135671033,
ytick style={color=black},
mark size=1,
width =\linewidth, 
height = 0.65*\linewidth
]
\addplot [draw=c1, fill=c1, mark=o, only marks]
table{%
x  y
0.1 0.999924022470059
0.2 0.9999265354598
0.25 0.999911779705014
0.3 0.999865781226853
0.4 0.999854598251404
0.5 0.999545374483015
0.6 0.999394180852329
0.7 0.999335423582628
0.75 0.999136376813519
0.8 0.999113249037908
0.9 0.998859194084344
1 0.998381365518054
1.1 0.998137834000742
1.2 0.997679260242739
1.25 0.997719849835131
1.4 0.996947742072037
1.5 0.996731949196309
1.6 0.996255986521746
1.7 0.995652303378198
1.75 0.995434882455833
1.8 0.995198230676519
1.9 0.994705858528655
2 0.994197869530671
2.1 0.993501935368055
2.2 0.993148508174522
2.25 0.99260303245699
2.3 0.992399255048476
2.3 0.992399255048476
2.4 0.99198842573019
2.5 0.991173998059529
2.6 0.990629015363984
2.7 0.99011390164071
2.75 0.989391718617101
2.8 0.989516054734701
2.9 0.988884412157464
3 0.987516546218509
3.1 0.98683778928989
3.2 0.986074650379576
3.25 0.98593296311583
3.3 0.985714996808797
3.4 0.984971781442856
3.5 0.983972156102084
3.6 0.983268052351912
3.7 0.982503792044402
3.75 0.981822833820323
3.8 0.981702211658601
3.9 0.981331901722868
4 0.980490501646336
4.1 0.979472154292166
4.2 0.978790396816198
4.25 0.978651152880428
4.3 0.978046007817905
4.4 0.977414560657856
4.5 0.976911114834839
4.6 0.975631710974177
4.7 0.974568488222116
4.75 0.97425789208788
4.8 0.974115819016961
4.9 0.973027827840941
5 0.972511822925215
5.25 0.970403245857965
5.5 0.969061934998307
5.75 0.967534543685753
6 0.965799037327236
6.25 0.964137312684246
6.5 0.963327472649565
6.75 0.961567640113805
7 0.96079698162849
7.25 0.959838041340581
7.5 0.958904050223959
7.75 0.958591083519592
8 0.957513539475852
8.25 0.956573185986441
8.5 0.955991042673364
8.75 0.955955160819951
9 0.955510417631882
9.25 0.955503237776611
9.5 0.955419700621846
9.75 0.955128529164552
10 0.954696280357008
10.25 0.954271801995136
10.5 0.954332390273906
10.75 0.954267996169302
11 0.954667501213132
11.25 0.95520664450238
11.5 0.955009260392798
11.75 0.955052149643205
12 0.954616561675725
12.25 0.954762593370389
12.5 0.95495836591798
12.75 0.95576323392039
13 0.955381620631046
13.25 0.95634097860188
13.5 0.955792533079931
13.75 0.955688821100123
14 0.956076549913744
14.25 0.956549591256112
14.5 0.957142076215365
14.75 0.958279374716224
15 0.958748932089414
15.25 0.958593172028664
15.5 0.958563340683626
15.75 0.960604508892263
16 0.961138456751951
16.25 0.962384188322429
16.5 0.963412363997328
16.75 0.964838794813757
17 0.965144396666414
17.25 0.966713727120818
17.5 0.968551531923784
17.75 0.969525245814461
18 0.971383084228807
18.25 0.973098952619186
18.5 0.97466658983429
18.75 0.97700369839992
19 0.978425594422306
19.25 0.979989178123493
19.5 0.981813808494213
19.75 0.983917026260331
20 0.985794277971866
};
\addplot [
  draw=c2,
  fill=c2,
  mark=x,
  only marks
]
table{%
x  y
0.1 1.00004273558581
0.2 0.999814421633299
0.3 0.999793386423507
0.4 0.99980927257291
0.5 0.999621458551089
0.6 0.999537777765303
0.7 0.999269859461762
0.8 0.999166113168338
0.9 0.998898485754368
1 0.9982380803605
1.1 0.997939185610932
1.2 0.997752637938528
1.3 0.997453019831099
1.4 0.997057836298118
1.5 0.996430685555453
1.6 0.996086201113156
1.7 0.995587524063577
1.8 0.995228627289901
1.9 0.99435947779336
2 0.994438578885452
2.1 0.993425878513494
2.2 0.99315459574174
2.3 0.9922403482222
2.4 0.991738556838991
2.5 0.991322958471822
2.6 0.990153991108062
2.7 0.98965482976823
2.8 0.9889365035641
2.9 0.98844808354424
3 0.987295458379659
3.1 0.986880139123544
3.2 0.986106803386428
3.3 0.985239663095875
3.4 0.984296522107508
3.5 0.983396493340983
3.6 0.982833603544777
3.7 0.982658010398332
3.8 0.981769239226172
3.9 0.980420693324012
4 0.980047220126914
4.1 0.979636953289062
4.2 0.978440389333702
4.3 0.977129253711187
4.4 0.977050422154811
4.5 0.976520440651408
4.6 0.975449572331949
4.7 0.974488846004635
4.8 0.974600519349106
4.9 0.973231783240233
5.2 0.971405111896078
5.4 0.969776530411138
5.6 0.96783379572022
5.8 0.966546241809953
6 0.965986928857969
6.2 0.964955214947986
6.4 0.963406066479205
6.6 0.962630501944407
6.8 0.960983033052032
7 0.959945212153019
7.2 0.959926612566008
7.4 0.959279987013426
7.6 0.957821961277929
7.8 0.957076932208192
8 0.956620006664987
8.2 0.956307934674164
8.4 0.956203494023377
8.6 0.955998553340842
8.8 0.956073265138447
9 0.955436915478839
9.2 0.955350212827988
9.4 0.955271061437004
9.6 0.954438383244055
9.8 0.954552002122562
10 0.954348743754094
10.2 0.953772598897841
10.4 0.954965005033504
10.6 0.953525399715829
10.8 0.952561737603247
11 0.954488320643465
11.2 0.953823625619815
11.4 0.954667093604239
11.6 0.954539163232741
11.8 0.954291338307883
12 0.954263404428174
12.2 0.953981526605268
12.4 0.955121140915562
12.6 0.955143513065586
12.8 0.954496663291347
13 0.954504657108951
13.2 0.955670736168954
13.4 0.955721529308193
13.6 0.955153160547476
13.8 0.955577435024528
14 0.95633809823467
14.2 0.95691921914914
14.4 0.957649625570683
14.6 0.957123154439261
14.8 0.957538505282965
15 0.957825631507807
15.2 0.957294384252731
15.4 0.9588900684306
15.6 0.959040687682585
15.8 0.958762354170819
16 0.960438005764873
16.2 0.962137080924514
16.4 0.961906377991976
16.6 0.962711168697195
16.8 0.963598167513563
17 0.96532604760529
17.2 0.966241963713086
17.4 0.966054441803917
17.6 0.96657298805083
17.8 0.969339005965222
18 0.97162471905426
18.2 0.97272200384982
18.4 0.973715337622528
18.6 0.974872946830473
18.8 0.977330170641752
19 0.976948182819261
19.2 0.979446356576863
19.4 0.980632037333613
19.6 0.981900332422848
19.8 0.984572540137959
};
\addplot [semithick, c3]
table {%
0.1 0.999984322253531
0.2 0.999937316048458
0.25 0.999902087993421
0.3 0.999859062423781
0.4 0.999749696232385
0.5 0.999609405823361
0.6 0.999438432598445
0.7 0.999237070443749
0.75 0.99912509954066
0.8 0.999005665038244
0.9 0.998744613040839
1 0.998454361158164
1.1 0.998135405095548
1.2 0.997788288393915
1.25 0.997604352428722
1.4 0.99701197866293
1.5 0.99658409989175
1.6 0.996130685926211
1.7 0.995652498276588
1.75 0.995404362911638
1.8 0.995150337106013
1.9 0.994625039370295
2 0.994077476875803
2.1 0.993508554260528
2.2 0.992919206903649
2.25 0.992617174129473
2.3 0.992310398769103
2.3 0.992310398769103
2.4 0.991683120188853
2.5 0.991038385591691
2.6 0.99037723118357
2.7 0.989700712585584
2.75 0.989357026316278
2.8 0.989009902435811
2.9 0.98830588796137
3 0.987589768527063
3.1 0.986862653167096
3.2 0.986125658106365
3.25 0.985753805863718
3.3 0.985379904277848
3.4 0.984626514842655
3.5 0.983866612719245
3.6 0.983101318128335
3.7 0.982331746159953
3.75 0.981945702629522
3.8 0.981559004369023
3.9 0.9807841904058
4 0.98000838968738
4.1 0.979232673116355
4.2 0.978458094852616
4.25 0.978071556897716
4.3 0.977685690144076
4.4 0.976916473222
4.5 0.976151435266367
4.6 0.975391542446561
4.7 0.974637734042424
4.75 0.974263397225317
4.8 0.973890920650495
4.9 0.973151982480018
5 0.972421767743045
5.25 0.97063965615509
5.5 0.96892881138133
5.75 0.967299502778876
6 0.965760428420085
6.25 0.964318652091784
6.5 0.962979569793162
6.75 0.961746905975175
7 0.960622739223592
7.25 0.959607556551009
7.5 0.958700334940377
7.75 0.957898648283958
8 0.957198797396691
8.25 0.956595960360986
8.5 0.956084360089292
8.75 0.95565744567886
9 0.955308083886447
9.25 0.955028756874138
9.5 0.954811762275025
9.75 0.954649411601136
10 0.954534223066697
10.25 0.954459105026463
10.5 0.954417526429264
10.75 0.954403670956914
11 0.954412571853028
11.25 0.954440224838352
11.5 0.954483676950941
11.75 0.954541089632261
12 0.954611774894168
12.25 0.954696203936569
12.5 0.954795988130704
12.75 0.954913832827626
13 0.955053464984911
13.25 0.955219536116504
13.5 0.955417502551073
13.75 0.955653485424212
14 0.955934113221046
14.25 0.956266350021176
14.5 0.956657312871512
14.75 0.957114081919665
15 0.957643507078036
15.25 0.958252015054501
15.5 0.958945420579272
15.75 0.959728745579925
16 0.960606049909829
16.25 0.961580277022944
16.5 0.962653117714371
16.75 0.963824894717154
17 0.965094470567682
17.25 0.966459180732235
17.5 0.967914793533262
17.75 0.969455497934343
18 0.97107391974594
18.25 0.972761166308634
18.5 0.974506899205412
18.75 0.976299434058091
19 0.978125865983579
19.25 0.979972218831163
19.5 0.981823615899892
19.75 0.983664469452122
20 0.985478686001584
};

\end{axis}
\node[anchor=south west] at ({$(current bounding box.south west)!0.05!(current bounding box.south east)$}|-{$(current bounding box.south west)!0.2!(current bounding box.north west)$})
{
\begin{axis}[
width=0.7\linewidth,
tick align=outside,
tick pos=left,
x grid style={darkgray176},
xtick style={color=black},
y grid style={darkgray176},
width =0.8\linewidth, 
height = 0.47*\linewidth,
ytick style={color=black},
]
\addplot [draw=c1, fill=c1, mark=o, only marks]
table{%
x  y
0.1 0.999924022470059
0.2 0.9999265354598
0.25 0.999911779705014
0.3 0.999865781226853
0.4 0.999854598251404
0.5 0.999545374483015
0.6 0.999394180852329
0.7 0.999335423582628
0.75 0.999136376813519
0.8 0.999113249037908
0.9 0.998859194084344
1 0.998381365518054
1.1 0.998137834000742
1.2 0.997679260242739
1.25 0.997719849835131
1.4 0.996947742072037
1.5 0.996731949196309
1.6 0.996255986521746
1.7 0.995652303378198
1.75 0.995434882455833
1.8 0.995198230676519
1.9 0.994705858528655
2 0.994197869530671
2.1 0.993501935368055
2.2 0.993148508174522
2.25 0.99260303245699
2.3 0.992399255048476
2.3 0.992399255048476
2.4 0.99198842573019
2.5 0.991173998059529
2.6 0.990629015363984
2.7 0.99011390164071
2.75 0.989391718617101
2.8 0.989516054734701
2.9 0.988884412157464
3 0.987516546218509
3.1 0.98683778928989
3.2 0.986074650379576
3.25 0.98593296311583
3.3 0.985714996808797
3.4 0.984971781442856
3.5 0.983972156102084
3.6 0.983268052351912
3.7 0.982503792044402
3.75 0.981822833820323
3.8 0.981702211658601
3.9 0.981331901722868
4 0.980490501646336
4.1 0.979472154292166
4.2 0.978790396816198
4.25 0.978651152880428
4.3 0.978046007817905
4.4 0.977414560657856
4.5 0.976911114834839
4.6 0.975631710974177
4.7 0.974568488222116
4.75 0.97425789208788
4.8 0.974115819016961
4.9 0.973027827840941
5 0.972511822925215
5.25 0.970403245857965
5.5 0.969061934998307
5.75 0.967534543685753
6 0.965799037327236
6.25 0.964137312684246
6.5 0.963327472649565
6.75 0.961567640113805
7 0.96079698162849
7.25 0.959838041340581
7.5 0.958904050223959
7.75 0.958591083519592
8 0.957513539475852
8.25 0.956573185986441
8.5 0.955991042673364
8.75 0.955955160819951
9 0.955510417631882
9.25 0.955503237776611
9.5 0.955419700621846
9.75 0.955128529164552
10 0.954696280357008
10.25 0.954271801995136
10.5 0.954332390273906
10.75 0.954267996169302
11 0.954667501213132
11.25 0.95520664450238
11.5 0.955009260392798
11.75 0.955052149643205
12 0.954616561675725
12.25 0.954762593370389
12.5 0.95495836591798
12.75 0.95576323392039
13 0.955381620631046
13.25 0.95634097860188
13.5 0.955792533079931
13.75 0.955688821100123
14 0.956076549913744
14.25 0.956549591256112
14.5 0.957142076215365
14.75 0.958279374716224
15 0.958748932089414
15.25 0.958593172028664
15.5 0.958563340683626
15.75 0.960604508892263
16 0.961138456751951
16.25 0.962384188322429
16.5 0.963412363997328
16.75 0.964838794813757
17 0.965144396666414
17.25 0.966713727120818
17.5 0.968551531923784
17.75 0.969525245814461
18 0.971383084228807
18.25 0.973098952619186
18.5 0.97466658983429
18.75 0.97700369839992
19 0.978425594422306
19.25 0.979989178123493
19.5 0.981813808494213
19.75 0.983917026260331
20 0.985794277971866
};
\addplot [
  draw=c2,
  fill=c2,
  mark=x,
  only marks
]
table{%
x  y
0.1 1.00004273558581
0.2 0.999814421633299
0.3 0.999793386423507
0.4 0.99980927257291
0.5 0.999621458551089
0.6 0.999537777765303
0.7 0.999269859461762
0.8 0.999166113168338
0.9 0.998898485754368
1 0.9982380803605
1.1 0.997939185610932
1.2 0.997752637938528
1.3 0.997453019831099
1.4 0.997057836298118
1.5 0.996430685555453
1.6 0.996086201113156
1.7 0.995587524063577
1.8 0.995228627289901
1.9 0.99435947779336
2 0.994438578885452
2.1 0.993425878513494
2.2 0.99315459574174
2.3 0.9922403482222
2.4 0.991738556838991
2.5 0.991322958471822
2.6 0.990153991108062
2.7 0.98965482976823
2.8 0.9889365035641
2.9 0.98844808354424
3 0.987295458379659
3.1 0.986880139123544
3.2 0.986106803386428
3.3 0.985239663095875
3.4 0.984296522107508
3.5 0.983396493340983
3.6 0.982833603544777
3.7 0.982658010398332
3.8 0.981769239226172
3.9 0.980420693324012
4 0.980047220126914
4.1 0.979636953289062
4.2 0.978440389333702
4.3 0.977129253711187
4.4 0.977050422154811
4.5 0.976520440651408
4.6 0.975449572331949
4.7 0.974488846004635
4.8 0.974600519349106
4.9 0.973231783240233
5.2 0.971405111896078
5.4 0.969776530411138
5.6 0.96783379572022
5.8 0.966546241809953
6 0.965986928857969
6.2 0.964955214947986
6.4 0.963406066479205
6.6 0.962630501944407
6.8 0.960983033052032
7 0.959945212153019
7.2 0.959926612566008
7.4 0.959279987013426
7.6 0.957821961277929
7.8 0.957076932208192
8 0.956620006664987
8.2 0.956307934674164
8.4 0.956203494023377
8.6 0.955998553340842
8.8 0.956073265138447
9 0.955436915478839
9.2 0.955350212827988
9.4 0.955271061437004
9.6 0.954438383244055
9.8 0.954552002122562
10 0.954348743754094
10.2 0.953772598897841
10.4 0.954965005033504
10.6 0.953525399715829
10.8 0.952561737603247
11 0.954488320643465
11.2 0.953823625619815
11.4 0.954667093604239
11.6 0.954539163232741
11.8 0.954291338307883
12 0.954263404428174
12.2 0.953981526605268
12.4 0.955121140915562
12.6 0.955143513065586
12.8 0.954496663291347
13 0.954504657108951
13.2 0.955670736168954
13.4 0.955721529308193
13.6 0.955153160547476
13.8 0.955577435024528
14 0.95633809823467
14.2 0.95691921914914
14.4 0.957649625570683
14.6 0.957123154439261
14.8 0.957538505282965
15 0.957825631507807
15.2 0.957294384252731
15.4 0.9588900684306
15.6 0.959040687682585
15.8 0.958762354170819
16 0.960438005764873
16.2 0.962137080924514
16.4 0.961906377991976
16.6 0.962711168697195
16.8 0.963598167513563
17 0.96532604760529
17.2 0.966241963713086
17.4 0.966054441803917
17.6 0.96657298805083
17.8 0.969339005965222
18 0.97162471905426
18.2 0.97272200384982
18.4 0.973715337622528
18.6 0.974872946830473
18.8 0.977330170641752
19 0.976948182819261
19.2 0.979446356576863
19.4 0.980632037333613
19.6 0.981900332422848
19.8 0.984572540137959
};
\addplot [semithick, c3]
table {%
0.1 0.999984322253531
0.2 0.999937316048458
0.25 0.999902087993421
0.3 0.999859062423781
0.4 0.999749696232385
0.5 0.999609405823361
0.6 0.999438432598445
0.7 0.999237070443749
0.75 0.99912509954066
0.8 0.999005665038244
0.9 0.998744613040839
1 0.998454361158164
1.1 0.998135405095548
1.2 0.997788288393915
1.25 0.997604352428722
1.4 0.99701197866293
1.5 0.99658409989175
1.6 0.996130685926211
1.7 0.995652498276588
1.75 0.995404362911638
1.8 0.995150337106013
1.9 0.994625039370295
2 0.994077476875803
2.1 0.993508554260528
2.2 0.992919206903649
2.25 0.992617174129473
2.3 0.992310398769103
2.3 0.992310398769103
2.4 0.991683120188853
2.5 0.991038385591691
2.6 0.99037723118357
2.7 0.989700712585584
2.75 0.989357026316278
2.8 0.989009902435811
2.9 0.98830588796137
3 0.987589768527063
3.1 0.986862653167096
3.2 0.986125658106365
3.25 0.985753805863718
3.3 0.985379904277848
3.4 0.984626514842655
3.5 0.983866612719245
3.6 0.983101318128335
3.7 0.982331746159953
3.75 0.981945702629522
3.8 0.981559004369023
3.9 0.9807841904058
4 0.98000838968738
4.1 0.979232673116355
4.2 0.978458094852616
4.25 0.978071556897716
4.3 0.977685690144076
4.4 0.976916473222
4.5 0.976151435266367
4.6 0.975391542446561
4.7 0.974637734042424
4.75 0.974263397225317
4.8 0.973890920650495
4.9 0.973151982480018
5 0.972421767743045
5.25 0.97063965615509
5.5 0.96892881138133
5.75 0.967299502778876
6 0.965760428420085
6.25 0.964318652091784
6.5 0.962979569793162
6.75 0.961746905975175
7 0.960622739223592
7.25 0.959607556551009
7.5 0.958700334940377
7.75 0.957898648283958
8 0.957198797396691
8.25 0.956595960360986
8.5 0.956084360089292
8.75 0.95565744567886
9 0.955308083886447
9.25 0.955028756874138
9.5 0.954811762275025
9.75 0.954649411601136
10 0.954534223066697
10.25 0.954459105026463
10.5 0.954417526429264
10.75 0.954403670956914
11 0.954412571853028
11.25 0.954440224838352
11.5 0.954483676950941
11.75 0.954541089632261
12 0.954611774894168
12.25 0.954696203936569
12.5 0.954795988130704
12.75 0.954913832827626
13 0.955053464984911
13.25 0.955219536116504
13.5 0.955417502551073
13.75 0.955653485424212
14 0.955934113221046
14.25 0.956266350021176
14.5 0.956657312871512
14.75 0.957114081919665
15 0.957643507078036
15.25 0.958252015054501
15.5 0.958945420579272
15.75 0.959728745579925
16 0.960606049909829
16.25 0.961580277022944
16.5 0.962653117714371
16.75 0.963824894717154
17 0.965094470567682
17.25 0.966459180732235
17.5 0.967914793533262
17.75 0.969455497934343
18 0.97107391974594
18.25 0.972761166308634
18.5 0.974506899205412
18.75 0.976299434058091
19 0.978125865983579
19.25 0.979972218831163
19.5 0.981823615899892
19.75 0.983664469452122
20 0.985478686001584
};
\end{axis}
};

\end{tikzpicture}

%% file: mainexpvstau1.0.tex
\begin{tikzpicture}

\begin{axis}[
tick align=outside,
tick pos=left,
xticklabel=\empty,
xmin=-0.895, xmax=20.995,
y grid style={darkgray176},
ylabel={$\braket{O}$, $p=0.001$},
ymin=-0.0871216912978824, ymax=1.16469135671033,
ytick style={color=black},
mark size=1,
width =\linewidth, 
height = 0.65*\linewidth
]
\addplot [draw=c1, fill=c1, mark=o, only marks]
table{%
x  y
0.1 0.999808347345181
0.2 1.00020201979863
0.25 0.999595155376648
0.3 0.999595865024691
0.4 1.00016901148727
0.5 0.99961015173215
0.6 0.999629463776136
0.7 0.999452214412409
0.75 0.999266916430586
0.8 0.999120623930955
0.9 0.998784424334304
1 0.998908351324127
1.1 0.998258784778531
1.2 0.997987739956705
1.25 0.997772675048787
1.4 0.9969025075854
1.5 0.996540480802629
1.6 0.996527858876526
1.7 0.995691331182357
1.75 0.995347621581486
1.8 0.995394904179688
1.9 0.994745868419234
2 0.994401610445927
2.1 0.993827905014085
2.2 0.99331385008157
2.25 0.992939018492288
2.3 0.992003314808633
2.3 0.992003314808633
2.4 0.992422636311866
2.5 0.991447067212358
2.6 0.990497758970759
2.7 0.989885486234895
2.75 0.989608843116952
2.8 0.989437426986496
2.9 0.988908290757953
3 0.987537723784141
3.1 0.987056623900192
3.2 0.986422396814482
3.25 0.986065555469907
3.3 0.985823622414831
3.4 0.985165675583141
3.5 0.98445379445884
3.6 0.98355277117105
3.7 0.982773364827906
3.75 0.982158974936564
3.8 0.981367359518001
3.9 0.981288412359387
4 0.97997230351525
4.1 0.979696968770756
4.2 0.979056097852616
4.25 0.978919449331443
4.3 0.977797664882809
4.4 0.977317865288751
4.5 0.976863880380256
4.6 0.975318206104318
4.7 0.975366614597433
4.75 0.974177073357167
4.8 0.973953972796376
4.9 0.973118487913227
5 0.972136975340304
5.25 0.971045532180379
5.5 0.969122600648514
5.75 0.967248991367051
6 0.96659342391881
6.25 0.964991347944822
6.5 0.963016776415196
6.75 0.962411697926026
7 0.960477653050595
7.25 0.961156467772623
7.5 0.959209471565761
7.75 0.958275857830135
8 0.957803116908302
8.25 0.956827222087653
8.5 0.956116405265811
8.75 0.956151535433532
9 0.955995282245294
9.25 0.955471593234452
9.5 0.955847002217905
9.75 0.95487693490566
10 0.955450088672238
10.25 0.955106245198154
10.5 0.955574992426028
10.75 0.955145641894603
11 0.955544458297675
11.25 0.9554798167342
11.5 0.955348418206938
11.75 0.954759387678649
12 0.955896759719311
12.25 0.955522649600535
12.5 0.955531483029222
12.75 0.955912344992378
13 0.955515923775873
13.25 0.956411114593255
13.5 0.956690005069368
13.75 0.956645730143224
14 0.957349575498577
14.25 0.957363746509597
14.5 0.958633821502082
14.75 0.958235151934605
15 0.959866967417858
15.25 0.960615429168243
15.5 0.960072928275912
15.75 0.961089648459594
16 0.961584925865873
16.25 0.963458128164048
16.5 0.964360106905488
16.75 0.965386521764824
17 0.966187431647821
17.25 0.967854083459546
17.5 0.968657539391641
17.75 0.970820217322081
18 0.972379089231216
18.25 0.97362059828808
18.5 0.974977597701416
18.75 0.977609101411988
19 0.979148680018976
19.25 0.981660767908128
19.5 0.983298532704672
19.75 0.984985763598736
20 0.987100961086268
};
\addplot [
  draw=c2,
  fill=c2,
  mark=x,
  only marks
]
table{%
x  y
0.1 0.999639567749171
0.2 1.00030932924286
0.3 1.00008683687169
0.4 0.999501421998206
0.5 1.0001695817252
0.6 0.999238360858551
0.7 0.999223524365462
0.8 0.998356912229116
0.9 1.00058488663982
1 0.998748843385313
1.1 0.998128102667117
1.2 0.996859887278973
1.3 0.997769625360024
1.4 0.996346637702163
1.5 0.997146655612708
1.6 0.997314707198654
1.7 0.996959194094995
1.8 0.99348956396958
1.9 0.994523268916816
2 0.99471173260295
2.1 0.995025130426337
2.2 0.993734837370389
2.3 0.991340234242695
2.4 0.992735930592354
2.5 0.990307850764606
2.6 0.990750407323676
2.7 0.992753828560538
2.8 0.993151572150835
2.9 0.988712887077446
3 0.988470595082924
3.1 0.986817838201706
3.2 0.986173606488159
3.3 0.985518339465486
3.4 0.984728768547503
3.5 0.98246349040704
3.6 0.982082019298034
3.7 0.981792375422053
3.8 0.983364126066071
3.9 0.978965959274793
4 0.9838775538303
4.1 0.980180997358902
4.2 0.978568496152865
4.3 0.980023651715511
4.4 0.976237071876653
4.5 0.978245567065074
4.6 0.974641867568665
4.7 0.978348845183992
4.8 0.976490919899992
4.9 0.972078671529282
5.2 0.964864778615714
5.4 0.970115158027879
5.6 0.968902769553706
5.8 0.964015640805437
6 0.959893125327932
6.2 0.973324293549276
6.4 0.962720208012627
6.6 0.967012800796431
6.8 0.961928574283419
7 0.956445521463901
7.2 0.964204651536118
7.4 0.962941761836273
7.6 0.962626102459034
7.8 0.95351349089719
8 0.94673970888999
8.2 0.947650948802216
8.4 0.963034174603005
8.6 0.959255548876488
8.8 0.954097866656648
9 0.961898574340677
9.2 0.950693239522526
9.4 0.96335000664142
9.6 0.950883265221753
9.8 0.958504370283807
10 0.952513165740583
10.2 0.952269032395861
10.4 0.947159831680101
10.6 0.958860009269096
10.8 0.954039777550218
11 0.95790744001511
11.2 0.955645959546264
11.4 0.969867161307729
11.6 0.946304434340645
11.8 0.947290543487977
12 0.952251723046473
12.2 0.952973812505976
12.4 0.934733455143355
12.6 0.951615212807033
12.8 0.961383332148917
13 0.936065508159433
13.2 0.95024255938438
13.4 0.934107463625245
13.6 0.966026398545195
13.8 0.950919625195925
14 0.957463704771662
14.2 0.967983993624059
14.4 0.944343613836493
14.6 0.93151116569598
14.8 0.948212432608225
15 0.943844640486939
15.2 0.950680282336703
15.4 0.944296582030058
15.6 0.927488213342431
15.8 0.96001601049297
16 0.981999475620925
16.2 0.973948442813417
16.4 0.971109991767467
16.6 0.958818567942142
16.8 0.933669886836001
17 0.914092367994302
17.2 0.969020120103341
17.4 0.966775299339752
17.6 0.987395036862848
17.8 0.990955501856938
18 0.969336417942914
18.2 0.926704539486377
18.4 0.980464351114038
18.6 0.952388352318606
18.8 0.958823913273756
19 0.936895906498282
19.2 0.900373236767771
19.4 0.975442509953557
19.6 0.968784949368569
19.8 1.03860281744968
};
\addplot [semithick,c3]
table {%
0.1 0.999984322253531
0.2 0.999937316048458
0.25 0.999902087993421
0.3 0.999859062423781
0.4 0.999749696232385
0.5 0.999609405823361
0.6 0.999438432598445
0.7 0.999237070443749
0.75 0.99912509954066
0.8 0.999005665038244
0.9 0.998744613040839
1 0.998454361158164
1.1 0.998135405095548
1.2 0.997788288393915
1.25 0.997604352428722
1.4 0.99701197866293
1.5 0.99658409989175
1.6 0.996130685926211
1.7 0.995652498276588
1.75 0.995404362911638
1.8 0.995150337106013
1.9 0.994625039370295
2 0.994077476875803
2.1 0.993508554260528
2.2 0.992919206903649
2.25 0.992617174129473
2.3 0.992310398769103
2.3 0.992310398769103
2.4 0.991683120188853
2.5 0.991038385591691
2.6 0.99037723118357
2.7 0.989700712585584
2.75 0.989357026316278
2.8 0.989009902435811
2.9 0.98830588796137
3 0.987589768527063
3.1 0.986862653167096
3.2 0.986125658106365
3.25 0.985753805863718
3.3 0.985379904277848
3.4 0.984626514842655
3.5 0.983866612719245
3.6 0.983101318128335
3.7 0.982331746159953
3.75 0.981945702629522
3.8 0.981559004369023
3.9 0.9807841904058
4 0.98000838968738
4.1 0.979232673116355
4.2 0.978458094852616
4.25 0.978071556897716
4.3 0.977685690144076
4.4 0.976916473222
4.5 0.976151435266367
4.6 0.975391542446561
4.7 0.974637734042424
4.75 0.974263397225317
4.8 0.973890920650495
4.9 0.973151982480018
5 0.972421767743045
5.25 0.97063965615509
5.5 0.96892881138133
5.75 0.967299502778876
6 0.965760428420085
6.25 0.964318652091784
6.5 0.962979569793162
6.75 0.961746905975175
7 0.960622739223592
7.25 0.959607556551009
7.5 0.958700334940377
7.75 0.957898648283958
8 0.957198797396691
8.25 0.956595960360986
8.5 0.956084360089292
8.75 0.95565744567886
9 0.955308083886447
9.25 0.955028756874138
9.5 0.954811762275025
9.75 0.954649411601136
10 0.954534223066697
10.25 0.954459105026463
10.5 0.954417526429264
10.75 0.954403670956914
11 0.954412571853028
11.25 0.954440224838352
11.5 0.954483676950941
11.75 0.954541089632261
12 0.954611774894168
12.25 0.954696203936569
12.5 0.954795988130704
12.75 0.954913832827626
13 0.955053464984911
13.25 0.955219536116504
13.5 0.955417502551073
13.75 0.955653485424212
14 0.955934113221046
14.25 0.956266350021176
14.5 0.956657312871512
14.75 0.957114081919665
15 0.957643507078036
15.25 0.958252015054501
15.5 0.958945420579272
15.75 0.959728745579925
16 0.960606049909829
16.25 0.961580277022944
16.5 0.962653117714371
16.75 0.963824894717154
17 0.965094470567682
17.25 0.966459180732235
17.5 0.967914793533262
17.75 0.969455497934343
18 0.97107391974594
18.25 0.972761166308634
18.5 0.974506899205412
18.75 0.976299434058091
19 0.978125865983579
19.25 0.979972218831163
19.5 0.981823615899892
19.75 0.983664469452122
20 0.985478686001584
};
\end{axis}
\node[anchor=south west] at ({$(current bounding box.south west)!0.05!(current bounding box.south east)$}|-{$(current bounding box.south west)!0.2!(current bounding box.north west)$})
{
\begin{axis}[
tick align=outside,
tick pos=left,
xmin=-0.895, xmax=20.995,
y grid style={darkgray176},
ymin=0.893461757733676, ymax=1.04551429648377,
ytick style={color=black},
mark size=1,
width =0.8\linewidth, 
height = 0.47*\linewidth,
]
\addplot [draw=c1, fill=c1, mark=o, only marks]
table{%
x  y
0.1 0.999808347345181
0.2 1.00020201979863
0.25 0.999595155376648
0.3 0.999595865024691
0.4 1.00016901148727
0.5 0.99961015173215
0.6 0.999629463776136
0.7 0.999452214412409
0.75 0.999266916430586
0.8 0.999120623930955
0.9 0.998784424334304
1 0.998908351324127
1.1 0.998258784778531
1.2 0.997987739956705
1.25 0.997772675048787
1.4 0.9969025075854
1.5 0.996540480802629
1.6 0.996527858876526
1.7 0.995691331182357
1.75 0.995347621581486
1.8 0.995394904179688
1.9 0.994745868419234
2 0.994401610445927
2.1 0.993827905014085
2.2 0.99331385008157
2.25 0.992939018492288
2.3 0.992003314808633
2.3 0.992003314808633
2.4 0.992422636311866
2.5 0.991447067212358
2.6 0.990497758970759
2.7 0.989885486234895
2.75 0.989608843116952
2.8 0.989437426986496
2.9 0.988908290757953
3 0.987537723784141
3.1 0.987056623900192
3.2 0.986422396814482
3.25 0.986065555469907
3.3 0.985823622414831
3.4 0.985165675583141
3.5 0.98445379445884
3.6 0.98355277117105
3.7 0.982773364827906
3.75 0.982158974936564
3.8 0.981367359518001
3.9 0.981288412359387
4 0.97997230351525
4.1 0.979696968770756
4.2 0.979056097852616
4.25 0.978919449331443
4.3 0.977797664882809
4.4 0.977317865288751
4.5 0.976863880380256
4.6 0.975318206104318
4.7 0.975366614597433
4.75 0.974177073357167
4.8 0.973953972796376
4.9 0.973118487913227
5 0.972136975340304
5.25 0.971045532180379
5.5 0.969122600648514
5.75 0.967248991367051
6 0.96659342391881
6.25 0.964991347944822
6.5 0.963016776415196
6.75 0.962411697926026
7 0.960477653050595
7.25 0.961156467772623
7.5 0.959209471565761
7.75 0.958275857830135
8 0.957803116908302
8.25 0.956827222087653
8.5 0.956116405265811
8.75 0.956151535433532
9 0.955995282245294
9.25 0.955471593234452
9.5 0.955847002217905
9.75 0.95487693490566
10 0.955450088672238
10.25 0.955106245198154
10.5 0.955574992426028
10.75 0.955145641894603
11 0.955544458297675
11.25 0.9554798167342
11.5 0.955348418206938
11.75 0.954759387678649
12 0.955896759719311
12.25 0.955522649600535
12.5 0.955531483029222
12.75 0.955912344992378
13 0.955515923775873
13.25 0.956411114593255
13.5 0.956690005069368
13.75 0.956645730143224
14 0.957349575498577
14.25 0.957363746509597
14.5 0.958633821502082
14.75 0.958235151934605
15 0.959866967417858
15.25 0.960615429168243
15.5 0.960072928275912
15.75 0.961089648459594
16 0.961584925865873
16.25 0.963458128164048
16.5 0.964360106905488
16.75 0.965386521764824
17 0.966187431647821
17.25 0.967854083459546
17.5 0.968657539391641
17.75 0.970820217322081
18 0.972379089231216
18.25 0.97362059828808
18.5 0.974977597701416
18.75 0.977609101411988
19 0.979148680018976
19.25 0.981660767908128
19.5 0.983298532704672
19.75 0.984985763598736
20 0.987100961086268
};
\addplot [
  draw=c2,
  fill=c2,
  mark=x,
  only marks
]
table{%
x  y
0.1 0.999639567749171
0.2 1.00030932924286
0.3 1.00008683687169
0.4 0.999501421998206
0.5 1.0001695817252
0.6 0.999238360858551
0.7 0.999223524365462
0.8 0.998356912229116
0.9 1.00058488663982
1 0.998748843385313
1.1 0.998128102667117
1.2 0.996859887278973
1.3 0.997769625360024
1.4 0.996346637702163
1.5 0.997146655612708
1.6 0.997314707198654
1.7 0.996959194094995
1.8 0.99348956396958
1.9 0.994523268916816
2 0.99471173260295
2.1 0.995025130426337
2.2 0.993734837370389
2.3 0.991340234242695
2.4 0.992735930592354
2.5 0.990307850764606
2.6 0.990750407323676
2.7 0.992753828560538
2.8 0.993151572150835
2.9 0.988712887077446
3 0.988470595082924
3.1 0.986817838201706
3.2 0.986173606488159
3.3 0.985518339465486
3.4 0.984728768547503
3.5 0.98246349040704
3.6 0.982082019298034
3.7 0.981792375422053
3.8 0.983364126066071
3.9 0.978965959274793
4 0.9838775538303
4.1 0.980180997358902
4.2 0.978568496152865
4.3 0.980023651715511
4.4 0.976237071876653
4.5 0.978245567065074
4.6 0.974641867568665
4.7 0.978348845183992
4.8 0.976490919899992
4.9 0.972078671529282
5.2 0.964864778615714
5.4 0.970115158027879
5.6 0.968902769553706
5.8 0.964015640805437
6 0.959893125327932
6.2 0.973324293549276
6.4 0.962720208012627
6.6 0.967012800796431
6.8 0.961928574283419
7 0.956445521463901
7.2 0.964204651536118
7.4 0.962941761836273
7.6 0.962626102459034
7.8 0.95351349089719
8 0.94673970888999
8.2 0.947650948802216
8.4 0.963034174603005
8.6 0.959255548876488
8.8 0.954097866656648
9 0.961898574340677
9.2 0.950693239522526
9.4 0.96335000664142
9.6 0.950883265221753
9.8 0.958504370283807
10 0.952513165740583
10.2 0.952269032395861
10.4 0.947159831680101
10.6 0.958860009269096
10.8 0.954039777550218
11 0.95790744001511
11.2 0.955645959546264
11.4 0.969867161307729
11.6 0.946304434340645
11.8 0.947290543487977
12 0.952251723046473
12.2 0.952973812505976
12.4 0.934733455143355
12.6 0.951615212807033
12.8 0.961383332148917
13 0.936065508159433
13.2 0.95024255938438
13.4 0.934107463625245
13.6 0.966026398545195
13.8 0.950919625195925
14 0.957463704771662
14.2 0.967983993624059
14.4 0.944343613836493
14.6 0.93151116569598
14.8 0.948212432608225
15 0.943844640486939
15.2 0.950680282336703
15.4 0.944296582030058
15.6 0.927488213342431
15.8 0.96001601049297
16 0.981999475620925
16.2 0.973948442813417
16.4 0.971109991767467
16.6 0.958818567942142
16.8 0.933669886836001
17 0.914092367994302
17.2 0.969020120103341
17.4 0.966775299339752
17.6 0.987395036862848
17.8 0.990955501856938
18 0.969336417942914
18.2 0.926704539486377
18.4 0.980464351114038
18.6 0.952388352318606
18.8 0.958823913273756
19 0.936895906498282
19.2 0.900373236767771
19.4 0.975442509953557
19.6 0.968784949368569
19.8 1.03860281744968
};
\addplot [semithick,c3]
table {%
0.1 0.999984322253531
0.2 0.999937316048458
0.25 0.999902087993421
0.3 0.999859062423781
0.4 0.999749696232385
0.5 0.999609405823361
0.6 0.999438432598445
0.7 0.999237070443749
0.75 0.99912509954066
0.8 0.999005665038244
0.9 0.998744613040839
1 0.998454361158164
1.1 0.998135405095548
1.2 0.997788288393915
1.25 0.997604352428722
1.4 0.99701197866293
1.5 0.99658409989175
1.6 0.996130685926211
1.7 0.995652498276588
1.75 0.995404362911638
1.8 0.995150337106013
1.9 0.994625039370295
2 0.994077476875803
2.1 0.993508554260528
2.2 0.992919206903649
2.25 0.992617174129473
2.3 0.992310398769103
2.3 0.992310398769103
2.4 0.991683120188853
2.5 0.991038385591691
2.6 0.99037723118357
2.7 0.989700712585584
2.75 0.989357026316278
2.8 0.989009902435811
2.9 0.98830588796137
3 0.987589768527063
3.1 0.986862653167096
3.2 0.986125658106365
3.25 0.985753805863718
3.3 0.985379904277848
3.4 0.984626514842655
3.5 0.983866612719245
3.6 0.983101318128335
3.7 0.982331746159953
3.75 0.981945702629522
3.8 0.981559004369023
3.9 0.9807841904058
4 0.98000838968738
4.1 0.979232673116355
4.2 0.978458094852616
4.25 0.978071556897716
4.3 0.977685690144076
4.4 0.976916473222
4.5 0.976151435266367
4.6 0.975391542446561
4.7 0.974637734042424
4.75 0.974263397225317
4.8 0.973890920650495
4.9 0.973151982480018
5 0.972421767743045
5.25 0.97063965615509
5.5 0.96892881138133
5.75 0.967299502778876
6 0.965760428420085
6.25 0.964318652091784
6.5 0.962979569793162
6.75 0.961746905975175
7 0.960622739223592
7.25 0.959607556551009
7.5 0.958700334940377
7.75 0.957898648283958
8 0.957198797396691
8.25 0.956595960360986
8.5 0.956084360089292
8.75 0.95565744567886
9 0.955308083886447
9.25 0.955028756874138
9.5 0.954811762275025
9.75 0.954649411601136
10 0.954534223066697
10.25 0.954459105026463
10.5 0.954417526429264
10.75 0.954403670956914
11 0.954412571853028
11.25 0.954440224838352
11.5 0.954483676950941
11.75 0.954541089632261
12 0.954611774894168
12.25 0.954696203936569
12.5 0.954795988130704
12.75 0.954913832827626
13 0.955053464984911
13.25 0.955219536116504
13.5 0.955417502551073
13.75 0.955653485424212
14 0.955934113221046
14.25 0.956266350021176
14.5 0.956657312871512
14.75 0.957114081919665
15 0.957643507078036
15.25 0.958252015054501
15.5 0.958945420579272
15.75 0.959728745579925
16 0.960606049909829
16.25 0.961580277022944
16.5 0.962653117714371
16.75 0.963824894717154
17 0.965094470567682
17.25 0.966459180732235
17.5 0.967914793533262
17.75 0.969455497934343
18 0.97107391974594
18.25 0.972761166308634
18.5 0.974506899205412
18.75 0.976299434058091
19 0.978125865983579
19.25 0.979972218831163
19.5 0.981823615899892
19.75 0.983664469452122
20 0.985478686001584
};
\end{axis}
};

\end{tikzpicture}

%% file: mainexpvstau10.0.tex

\begin{tikzpicture}

\begin{axis}[
mark size=1,
tick align=outside,
tick pos=left,
x grid style={darkgray176},
xlabel={simulation time $\tau$},
xmin=-0.895, xmax=20.995,
y grid style={darkgray176},
ylabel={$\braket{O}$, $p=0.01$},
ymin=-0.0871216912978824, ymax=1.16469135671033,
ytick style={color=black},
width =\linewidth, 
height = 0.65*\linewidth,
legend style={at={(axis cs:10,0.4)},anchor=south west},
]
\addplot [draw=c1, fill=c1, mark=o, only marks]
table{%
x  y
0.1 1.00714727566774
0.2 1.00617277739028
0.25 1.00647639094306
0.3 1.00549714350543
0.4 1.00449546484134
0.5 1.00596981743879
0.6 1.00584062856152
0.7 1.00616928510222
0.75 1.00646307292237
0.8 1.00398950502193
0.9 1.00615186139401
1 1.00520299638672
1.1 1.00433828100642
1.2 1.00445212317456
1.25 1.00454299099602
1.4 1.00528613738243
1.5 1.00360583005872
1.6 1.00375652897194
1.7 1.00373854596162
1.75 1.00475036333081
1.8 1.00180830745546
1.9 1.00233212602512
2 1.00178475960451
2.1 1.00287417722638
2.2 1.00213357382416
2.25 1.00293542425087
2.3 1.0011554368794
2.3 1.0011554368794
2.4 1.00120493087476
2.5 1.00180984719809
2.6 1.00185263690091
2.7 1.00055838129829
2.75 0.999849632601197
2.8 0.999033388151758
2.9 0.999476890980975
3 0.998468594353243
3.1 1.0001144051598
3.2 0.996894107542831
3.25 0.998001236909055
3.3 0.996178661142324
3.4 0.995445892584782
3.5 0.994691443740652
3.6 0.997096832475159
3.7 0.99286495078517
3.75 0.993208379880764
3.8 0.992794719466962
3.9 0.992105305691847
4 0.991771838633018
4.1 0.991624124231169
4.2 0.99135443586247
4.25 0.990343446622527
4.3 0.989822183537255
4.4 0.989682904545917
4.5 0.98789680098717
4.6 0.990335116730112
4.7 0.989063668742973
4.75 0.987396858057079
4.8 0.986686831344655
4.9 0.987461208359122
5 0.988517873292424
5.25 0.985651126595955
5.5 0.982478014893685
5.75 0.983827847881529
6 0.981818534790821
6.25 0.98146432063841
6.5 0.981260767277176
6.75 0.981103593656694
7 0.982040626294196
7.25 0.977818655513538
7.5 0.98103074980068
7.75 0.980647872909582
8 0.979390878859633
8.25 0.979007040778643
8.5 0.979641725584492
8.75 0.983514584211362
9 0.978594318498014
9.25 0.984318573603005
9.5 0.984740078268584
9.75 0.984699374436939
10 0.980461762668713
10.25 0.978536709479948
10.5 0.980664332963375
10.75 0.982927034758142
11 0.977784010027477
11.25 0.983539817360077
11.5 0.975831483228489
11.75 0.982353357317986
12 0.979328093704801
12.25 0.982494506056115
12.5 0.986082236442038
12.75 0.985039735952703
13 0.978309322967669
13.25 0.984842020107838
13.5 0.986914525879684
13.75 0.988573967043355
14 0.988821046311412
14.25 0.985363268535045
14.5 0.994746255948131
14.75 0.991801182350385
15 0.997358698059843
15.25 0.993182151676033
15.5 0.995579956474824
15.75 0.991320079613172
16 0.996339214826259
16.25 0.998883378060386
16.5 1.00099191540918
16.75 1.00286492263446
17 1.00366292320332
17.25 0.991292104679428
17.5 1.01193513208499
17.75 1.01043273061295
18 1.00265486771738
18.25 1.01812014160027
18.5 1.01526641233211
18.75 1.01284763572515
19 1.01511458844463
19.25 1.01819720816309
19.5 1.02522733154044
19.75 1.02254468378026
20 1.02753343094176
};
\addplot [
  draw=c2,
  fill=c2,
  mark=x,
  only marks
]
table{%
x  y
0.1 1.00324206735688
0.2 1.00510557361019
0.3 1.00877749525138
0.4 1.0101392094352
0.5 1.00965269340038
0.6 1.00933101706773
0.7 1.01088533321773
0.8 1.01576951466925
0.9 1.02457518711634
1 1.02308325453883
1.1 1.01560175608894
1.2 1.02132549611129
1.3 1.0060554617784
1.4 1.00005648950524
1.5 1.0378295755097
1.6 0.978549631099811
1.7 1.02024221770547
1.8 1.05504713308029
1.9 1.10779076361905
2 0.969176870450375
2.1 1.01582596824835
2.2 0.913560648969476
2.3 1.07563779553159
2.4 0.115504
2.5 0.664758490164313
2.6 0.720205463828608
2.7 0.954415564869694
2.8 0.358799908743184
2.9 0.0630376
3 0.0514924
3.1 0.503203767871775
3.2 0.0457664
3.3 0.367195010879152
3.4 0.0319316
3.5 0.027754
3.6 0.0809031237961048
3.7 0.0232232
3.8 0.0512300597291939
3.9 0.0159236
4 0.0168632
4.1 0.0115708
4.2 0.0116104
4.3 0.0137994984732741
4.4 0.0091036
4.5 0.0176179212096261
4.6 0.0197820016482098
4.7 0.0126846227919139
4.8 0.0083088
4.9 0.0010096
5.2 0.0426487791674688
5.4 0.00375838336306736
5.6 0.0122516534998129
5.8 0.000583150855948498
6 4.22589547886501e-21
6.2 0.000735326758833367
6.4 0.002656906230733
6.6 0.0026984
6.8 0.000130781446234735
7 0.0040244
7.2 0.0003556
7.4 -0.000296037134617434
7.6 0.00199647230700041
7.8 -0.0011844
8 -0.00119794296687819
8.2 -0.0265644454336191
8.4 0.0018964
8.6 -0.00116327267394262
8.8 -0.0031852
9 -0.000649896436148147
9.2 0.0008676
9.4 0.0023096
9.6 6.85491690238626e-07
9.8 0.000100076765371171
10 -0.0012844
10.2 -0.000365023105021196
10.4 0.0028228
10.6 -0.000151109394550728
10.8 0.0032756
11 9.94372448218628e-21
11.2 0.000224516723838204
11.4 -0.000183667189241641
11.6 8.84e-05
11.8 -0.000627503132698553
12 -0.0048868
12.2 -0.0302210982066001
12.4 -0.0028004
12.6 0.0029744
12.8 -0.003308
13 -0.002006
13.2 0.000457418434957321
13.4 0.000324447544954076
13.6 0.0733172496219894
13.8 -0.0017688
14 0.00354625440815416
14.2 0.001308
14.4 -0.0112947674109125
14.6 0.0009768
14.8 -0.0014744
15 -0.0016692
15.2 -0.0003052
15.4 -4.6707288145636e-05
15.6 0.0020144
15.8 -0.00396
16 -0.00059461126701584
16.2 0.00124237372409793
16.4 -0.000346
16.6 -0.0028132
16.8 0.0037892
17 -0.0024532
17.2 -0.0001836
17.4 0.0056196
17.6 -0.0009972
17.8 -0.0018248
18 -0.0016908
18.2 0.000929415958526693
18.4 0.0014956
18.6 -0.0014156
18.8 0.002262
19 -0.003196
19.2 -0.0044056
19.4 0.00152846548652277
19.6 0.004436
19.8 0.00101915710902652
};
\addplot [semithick, c3]
table {%
0.1 0.999984322253531
0.2 0.999937316048458
0.25 0.999902087993421
0.3 0.999859062423781
0.4 0.999749696232385
0.5 0.999609405823361
0.6 0.999438432598445
0.7 0.999237070443749
0.75 0.99912509954066
0.8 0.999005665038244
0.9 0.998744613040839
1 0.998454361158164
1.1 0.998135405095548
1.2 0.997788288393915
1.25 0.997604352428722
1.4 0.99701197866293
1.5 0.99658409989175
1.6 0.996130685926211
1.7 0.995652498276588
1.75 0.995404362911638
1.8 0.995150337106013
1.9 0.994625039370295
2 0.994077476875803
2.1 0.993508554260528
2.2 0.992919206903649
2.25 0.992617174129473
2.3 0.992310398769103
2.3 0.992310398769103
2.4 0.991683120188853
2.5 0.991038385591691
2.6 0.99037723118357
2.7 0.989700712585584
2.75 0.989357026316278
2.8 0.989009902435811
2.9 0.98830588796137
3 0.987589768527063
3.1 0.986862653167096
3.2 0.986125658106365
3.25 0.985753805863718
3.3 0.985379904277848
3.4 0.984626514842655
3.5 0.983866612719245
3.6 0.983101318128335
3.7 0.982331746159953
3.75 0.981945702629522
3.8 0.981559004369023
3.9 0.9807841904058
4 0.98000838968738
4.1 0.979232673116355
4.2 0.978458094852616
4.25 0.978071556897716
4.3 0.977685690144076
4.4 0.976916473222
4.5 0.976151435266367
4.6 0.975391542446561
4.7 0.974637734042424
4.75 0.974263397225317
4.8 0.973890920650495
4.9 0.973151982480018
5 0.972421767743045
5.25 0.97063965615509
5.5 0.96892881138133
5.75 0.967299502778876
6 0.965760428420085
6.25 0.964318652091784
6.5 0.962979569793162
6.75 0.961746905975175
7 0.960622739223592
7.25 0.959607556551009
7.5 0.958700334940377
7.75 0.957898648283958
8 0.957198797396691
8.25 0.956595960360986
8.5 0.956084360089292
8.75 0.95565744567886
9 0.955308083886447
9.25 0.955028756874138
9.5 0.954811762275025
9.75 0.954649411601136
10 0.954534223066697
10.25 0.954459105026463
10.5 0.954417526429264
10.75 0.954403670956914
11 0.954412571853028
11.25 0.954440224838352
11.5 0.954483676950941
11.75 0.954541089632261
12 0.954611774894168
12.25 0.954696203936569
12.5 0.954795988130704
12.75 0.954913832827626
13 0.955053464984911
13.25 0.955219536116504
13.5 0.955417502551073
13.75 0.955653485424212
14 0.955934113221046
14.25 0.956266350021176
14.5 0.956657312871512
14.75 0.957114081919665
15 0.957643507078036
15.25 0.958252015054501
15.5 0.958945420579272
15.75 0.959728745579925
16 0.960606049909829
16.25 0.961580277022944
16.5 0.962653117714371
16.75 0.963824894717154
17 0.965094470567682
17.25 0.966459180732235
17.5 0.967914793533262
17.75 0.969455497934343
18 0.97107391974594
18.25 0.972761166308634
18.5 0.974506899205412
18.75 0.976299434058091
19 0.978125865983579
19.25 0.979972218831163
19.5 0.981823615899892
19.75 0.983664469452122
20 0.985478686001584
};
\addlegendentry{$\braket{E_O}$ QSP}
\addlegendentry{$\braket{E_O}$ Trotter}
\addlegendentry{$\braket{O}_{ideal}$}
\end{axis}

\end{tikzpicture}

%% file: biasvstau_p0.1.tex
\begin{tikzpicture}

\begin{axis}[
legend style={at={(axis cs:19, 0.075)},anchor=north east},
tick align=outside,
tick pos=left,
x grid style={darkgray176},
xticklabel=\empty,
xmin=-0.895, xmax=20.995,
xtick style={color=black},
y grid style={darkgray176},
ylabel={$\left|b_{E_O}\right|$, $p=0.0001$},
ytick style={color=black},
mark size=1,
width =\linewidth, 
height = 0.65*\linewidth
]
\addplot [semithick, c1, mark=o]
table {%
0.1  6.02997834714403e-05
0.2  1.07805886578882e-05
0.25 9.69171159237447e-06
0.3 6.71880307168138e-06
0.4 0.000104902019018471
0.5  6.40313403462711e-05
0.6  4.42517461156644e-05
0.7 9.83531388791814e-05
0.75 1.12772728586563e-05
0.8 0.000107583999663596
0.9 0.000114581043505146
1  7.29956401099852e-05
1.1 2.42890519364103e-06
1.2  0.000109028151175639
1.25 0.00011549740640926
1.4  6.423659089283e-05
1.5 0.000147849304558356
1.6 0.000125300595534727
1.7  1.94898390204479e-07
1.75 3.05195441945294e-05
1.8 4.78935705058303e-05
1.9 8.08191583593354e-05
2 0.000120392654868517
2.1  6.61889247310921e-06
2.2 0.000229301270872839
2.25  1.41416724832588e-05
2.3 8.88562793732728e-05
2.3 8.88562793732728e-05
2.4 0.000305305541337231
2.5 0.000135612467837998
2.6 0.000251784180413828
2.7 0.000413189055125995
2.75 3.46923008230915e-05
2.8 0.000506152298889728
2.9 0.000578524196094832
3 7.32223085546657e-05
3.1 2.48638772059051e-05
3.2 5.10077267893427e-05
3.25 0.000179157252112594
3.3 0.000335092530948766
3.4 0.000345266600200511
3.5 0.000105543382839079
3.6 0.000166734223577047
3.7 0.000172045884449301
3.75 0.000122868809198762
3.8 0.00014320728957784
3.9 0.000547711317067989
4 0.000482111958956088
4.1 0.000239481175811185
4.2 0.000332301963582804
4.25 0.000579595982712622
4.3 0.000360317673828825
4.4 0.000498087435855665
4.5 0.000759679568471561
4.6 0.000240168527616103
4.7 6.92458203077617e-05
4.75 5.50513743757808e-06
4.8 0.000224898366466553
4.9 0.000124154639077223
5 9.00551821706452e-05
5.25 0.000236410297125067
5.5 0.000133123616976816
5.75 0.00023504090687676
6 3.86089071514606e-05
6.25 0.000181339407537506
6.5 0.000347902856403071
6.75 0.000179265861369293
7 0.000174242404898162
7.25 0.000230484789572549
7.5 0.000203715283581873
7.75 0.000692435235633893
8 0.000314742079161157
8.25 2.27743745454534e-05
8.5 9.33174159279959e-05
8.75 0.00029771514109056
9 0.000202333745435213
9.25 0.000474480902472818
9.5 0.000607938346821157
9.75 0.000479117563415565
10 0.000162057290311823
10.25 0.000187303031327191
10.5 8.5136155358434e-05
10.75 0.000135674787611895
11 0.000254929360103517
11.25 0.000766419664027462
11.5 0.000525583441857269
11.75 0.000511060010944253
12 4.78678155690915e-06
12.25 6.6389433820313e-05
12.5 0.00016237778727668
12.75 0.000849401092764213
13 0.000328155646135198
13.25 0.0011214424853756
13.5 0.000375030528858233
13.75 3.53356759100754e-05
14 0.000142436692698422
14.25 0.000283241234935616
14.5 0.000484763343853389
14.75 0.00116529279655875
15 0.00110542501137711
15.25 0.00034115697416337
15.5 0.00038207989564687
15.75 0.000875763312337963
16 0.000532406842121302
16.25 0.000803911299485205
16.5 0.000759246282957426
16.75 0.00101390009660318
17 4.99260987318895e-05
17.25 0.000254546388582577
17.5 0.000636738390522695
17.75 6.97478801175988e-05
18 0.000309164482867108
18.25 0.000337786310552146
18.5 0.000159690628878106
18.75 0.000704264341828242
19 0.000299728438726943
19.25 1.69592923301431e-05
19.5 9.80740567924077e-06
19.75 0.000252556808209126
20 0.000315591970281193
};
\addplot [semithick, c2, mark=x]
table {%
0.1 5.84154831299033e-05
0.2 0.00012288581635933
0.3 6.56566719688945e-05
0.4 5.96106540909913e-05
0.5 1.21062462905552e-05
0.6 9.94220640029297e-05
0.7 3.28934111745172e-05
0.8 0.000160584070702519
0.9 0.000154044177341772
1 0.000216069920039752
1.1 0.000195965396964759
1.2 3.53494648673847e-05
1.3 3.97701495686853e-05
1.4 4.62630531952168e-05
1.5 0.000152951642757748
1.6 4.39616486452854e-05
1.7 6.43875261000515e-05
1.8 7.8943294241629e-05
1.9 0.000264839299477404
2 0.000361896034431264
2.1 8.18075638732152e-05
2.2 0.000236333416244849
2.3 6.90275163812704e-05
2.4 5.65400068374133e-05
2.5 0.000285758249423784
2.6 0.000221971197933879
2.7 4.45291293514893e-05
2.8 7.19592669803371e-05
2.9 0.000143722013011938
3 0.000292696182039864
3.1 1.9187967262746e-05
3.2 1.70643532125814e-05
3.3 0.000138362348305665
3.4 0.000328025522044584
3.5 0.00046806407036537
3.6 0.000265571660712172
3.7 0.000328494103580934
3.8 0.000212550802322853
3.9 0.00036109610535151
4 4.13152161569119e-05
4.1 0.000406847340504779
4.2 1.50575419793908e-05
4.3 0.0005537093965301
4.4 0.000136753117012223
4.5 0.000371884649843501
4.6 6.09820145159867e-05
4.7 0.000145865402640832
4.8 0.000712689346815232
4.9 8.29568015662252e-05
5.2 0.000417714787051882
5.4 0.000176064562726452
5.6 0.000429453247239864
5.8 0.000434367553805437
6 0.000230175149614209
6.2 0.000360000282978001
6.4 9.27744016803178e-05
6.6 0.000160707155988482
6.8 0.000526525115104093
7 0.000673733170656665
7.2 0.000128488906850133
7.4 0.000233345706835086
7.6 0.000541500160123332
7.8 0.000670070555077618
8 0.000575179003643012
8.2 0.000397548705498796
8.4 7.14834481960747e-05
8.6 9.81363388683487e-05
8.8 0.000494984639960006
9 0.0001320754770614
9.2 0.000273986485707312
9.4 0.000382566620259928
9.6 0.000299307184797315
9.8 6.79097441883547e-05
10 0.000182629654910005
10.2 0.00069586508156938
10.4 0.000537202424001459
10.6 0.000880889899449011
10.8 0.00183941524770892
11 7.8332176835505e-05
11.2 0.000607170048045003
11.4 0.000205088551766819
11.6 3.66630505003185e-05
11.8 0.000260293825673497
12 0.000345822432433196
12.2 0.000694058690528521
12.4 0.000369664711533546
12.6 0.000305391166716462
12.8 0.000440513726951308
13 0.000546043535795571
13.2 0.000489614031718322
13.4 0.000390328186505196
13.6 0.000350821610447971
13.8 0.000125309643176363
14 0.000407149468185652
14.2 0.000727041457276645
14.4 0.0011595480073664
14.6 0.00029485919895722
14.8 0.000328090213993493
15 0.000185736186680829
15.2 0.000825620880197442
15.4 0.000236305685725413
15.6 0.000203188146566835
15.8 0.00113033049500655
16 0.000164103130851356
16.2 0.000763482388653869
16.4 0.000301711901429025
16.6 0.000394798259956142
16.8 0.000468861268203669
17 0.000235584294390412
17.2 6.71357021615471e-05
17.4 0.00126359580498614
17.6 0.00194437109615408
17.8 0.000430408516656655
18 0.000554528780971397
18.2 0.000306948429822707
18.4 8.34481952396438e-05
18.6 0.000342649399800554
18.8 0.000670991658183184
19 0.00117456629808432
19.2 0.0001527389760001
19.4 0.000448600435348001
19.6 0.000659154991806687
19.8 0.000545025907943475
};
\addplot [semithick, c3]
table {%
0.1 0.08
0.2 0.08
0.3 0.08
0.4 0.08
0.5 0.08
0.6 0.08
0.7 0.08
0.8 0.08
0.9 0.08
1 0.08
1.1 0.08
1.2 0.08
1.3 0.08
1.4 0.08
1.5 0.08
1.6 0.08
1.7 0.08
1.8 0.08
1.9 0.08
2 0.08
2.1 0.08
2.2 0.08
2.3 0.08
2.4 0.08
2.5 0.08
2.6 0.08
2.7 0.08
2.8 0.08
2.9 0.08
3 0.08
3.1 0.08
3.2 0.08
3.3 0.08
3.4 0.08
3.5 0.08
3.6 0.08
3.7 0.08
3.8 0.08
3.9 0.08
4 0.08
4.1 0.08
4.2 0.08
4.3 0.08
4.4 0.08
4.5 0.08
4.6 0.08
4.7 0.08
4.8 0.08
4.9 0.08
5.2 0.08
5.4 0.08
5.6 0.08
5.8 0.08
6 0.08
6.2 0.08
6.4 0.08
6.6 0.08
6.8 0.08
7 0.08
7.2 0.08
7.4 0.08
7.6 0.08
7.8 0.08
8 0.08
8.2 0.08
8.4 0.08
8.6 0.08
8.8 0.08
9 0.08
9.2 0.08
9.4 0.08
9.6 0.08
9.8 0.08
10 0.08
10.2 0.08
10.4 0.08
10.6 0.08
10.8 0.08
11 0.08
11.2 0.08
11.4 0.08
11.6 0.08
11.8 0.08
12 0.08
12.2 0.08
12.4 0.08
12.6 0.08
12.8 0.08
13 0.08
13.2 0.08
13.4 0.08
13.6 0.08
13.8 0.08
14 0.08
14.2 0.08
14.4 0.08
14.6 0.08
14.8 0.08
15 0.08
15.2 0.08
15.4 0.08
15.6 0.08
15.8 0.08
16 0.08
16.2 0.08
16.4 0.08
16.6 0.08
16.8 0.08
17 0.08
17.2 0.08
17.4 0.08
17.6 0.08
17.8 0.08
18 0.08
18.2 0.08
18.4 0.08
18.6 0.08
18.8 0.08
19 0.08
19.2 0.08
19.4 0.08
19.6 0.08
19.8 0.08
};

\addlegendentry{$\left|b_{E_O}\right|$ QSP}
\addlegendentry{$\left|b_{E_O}\right|$ Trotter}
\addlegendentry{$8\cdot 10^{-2}$}
\end{axis}

\end{tikzpicture}

%% file: biasvstau_p1.0.tex
\begin{tikzpicture}

\begin{axis}[
tick align=outside,
tick pos=left,
xticklabel=\empty,
x grid style={darkgray176},
xlabel={simulation time $\tau$},
xmin=-0.895, xmax=20.995,
xtick style={color=black},
y grid style={darkgray176},
ylabel={$\left|b_{E_O}\right|$, $p=0.001$},
ytick style={color=black},
mark size=1,
width =\linewidth, 
height = 0.65*\linewidth
]
\addplot [semithick, c1, mark=o]
table {%
0.1 0.000175974908350085
0.2 0.000264703750175732
0.25 0.00030693261677317
0.3 0.000263197399089865
0.4 0.000419315254885588
0.5 7.45908789068217e-07
0.6 0.000191031177691037
0.7 0.00021514396866007
0.75 0.00014181688992565
0.8 0.000114958892710626
0.9 3.98112934651973e-05
1 0.000453990165963702
1.1 0.000123379682982816
1.2 0.000199451562790665
1.25 0.000168322620065275
1.4 0.000109471077530454
1.5 4.36190891209831e-05
1.6 0.000397172950314495
1.7 3.88329057694481e-05
1.75 5.67413301519881e-05
1.8 0.000244567073674373
1.9 0.000120829048938398
2 0.000324133570124285
2.1 0.000319350753556491
2.2 0.000394643177921594
2.25 0.000321844362814905
2.3 0.000307083960470078
2.3 0.000307083960470078
2.4 0.000739516123012485
2.5 0.000408681620667317
2.6 0.00012052778718874
2.7 0.000184773649311354
2.75 0.000251816800674098
2.8 0.000427524550685399
2.9 0.00060240279658319
3 5.20447429227522e-05
3.1 0.000193970733095794
3.2 0.000296738708117306
3.25 0.000311749606189404
3.3 0.000443718136983184
3.4 0.000539160740486211
3.5 0.000587181739594822
3.6 0.000451453042715433
3.7 0.00044161866795267
3.75 0.000213272307041601
3.8 0.000191644851021411
3.9 0.000504221953587014
4 3.6086172129246e-05
4.1 0.000464295654400759
4.2 0.000598003000000236
4.25 0.000847892433726849
4.3 0.000111974738733034
4.4 0.000401392066751272
4.5 0.000712445113889348
4.6 7.33363422424915e-05
4.7 0.000728880555009748
4.75 8.63238681498846e-05
4.8 6.3052145881648e-05
4.9 3.34945667910258e-05
5 -0.00028479240274093
5.25 0.000405876025288743
5.5 0.000193789267183719
5.75 5.05114118246608e-05
6 0.000832995498725175
6.25 0.000672695853038308
6.5 3.72066220336587e-05
6.75 0.000664791950851229
7 0.000145086172996578
7.25 0.00154891122161416
7.5 0.000509136625383988
7.75 0.000377209546176482
8 0.000604319511610996
8.25 0.00023126172666732
8.5 3.20451765190022e-05
8.75 0.000494089754671689
9 0.000687198358847452
9.25 0.000442836360313725
9.5 0.00103523994288002
9.75 0.000227523304523469
10 0.000915865605541244
10.25 0.000647140171691318
10.5 0.0011574659967637
10.75 0.000741970937688796
11 0.00113188644464701
11.25 0.00103959189584768
11.5 0.000864741255997226
11.75 0.000218298046388066
12 0.00128498482514283
12.25 0.000826445663965614
12.5 0.000735494898518096
12.75 0.000998512164752663
13 0.000462458790962295
13.25 0.00119157847675144
13.5 0.00127250251829492
13.75 0.00099224471901127
14 0.00141546227753098
14.25 0.00109739648842133
14.5 0.00197650863057042
14.75 0.00112107001493933
15 0.00222346033982146
15.25 0.00236341411374252
15.5 0.00112750769663938
15.75 0.00136090287966917
16 0.000978875956043068
16.25 0.00187785114110384
16.5 0.00170698919111667
16.75 0.00156162704767038
17 0.00109296108013834
17.25 0.00139490272731047
17.5 0.000742745858379412
17.75 0.00136471938773819
18 0.00130516948527604
18.25 0.000859431979446157
18.5 0.000470698496003785
18.75 0.00130966735389693
19 0.00102281403539684
19.25 0.00168854907696536
19.5 0.00147491680477951
19.75 0.00132129414661453
20 0.00162227508468349
};
\addplot [semithick, c2, mark=x]
table {%
0.1 0.000344752353507838
0.2 0.000372021793201616
0.3 0.000227793776209184
0.4 0.000248239920612625
0.5 0.000560229420404568
0.6 0.000199994842749796
0.7 1.34416851256258e-05
0.8 0.000648616868519469
0.9 0.00184044506278946
1 0.000294693104773147
1.1 7.04834077980276e-06
1.2 0.00092810012442146
1.3 0.000356375678493781
1.4 0.000664935542759704
1.5 0.000563018414497241
1.6 0.00118454443685179
1.7 0.00130728250531875
1.8 0.00166012002607985
1.9 0.000101048176020746
2 0.000635049751929406
2.1 0.00151744434896994
2.2 0.00081657504489363
2.3 -0.000969141495886228
2.4 0.00105391376020025
2.5 0.000729349457792128
2.6 0.00037444501767947
2.7 0.00305446966295564
2.8 0.00414310931975481
2.9 0.000408525546217953
3 0.000882440521225325
3.1 4.3112954575153e-05
3.2 4.97387485188794e-05
3.3 0.000140314021305987
3.4 0.000104220917950704
3.5 0.00140106700430909
3.6 0.00101715590745544
3.7 0.000537140872698849
3.8 0.00180743764222224
3.9 0.0018158301545711
4 0.00387164891954228
4.1 0.0009508914103451
4.2 0.000113049277183253
4.3 0.00234068860779391
4.4 0.00067659716114532
4.5 0.00209701106351001
4.6 0.000746722748767747
4.7 0.00371413377671592
4.8 0.00260308989770164
4.9 0.00107015490938434
5.2 0.00612261849331186
5.4 0.000514692179468312
5.6 0.000639520586246167
5.8 0.00296496855832118
6 0.00586362838042309
6.2 0.00872907888426833
6.4 0.000778632868258855
6.6 0.0045430060080125
6.8 0.000419016116283366
7 0.00417342385977393
7.2 0.00440652787695928
7.4 0.00389512052968177
7.6 0.00426264102098206
7.8 0.0042335118660799
8 0.01045547677864
8.2 0.00905453457744676
8.4 0.00675919713143169
8.6 0.00335513187451397
8.8 0.00148041384183917
9 0.00659373433889965
9.2 0.00438298681975513
9.4 0.00846151182467647
9.6 0.00385442520709989
9.8 0.00388445841705642
10 0.00201820766842142
10.2 0.00219943158354963
10.4 0.00726797092940201
10.6 0.00445371965381813
10.8 0.000361375300737654
11 0.00349745154848069
11.2 0.00121516387840415
11.4 0.015405156255257
11.6 0.00819806584159599
11.8 0.00726108864557962
12 0.00235750381413447
12.2 0.00170177278981998
12.4 0.0200180210606734
12.6 0.00322290909183609
12.8 0.00644615513061908
13 0.0189851924853141
13.2 0.00493856275285642
13.4 0.0212237374964431
13.6 0.0105224163872706
13.8 0.00478311947177912
14 0.00153275600517833
14.2 0.0117918159321961
14.4 0.0121464637268229
14.6 0.0253171295443234
14.8 0.00899798246074679
15 0.0137952548341868
15.2 0.00743972279622551
15.4 0.0143571807148172
15.6 0.0317556624867207
15.8 0.00012332582714436
16 0.0213973667252001
16.2 0.0125748442775572
16.4 0.00890190187406192
16.6 0.00428739901500907
16.8 0.0303971419457657
17 0.0509980953165968
17.2 0.00284529209241702
17.4 0.000542738269151521
17.6 0.0188776777158634
17.8 0.0211860873750598
18 0.00173377233037519
18.2 0.0457105159336195
18.4 0.00666556529627005
18.6 0.0228272439116678
18.8 0.0178352657098124
19 0.0412268426190635
19.2 0.0792258587850919
19.4 0.00563812781540407
19.6 0.0137745380460864
19.8 0.0545753032196619
};
\addplot [semithick, c3]
table {%
0.1 0.08
0.2 0.08
0.3 0.08
0.4 0.08
0.5 0.08
0.6 0.08
0.7 0.08
0.8 0.08
0.9 0.08
1 0.08
1.1 0.08
1.2 0.08
1.3 0.08
1.4 0.08
1.5 0.08
1.6 0.08
1.7 0.08
1.8 0.08
1.9 0.08
2 0.08
2.1 0.08
2.2 0.08
2.3 0.08
2.4 0.08
2.5 0.08
2.6 0.08
2.7 0.08
2.8 0.08
2.9 0.08
3 0.08
3.1 0.08
3.2 0.08
3.3 0.08
3.4 0.08
3.5 0.08
3.6 0.08
3.7 0.08
3.8 0.08
3.9 0.08
4 0.08
4.1 0.08
4.2 0.08
4.3 0.08
4.4 0.08
4.5 0.08
4.6 0.08
4.7 0.08
4.8 0.08
4.9 0.08
5.2 0.08
5.4 0.08
5.6 0.08
5.8 0.08
6 0.08
6.2 0.08
6.4 0.08
6.6 0.08
6.8 0.08
7 0.08
7.2 0.08
7.4 0.08
7.6 0.08
7.8 0.08
8 0.08
8.2 0.08
8.4 0.08
8.6 0.08
8.8 0.08
9 0.08
9.2 0.08
9.4 0.08
9.6 0.08
9.8 0.08
10 0.08
10.2 0.08
10.4 0.08
10.6 0.08
10.8 0.08
11 0.08
11.2 0.08
11.4 0.08
11.6 0.08
11.8 0.08
12 0.08
12.2 0.08
12.4 0.08
12.6 0.08
12.8 0.08
13 0.08
13.2 0.08
13.4 0.08
13.6 0.08
13.8 0.08
14 0.08
14.2 0.08
14.4 0.08
14.6 0.08
14.8 0.08
15 0.08
15.2 0.08
15.4 0.08
15.6 0.08
15.8 0.08
16 0.08
16.2 0.08
16.4 0.08
16.6 0.08
16.8 0.08
17 0.08
17.2 0.08
17.4 0.08
17.6 0.08
17.8 0.08
18 0.08
18.2 0.08
18.4 0.08
18.6 0.08
18.8 0.08
19 0.08
19.2 0.08
19.4 0.08
19.6 0.08
19.8 0.08
};

\end{axis}

\end{tikzpicture}

%% file: biasvstau_p10.0.tex
\begin{tikzpicture}
\begin{axis}[
mark size=1,
tick align=outside,
tick pos=left,
x grid style={darkgray176},
xlabel={simulation time $\tau$},
xmin=-0.895, xmax=20.995,
xtick style={color=black},
y grid style={darkgray176},
ylabel={ $\left|b_{E_O}\right|$, $p=0.01$},
ytick style={color=black},
width =\linewidth, 
height = 0.65*\linewidth
]
\addplot[semithick, c1, mark=o]
table {%
0.1 0.00716295341420714
0.2 0.00623546134182118
0.25 0.00657430294963712
0.3 0.00563808108165142
0.4 0.00474576860895393
0.5 0.00636041161543399
0.6 0.00640219596307212
0.7 0.00693221465846616
0.75 0.00733797338171294
0.8 0.00498383998368357
0.9 0.00740724835316731
1 0.00674863522855174
1.1 0.00620287591087598
1.2 0.00666383478064558
1.25 0.00693863856729382
1.4 0.00827415871950465
1.5 0.00702173016697027
1.6 0.00762584304573211
1.7 0.00808604768503574
1.75 0.00934600041917655
1.8 0.00665797034944171
1.9 0.007707086654825
2 0.00770728272870647
2.1 0.00936562296585108
2.2 0.00921436692051392
2.25 0.0103182501213979
2.3 0.00884503811030146
2.3 0.00884503811030146
2.4 0.00952181068590885
2.5 0.0107714616063959
2.6 0.011475405717341
2.7 0.0108576687127075
2.75 0.0104926062849183
2.8 0.0100234857159468
2.9 0.0111710030196055
3 0.0108788258261791
3.1 0.0132517519926993
3.2 0.0107684494364662
3.25 0.012247431045337
3.3 0.0107987568644755
3.4 0.0108193777421265
3.5 0.010824831021407
3.6 0.0139955143468243
3.7 0.0105332046252166
3.75 0.0112626772512423
3.8 0.0112357150979386
3.9 0.0113211152860465
4 0.0117634489456381
4.1 0.0123914511148142
4.2 0.0128963410098548
4.25 0.0122718897248114
4.3 0.0121364933931786
4.4 0.012766431323917
4.5 0.0117453657208031
4.6 0.0149435742835511
4.7 0.0144259347005495
4.75 0.0131334608317621
4.8 0.0127959106941604
4.9 0.014309225879104
5 0.0160961055493791
5.25 0.015011470440865
5.5 0.0135492035123544
5.75 0.0165283451026532
6 0.0160581063707361
6.25 0.0171456685466258
6.5 0.018281197484014
6.75 0.0193566876815194
7 0.0214178870706037
7.25 0.0182110989625289
7.5 0.022330414860303
7.75 0.022749224625624
8 0.0221920814629426
8.25 0.0224110804176572
8.5 0.0235573654952004
8.75 0.027857138532502
9 0.0232862346115679
9.25 0.0292898167288673
9.5 0.0299283159935595
9.75 0.0300499628358027
10 0.0259275396020168
10.25 0.0240776044534853
10.5 0.026246806534111
10.75 0.0285233638012279
11 0.0233714381744489
11.25 0.0290995925217241
11.5 0.0213478062775476
11.75 0.0278122676857255
12 0.0247163188106333
12.25 0.027798302119546
12.5 0.0312862483113342
12.75 0.0301259031250776
13 0.0232558579827585
13.25 0.0296224839913343
13.5 0.0314970233286114
13.75 0.0329204816191421
14 0.0328869330903659
14.25 0.0290969185138689
14.5 0.0380889430766197
14.75 0.0346871004307194
15 0.0397151909818061
15.25 0.034930136621532
15.5 0.0366345358955511
15.75 0.031591334033247
16 0.0357331649164292
16.25 0.0373031010374427
16.5 0.0383387976948061
16.75 0.0390400279173062
17 0.0385684526356365
17.25 0.0248329239471924
17.5 0.0440203385517327
17.75 0.0409772326786028
18 0.0315809479714391
18.25 0.0453589752916361
18.5 0.0407595131266988
18.75 0.0365482016670569
19 0.0369887224610554
19.25 0.0382249893319258
19.5 0.0434037156405457
19.75 0.0388802143281356
20 0.0420547449401799
};
\addplot[semithick, c2, mark=x]
table {%
0.1 0.00325774725419825
0.2 0.0051682661605289
0.3 0.00891845215590426
0.4 0.0103895475163771
0.5 0.0100433410955838
0.6 0.00989266136643396
0.7 0.0116483671671435
0.8 0.0167639855716172
0.9 0.0258307455393189
1 0.0246291042582854
1.1 0.0174666050810388
1.2 0.023537508707893
1.3 0.00864221209686722
1.4 0.00304491626031955
1.5 0.0412459383114885
1.6 0.0175805316619911
1.7 0.0245903061157896
1.8 0.0598974490846307
1.9 0.113166446526209
2 0.0248998124006456
2.1 0.0223182821709811
2.2 0.0793576133560194
2.3 0.0833284197930119
2.4 0.876178016832153
2.5 0.326278710058086
2.6 0.270170498477388
2.7 0.0352837940278877
2.8 0.630208554087896
2.9 0.925266761531228
3 0.936095754561699
3.1 0.483657183284507
3.2 0.94035746773964
3.3 0.618183014565028
3.4 0.952692947629552
3.5 0.956110557411349
3.6 0.902196051409384
3.7 0.959106316294752
3.8 0.930326628694655
3.9 0.964858189429364
4 0.963142704910758
4.1 0.967659305948557
4.2 0.966845046875682
4.3 0.963883464634443
4.4 0.967810069037799
4.5 0.958530634791938
4.6 0.955606588669223
4.7 0.961950088615362
4.8 0.96557903000229
4.9 0.972139226438666
5.2 0.928338617941557
5.4 0.965842082485344
5.6 0.956011595467647
5.8 0.96639745850781
6 0.965756753708355
6.2 0.963859887906175
6.4 0.960841934650153
6.6 0.959771394788419
6.8 0.961378776720901
7 0.956594545323675
7.2 0.959442523659158
7.4 0.959342678441209
7.6 0.956366989131052
7.8 0.95893140276327
8 0.958393128635508
8.2 0.983269928813282
8.4 0.954378577471574
8.6 0.957063689675917
8.8 0.958763480498487
9 0.955954736437925
9.2 0.954208626342281
9.4 0.952578894816744
9.6 0.954737004937163
9.8 0.954519835101379
10 0.955815773409004
10.2 0.954833487084432
10.4 0.951605002609503
10.6 0.954557399009829
10.8 0.951125552850955
11 0.95440998846663
11.2 0.954206278944022
11.4 0.954645672241714
11.6 0.954414100182241
11.8 0.955179135266255
12 0.959496026860607
12.2 0.984896683502396
12.4 0.957551876204028
12.6 0.951863721898869
12.8 0.958245177018298
13 0.957056700644747
13.2 0.954723703702279
13.4 0.955006753576734
13.6 0.882186732535935
13.8 0.957471544667704
14 0.95238469435833
14.2 0.954884177691863
14.4 0.967784844974229
14.6 0.955851495240304
14.8 0.958684815068972
15 0.959309095321126
15.2 0.958425205132929
15.4 0.958700470033021
15.6 0.957229475829152
15.8 0.963852684665825
16 0.96119672016274
16.2 0.960131224811762
16.4 0.962554089893405
16.6 0.965919166957151
16.8 0.960277828781766
17 0.967543663310899
17.2 0.966358428010924
17.4 0.961698437608904
17.6 0.969514559146984
17.8 0.971594214481878
18 0.972760990273289
18.2 0.97148563946147
18.4 0.972303185817768
18.6 0.976631196230274
18.8 0.974397178983568
19 0.981318749117346
19.2 0.984004695552863
19.4 0.979552172282438
19.6 0.978123487414655
19.8 0.983008357120989
};
\addplot [semithick, c3]
table {%
0.1 0.08
0.2 0.08
0.3 0.08
0.4 0.08
0.5 0.08
0.6 0.08
0.7 0.08
0.8 0.08
0.9 0.08
1 0.08
1.1 0.08
1.2 0.08
1.3 0.08
1.4 0.08
1.5 0.08
1.6 0.08
1.7 0.08
1.8 0.08
1.9 0.08
2 0.08
2.1 0.08
2.2 0.08
2.3 0.08
2.4 0.08
2.5 0.08
2.6 0.08
2.7 0.08
2.8 0.08
2.9 0.08
3 0.08
3.1 0.08
3.2 0.08
3.3 0.08
3.4 0.08
3.5 0.08
3.6 0.08
3.7 0.08
3.8 0.08
3.9 0.08
4 0.08
4.1 0.08
4.2 0.08
4.3 0.08
4.4 0.08
4.5 0.08
4.6 0.08
4.7 0.08
4.8 0.08
4.9 0.08
5.2 0.08
5.4 0.08
5.6 0.08
5.8 0.08
6 0.08
6.2 0.08
6.4 0.08
6.6 0.08
6.8 0.08
7 0.08
7.2 0.08
7.4 0.08
7.6 0.08
7.8 0.08
8 0.08
8.2 0.08
8.4 0.08
8.6 0.08
8.8 0.08
9 0.08
9.2 0.08
9.4 0.08
9.6 0.08
9.8 0.08
10 0.08
10.2 0.08
10.4 0.08
10.6 0.08
10.8 0.08
11 0.08
11.2 0.08
11.4 0.08
11.6 0.08
11.8 0.08
12 0.08
12.2 0.08
12.4 0.08
12.6 0.08
12.8 0.08
13 0.08
13.2 0.08
13.4 0.08
13.6 0.08
13.8 0.08
14 0.08
14.2 0.08
14.4 0.08
14.6 0.08
14.8 0.08
15 0.08
15.2 0.08
15.4 0.08
15.6 0.08
15.8 0.08
16 0.08
16.2 0.08
16.4 0.08
16.6 0.08
16.8 0.08
17 0.08
17.2 0.08
17.4 0.08
17.6 0.08
17.8 0.08
18 0.08
18.2 0.08
18.4 0.08
18.6 0.08
18.8 0.08
19 0.08
19.2 0.08
19.4 0.08
19.6 0.08
19.8 0.08
};
\end{axis}

\end{tikzpicture}

%% file: mainexpvstauprecise4q.tex
\begin{tikzpicture}

\begin{axis}[
tick align=outside,
tick pos=left,
x grid style={darkgray176},
xticklabel=\empty,
xmin=-0.895, xmax=20.995,
xtick style={color=black},
y grid style={darkgray176},
ylabel={$\braket{O}$},
ymin=0.951796339566437, ymax=1.00227898809577,
ytick style={color=black},
mark size=1,
width =\linewidth, 
height = 0.65*\linewidth
]
\addplot [
  draw=c4,
  fill=c4,
  mark=*,
  only marks
]
table{%
x  y
0.1 0.999906984554111
0.2 0.999888628243547
0.3 0.999920898218107
0.4 0.999734950909632
0.6 0.999534728159085
0.7 0.999316507497087
0.8 0.999091749496072
0.9 0.998689559673756
1.1 0.998050557299041
1.2 0.997781330018956
2.3 0.992256281022811
1.4 0.996885529559819
1.6 0.996138036373307
1.7 0.995647277886036
1.8 0.995041699700566
1.9 0.994619350121401
2.1 0.993449426104197
2.2 0.993005065372147
2.3 0.992256281022811
2.4 0.991571540580027
2.6 0.990412872171851
2.7 0.989802498359285
2.8 0.988932950298551
2.9 0.988043648134844
3.1 0.986861203878498
3.2 0.986164156641771
3.3 0.985509542679343
3.4 0.984757023662438
3.6 0.983292338288543
3.7 0.982433750662943
3.8 0.981728069888375
3.9 0.980898010259301
4.1 0.979327147552488
4.2 0.978748975679891
4.3 0.977628891229139
4.4 0.976738142440546
4.6 0.97554634572984
4.7 0.974559431308724
4.8 0.974082499066394
4.9 0.973043212709274
5.25 0.970597184953824
5.5 0.968654915178003
5.75 0.967388700840801
6 0.965437224395334
6.25 0.9644794321992
6.5 0.962984318780432
6.75 0.961762694003782
7 0.960808874030845
7.25 0.959597968119803
7.5 0.958954039761167
7.75 0.958066742214087
8 0.957059050271508
8.25 0.956264230667805
8.5 0.95593222968931
8.75 0.955839759177786
9 0.95542797698981
9.25 0.955189768627105
9.5 0.954746807083104
9.75 0.954292494200696
10 0.954306730486332
10.25 0.954898992364336
10.5 0.954367821094968
10.75 0.954227088537122
11 0.954248334216879
11.25 0.95409100540868
11.5 0.954515285666599
11.75 0.95474200933849
12 0.954617051709271
12.25 0.954922580514357
12.5 0.954744528893434
12.75 0.954856088129795
13 0.955078791212621
13.25 0.95511136884002
13.5 0.95509979735963
13.75 0.955604828551707
14 0.955969599887685
14.25 0.956288690371654
14.5 0.956715979997326
14.75 0.95708868773126
15.25 0.958405413737228
15.5 0.959125026104318
15.75 0.960215645721586
16 0.960339797340811
16.25 0.961360538384569
16.5 0.962640192365879
16.75 0.963598895013855
17 0.965012969630341
17.25 0.96613717388442
17.5 0.967941528589066
17.75 0.969346900682033
18 0.97107393009856
18.25 0.97277584851466
18.5 0.97422126933908
18.75 0.976465148560798
19 0.977998377032243
19.25 0.980020912989682
19.5 0.981837432655336
19.75 0.984070226549029
20 0.985476048854293
};
\addplot [semithick, c3]
table {%
0.1 0.999984322253531
0.2 0.999937316048458
0.3 0.999859062423781
0.4 0.999749696232385
0.6 0.999438432598445
0.7 0.999237070443749
0.8 0.999005665038244
0.9 0.998744613040839
1.1 0.998135405095548
1.2 0.997788288393915
2.3 0.992310398769103
1.4 0.99701197866293
1.6 0.996130685926211
1.7 0.995652498276588
1.8 0.995150337106013
1.9 0.994625039370295
2.1 0.993508554260528
2.2 0.992919206903649
2.3 0.992310398769103
2.4 0.991683120188853
2.6 0.99037723118357
2.7 0.989700712585584
2.8 0.989009902435811
2.9 0.98830588796137
3.1 0.986862653167096
3.2 0.986125658106365
3.3 0.985379904277848
3.4 0.984626514842655
3.6 0.983101318128335
3.7 0.982331746159953
3.8 0.981559004369023
3.9 0.9807841904058
4.1 0.979232673116355
4.2 0.978458094852616
4.3 0.977685690144076
4.4 0.976916473222
4.6 0.975391542446561
4.7 0.974637734042424
4.8 0.973890920650495
4.9 0.973151982480018
5.25 0.97063965615509
5.5 0.96892881138133
5.75 0.967299502778876
6 0.965760428420085
6.25 0.964318652091784
6.5 0.962979569793162
6.75 0.961746905975175
7 0.960622739223592
7.25 0.959607556551009
7.5 0.958700334940377
7.75 0.957898648283958
8 0.957198797396691
8.25 0.956595960360986
8.5 0.956084360089292
8.75 0.95565744567886
9 0.955308083886447
9.25 0.955028756874138
9.5 0.954811762275025
9.75 0.954649411601136
10 0.954534223066697
10.25 0.954459105026463
10.5 0.954417526429264
10.75 0.954403670956914
11 0.954412571853028
11.25 0.954440224838352
11.5 0.954483676950941
11.75 0.954541089632261
12 0.954611774894168
12.25 0.954696203936569
12.5 0.954795988130704
12.75 0.954913832827626
13 0.955053464984911
13.25 0.955219536116504
13.5 0.955417502551073
13.75 0.955653485424212
14 0.955934113221046
14.25 0.956266350021176
14.5 0.956657312871512
14.75 0.957114081919665
15.25 0.958252015054501
15.5 0.958945420579272
15.75 0.959728745579925
16 0.960606049909829
16.25 0.961580277022944
16.5 0.962653117714371
16.75 0.963824894717154
17 0.965094470567682
17.25 0.966459180732235
17.5 0.967914793533262
17.75 0.969455497934343
18 0.97107391974594
18.25 0.972761166308634
18.5 0.974506899205412
18.75 0.976299434058091
19 0.978125865983579
19.25 0.979972218831163
19.5 0.981823615899892
19.75 0.983664469452122
20 0.985478686001584
};

\addlegendentry{$\braket{E_O}$ $4$-qubit QSP}
\addlegendentry{$\braket{O}_{ideal}$}
\end{axis}

\end{tikzpicture}

%% file: mainexpvstauprecise6q.tex
\begin{tikzpicture}

\begin{axis}[
tick align=outside,
tick pos=left,
x grid style={darkgray176},
xmin=-0.895, xmax=20.995,
xticklabel=\empty,
xtick style={color=black},
y grid style={darkgray176},
ylabel={$\braket{O}$},
ymin=0.967644089299371, ymax=1.0015243340441,
ytick style={color=black},
mark size=1,
width =\linewidth, 
height = 0.65*\linewidth
]
\addplot [
  draw=c5,
  fill=c5,
  mark=star,
  only marks
]
table{%
x  y
0.1 0.999962076487988
0.2 0.999863592249848
0.3 0.999788705396519
0.4 0.999741873104754
0.6 0.999278100422971
0.7 0.999250243437232
0.8 0.999099593537475
0.9 0.998867355958858
1.1 0.998193012733195
1.2 0.997868471629905
2.3 0.992473471927722
1.4 0.996953223897276
1.6 0.996196708812997
1.7 0.995528590683209
1.8 0.995282675620164
1.9 0.994794192250143
2.1 0.993401069486224
2.2 0.993017035189914
2.3 0.992473471927722
2.4 0.992043999188969
2.6 0.990488223860911
2.7 0.99007713474514
2.8 0.989417352609459
2.9 0.988394651590196
3.1 0.987426190652609
3.2 0.986765890675691
3.3 0.986313459466845
3.4 0.98541870994921
3.6 0.984105882947824
3.7 0.983310917903949
3.8 0.982735787370024
3.9 0.982401373662282
4.1 0.980712217028389
4.2 0.980149160481884
4.3 0.979410231257224
4.4 0.979041079735063
4.6 0.977710823652212
4.7 0.97704975399449
4.8 0.976750302218886
4.9 0.976160658213134
5.25 0.974259627283337
5.5 0.972923985808332
5.75 0.972231146971043
6 0.971279518708752
6.25 0.970444740312919
6.5 0.970101101149249
6.75 0.969441076308569
7 0.969777560662951
7.25 0.969297028951486
7.5 0.969193707017764
7.75 0.969184100424131
8 0.969351634669721
8.25 0.969853732349908
8.5 0.970470466706857
8.75 0.970417971501847
9 0.97093823542344
9.25 0.97161703756828
9.5 0.972035050551009
9.75 0.972338555660666
10 0.973100612643245
10.25 0.973855415457333
10.5 0.97417624025873
10.75 0.974755980031186
11 0.974539921085344
11.25 0.975508531674456
11.5 0.975536048445565
11.75 0.975258903446641
12 0.975617420367273
12.25 0.975355373972787
12.5 0.975001994695722
12.75 0.97533564341775
13 0.975077753679343
13.25 0.974485590301741
13.5 0.974760330229684
13.75 0.973754397567578
14 0.973502775445608
14.25 0.973328477525204
14.5 0.972779571850066
14.75 0.971696244109809
15.25 0.971532109670732
15.5 0.970786957570695
15.75 0.9702209191413
16 0.970708391826775
16.25 0.970435714063014
16.5 0.970347408102304
16.75 0.970299085353946
17 0.970730523750186
17.25 0.970824477967576
17.5 0.971364778820502
17.75 0.971919347481197
18 0.972434548979634
18.25 0.972991228546822
18.5 0.974349200426843
18.75 0.975001199160577
19 0.976179630594889
19.25 0.977526333502331
19.5 0.978659528017677
19.75 0.980276976992077
20 0.981560304628915
};
\addplot [semithick, c3]
table {%
0.1 0.999984322919342
0.2 0.99993732669744
0.3 0.999859116302136
0.4 0.999749866372792
0.6 0.999439291886414
0.7 0.999238659919064
0.8 0.999008371774317
0.9 0.998748939939292
1.1 0.998145014745892
1.2 0.997801861246161
2.3 0.992485329476364
1.4 0.997036968575635
1.6 0.996173013367103
1.7 0.995706229386173
1.8 0.995217588845861
1.9 0.994708159549164
2.1 0.993631414467699
2.2 0.993066434042418
2.3 0.992485329476364
2.4 0.991889350032809
2.6 0.99065789448704
2.7 0.990025037808803
2.8 0.989382538296018
2.9 0.988731745713289
3.1 0.98741072571382
3.2 0.986743232604209
3.3 0.986072908002456
3.4 0.985401115361374
3.6 0.984058537451234
3.7 0.983390426442256
3.8 0.982726189134995
3.9 0.982067116018345
4.1 0.980769498407818
4.2 0.98013339959682
4.3 0.979507350292508
4.4 0.978892487556741
4.6 0.9777006704252
4.7 0.977125782736866
4.8 0.976566210131885
4.9 0.976022867569172
5.25 0.974260964520918
5.5 0.97314818804521
5.75 0.97216712668264
6 0.971324223702447
6.25 0.970623467115943
6.5 0.970066372673323
6.75 0.969652010964895
7 0.969377077764297
7.25 0.969236005949882
7.5 0.969221116576396
7.75 0.969322805958359
8 0.969529764982402
8.25 0.969829226301034
8.5 0.970207234584443
8.75 0.970648934628482
9 0.971138871841153
9.25 0.971661299460405
9.5 0.972200486793834
9.75 0.972741022814958
10 0.973268109598261
10.25 0.973767840321205
10.5 0.974227456899752
10.75 0.974635582746779
11 0.974982426641316
11.25 0.975259954260934
11.5 0.975462024549322
11.75 0.975584488754863
12 0.975625250672105
12.25 0.97558428733458
12.5 0.975463630131892
12.75 0.975267307044289
13 0.975001247391448
13.25 0.974673151166838
13.5 0.974292325662917
13.75 0.973869492674011
14 0.973416570082421
14.25 0.972946432079359
14.5 0.972472652636873
14.75 0.97200923712309
15.25 0.971170023706187
15.5 0.97082191403391
15.75 0.970539006749233
16 0.97033338054914
16.25 0.970215970700178
16.5 0.970196357537369
16.75 0.970282580616649
17 0.970480981649119
17.25 0.970796078762293
17.5 0.971230474010434
17.75 0.971784795407167
18 0.972457674093119
18.25 0.973245756593622
18.5 0.974143751480229
18.75 0.975144509137765
19 0.97623913276757
19.25 0.977417118237557
19.5 0.978666519929177
19.75 0.979974139337045
20 0.981325732853615
};

\addlegendentry{$\braket{E_O}$ $6$-qubit QSP}
\addlegendentry{$\braket{O}_{ideal}$}
\end{axis}

\end{tikzpicture}

%% file: biasvstau_precise.tex
\begin{tikzpicture}

\begin{axis}[
legend style={at={(axis cs:0.5,0.0005)},anchor=south west},
tick align=outside,
tick pos=left,
x grid style={darkgray176},
xlabel={simulation time $\tau$},
xmin=-0.895, xmax=20.995,
xtick style={color=black},
y grid style={darkgray176},
ylabel={ $\left|b_{E_O}\right|$},
ytick style={color=black},
mark size=1,
width =\linewidth, 
height = 0.75*\linewidth
]
\addplot [semithick, c4, mark=*]
table {%
0.1 7.7337699419866e-05
0.2 4.86878049106299e-05
0.3 6.18357943255043e-05
0.4 1.47453227536953e-05
0.6 9.62955606399474e-05
0.7 7.94370533374611e-05
0.8 8.6084457827651e-05
0.9 5.50533670822251e-05
1.1 8.48477965073036e-05
1.2 6.95837495912066e-06
2.3 5.41177462916043e-05
1.4 0.000126449103111437
1.6 7.35044709598842e-06
1.7 5.22039055206314e-06
1.8 0.000108637405447642
1.9 5.68924889410294e-06
2.1 5.91281563314672e-05
2.2 8.58584684981789e-05
2.3 5.41177462916043e-05
2.4 0.000111579608826573
2.6 3.5640988280794e-05
2.7 0.000101785773701346
2.8 7.69521372594451e-05
2.9 0.000262239826525179
3.1 1.44928859791005e-06
3.2 3.84985354062728e-05
3.3 0.00012963840149427
3.4 0.000130508819782493
3.6 0.000191020160207556
3.7 0.00010200450299036
3.8 0.000169065519352363
3.9 0.000113819853500785
4.1 9.44744361329741e-05
4.2 0.000290880827275464
4.3 5.67989149370396e-05
4.4 0.000178330781453684
4.6 0.000154803283279104
4.7 7.83027336999087e-05
4.8 0.000191578415898785
4.9 0.000108769770744144
5.25 4.24712012661033e-05
5.5 0.000273896203327517
5.75 8.91980619253108e-05
6 0.000323204024750612
6.25 0.000160780107416381
6.5 4.7489872697648e-06
6.75 1.57880286069112e-05
7 0.000186134807253491
7.25 9.58843120568709e-06
7.5 0.000253704820790346
7.75 0.000168093930129287
8 0.000139747125182499
8.25 0.000331729693180738
8.5 0.000152130399981631
8.75 0.000182313498926168
9 0.000119893103363022
9.25 0.000161011752967455
9.5 6.49551919201796e-05
9.75 0.000356917400440304
10 0.000227492580364919
10.25 0.000439887337873257
10.5 4.97053342967124e-05
10.75 0.000176582419792015
11 0.000164237636149478
11.25 0.000349219429672809
11.5 3.16087156581046e-05
11.75 0.000200919706229485
12 5.27681510276157e-06
12.25 0.000226376577788234
12.5 5.14592372696798e-05
12.75 5.77446978302643e-05
13 2.53262277108979e-05
13.25 0.000108167276484106
13.5 0.000317705191442696
13.75 4.8656872505326e-05
14 3.54866666387732e-05
14.25 2.23403504782116e-05
14.5 5.86671258142601e-05
14.75 2.53941884048592e-05
15.25 0.000153398682726902
15.5 0.000179605525045257
15.75 0.000486900141661306
16 0.000266252569018799
16.25 0.000219738638374434
16.5 1.29253484915104e-05
16.75 0.000225999703298729
17 8.15009373416054e-05
17.25 0.00032200684781547
17.5 2.67350558046386e-05
17.75 0.000108597252309495
18 1.03526203165671e-08
18.25 1.46822060261842e-05
18.5 0.000285629866332182
18.75 0.000165714502706749
19 0.000127488951336296
19.25 4.86941585197131e-05
19.5 1.3816755444207e-05
19.75 0.000405757096907156
20 2.63714729165088e-06
};
\addplot [semithick, c5, mark=star]
table {%
0.1 2.22464313538451e-05
0.2 7.37344475914004e-05
0.3 7.04109056174351e-05
0.4 7.99326803802014e-06
0.6 0.000161191463442956
0.7 1.15835181682122e-05
0.8 9.12217631584555e-05
0.9 0.000118416019565637
1.1 4.79979873027059e-05
1.2 6.66103837445586e-05
2.3 1.18575486423023e-05
1.4 8.37446783591433e-05
1.6 2.36954458940897e-05
1.7 0.0001776387029645
1.8 6.50867743031691e-05
1.9 8.60327009790662e-05
2.1 0.000230344981475006
2.2 4.9398852503546e-05
2.3 1.18575486423023e-05
2.4 0.000154649156159925
2.6 0.000169670626129048
2.7 5.20969363368984e-05
2.8 3.48143134403234e-05
2.9 0.000337094123093795
3.1 1.54649387895667e-05
3.2 2.26580714826063e-05
3.3 0.000240551464389172
3.4 1.75945878363448e-05
3.6 4.73454965895792e-05
3.7 7.95085383074046e-05
3.8 9.59823502888479e-06
3.9 0.000334257643937685
4.1 5.72813794286198e-05
4.2 1.57608850649416e-05
4.3 9.71190352835327e-05
4.4 0.000148592178322415
4.6 1.0153227011811e-05
4.7 7.60287423752226e-05
4.8 0.000184092087001542
4.9 0.000137790643962021
5.25 1.33723758111159e-06
5.5 0.000224202236877691
5.75 6.40202884034657e-05
6 4.47049936945776e-05
6.25 0.000178726803024221
6.5 3.47284759253386e-05
6.75 0.000210934656325268
7 0.000400482898654464
7.25 6.10230016045987e-05
7.5 2.74095586316747e-05
7.75 0.000138705534227235
8 0.000178130312680769
8.25 2.45060488737181e-05
8.5 0.000263232122413304
8.75 0.000230963126635064
9 0.000200636417712508
9.25 4.42618921258076e-05
9.5 0.000165436242824946
9.75 0.00040246715429193
10 0.000167496955016277
10.25 8.75751361282262e-05
10.5 5.12166410224779e-05
10.75 0.000120397284406937
11 0.000442505555971695
11.25 0.000248577413521467
11.5 7.40238962426698e-05
11.75 0.000325585308222132
12 7.83030483197944e-06
12.25 0.000228913361792649
12.5 0.000461635436169838
12.75 6.83363734614861e-05
13 7.65062878954614e-05
13.25 0.000187560865096859
13.5 0.000468004566767211
13.75 0.000115095106432594
14 8.62053631873749e-05
14.25 0.000382045445844637
14.5 0.000306919213192924
14.75 0.000312993013281582
15.25 0.000362085964544567
15.5 3.49564632150523e-05
15.75 0.00031808760793306
16 0.00037501127763484
16.25 0.000219743362836256
16.5 0.000151050564934807
16.75 1.65047372967519e-05
17 0.000249542101067424
17.25 2.83992052833026e-05
17.5 0.000134304810067265
17.75 0.000134552074029903
18 2.31251134847898e-05
18.25 0.000254528046799818
18.5 0.000205448946613118
18.75 0.000143309977188255
19 5.95021726809142e-05
19.25 0.000109215264774476
19.5 6.99191150010936e-06
19.75 0.000302837655031585
20 0.000234571775300063
};
\addplot [semithick, c3]
table {%
0.1 0.0008
0.2 0.0008
0.3 0.0008
0.4 0.0008
0.6 0.0008
0.7 0.0008
0.8 0.0008
0.9 0.0008
1.1 0.0008
1.2 0.0008
2.3 0.0008
1.4 0.0008
1.6 0.0008
1.7 0.0008
1.8 0.0008
1.9 0.0008
2.1 0.0008
2.2 0.0008
2.3 0.0008
2.4 0.0008
2.6 0.0008
2.7 0.0008
2.8 0.0008
2.9 0.0008
3.1 0.0008
3.2 0.0008
3.3 0.0008
3.4 0.0008
3.6 0.0008
3.7 0.0008
3.8 0.0008
3.9 0.0008
4.1 0.0008
4.2 0.0008
4.3 0.0008
4.4 0.0008
4.6 0.0008
4.7 0.0008
4.8 0.0008
4.9 0.0008
5.25 0.0008
5.5 0.0008
5.75 0.0008
6 0.0008
6.25 0.0008
6.5 0.0008
6.75 0.0008
7 0.0008
7.25 0.0008
7.5 0.0008
7.75 0.0008
8 0.0008
8.25 0.0008
8.5 0.0008
8.75 0.0008
9 0.0008
9.25 0.0008
9.5 0.0008
9.75 0.0008
10 0.0008
10.25 0.0008
10.5 0.0008
10.75 0.0008
11 0.0008
11.25 0.0008
11.5 0.0008
11.75 0.0008
12 0.0008
12.25 0.0008
12.5 0.0008
12.75 0.0008
13 0.0008
13.25 0.0008
13.5 0.0008
13.75 0.0008
14 0.0008
14.25 0.0008
14.5 0.0008
14.75 0.0008
15.25 0.0008
15.5 0.0008
15.75 0.0008
16 0.0008
16.25 0.0008
16.5 0.0008
16.75 0.0008
17 0.0008
17.25 0.0008
17.5 0.0008
17.75 0.0008
18 0.0008
18.25 0.0008
18.5 0.0008
18.75 0.0008
19 0.0008
19.25 0.0008
19.5 0.0008
19.75 0.0008
20 0.0008
};

\addlegendentry{$\left|b_{E_O}\right|$, $N=4$}
\addlegendentry{$\left|b_{E_O}\right|$, $N=6$ }
\addlegendentry{$8\cdot 10^{-4}$}
\end{axis}

\end{tikzpicture}

%% file: mainexpvstaulong0.1.tex
\begin{tikzpicture}

\begin{axis}[
legend style={at={(axis cs:390,0.94)},anchor=south east},
tick align=outside,
tick pos=left,
x grid style={darkgray176},
xticklabel=\empty,
xmin=6.75, xmax=408.25,
xtick style={color=black},
y grid style={darkgray176},
ylabel={$\braket{O}$},
ytick style={color=black},
width =\linewidth, 
height = 0.65*\linewidth
]
\addplot [
  draw=c4,
  fill=c4,
  mark=*,
  mark size=1,
  only marks
]
table{%
x  y
25 0.995863323106836
50 0.988817015899282
75 0.986856743153999
100 0.991058279321998
125 0.992927692991226
150 0.990689876508455
175 0.98341098212019
200 0.984761396002753
225 0.987866276756411
250 0.984082060430621
275 0.978698977416088
300 0.967342743251377
325 0.970044146578197
350 0.98006649677807
375 0.982872843553438
390 0.98663829465227
};
\addplot [semithick, c3,  mark=o,  only marks]
table {%
25 0.996079902617241
50 0.988893924866859
75 0.986387197485224
100 0.989943145886222
125 0.992866103126651
150 0.98983905197017
175 0.984712810675883
200 0.984043334672198
225 0.986446305826655
250 0.984767491339003
275 0.97642010064289
300 0.968123225863805
325 0.969083709348857
350 0.978077917447031
375 0.982311735321907
390 0.985421979615637
};
\addplot [semithick, c3]
table {%
25 0.996079902617241
25.7314629258517 0.992356811259673
26.4629258517034 0.987820119272125
27.1943887775551 0.98288320486723
27.9258517034068 0.977965710393942
28.6573146292585 0.973443693170157
29.3887775551102 0.96960761119979
30.1202404809619 0.966634064808962
30.8517034068136 0.964575088434133
31.5831663326653 0.963366057626526
32.314629258517 0.962850384725129
33.0460921843687 0.962816595490779
33.7775551102204 0.963041529456536
34.5090180360721 0.963332582846153
35.2404809619239 0.963562231057834
35.9719438877756 0.963689451106867
36.7034068136273 0.963764863880606
37.434869739479 0.963919060728153
38.1663326653307 0.964336244436222
38.8977955911824 0.965217594130224
39.6292585170341 0.966740328259028
40.3607214428858 0.969019080505011
41.0921843687375 0.972075850528038
41.8236472945892 0.975823513858362
42.5551102204409 0.980065863890704
43.2865731462926 0.984514698881551
44.0180360721443 0.988821902325147
44.749498997996 0.992622159216078
45.4809619238477 0.995580241689098
46.2124248496994 0.99743595337185
46.9438877755511 0.998039998154778
47.6753507014028 0.997375246986989
48.4068136272545 0.995559967712715
49.1382765531062 0.992832261159299
49.8697394789579 0.989517804338691
50.6012024048096 0.985985583352244
51.3326653306613 0.98259817834475
52.064128256513 0.979664021325042
52.7955911823647 0.977398733286804
53.5270541082164 0.97590120674213
54.2585170340681 0.975147771919226
54.9899799599198 0.975004955660906
55.7214428857715 0.975258474548096
56.4529058116232 0.975653655088622
57.1843687374749 0.975940817669766
57.9158316633266 0.975918530223961
58.6472945891784 0.975468095675157
59.3787575150301 0.974574080439863
60.1102204408818 0.97332787650408
60.8416833667335 0.971913879585854
61.5731462925852 0.970580481288334
62.3046092184369 0.969600343781306
63.0360721442886 0.969226035349245
63.7675350701403 0.969647828340865
64.498997995992 0.970960188712813
65.2304609218437 0.973142242222577
65.9619238476954 0.976055445898554
66.6933867735471 0.979459107314439
67.4248496993988 0.983041653295252
68.1563126252505 0.986463072113955
68.8877755511022 0.989402137551392
69.6192384769539 0.99160118080253
70.3507014028056 0.992901471840512
71.0821643286573 0.993263685459993
71.813627254509 0.992770248016892
72.5450901803607 0.991609217640727
73.2765531062124 0.99004227449721
74.0080160320641 0.988361905225074
74.7394789579158 0.986844546167167
75.4709418837675 0.985707037544635
76.2024048096192 0.985073157563767
76.9338677354709 0.984955368814177
77.6653306613226 0.985254502509148
78.3967935871743 0.985777322612519
79.128256513026 0.986269185702982
79.8597194388778 0.986456748869077
80.5911823647295 0.986094196453734
81.3226452905812 0.98500595237153
82.0541082164329 0.983119371594916
82.7855711422846 0.980482375296627
83.5170340681363 0.977263197695013
84.248496993988 0.973732040689282
84.9799599198397 0.970227116413558
85.7114228456914 0.96710991306053
86.4428857715431 0.964716193884337
87.1743486973948 0.9633099651096
87.9058116232465 0.9630472868927
88.6372745490982 0.963955375101086
89.3687374749499 0.965930143497836
90.1002004008016 0.968752509649996
90.8316633266533 0.972120878655446
91.563126252505 0.975694700517313
92.2945891783567 0.979142289086636
93.0260521042084 0.98218548059927
93.7575150300601 0.984634299355537
94.4889779559118 0.986406482657833
95.2204408817635 0.987529206567501
95.9519038076152 0.988123227430282
96.6833667334669 0.988372437515738
97.4148296593186 0.988484083365097
98.1462925851703 0.988646273057783
98.877755511022 0.988989716861611
99.6092184368737 0.989559889610328
100.340681362725 0.99030411893344
101.072144288577 0.99107576528897
101.803607214429 0.991655022385811
102.535070140281 0.991783313483132
103.266533066132 0.991206152670253
103.997995991984 0.989717978658271
104.729458917836 0.98720205055841
105.460921843687 0.983659094467398
106.192384769539 0.979219941647114
106.923847695391 0.974139701019996
107.655310621242 0.968773738823047
108.386773547094 0.963538494152271
109.118236472946 0.958862514146762
109.849699398798 0.955134663554939
110.581162324649 0.952656978471237
111.312625250501 0.951608986422706
112.044088176353 0.952028590502316
112.775551102204 0.953812084353715
113.507014028056 0.95673293886369
114.238476953908 0.960476160364571
114.96993987976 0.964682726224649
115.701402805611 0.96899721907889
116.432865731463 0.973111506304324
117.164328657315 0.976798157097587
117.895791583166 0.979929083533194
118.627254509018 0.982477319278325
119.35871743487 0.984502513191173
120.090180360721 0.98612320392339
120.821643286573 0.987480893529964
121.553106212425 0.988702086670216
122.284569138277 0.989864665370164
123.016032064128 0.990974219501709
123.74749498998 0.99195436967822
124.478957915832 0.992652930834523
125.210420841683 0.992863281228076
125.941883767535 0.992357878149196
126.673346693387 0.990928855842956
127.404809619238 0.988429365241906
128.13627254509 0.98480898794011
128.867735470942 0.980137263060497
129.599198396794 0.974611031092045
130.330661322645 0.968543690125389
131.062124248497 0.962337213434705
131.793587174349 0.956440454584552
132.5250501002 0.951299428448476
133.256513026052 0.947306543590777
133.987975951904 0.944755958698062
134.719438877756 0.94381131393198
135.450901803607 0.944490207742095
136.182364729459 0.946667268949081
136.913827655311 0.950094927533562
137.645290581162 0.954438451052724
138.376753507014 0.959319868708724
139.108216432866 0.964364321098464
139.839679358717 0.969242273756698
140.571142284569 0.973701887765758
141.302605210421 0.977587487701047
142.034068136273 0.98084224288114
142.765531062124 0.983495561522481
143.496993987976 0.985637953636979
144.228456913828 0.987387937155059
144.959919839679 0.988856692283312
145.691382765531 0.990116449574374
146.422845691383 0.991177976465235
147.154308617234 0.991981077716967
147.885771543086 0.99239994309082
148.617234468938 0.992262761816925
149.34869739479 0.991382646963182
150.080160320641 0.989594956241667
150.811623246493 0.986794891280982
151.543086172345 0.982969025139676
152.274549098196 0.978215210224213
153.006012024048 0.972747041985086
153.7374749499 0.96688142298028
154.468937875751 0.961010398641969
155.200400801603 0.95556088844957
155.931863727455 0.950947817108442
156.663326653307 0.947527167133975
157.394789579158 0.945555485716195
158.12625250501 0.94516141114892
158.857715430862 0.946333015961238
159.589178356713 0.948922485870029
160.320641282565 0.952667214639262
161.052104208417 0.957224145925196
161.783567134269 0.962212439134329
162.51503006012 0.967258501009959
163.246492985972 0.972037230210681
163.977955911824 0.976303981206684
164.709418837675 0.979913171854866
165.440881763527 0.982821444829144
166.172344689379 0.985075577610129
166.90380761523 0.986787600097416
167.635270541082 0.988101492308946
168.366733466934 0.989157100283732
169.098196392786 0.990057311902206
169.829659318637 0.990843986396276
170.561122244489 0.991486693134163
171.292585170341 0.991886199811156
172.024048096192 0.991892193884665
172.755511022044 0.991332329127172
173.486973947896 0.990047763838847
174.218436873747 0.987929224104429
174.949899799599 0.984947473047365
175.681362725451 0.981172913484928
176.412825651303 0.976780744726741
177.144288577154 0.972040345035459
177.875751503006 0.967289989118348
178.607214428858 0.962900250728368
179.338677354709 0.959231149284446
180.070140280561 0.956589039219649
180.801603206413 0.955189298595254
181.533066132265 0.955130065330538
182.264529058116 0.956380726788721
182.995991983968 0.95878681366297
183.72745490982 0.962090664005774
184.458917835671 0.965965016175589
185.190380761523 0.970054861345948
185.921843687375 0.974021696324441
186.653306613226 0.977583949850068
187.384769539078 0.980547889137598
188.11623246493 0.982824700713455
188.847695390782 0.984431501042705
189.579158316633 0.985476472107766
190.310621242485 0.986130761189194
191.042084168337 0.986591839322155
191.773547094188 0.987044334911305
192.50501002004 0.987624717358327
193.236472945892 0.988395530667391
193.967935871743 0.989333277241966
194.699398797595 0.990331796347291
195.430861723447 0.991220450092395
196.162324649299 0.991794043256522
196.89378757515 0.991849547559199
197.625250501002 0.991223660644606
198.356713426854 0.989825145621707
199.088176352705 0.987656751106614
199.819639278557 0.984823143695311
200.551102204409 0.981523426862576
201.282565130261 0.978029145332137
202.014028056112 0.974650845131788
202.745490981964 0.971697974112399
203.476953907816 0.969437931503365
204.208416833667 0.968060266686616
204.939879759519 0.967651352877459
205.671342685371 0.968183400093298
206.402805611222 0.969519611906413
207.134268537074 0.971434912749415
207.865731462926 0.973649319529754
208.597194388778 0.975869062607046
209.328657314629 0.977829298510791
210.060120240481 0.979331927238727
210.791583166333 0.98027271596776
211.523046092184 0.980653556449761
212.254509018036 0.980577999941054
212.985971943888 0.980230849317775
213.717434869739 0.979845108282644
214.448897795591 0.979661573142466
215.180360721443 0.979887476860987
215.911823647295 0.98066068241189
216.643286573146 0.982024976738216
217.374749498998 0.983920214295195
218.10621242485 0.986188710584911
218.837675350701 0.988596775483774
219.569138276553 0.990867993824648
220.300601202405 0.992723141359733
221.032064128257 0.993920701519021
221.763527054108 0.994291930161893
222.49498997996 0.993765279730895
223.226452905812 0.992376601124637
223.957915831663 0.990263657340363
224.689378757515 0.987645809931705
225.420841683367 0.984791951130766
226.152304609218 0.981981533741475
226.88376753507 0.979464629337412
227.615230460922 0.977427142853198
228.346693386774 0.975966569726773
229.078156312625 0.975082083156048
229.809619238477 0.974680507685377
230.541082164329 0.974597210842717
231.27254509018 0.974628532244276
232.004008016032 0.974570472723339
232.735470941884 0.974257312530409
233.466933867735 0.973593806744498
234.198396793587 0.972575629485414
234.929859719439 0.971294635810523
235.661322645291 0.969927959688129
236.392785571142 0.968712555898673
237.124248496994 0.967909096654596
237.855711422846 0.967760785279195
238.587174348697 0.968453405912996
239.318637274549 0.970082700891426
240.050100200401 0.972634024717881
240.781563126252 0.975977368083015
241.513026052104 0.979878572202928
242.244488977956 0.984025200766933
242.975951903808 0.98806343645724
243.707414829659 0.99164080732338
244.438877755511 0.994448733581941
245.170340681363 0.996258926957643
245.901803607214 0.996948570631565
246.633266533066 0.996510842569922
247.364729458918 0.995049497506785
248.09619238477 0.992758588795692
248.827655310621 0.989890639654575
249.559118236473 0.986718315847449
250.290581162325 0.983495619965807
251.022044088176 0.980424645847541
251.753507014028 0.977632977146284
252.48496993988 0.975165029173726
253.216432865731 0.972988308175258
253.947895791583 0.971013086911758
254.679358717435 0.969121788931543
255.410821643287 0.967202807317569
256.142284569138 0.96518281196726
256.87374749499 0.963051917318799
257.605210420842 0.960877312772566
258.336673346693 0.958802873756777
259.068136272545 0.957034543554133
259.799599198397 0.955813536335435
260.531062124248 0.955381315263726
261.2625250501 0.955941575308123
261.993987975952 0.957624944634834
262.725450901804 0.960461766318512
263.456913827655 0.964367202894247
264.188376753507 0.969141186066927
264.919839679359 0.974483651345707
265.65130260521 0.980023335360614
266.382765531062 0.985356468251546
267.114228456914 0.990090240475672
267.845691382765 0.993885182459548
268.577154308617 0.996490696411884
269.308617234469 0.99776893447654
270.040080160321 0.9977039075397
270.771543086172 0.99639489209027
271.503006012024 0.994035543624217
272.234468937876 0.990882248992719
272.965931863727 0.987216806738654
273.697394789579 0.983309254152466
274.428857715431 0.97938644388007
275.160320641283 0.975610857867097
275.891783567134 0.972072334903038
276.623246492986 0.968793199037598
277.354709418838 0.965745083177264
278.086172344689 0.962873903631442
278.817635270541 0.960128238392758
279.549098196393 0.957485955845632
280.280561122244 0.954974356459833
281.012024048096 0.952680225726523
281.743486973948 0.950747851477986
282.4749498998 0.949364972693985
283.206412825651 0.948738520978431
283.937875751503 0.949063628006573
284.669338677355 0.950490485273871
285.400801603206 0.953094104940372
286.132264529058 0.956851771392673
286.86372745491 0.961632011396858
287.595190380761 0.96719735908201
288.326653306613 0.97322125110749
289.058116232465 0.979317327413492
289.789579158317 0.985077540807161
290.521042084168 0.99011409244216
291.25250501002 0.994099547953338
291.983967935872 0.996799681053342
292.715430861723 0.998094629085238
293.446893787575 0.99798567390961
294.178356713427 0.99658710392594
294.909819639279 0.994104816791069
295.64128256513 0.990805225804511
296.372745490982 0.986979331932535
297.104208416834 0.982907326034453
297.835671342685 0.978828742207332
298.567134268537 0.974922086358091
299.298597194389 0.971296223675969
300.03006012024 0.967993905371421
300.761523046092 0.965005948548381
301.492985971944 0.962293024021347
302.224448897796 0.959810961363318
302.955911823647 0.95753507024457
303.687374749499 0.955479234346493
304.418837675351 0.953706406473911
305.150300601202 0.952328495295289
305.881763527054 0.95149530092158
306.613226452906 0.951373902413143
307.344689378757 0.952121479906287
308.076152304609 0.953855730271419
308.807615230461 0.956627612272589
309.539078156313 0.96040101511082
310.270541082164 0.96504306770415
311.002004008016 0.970327299383
311.733466933868 0.975949943678977
312.464929859719 0.981557647092377
313.196392785571 0.986783039141919
313.927855711423 0.991283345728079
314.659318637275 0.994776704852216
315.390781563126 0.997071161557127
316.122244488978 0.998082419597861
316.85370741483 0.99783811839505
317.585170340681 0.996468398119527
318.316633266533 0.994184487774743
319.048096192385 0.991248696594477
319.779559118236 0.987940274887971
320.511022044088 0.984522006021463
321.24248496994 0.981212077538636
321.973947895792 0.978164838754825
322.705410821643 0.97546264447661
323.436873747495 0.973119318362561
324.168336673347 0.971094073466085
324.899799599198 0.96931322545569
325.63126252505 0.96769592363127
326.362725450902 0.966179558811305
327.094188376753 0.964740574949896
327.825651302605 0.963407123053536
328.557114228457 0.962261271503672
329.288577154309 0.961430155798457
330.02004008016 0.961067267464284
330.751503006012 0.961326758289263
331.482965931864 0.962334887951144
332.214428857715 0.964163343169787
332.945891783567 0.966808981388582
333.677354709419 0.970183612329401
334.408817635271 0.974115874983088
335.140280561122 0.978365356915077
335.871743486974 0.982647162181294
336.603206412826 0.986663490222584
337.334669338677 0.990137707784997
338.066132264529 0.992846039989597
338.797595190381 0.994642407147164
339.529058116232 0.995472998005625
340.260521042084 0.99537870701314
340.991983967936 0.994485324654783
341.723446893788 0.992983094363571
342.454909819639 0.991098700258013
343.186372745491 0.989063742681145
343.917835671343 0.987084176633479
344.649298597194 0.985314988732843
345.380761523046 0.983843601695863
346.112224448898 0.982684221486123
346.843687374749 0.98178374358835
347.575150300601 0.981038123626341
348.306613226453 0.98031653536523
349.038076152305 0.979489427071714
349.769539078156 0.978455948438564
350.501002004008 0.977166281429519
351.23246492986 0.975635188233068
351.963927855711 0.973944484443479
352.695390781563 0.972233938626718
353.426853707415 0.970681993876246
354.158316633267 0.969479379551742
354.889779559118 0.968799843888054
355.62124248497 0.96877269343691
356.352705410822 0.969461504806177
357.084168336673 0.97085234823981
357.815631262525 0.97285332210806
358.547094188377 0.975305412253779
359.278557114228 0.978002955583665
360.01002004008 0.980720570718909
360.741482965932 0.98324251809426
361.472945891784 0.985390173281593
362.204408817635 0.987043649529493
362.935871743487 0.988154512039834
363.667334669339 0.988747845330648
364.39879759519 0.988913481446978
365.130260521042 0.988787765084088
365.861723446894 0.988528615007539
366.593186372745 0.988287650679732
367.324649298597 0.988183638367022
368.056112224449 0.988281381868966
368.787575150301 0.988579429052618
369.519038076152 0.989008669239291
370.250501002004 0.989442233477508
370.981963927856 0.989715329568252
371.713426853707 0.989652033968867
372.444889779559 0.98909489890398
373.176352705411 0.987932725205334
373.907815631263 0.986122099737057
374.639278557114 0.983699266981135
375.370741482966 0.980780433213187
376.102204408818 0.97755042499132
376.833667334669 0.974241429998756
377.565130260521 0.971105038780642
378.296593186373 0.968381749180002
379.028056112224 0.966272364895709
379.759519038076 0.964915307818644
380.490981963928 0.964372871285968
381.22244488978 0.964628048642223
381.953907815631 0.96559200137442
382.685370741483 0.96712071160893
383.416833667335 0.969038097983622
384.148296593186 0.971162019731429
384.879759519038 0.973329252629522
385.61122244489 0.975415731719686
386.342685370741 0.977349094396594
387.074148296593 0.979111733829688
387.805611222445 0.980734035720038
388.537074148297 0.982279019614823
389.268537074148 0.983821008570519
390 0.985421979615637
};

\addlegendentry{$\braket{E_o}$, $p=0.0001$}
\addlegendentry{$\braket{O}_{ideal}$}
\end{axis}

\end{tikzpicture}

%% file: mainexpvstaulong1.0.tex
\begin{tikzpicture}

\begin{axis}[
legend style={at={(axis cs:390,1.02)},anchor=north east},
tick align=outside,
tick pos=left,
x grid style={darkgray176},
xmin=6.75, xmax=408.25,
xtick style={color=black},
y grid style={darkgray176},
ylabel={$\braket{O}$},
xticklabel=\empty,
ytick style={color=black},
width =\linewidth, 
height = 0.65*\linewidth
]
\addplot [
  draw=c5,
  fill=c5,
  mark=star,
  only marks
]
table{%
x  y
25 0.99556218949386
50 0.993761338244986
75 0.983073392169312
100 1.00083551150125
125 1.00333788694662
150 0.99593281285961
175 0.984266277748168
200 0.97892140352941
225 0.992482276479411
250 0.999727816397137
275 0.970910792019401
300 0.995170926345506
325 0.96379216644353
350 0.98434422111076
375 0.980780044717271
390 1.01360480546653
};
\addplot [semithick, c3, mark=o,  only marks]
table {%
25 0.996079902617241
50 0.988893924866859
75 0.986387197485224
100 0.989943145886222
125 0.992866103126651
150 0.98983905197017
175 0.984712810675883
200 0.984043334672198
225 0.986446305826655
250 0.984767491339003
275 0.97642010064289
300 0.968123225863805
325 0.969083709348857
350 0.978077917447031
375 0.982311735321907
390 0.985421979615637
};

\addplot [semithick, c3]
table {%
25 0.996079902617241
25.7314629258517 0.992356811259673
26.4629258517034 0.987820119272125
27.1943887775551 0.98288320486723
27.9258517034068 0.977965710393942
28.6573146292585 0.973443693170157
29.3887775551102 0.96960761119979
30.1202404809619 0.966634064808962
30.8517034068136 0.964575088434133
31.5831663326653 0.963366057626526
32.314629258517 0.962850384725129
33.0460921843687 0.962816595490779
33.7775551102204 0.963041529456536
34.5090180360721 0.963332582846153
35.2404809619239 0.963562231057834
35.9719438877756 0.963689451106867
36.7034068136273 0.963764863880606
37.434869739479 0.963919060728153
38.1663326653307 0.964336244436222
38.8977955911824 0.965217594130224
39.6292585170341 0.966740328259028
40.3607214428858 0.969019080505011
41.0921843687375 0.972075850528038
41.8236472945892 0.975823513858362
42.5551102204409 0.980065863890704
43.2865731462926 0.984514698881551
44.0180360721443 0.988821902325147
44.749498997996 0.992622159216078
45.4809619238477 0.995580241689098
46.2124248496994 0.99743595337185
46.9438877755511 0.998039998154778
47.6753507014028 0.997375246986989
48.4068136272545 0.995559967712715
49.1382765531062 0.992832261159299
49.8697394789579 0.989517804338691
50.6012024048096 0.985985583352244
51.3326653306613 0.98259817834475
52.064128256513 0.979664021325042
52.7955911823647 0.977398733286804
53.5270541082164 0.97590120674213
54.2585170340681 0.975147771919226
54.9899799599198 0.975004955660906
55.7214428857715 0.975258474548096
56.4529058116232 0.975653655088622
57.1843687374749 0.975940817669766
57.9158316633266 0.975918530223961
58.6472945891784 0.975468095675157
59.3787575150301 0.974574080439863
60.1102204408818 0.97332787650408
60.8416833667335 0.971913879585854
61.5731462925852 0.970580481288334
62.3046092184369 0.969600343781306
63.0360721442886 0.969226035349245
63.7675350701403 0.969647828340865
64.498997995992 0.970960188712813
65.2304609218437 0.973142242222577
65.9619238476954 0.976055445898554
66.6933867735471 0.979459107314439
67.4248496993988 0.983041653295252
68.1563126252505 0.986463072113955
68.8877755511022 0.989402137551392
69.6192384769539 0.99160118080253
70.3507014028056 0.992901471840512
71.0821643286573 0.993263685459993
71.813627254509 0.992770248016892
72.5450901803607 0.991609217640727
73.2765531062124 0.99004227449721
74.0080160320641 0.988361905225074
74.7394789579158 0.986844546167167
75.4709418837675 0.985707037544635
76.2024048096192 0.985073157563767
76.9338677354709 0.984955368814177
77.6653306613226 0.985254502509148
78.3967935871743 0.985777322612519
79.128256513026 0.986269185702982
79.8597194388778 0.986456748869077
80.5911823647295 0.986094196453734
81.3226452905812 0.98500595237153
82.0541082164329 0.983119371594916
82.7855711422846 0.980482375296627
83.5170340681363 0.977263197695013
84.248496993988 0.973732040689282
84.9799599198397 0.970227116413558
85.7114228456914 0.96710991306053
86.4428857715431 0.964716193884337
87.1743486973948 0.9633099651096
87.9058116232465 0.9630472868927
88.6372745490982 0.963955375101086
89.3687374749499 0.965930143497836
90.1002004008016 0.968752509649996
90.8316633266533 0.972120878655446
91.563126252505 0.975694700517313
92.2945891783567 0.979142289086636
93.0260521042084 0.98218548059927
93.7575150300601 0.984634299355537
94.4889779559118 0.986406482657833
95.2204408817635 0.987529206567501
95.9519038076152 0.988123227430282
96.6833667334669 0.988372437515738
97.4148296593186 0.988484083365097
98.1462925851703 0.988646273057783
98.877755511022 0.988989716861611
99.6092184368737 0.989559889610328
100.340681362725 0.99030411893344
101.072144288577 0.99107576528897
101.803607214429 0.991655022385811
102.535070140281 0.991783313483132
103.266533066132 0.991206152670253
103.997995991984 0.989717978658271
104.729458917836 0.98720205055841
105.460921843687 0.983659094467398
106.192384769539 0.979219941647114
106.923847695391 0.974139701019996
107.655310621242 0.968773738823047
108.386773547094 0.963538494152271
109.118236472946 0.958862514146762
109.849699398798 0.955134663554939
110.581162324649 0.952656978471237
111.312625250501 0.951608986422706
112.044088176353 0.952028590502316
112.775551102204 0.953812084353715
113.507014028056 0.95673293886369
114.238476953908 0.960476160364571
114.96993987976 0.964682726224649
115.701402805611 0.96899721907889
116.432865731463 0.973111506304324
117.164328657315 0.976798157097587
117.895791583166 0.979929083533194
118.627254509018 0.982477319278325
119.35871743487 0.984502513191173
120.090180360721 0.98612320392339
120.821643286573 0.987480893529964
121.553106212425 0.988702086670216
122.284569138277 0.989864665370164
123.016032064128 0.990974219501709
123.74749498998 0.99195436967822
124.478957915832 0.992652930834523
125.210420841683 0.992863281228076
125.941883767535 0.992357878149196
126.673346693387 0.990928855842956
127.404809619238 0.988429365241906
128.13627254509 0.98480898794011
128.867735470942 0.980137263060497
129.599198396794 0.974611031092045
130.330661322645 0.968543690125389
131.062124248497 0.962337213434705
131.793587174349 0.956440454584552
132.5250501002 0.951299428448476
133.256513026052 0.947306543590777
133.987975951904 0.944755958698062
134.719438877756 0.94381131393198
135.450901803607 0.944490207742095
136.182364729459 0.946667268949081
136.913827655311 0.950094927533562
137.645290581162 0.954438451052724
138.376753507014 0.959319868708724
139.108216432866 0.964364321098464
139.839679358717 0.969242273756698
140.571142284569 0.973701887765758
141.302605210421 0.977587487701047
142.034068136273 0.98084224288114
142.765531062124 0.983495561522481
143.496993987976 0.985637953636979
144.228456913828 0.987387937155059
144.959919839679 0.988856692283312
145.691382765531 0.990116449574374
146.422845691383 0.991177976465235
147.154308617234 0.991981077716967
147.885771543086 0.99239994309082
148.617234468938 0.992262761816925
149.34869739479 0.991382646963182
150.080160320641 0.989594956241667
150.811623246493 0.986794891280982
151.543086172345 0.982969025139676
152.274549098196 0.978215210224213
153.006012024048 0.972747041985086
153.7374749499 0.96688142298028
154.468937875751 0.961010398641969
155.200400801603 0.95556088844957
155.931863727455 0.950947817108442
156.663326653307 0.947527167133975
157.394789579158 0.945555485716195
158.12625250501 0.94516141114892
158.857715430862 0.946333015961238
159.589178356713 0.948922485870029
160.320641282565 0.952667214639262
161.052104208417 0.957224145925196
161.783567134269 0.962212439134329
162.51503006012 0.967258501009959
163.246492985972 0.972037230210681
163.977955911824 0.976303981206684
164.709418837675 0.979913171854866
165.440881763527 0.982821444829144
166.172344689379 0.985075577610129
166.90380761523 0.986787600097416
167.635270541082 0.988101492308946
168.366733466934 0.989157100283732
169.098196392786 0.990057311902206
169.829659318637 0.990843986396276
170.561122244489 0.991486693134163
171.292585170341 0.991886199811156
172.024048096192 0.991892193884665
172.755511022044 0.991332329127172
173.486973947896 0.990047763838847
174.218436873747 0.987929224104429
174.949899799599 0.984947473047365
175.681362725451 0.981172913484928
176.412825651303 0.976780744726741
177.144288577154 0.972040345035459
177.875751503006 0.967289989118348
178.607214428858 0.962900250728368
179.338677354709 0.959231149284446
180.070140280561 0.956589039219649
180.801603206413 0.955189298595254
181.533066132265 0.955130065330538
182.264529058116 0.956380726788721
182.995991983968 0.95878681366297
183.72745490982 0.962090664005774
184.458917835671 0.965965016175589
185.190380761523 0.970054861345948
185.921843687375 0.974021696324441
186.653306613226 0.977583949850068
187.384769539078 0.980547889137598
188.11623246493 0.982824700713455
188.847695390782 0.984431501042705
189.579158316633 0.985476472107766
190.310621242485 0.986130761189194
191.042084168337 0.986591839322155
191.773547094188 0.987044334911305
192.50501002004 0.987624717358327
193.236472945892 0.988395530667391
193.967935871743 0.989333277241966
194.699398797595 0.990331796347291
195.430861723447 0.991220450092395
196.162324649299 0.991794043256522
196.89378757515 0.991849547559199
197.625250501002 0.991223660644606
198.356713426854 0.989825145621707
199.088176352705 0.987656751106614
199.819639278557 0.984823143695311
200.551102204409 0.981523426862576
201.282565130261 0.978029145332137
202.014028056112 0.974650845131788
202.745490981964 0.971697974112399
203.476953907816 0.969437931503365
204.208416833667 0.968060266686616
204.939879759519 0.967651352877459
205.671342685371 0.968183400093298
206.402805611222 0.969519611906413
207.134268537074 0.971434912749415
207.865731462926 0.973649319529754
208.597194388778 0.975869062607046
209.328657314629 0.977829298510791
210.060120240481 0.979331927238727
210.791583166333 0.98027271596776
211.523046092184 0.980653556449761
212.254509018036 0.980577999941054
212.985971943888 0.980230849317775
213.717434869739 0.979845108282644
214.448897795591 0.979661573142466
215.180360721443 0.979887476860987
215.911823647295 0.98066068241189
216.643286573146 0.982024976738216
217.374749498998 0.983920214295195
218.10621242485 0.986188710584911
218.837675350701 0.988596775483774
219.569138276553 0.990867993824648
220.300601202405 0.992723141359733
221.032064128257 0.993920701519021
221.763527054108 0.994291930161893
222.49498997996 0.993765279730895
223.226452905812 0.992376601124637
223.957915831663 0.990263657340363
224.689378757515 0.987645809931705
225.420841683367 0.984791951130766
226.152304609218 0.981981533741475
226.88376753507 0.979464629337412
227.615230460922 0.977427142853198
228.346693386774 0.975966569726773
229.078156312625 0.975082083156048
229.809619238477 0.974680507685377
230.541082164329 0.974597210842717
231.27254509018 0.974628532244276
232.004008016032 0.974570472723339
232.735470941884 0.974257312530409
233.466933867735 0.973593806744498
234.198396793587 0.972575629485414
234.929859719439 0.971294635810523
235.661322645291 0.969927959688129
236.392785571142 0.968712555898673
237.124248496994 0.967909096654596
237.855711422846 0.967760785279195
238.587174348697 0.968453405912996
239.318637274549 0.970082700891426
240.050100200401 0.972634024717881
240.781563126252 0.975977368083015
241.513026052104 0.979878572202928
242.244488977956 0.984025200766933
242.975951903808 0.98806343645724
243.707414829659 0.99164080732338
244.438877755511 0.994448733581941
245.170340681363 0.996258926957643
245.901803607214 0.996948570631565
246.633266533066 0.996510842569922
247.364729458918 0.995049497506785
248.09619238477 0.992758588795692
248.827655310621 0.989890639654575
249.559118236473 0.986718315847449
250.290581162325 0.983495619965807
251.022044088176 0.980424645847541
251.753507014028 0.977632977146284
252.48496993988 0.975165029173726
253.216432865731 0.972988308175258
253.947895791583 0.971013086911758
254.679358717435 0.969121788931543
255.410821643287 0.967202807317569
256.142284569138 0.96518281196726
256.87374749499 0.963051917318799
257.605210420842 0.960877312772566
258.336673346693 0.958802873756777
259.068136272545 0.957034543554133
259.799599198397 0.955813536335435
260.531062124248 0.955381315263726
261.2625250501 0.955941575308123
261.993987975952 0.957624944634834
262.725450901804 0.960461766318512
263.456913827655 0.964367202894247
264.188376753507 0.969141186066927
264.919839679359 0.974483651345707
265.65130260521 0.980023335360614
266.382765531062 0.985356468251546
267.114228456914 0.990090240475672
267.845691382765 0.993885182459548
268.577154308617 0.996490696411884
269.308617234469 0.99776893447654
270.040080160321 0.9977039075397
270.771543086172 0.99639489209027
271.503006012024 0.994035543624217
272.234468937876 0.990882248992719
272.965931863727 0.987216806738654
273.697394789579 0.983309254152466
274.428857715431 0.97938644388007
275.160320641283 0.975610857867097
275.891783567134 0.972072334903038
276.623246492986 0.968793199037598
277.354709418838 0.965745083177264
278.086172344689 0.962873903631442
278.817635270541 0.960128238392758
279.549098196393 0.957485955845632
280.280561122244 0.954974356459833
281.012024048096 0.952680225726523
281.743486973948 0.950747851477986
282.4749498998 0.949364972693985
283.206412825651 0.948738520978431
283.937875751503 0.949063628006573
284.669338677355 0.950490485273871
285.400801603206 0.953094104940372
286.132264529058 0.956851771392673
286.86372745491 0.961632011396858
287.595190380761 0.96719735908201
288.326653306613 0.97322125110749
289.058116232465 0.979317327413492
289.789579158317 0.985077540807161
290.521042084168 0.99011409244216
291.25250501002 0.994099547953338
291.983967935872 0.996799681053342
292.715430861723 0.998094629085238
293.446893787575 0.99798567390961
294.178356713427 0.99658710392594
294.909819639279 0.994104816791069
295.64128256513 0.990805225804511
296.372745490982 0.986979331932535
297.104208416834 0.982907326034453
297.835671342685 0.978828742207332
298.567134268537 0.974922086358091
299.298597194389 0.971296223675969
300.03006012024 0.967993905371421
300.761523046092 0.965005948548381
301.492985971944 0.962293024021347
302.224448897796 0.959810961363318
302.955911823647 0.95753507024457
303.687374749499 0.955479234346493
304.418837675351 0.953706406473911
305.150300601202 0.952328495295289
305.881763527054 0.95149530092158
306.613226452906 0.951373902413143
307.344689378757 0.952121479906287
308.076152304609 0.953855730271419
308.807615230461 0.956627612272589
309.539078156313 0.96040101511082
310.270541082164 0.96504306770415
311.002004008016 0.970327299383
311.733466933868 0.975949943678977
312.464929859719 0.981557647092377
313.196392785571 0.986783039141919
313.927855711423 0.991283345728079
314.659318637275 0.994776704852216
315.390781563126 0.997071161557127
316.122244488978 0.998082419597861
316.85370741483 0.99783811839505
317.585170340681 0.996468398119527
318.316633266533 0.994184487774743
319.048096192385 0.991248696594477
319.779559118236 0.987940274887971
320.511022044088 0.984522006021463
321.24248496994 0.981212077538636
321.973947895792 0.978164838754825
322.705410821643 0.97546264447661
323.436873747495 0.973119318362561
324.168336673347 0.971094073466085
324.899799599198 0.96931322545569
325.63126252505 0.96769592363127
326.362725450902 0.966179558811305
327.094188376753 0.964740574949896
327.825651302605 0.963407123053536
328.557114228457 0.962261271503672
329.288577154309 0.961430155798457
330.02004008016 0.961067267464284
330.751503006012 0.961326758289263
331.482965931864 0.962334887951144
332.214428857715 0.964163343169787
332.945891783567 0.966808981388582
333.677354709419 0.970183612329401
334.408817635271 0.974115874983088
335.140280561122 0.978365356915077
335.871743486974 0.982647162181294
336.603206412826 0.986663490222584
337.334669338677 0.990137707784997
338.066132264529 0.992846039989597
338.797595190381 0.994642407147164
339.529058116232 0.995472998005625
340.260521042084 0.99537870701314
340.991983967936 0.994485324654783
341.723446893788 0.992983094363571
342.454909819639 0.991098700258013
343.186372745491 0.989063742681145
343.917835671343 0.987084176633479
344.649298597194 0.985314988732843
345.380761523046 0.983843601695863
346.112224448898 0.982684221486123
346.843687374749 0.98178374358835
347.575150300601 0.981038123626341
348.306613226453 0.98031653536523
349.038076152305 0.979489427071714
349.769539078156 0.978455948438564
350.501002004008 0.977166281429519
351.23246492986 0.975635188233068
351.963927855711 0.973944484443479
352.695390781563 0.972233938626718
353.426853707415 0.970681993876246
354.158316633267 0.969479379551742
354.889779559118 0.968799843888054
355.62124248497 0.96877269343691
356.352705410822 0.969461504806177
357.084168336673 0.97085234823981
357.815631262525 0.97285332210806
358.547094188377 0.975305412253779
359.278557114228 0.978002955583665
360.01002004008 0.980720570718909
360.741482965932 0.98324251809426
361.472945891784 0.985390173281593
362.204408817635 0.987043649529493
362.935871743487 0.988154512039834
363.667334669339 0.988747845330648
364.39879759519 0.988913481446978
365.130260521042 0.988787765084088
365.861723446894 0.988528615007539
366.593186372745 0.988287650679732
367.324649298597 0.988183638367022
368.056112224449 0.988281381868966
368.787575150301 0.988579429052618
369.519038076152 0.989008669239291
370.250501002004 0.989442233477508
370.981963927856 0.989715329568252
371.713426853707 0.989652033968867
372.444889779559 0.98909489890398
373.176352705411 0.987932725205334
373.907815631263 0.986122099737057
374.639278557114 0.983699266981135
375.370741482966 0.980780433213187
376.102204408818 0.97755042499132
376.833667334669 0.974241429998756
377.565130260521 0.971105038780642
378.296593186373 0.968381749180002
379.028056112224 0.966272364895709
379.759519038076 0.964915307818644
380.490981963928 0.964372871285968
381.22244488978 0.964628048642223
381.953907815631 0.96559200137442
382.685370741483 0.96712071160893
383.416833667335 0.969038097983622
384.148296593186 0.971162019731429
384.879759519038 0.973329252629522
385.61122244489 0.975415731719686
386.342685370741 0.977349094396594
387.074148296593 0.979111733829688
387.805611222445 0.980734035720038
388.537074148297 0.982279019614823
389.268537074148 0.983821008570519
390 0.985421979615637
};
\addlegendentry{$\braket{E_O}$, $p=0.001$}
\addlegendentry{$\braket{O}_{ideal}$}
\end{axis}

\end{tikzpicture}

%% file: biasvstau_long.tex
\begin{tikzpicture}

\begin{axis}[
legend style={at={(axis cs:390,0.07)},anchor=north east},
tick align=outside,
tick pos=left,
x grid style={darkgray176},
xlabel={simulation time $\tau$},
xmin=6.75, xmax=408.25,
xtick style={color=black},
y grid style={darkgray176},
ylabel={$\left| b_{E_O}\right|$},
ytick style={color=black},
mark size=1
]
\addplot [semithick, c4, mark=*]
table {%
25 0.000216579510405324
50 7.69089675766166e-05
75 0.000469545668775107
100 0.00111513343577563
125 6.15898645751622e-05
150 0.000850824538284556
175 0.0013018285556935
200 0.000718061330554409
225 0.00141997092975576
250 0.000685430908382578
275 0.00227887677319827
300 0.000780482612427469
325 0.000960437229339695
350 0.00198857933103891
375 0.000561108231531549
390 0.00121631503663244
};
\addplot [semithick, c5, mark=star]
table {%
25 0.000517713123380781
50 0.00486741337812691
75 0.00331380531591208
100 0.0108923656150298
125 0.0104717838199659
150 0.00609376088944003
175 0.000446532927715571
200 0.00512193114278847
225 0.00603597065275641
250 0.0149603250581342
275 0.00550930862348842
300 0.0270477004817014
325 0.00529154290532707
350 0.00626630366372927
375 0.00153169060463565
390 0.0281828258508879
};
\addplot [semithick, c3]
table {%
25 0.08
50 0.08
75 0.08
100 0.08
125 0.08
150 0.08
175 0.08
200 0.08
225 0.08
250 0.08
275 0.08
300 0.08
325 0.08
350 0.08
375 0.08
390 0.08
};

\addlegendentry{$\braket{E_o}$, $p=0.0001$}
\addlegendentry{$\braket{E_o}$, $p=0.001$}
\addlegendentry{$8\cdot 10^{-2}$}
\end{axis}

\end{tikzpicture}

%% file: discussion.tex
Herein, we have combined standard QSP and QEM protocols, providing a simple model to study the interplay of noise, precision, and depth trade-offs. Besides our spin-state simulations, we have adapted some of the general bounds on QEM cost to QSP and numerically explored at which noise level makes the expectation values and variance decay to constants, leading to unrecoverable expectation values. These limits suggest that for specific noise models and circuits, there exist regions where ZNE stops working even when provided with an unlimited number of samples.

While choosing the extrapolation method, we observed that the overwhelmingly preferable technique of extrapolation was exponential, as the noise accumulates in the circuit layerwise. We have found that for low noise levels, $p=0.0001, 0.001$, the ZNE-QSP protocol is successful in recovering expectation values of interest during the time intervals, $\tau\in[0.1, 20.0]$. The ZNE-QSP protocol struggles with higher initial noise, $0.01$, where the expectation value of interest is recovered reliably only for short-time intervals.

For noise levels attainable now or in the next few years, $p=0.0001, 0.001$, the MSE after QEM was successfully suppressed to the same precision as an algorithmic error $\mathcal{O}\left(10^{-4}\right)$ with our resources. For longer circuits, we expect that the precision of the QSP approximation will be constrained by the expected precision of the ZNE protocol.

Our long-time simulations show that ZNE performs successfully for small noise, $p = 0.0001$ and  $p=0.001$. We observe that long-time QSP is mitigated quite well in comparison to similar depth Trotter circuits (with a smaller simulation time), which may be attributed to the fact that QSP has built-in post-selection, so the QEM is working on already post-selected states.

This work has a major simplification from a real QSP circuit\textemdash we calculate and apply $e^{i\arccos(H)}$  defined from its matrix form rather than decomposing it into elementary gates. Thus, we are quantifying QEM with QSP without having to correct for the QEM performance of a given block encoding and extra ancilla qubit costs. Importantly, this means that the depth comparison between QSP and Trotter is not rigorous. Instead, our results are a best-case limit on QSP circuits, and our Trotter circuits highlight regimes where the standard technique does or does not succeed. The actual proof of performance would require full QSP circuits, optimization and decomposition of circuits for both simulation methods, and a more practical implementation of QEM.

Our simulations show some relative simulation times and low-precisions for which ZNE might succeed for QSP circuits. Exploration on near-term devices will require access to reasonably sized QSP oracles, by which we mean oracles in $\mathcal{O}(10)$, for which the total circuit length might still allow mitigation for longer times than Trotter simulations.

Thus, we have shown that QSP Hamiltonian simulations might be useful to consider for the NISQ era, as they might allow low-precision long-timescale simulations with relatively short circuit lengths, which results in better performance of error mitigation techniques, as sample cost is exponential in depth. One setting where such short-depth oracles might be natural is circuit benchmarking, for which early QSP circuits have already been explored \cite{Dong_2022}.

%% file: qsp-methods.tex
Herein, we summarize some well-known results on QSP and Hamiltonian simulation with QSP.
\subsection{Jacobi-Anger expansion}\label{sec:hsim_function}
We begin with the Jacobi-Anger expansion for $e^{i\tau x}$,
\begin{align}  e^{i\tau x}&=\cos(\tau x)+i\sin(\tau x)\\
    \cos(\tau x)&=J_0(t)+\sum_{k=1}^{\infty}(-1)^kJ_{2k}(\tau)\mathcal{T}_{2k}(x)\\
    \sin(\tau x)&=\sum_{k=1}^{\infty}(-1)^kJ_{2k+1}(\tau)\mathcal{T}_{2k+1}(x)
\end{align}
One can truncate the expansion at $R$ and bound the result on domain $\mathcal{D}_0=[-1, 1]$,
\begin{align}
    \left|\left|\right.\right.\cos(\tau x)&-J_0(\tau)-\sum_k(-1)^kJ_{2k}(\tau)\mathcal{T}_{2k}(x)\left.\left.\right|\right|_{\mathcal{D}_0}\nonumber\\
    &\leq 2\sum_{l=0}^{\infty}\left|J_{2R+2l+2}(\tau)\right|,\\
     \left|\left|\right.\right.\sin(\tau x)&-\sum_{k=1}^{\infty}(-1)^kJ_{2k+1}(\tau)\mathcal{T}_{2k+1}(x)\left.\left.\right|\right|_{\mathcal{D}_0}\nonumber\\
    &\leq 2\sum_{l=0}^{\infty}\left|J_{2R+2l+3}(\tau)\right|.
\end{align}
Following \cite{gilyen_quantum_2019},
\begin{align}
    \left|\left|\cos(\tau x)-J_0(\tau)-\sum_{k=0}^R(-1)^kJ_{2k}(\tau)\mathcal{T}_{2k}(x)\right|\right|_{\mathcal{D}_0} &\leq \epsilon\\
    \left|\left|\sin(\tau x)-\sum_{k=0}^R(-1)^kJ_{2k+1}(\tau)\mathcal{T}_{2k+1}(x)\right|\right|_{\mathcal{D}_0} &\leq \epsilon
\end{align}
whenever
\begin{equation}\label{eq;lambeth_thing_for_bound}
    \frac{4}{3\sqrt{\pi}}\left(\frac{e|\tau|}{4(R+1)}\right)^{2(R+1)}\leq\epsilon.
\end{equation}
Furthermore,  \cite{gilyen_quantum_2019} shows that $R+1$ can be upper-bound by a solution to
\begin{equation}
    \left(\frac{e|\tau|}{2r}\right)^r=\epsilon,
\end{equation}
which is related to the W-Lambeth function. Then 
\begin{equation}
    2(R+1)=\max\left(\lceil r(\frac{e|\tau|}{2}, \epsilon)\rceil, \left|\tau\right|\right)
\end{equation}
and then the highest degree of the truncated sums will be $2R+1$. Again following \cite{gilyen_quantum_2019}, which itself adapts \cite{low_hamiltonian_2019}, we bound $r$ with \cref{lemma:hsdegreebound}.
\begin{lemma}[Bound on $r$, paraphrased from lemma 59 \cite{gilyen_quantum_2019}]\label{lemma:hsdegreebound}
    For $\tau\in\mathbb{R}_+$ and $\epsilon\in(0, 1)$, an upper-bound on the solution to $r(\tau, \epsilon)$, $\Tilde{r}(\tau, \epsilon)$ is
    \begin{equation}\label{eq:rupperbound}
        \tilde{r}(\tau, \epsilon)=\begin{cases}
            \lceil e\tau\rceil & \tau> \frac{ln(1/\epsilon)}{e}\\
             \lceil \frac{4\log(1/\epsilon)}{\log\left(e+\frac{1}{\tau}\log(1/\epsilon)\right)}\rceil & \tau\leq \frac{ln(1/\epsilon)}{e}\\
        \end{cases}
    \end{equation}
    Moreover, for all $q\in\mathbb{R}_+$,
    \begin{equation}
        r(\tau, \epsilon)<e^q\tau+\frac{ln(1/\epsilon)}{q}.
    \end{equation}
\end{lemma}
Note that we have modified the lemma statement from \cite{gilyen_quantum_2019} to obtain a more concise lower bound for $r(\tau, \epsilon)$.

To coax this expansion into the correct form for \cref{lemma:QSP_Haah}, we use relations $\cos(j\theta)=\frac{e^{i\theta}+e^{-i\theta}}{2}, \sin(j\theta)=\frac{e^{i\theta}-e^{-i\theta}}{2}$ to define Laurent polynomials $\mathcal{A}(z), \mathcal{B}(z)$ which are identical to the Jacobi-Anger truncation for $z\in U(1)$, and are respectively reciprocal and anti-reciprocal. We also introduce a sub-normalization of $\frac{1}{\sqrt{2}}$ to make the QSP-processing easier, and finally, \cref{eq:laurent_HSpoly} is the polynomial used in our QSP protocol.
\begin{align}\label{eq:laurent_HSpoly}
    \mathcal{P}_{HS}&=\mathcal{A}(z)-i\mathcal{B}(z)\\
    \mathcal{A}(z)&=\frac{1}{\sqrt{2}}\left(J_0(t)+\sum_{k=-R}^{R}(-1)^k\frac{J_{2k}(\tau)}{2}z^{2k}\right)\\
    \mathcal{B}(z)&=\frac{1}{\sqrt{2}}\sum_{-R-1}(-1)^k\frac{J_{2k+1}(\tau)}{2}z^{2k+1}
\end{align}

\subsection{QSP Hamiltonian Simulation}
We begin by defining with Hamiltonian $H$ with eigendecomposition $H=Q\text{Diag}\left(\{\lambda_j\}\right)Q^{\dag}$, where column $j$ of $Q$ defines eigenvector $\ket{\lambda_j}$ and  $\{\lambda_j\}$ is the set of all eigenvalues. 
We then construct our oracle, 
\begin{equation}
    U=Q\text{Diag}_j\left(\{e^{i\arccos(\lambda_j})\}\right)Q^{\dag}
\end{equation}
This object is unitary and has eigenvalues $e^{i\arccos(\lambda_j)}$, so it is a suitable QSP oracle. In QSP, our polynomial $\mathcal{P}_{HS}$ will be applied to each $e^{i\arccos(\lambda_j)}$. 

\begin{lemma}[Hamiltonian Simulation with Laurent QSP, simplified from \cite{Haah2019product, gilyen_quantum_2019}]\label{lemma:qsp_hsim}
    Given access to $U_H$, a $(1,0,  0)$  block-encoding\footnote{That is, a block encoding with subnormalization $1$, $0$ ancilla qubits, and precision $\epsilon_{BE}=0$} the elements of Hamiltonian $H\in \mathbb{C}^{2^h\otimes 2^h}$, we can prepare a $(\frac{1}{\sqrt{2}}, 1, \epsilon)$ block-encoding of $e^{i\tau H}$, $\tau\in\mathbb{R}_+$ on a quantum device within precision $\epsilon$ using $h+1$ qubits
\end{lemma}
\begin{proof}
For consistency with existing proofs, we are treating $e^{i\arccos(H)}$ as a completely precise block encoding using no ancilla qubits. We choose to instead approximate $\frac{1}{\sqrt{2}}e^{i\tau x}$ for pre-processing convenience. Recall that $\cos(\tau x)-i \sin(\tau x)$ can be approximated by \cref{eq:laurent_HSpoly} up to some $\epsilon_{approx}$, fulfilling the conditions in \cref{lemma:QSP_Haah}.

The QSP circuit prepares  $U_{QSP, H}=\frac{1}{\sqrt{2}}\ket{+}\bra{+}\otimes U_{e^{i\tau H'}} +U^{\perp}$, where $\braket{+|U_{prep}|+}=0$, up to error $\epsilon_{QSP}$. This error is bound by the degree choice \cite{gilyen_quantum_2019, low_hamiltonian_2019} and pre-processing method  \cite{Haah2019product, skelton2025hitchhikersguideqsppreprocessing}. Stipulating $\epsilon=\epsilon_{QSP}$ and $\epsilon_{coeff}=\epsilon_{QSP}/C(n)$ ensures that the circuit will be bound within error.

QSP approximately block-encodes $e^{i\tau H'}$ with query complexity $2n$ and $2n+1$ controlled rotations, which are usually further decomposed into $1$-controlled operations and single-qubit gates. Irregardless, the query complexity of QSP scales with $n=\mathcal{O}\left(2\lceil \tilde{r}({|\tau|},{\frac{\epsilon_{approx}}{C(n)}})\rceil\right)$, and uses one ancilla qubit. 
\end{proof}

We use QSP-processing from \cite{skelton2024harmlessmethodsqspprocessinglaurent} to obtain the QSP parameters $\{P_j,E_0\}$, defining the QSP circuit $U_{QSP}$. The joint operator error of the polynomial approximation and decomposition step means we can write
\begin{equation}
    U_{QSP}=\begin{bmatrix}
    P_{real}(U)+iQ_{im}(U) & Q_{real}(U)-iP_{im}(U)\\
    Q_{real}(U)+iP_{im}(U) & P_{real}(U)-iQ_{im}(U).\\
\end{bmatrix}
\end{equation}
for some polynomial $P$ which is $\epsilon_{QSP}$-close to $\mathcal{P}_{HS}$ and complimentary polynomial $Q$.
Then, when we simulate the evolution 
$U_{QSP}\rho U_{QSP}^{\dag}$ for some (possibly mixed) state $\rho$ and project to the $\ket{+}\bra{+}$ subspace of the QSP-ancilla qubit $a$, we get $P(U)\rho P^{*}(U)$. In math, this process is 
\begin{align}
    \rho_{red}&=\textbf{Tr}_{a}\left(\ket{+}\bra{+}\otimes I_{syst}\cdot U_{QSP}\rho U_{QSP}^{\dag}\right).
\end{align}
With  $\rho=\ket{+}\bra{+}\otimes \ket{\lambda_j}\bra{\lambda_j}$, the result after post-selection is

\begin{equation}
\rho_{red}=\frac{P(U)\ket{\lambda_j}\bra{\lambda_j}P^{*}(U)}{\left|\left|P(U)\ket{\lambda_j}\bra{\lambda_j}P^{*}(U)\right|\right|}.
\end{equation}
When $\rho$ is built from some arbitrary density matrix in the state space, $\rho_{init}=\ket{+}\bra{+}\otimes \sum_jp_j\ket{\lambda_j}\bra{\lambda_j}$, the additivity of the trace ensures that after postselection we still arrive at
\begin{align}
\rho_{red}&=\sum p_j P(\lambda_j)\ket{\lambda_j}\bra{\lambda_j}P^{*}(\lambda_j)\\
&=P(U)\sum_jp_j\ket{\lambda_j}\bra{\lambda_j}P^{*}(U).
\end{align}

We then compute the expectation value with respect to any measurement operator $O$ as
\begin{equation}
    \text{Tr}\left(O\rho_{red}\right).
\end{equation} 
We can also classically build matrix $P(U)$ using the known eigendecomposition and compare the simulated expectation value to  $\text{Tr}\left(OP(U)\ket{\lambda_j}\bra{\lambda_j}P^{*}(U)\right).$

\subsection{Error and circuit depth}\label{sec:qsperrorcircuitdepth}
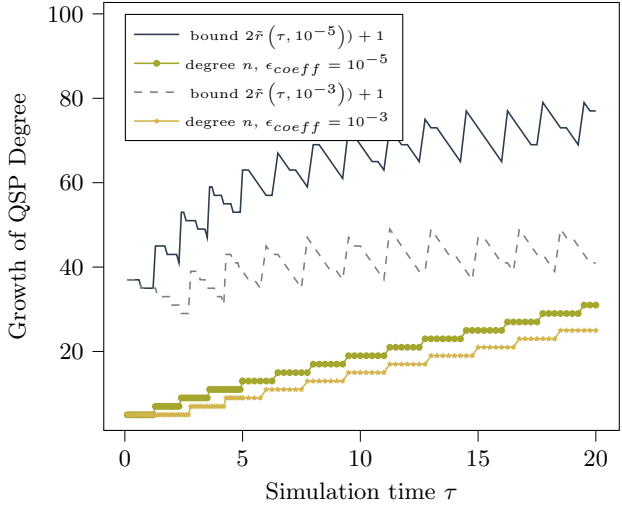
\begin{figure}
     \centering
        \input{neccdegvstau20}
     \caption{Comparison of the degree bound and the numerically sufficient degree. We compute the degree necessary to obtain $\epsilon_{approx}=\mathcal{O}\left(10^{-5}\right)$ for simulation times $\tau\in\{0.1, 0.2, 0.3, ... 5\}\cup\{5, 5.25,... 20\}$ or to obtain $\epsilon_{approx}=\mathcal{O}\left(10^{-3}\right)$ for simulation times $\tau\in\{0.25, 0.5, 0.75,...20\}$ and compare these against the degree bound computed using \cref{eq:rupperbound}.} 
     \label{fig:qspdegrees}
 \end{figure}

\begin{figure}
    \centering
    \input{fixeddepthsteps}
    \caption{Decay of error with simulation time at a fixed depth. With fixed-depth circuits, we no longer guarantee that the error in the approximation will be beneath our $\leq \epsilon_{QSP}$ threshold and numerically compute the full error. We examine the growth of error for $\tau\in\{0.1, 0.2,0.25...5\}$ and $\tau\in\{5, 5.25,...20\}$. We select circuit depths ranging from the depth necessary to achieve $\leq \epsilon_{QSP}$ error for the smallest $\tau$ to the depth necessary to achieve $\leq \epsilon_{QSP}$ error for the largest $\tau$, in this case, $d=5, 9, 13, 17, 21, 25, 31$. Note that due to numeric instabilities in the decomposition step, circuits corresponding to $\mathcal{O}(10^{-1})$ error are not physically feasible (cannot recover a unitary form of the QSP circuit). }
    \label{fig:growthoferrorvstau}
\end{figure}
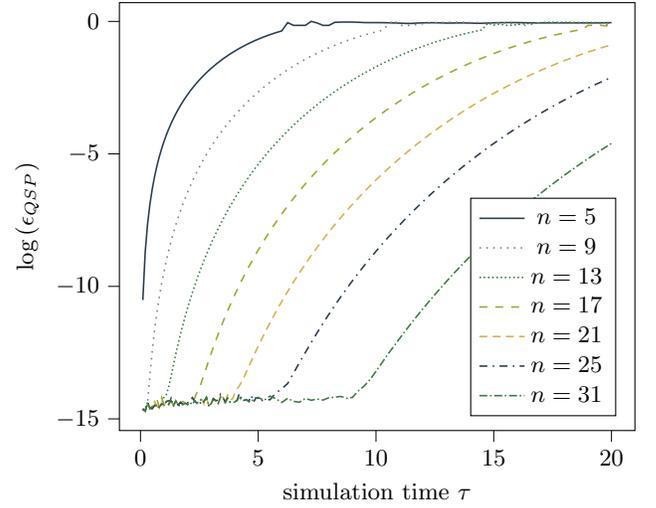

Recall that $n=n(t, \epsilon_{approx})$ is defined from \cref{lemma:qsp_hsim} and \cref{eq:rupperbound}, and then the relationship between our Hamiltonian simulation circuit depth and $\epsilon_{approx}$ is
 \begin{align}
      d_{qsp}\leq 4\lceil\tilde{r}\left(\left|\tau\right|, \epsilon_{approx}\right)\rceil+1+4\lceil\tilde{r}\left(\left|\tau\right|, \epsilon_{approx}\right)\rceil d_o.
 \end{align}
 For $\tau\leq \frac{ln(1/\epsilon)}{e}$, this is
 \begin{align}
     d_{qsp}&\leq  \left(4 \lceil \frac{4\log(1/\epsilon)}{\log\left(e+\frac{1}{\tau}\log(1/\epsilon)\right)}\rceil+2\right)\cdot\left(1+d_o\right)+1\label{eq:qspdepthhs}
 \end{align}
 and a similar expression holds for $\tau> \frac{ln(1/\epsilon)}{e}$. However, these bounds are not tight. Circuit depth is a major constraint, so we will instead determine the degree required to obtain error $\mathcal{O}\left(10^{-5}\right)$ or $\mathcal{O}\left(10^{-3}\right)$ numerically. We define $\epsilon_{approx}$ as the $l_{\infty}$-norm difference between $\frac{1}{\sqrt{2}}e^{-i\tau x}$ and $\mathcal{P}(e^{i\arccos(x)})$ within $x\in [-1, 1]$.  In \cref{fig:qspdegrees}, we determine the degree necessary to obtain such an approximation for each $\tau$. 
 In \cref{fig:growthoferrorvstau}, we show how $\epsilon_{coeff}$ increases numerically with $\tau$ for fixed depth circuits.  

By selecting $n$ numerically, we are able to reduce the circuit depths by an average factor of  $4.52$ for the more precise instances ($\epsilon_{coeff}=10^{-5}$) and  $4.10$ for the less precise instances, ($\epsilon_{coeff}=10^{-3}$). 

\begin{figure}
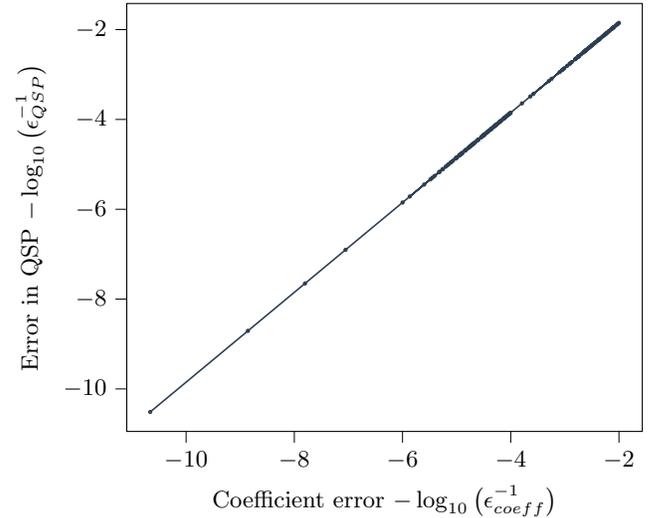

     \centering
        \include{epsivsepsi}
     \caption{Growth of error in the QSP decomposition step, shown on a log-log scale. Instead of using a bound on $C(n)$, we numerically determine the precision of the QSP circuit. We show that the error grows linearly for $\tau\in\{0.1, 0.2,0.25, 0.3...5\}\cup  \{5, 5.25,...20\}$.}
     \label{fig:growthoferror}
\end{figure}

In \cref{lemma:qsp_hsim}, the error in the noiseless QSP circuit arises from the limited precision of the functional approximation, $\epsilon_{approx}$. Technically, QSP circuits contain some additional error from the limited precision solution of the QSP-processing, and the full error is $\epsilon_{qsp}=C(n)\epsilon_{approx}$. For Laurent-QSP, this can be rigorously bound \cite{Haah2019product, skelton2025hitchhikersguideqsppreprocessing}, but in our implementation, we expect this error to grow linearly rather than the exponential worst case. We confirm that $\epsilon_{QSP}$ remains controlled by estimating the $l_{\infty}$ norm distance between $\bra{+, \lambda}U_{QSp}HU_{QSP}^{\dag}\ket{+, \lambda}$ against $\mathcal{P}(e^{i\arccos(x)})$, as shown in \cref{fig:growthoferror}

Because we numerically select $n$, it is more convenient to use the generic expression for the depth of a degree-$n$ QSP circuit whose oracle has some depth $d_o$,
\begin{equation}
    d_{qsp}(d_o)=2nd_o+1\label{eq:qspdepth}.
\end{equation}
This is the gate complexity when we assume we can implement generalized-control gates\footnote{ Note that each of these could be decomposed into a more standard form of control-zero or control-one gate with a constant overhead of single-qubit gates.} of the form \cref{eq:generalizedcontrolgate}.

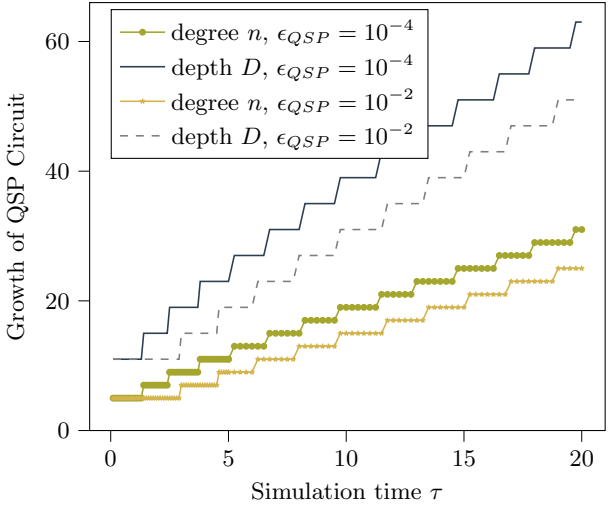
\begin{figure}
     \centering
       \input{depthvstau}
     \caption{Degrees and depths for $d_o=1$ necessary to obtain precision better than $\epsilon_{qsp}$ for simulation times $\tau\in\{0.1, 0.2, 0.3, ... 5\}$ (above) $\tau\in\{5, 5.25,... 20\}$ (below).}
     \label{fig:qspdepths}
 \end{figure}

In \cref{fig:qspdepths}, we show the circuit depths needed to reach $\epsilon_{QSP}$ precision on a noiseless circuit. All of our instances succeed with error $\epsilon_{QSP}\leq 1.405 \cdot 10^{-4}$ or $\epsilon_{QSP}\leq 1.408\cdot 10^{-2}$. All of our smaller benchmark, $\tau\in\{0.1, 0.2, 0.3, ... 5\}$, succeed with circuit depth under $D=40$, and $\tau\in\{5, 5.25,... 20\}$ requires up to $D=125$. However, we stress that this does not account for the substantial cost of $d_0$. 
Then finally with circuit noise, we have to account for the additional error $\epsilon_{noise}$. This is additive to $\epsilon_{qsp}$, through the simple relation
\begin{eqnarray}
    &\left|\trace{\ket{+}\bra{+} OU_{QSP, p}^{\dag}\rho U_{QSP, p}} \right.\nonumber\\
    &\left.-\trace{O\sum_{\lambda, \lambda'} e^{i\tau(\lambda'-\lambda)}\left(\ket{\lambda}\bra{\lambda}\right)^{\dag}\rho \ket{\lambda'}\bra{\lambda'}}\right| \\
    &\leq \left|\trace{\ket{+}\bra{+}OU_{QSP, p}^{\dag}\rho U_{QSP, p}}\right. \nonumber\\
    &\left.-\trace{\ket{+}\bra{+}OU_{QSP}^{\dag}\rho U_{QSP}}\right|\nonumber\\
    &+\left|\trace{\ket{+}\bra{+}OU_{QSP}^{\dag}\rho U_{QSP}}\right. \nonumber\\
    &\left.-\trace{O\sum_{\lambda, \lambda'} e^{i\tau(\lambda'-\lambda)}\left(\ket{\lambda}\bra{\lambda}\right)^{\dag}\rho \ket{\lambda'}\bra{\lambda'}}\right|\\
    &\leq \left|\trace{\ket{+}\bra{+}OU_{QSP, p}^{\dag}\rho U_{QSP, p}}\right. \nonumber\\
    &\left.-\trace{\ket{+}\bra{+}OU_{QSP}^{\dag}\rho U_{QSP}}\right|+\epsilon_{qsp}\\
    &=\epsilon_{noise}+\epsilon_{qsp}.
\end{eqnarray}

\subsection{Success Probability of Hamiltonian Simulation}
The success probability of the QSP protocol is the component of $U_{QSP}\rho_{init}U_{QSP}^{\dag}$ approximating $e^{i\arccos(H)}$. We have to post-select the correct QSP-ancillary measurement, so the success probability is
\begin{equation}
    p_{QSP}\approx {\left|\left|\frac{1}{2}e^{-i\tau H}\rho e^{i\tau H} \right|\right|}=\frac{1}{2}\label{eq:qsp_success},
\end{equation}

More rigorously, for any density matrix with decomposition $\rho=\sum_{l=0}^{N}c_l\ket{\lambda_l}\bra{\lambda_l}$, 
\begin{eqnarray}
    p_{QSP}&=Tr\left(\bra{+}\otimes I_{N}U_{QSP}\rho U_{QSP}^{\dag}\ket{+}\otimes I_{N}\right)\\
    &=\sum_{l}c_l|P(\lambda_l)|^2.
\end{eqnarray}

Recall that from \cref{lemma:qsp_hsim}, the distance between $\bra{+}\otimes I_{\text{rank}(U)}U_{QSP}\ket{+}\otimes I_{\text{rank}(U)}=P(U)$  and $\frac{1}{\sqrt{2}}e^{i\tau H}$ is bound. Specifically, we have the following operator norm bound for QSP circuits,
\begin{align}
    \left|\left|P(U)-\frac{1}{\sqrt{2}}e^{i\tau H}\right|\right|_{op}\leq \epsilon_{QSP},
\end{align}
for some $\epsilon_{QSP}\in (0, 1)$. After measuring the $QSP$-ancilla, we have the following state norm bound 
\begin{eqnarray}
    &\left|\left|P(H)\rho P(H)^{\dag}-e^{-i\tau H}\rho e^{i\tau H}\right|\right|_{op}\nonumber\\
    &\leq  \left|\left|P(H)\rho P(H)^{\dag}-P(H)\rho e^{i\tau H}\right|\right|_{op} \nonumber\\
    &\quad +  \left|\left|P(H)\rho e^{i\tau H}-e^{-i\tau H}\rho e^{i\tau H}\right|\right|_{op}\\
    &\leq \left|\left|P(H)\right|\right|_{op}\cdot  \left|\left|\rho\right|\right|_{op} \cdot \left|\left|P(H)^{\dag}-e^{i\tau H}\right|\right|_{op}\nonumber\\
    &\quad +  \left|\left|P(H)-e^{-1\tau H}\right|\right|_{op}\cdot  \left|\left|\rho\right|\right|_{op} \cdot \left|\left|e^{i\tau H}\right|\right|_{op}\\
    &\leq \left(\left|\left|P(H)\right|\right|_{op}+\left|\left|e^{i\tau H}\right|\right|_{op}\right)\left|\left|P(H)^{\dag}-e^{i\tau H}\right|\right|_{op}\\
    &\leq \left(2+\epsilon_{QSP}\right)\epsilon_{QSP}\\
    &\leq 3\epsilon_{QSP}\label{eq:qspsuccprob}
\end{eqnarray}
 Thus, when $\epsilon_{QSP}$ is very small, we have a good reason to accept \cref{eq:qsp_success}.

%% file: neccdegvstau20.tex
\begin{tikzpicture}

\begin{axis}[
legend style={at={(axis cs:0,100)},anchor=north west},
tick align=outside,
tick pos=left,
x grid style={darkgray176},
xlabel={Simulation time \(\displaystyle \tau\)},
xmin=-0.895, xmax=20.995,
xtick style={color=black},
y grid style={darkgray176},
ylabel={Growth of QSP Degree},
ymin=1.3, ymax=102.7,
ytick style={color=black}
]
\addplot [semithick, c1]
table {%
0.1 37
0.2 37
0.3 37
0.4 37
0.5 37
0.6 37
0.7 35
0.8 35
0.9 35
1 35
1.1 35
1.2 35
1.3 45
1.4 45
1.5 45
1.6 45
1.7 45
1.8 43
1.9 43
2 43
2.1 43
2.2 43
2.3 41
2.4 53
2.5 53
2.6 51
2.7 51
2.8 51
2.9 51
3 51
3.1 49
3.2 49
3.3 49
3.4 49
3.5 47
3.6 59
3.7 59
3.8 57
3.9 57
4 57
4.1 57
4.2 55
4.3 55
4.4 55
4.5 55
4.6 53
4.7 53
4.8 53
4.9 53
5 63
5 63
5.25 63
5.5 61
5.75 59
6 57
6.25 57
6.5 67
6.75 65
7 63
7.25 63
7.5 61
7.75 59
8 69
8.25 69
8.5 67
8.75 65
9 63
9.25 61
9.5 73
9.75 71
10 69
10.25 67
10.5 65
10.75 65
11 63
11.25 73
11.5 71
11.75 69
12 69
12.25 67
12.5 65
12.75 75
13 73
13.25 73
13.5 71
13.75 69
14 67
14.25 65
14.5 77
14.75 75
15 73
15.25 71
15.5 69
15.75 67
16 65
16.25 77
16.5 75
16.75 73
17 71
17.25 69
17.5 69
17.75 79
18 77
18.25 75
18.5 73
18.75 73
19 71
19.25 69
19.5 79
19.75 77
20 77
};
\addplot [semithick, c4, mark=*, mark size=1]
table {%
0.1 5
0.2 5
0.3 5
0.4 5
0.5 5
0.6 5
0.7 5
0.8 5
0.9 5
1 5
1.1 5
1.2 5
1.3 7
1.4 7
1.5 7
1.6 7
1.7 7
1.8 7
1.9 7
2 7
2.1 7
2.2 7
2.3 7
2.4 9
2.5 9
2.6 9
2.7 9
2.8 9
2.9 9
3 9
3.1 9
3.2 9
3.3 9
3.4 9
3.5 9
3.6 11
3.7 11
3.8 11
3.9 11
4 11
4.1 11
4.2 11
4.3 11
4.4 11
4.5 11
4.6 11
4.7 11
4.8 11
4.9 11
5 13
5 13
5.25 13
5.5 13
5.75 13
6 13
6.25 13
6.5 15
6.75 15
7 15
7.25 15
7.5 15
7.75 15
8 17
8.25 17
8.5 17
8.75 17
9 17
9.25 17
9.5 19
9.75 19
10 19
10.25 19
10.5 19
10.75 19
11 19
11.25 21
11.5 21
11.75 21
12 21
12.25 21
12.5 21
12.75 23
13 23
13.25 23
13.5 23
13.75 23
14 23
14.25 23
14.5 25
14.75 25
15 25
15.25 25
15.5 25
15.75 25
16 25
16.25 27
16.5 27
16.75 27
17 27
17.25 27
17.5 27
17.75 29
18 29
18.25 29
18.5 29
18.75 29
19 29
19.25 29
19.5 31
19.75 31
20 31
};
\addplot [semithick, c2, dashed]
table {%
0.1 37
0.2 37
0.3 37
0.4 37
0.5 37
0.6 37
0.7 35
0.8 35
0.9 35
1 35
1.1 35
1.2 35
1.3 35
1.4 33
1.5 33
1.6 33
1.7 33
1.8 33
1.9 33
2 31
2.1 31
2.2 31
2.3 31
2.4 29
2.5 29
2.6 29
2.7 29
2.8 39
2.9 39
3 39
3.1 39
3.2 37
3.3 37
3.4 37
3.5 35
3.6 35
3.7 35
3.8 35
3.9 33
4 33
4.1 33
4.2 31
4.3 43
4.4 43
4.5 43
4.6 41
4.7 41
4.8 41
4.9 39
5 39
5 39
5.25 37
5.5 37
5.75 35
6 45
6.25 43
6.5 43
6.75 41
7 39
7.25 37
7.5 35
7.75 47
8 45
8.25 43
8.5 41
8.75 39
9 39
9.25 37
9.5 47
9.75 45
10 45
10.25 43
10.5 41
10.75 39
11 37
11.25 49
11.5 47
11.75 45
12 43
12.25 41
12.5 39
12.75 39
13 49
13.25 47
13.5 45
13.75 43
14 43
14.25 41
14.5 39
14.75 37
15 47
15.25 47
15.5 45
15.75 43
16 41
16.25 41
16.5 39
16.75 49
17 47
17.25 45
17.5 43
17.75 43
18 41
18.25 39
18.5 49
18.75 47
19 47
19.25 45
19.5 43
19.75 41
20 41
};
\addplot [semithick, c5, mark=star, mark size=1,]
table {%
0.1 5
0.2 5
0.3 5
0.4 5
0.5 5
0.6 5
0.7 5
0.8 5
0.9 5
1 5
1.1 5
1.2 5
1.3 5
1.4 5
1.5 5
1.6 5
1.7 5
1.8 5
1.9 5
2 5
2.1 5
2.2 5
2.3 5
2.4 5
2.5 5
2.6 5
2.7 5
2.8 7
2.9 7
3 7
3.1 7
3.2 7
3.3 7
3.4 7
3.5 7
3.6 7
3.7 7
3.8 7
3.9 7
4 7
4.1 7
4.2 7
4.3 9
4.4 9
4.5 9
4.6 9
4.7 9
4.8 9
4.9 9
5 9
5 9
5.25 9
5.5 9
5.75 9
6 11
6.25 11
6.5 11
6.75 11
7 11
7.25 11
7.5 11
7.75 13
8 13
8.25 13
8.5 13
8.75 13
9 13
9.25 13
9.5 15
9.75 15
10 15
10.25 15
10.5 15
10.75 15
11 15
11.25 17
11.5 17
11.75 17
12 17
12.25 17
12.5 17
12.75 17
13 19
13.25 19
13.5 19
13.75 19
14 19
14.25 19
14.5 19
14.75 19
15 21
15.25 21
15.5 21
15.75 21
16 21
16.25 21
16.5 21
16.75 23
17 23
17.25 23
17.5 23
17.75 23
18 23
18.25 23
18.5 25
18.75 25
19 25
19.25 25
19.5 25
19.75 25
20 25
};
\addlegendentry{{\tiny bound $2\tilde{r}\left(\tau, 10^{-5}\right))+1$}}
\addlegendentry{{\tiny degree $n$, $\epsilon_{coeff}=10^{-5}$}}
\addlegendentry{{\tiny bound $2\tilde{r}\left(\tau, 10^{-3}\right))+1$}}
\addlegendentry{{\tiny degree $n$, $\epsilon_{coeff}=10^{-3}$}}
\end{axis}

\end{tikzpicture}

%% file: fixeddepthsteps.tex
\begin{tikzpicture}

\begin{axis}[
legend style={at={(axis cs:14,-14.9)},anchor=south west},
tick align=outside,
tick pos=left,
x grid style={darkgray176},
xlabel={simulation time \(\displaystyle \tau\)},
xmin=-0.895, xmax=20.995,
xtick style={color=black},
y grid style={darkgray176},
ylabel={\(\displaystyle \log\left(\epsilon_{QSP}\right)\)},
ymin=-15.440177419123, ymax=0.709987193212572,
ytick style={color=black},
mark size=1
]
\addplot [semithick, c1, solid]
table {%
0.1 -10.5131302993778
0.2 -8.70737946547059
0.3 -7.65151086579953
0.4 -6.90282877796809
0.5 -6.32259081857493
0.6 -5.84899744339381
0.7 -5.44908303001197
0.8 -5.10317018052219
0.9 -4.79856671551276
1 -4.52660665332788
1.1 -4.28110905377562
1.2 -4.05751029398562
1.3 -3.85234486556879
1.4 -3.66291913330279
1.5 -3.48709792128036
1.6 -3.32316011629825
1.7 -3.16969812413537
1.8 -3.02554610824451
1.9 -2.88972765671969
2 -2.76141688859393
2.1 -2.63990905955409
2.2 -2.52459801214476
2.3 -2.414958642816
2.4 -2.31053310336708
2.5 -2.21091982122477
2.6 -2.11576467462773
2.7 -2.02475383436483
2.8 -1.93760790814313
2.9 -1.85407711313589
3 -1.77393726743634
3.1 -1.69698643921783
3.2 -1.62304212824435
3.3 -1.55193888139934
3.4 -1.48352626445654
3.5 -1.41766712811095
3.6 -1.35423611851299
3.7 -1.29311839209893
3.8 -1.2342085020173
3.9 -1.17740942940016
4 -1.12263173746949
4.1 -1.06979283027371
4.2 -1.01881630092119
4.3 -0.969631356668784
4.4 -0.922172310258008
4.5 -0.876378128558503
4.6 -0.832192030952662
4.7 -0.789561131032602
4.8 -0.748436116125917
4.9 -0.708770959955467
5 -0.670522664399677
5 -0.670522664399672
5.25 -0.58084317695467
5.5 -0.499213631168596
5.75 -0.425157007510904
6 -0.358266748182742
6.25 -0.0444371996719006
6.5 -0.150484116762131
6.75 -0.150090068389503
7 -0.150057753930041
7.25 0.00202827701507732
7.5 -0.0581286387772557
7.75 -0.150492362042686
8 -0.150492776149763
8.25 -0.0351240188430184
8.5 -0.0270590638790042
8.75 -0.0241111982572249
9 -0.026152011169473
9.25 -0.0330230094268977
9.5 -0.0441106078344044
9.75 -0.0459837382630354
10 -0.0457213604058915
10.25 -0.0473378702883847
10.5 -0.0513490239687149
10.75 -0.0576311399275008
11 -0.0658250610424917
11.25 -0.075421665250917
11.5 -0.0769716557633951
11.75 -0.0630696790693881
12 -0.0539006500077306
12.25 -0.0496626076522263
12.5 -0.0503609568635914
12.75 -0.0557087777482286
13 -0.0638990940819165
13.25 -0.0605219759748804
13.5 -0.057465545344096
13.75 -0.0580460341746135
14 -0.0627396254858239
14.25 -0.0715516605536015
14.5 -0.0699557311448398
14.75 -0.0616027064210752
15 -0.0540084821786378
15.25 -0.0480510268794578
15.5 -0.0444355278940433
15.75 -0.0435664899065742
16 -0.045463673719012
16.25 -0.0492990456077447
16.5 -0.0529565573369244
16.75 -0.0546105817504693
17 -0.0548813603577568
17.25 -0.0548696008797123
17.5 -0.0549965579054327
17.75 -0.0552395722295282
18 -0.0554718167253953
18.25 -0.0556403777906607
18.5 -0.0557724753664075
18.75 -0.0559308249176515
19 -0.0560349983784662
19.25 -0.0556360164341302
19.5 -0.0539330937863766
19.75 -0.0508192166346898
20 -0.0476898239535235
};
\addplot [semithick, c2, dotted]
table {%
0.1 -14.6428177443205
0.2 -14.7060790276532
0.3 -14.3776442453529
0.4 -13.3840459476559
0.5 -12.4309448569764
0.6 -11.6410621530594
0.7 -10.9730327086251
0.8 -10.394506023074
0.9 -9.88451076925109
1 -9.42865304190391
1.1 -9.01663097436496
1.2 -8.64083031012581
1.3 -8.29547617870655
1.4 -7.97607841543502
1.5 -7.67907755250623
1.6 -7.40160425617085
1.7 -7.14131167082947
1.8 -6.89625574708061
1.9 -6.66480806294591
2 -6.44559103302131
2.1 -6.23742889344608
2.2 -6.03931012708846
2.3 -5.85035823967751
2.4 -5.66980875995324
2.5 -5.49699093951383
2.6 -5.33131304206714
2.7 -5.17225041415657
2.8 -5.01933572402159
2.9 -4.87215091333344
3 -4.730320516219
3.1 -4.59350607274334
3.2 -4.46140143035627
3.3 -4.33372876778687
3.4 -4.21023521369787
3.5 -4.09068995486237
3.6 -3.9748817523732
3.7 -3.86261679798244
3.8 -3.75371685648845
3.9 -3.64801764957194
4 -3.54536744426783
4.1 -3.44562581576877
4.2 -3.34866255944175
4.3 -3.25435673089956
4.4 -3.1625957964691
4.5 -3.07327487921917
4.6 -2.986296087878
4.7 -2.90156791800385
4.8 -2.8190047162321
4.9 -2.73852619982741
5 -2.66005702482833
5 -2.66005702482833
5.25 -2.47222062228078
5.5 -2.29551395395277
5.75 -2.12907532250999
6 -1.97215605321608
6.25 -1.82410224259865
6.5 -1.68434010221784
6.75 -1.55236408374245
7 -1.42772717888894
7.25 -1.31003293685209
7.5 -1.19892885041452
7.75 -1.09410084193876
8 -0.995268640052803
8.25 -0.902181882675481
8.5 -0.814616816030302
8.75 -0.732373485249653
9 -0.655273332062378
9.25 -0.583157130307247
9.5 -0.515883201637369
9.75 -0.453325862493262
10 -0.395374059712031
10.25 -0.341930156311588
10.5 -0.150509621047357
10.75 0.0563780386611017
11 -0.150514934159792
11.25 -0.150511925360324
11.5 -0.150513528313178
11.75 -0.111889692785664
12 -0.0881741460152175
12.25 -0.068541846554688
12.5 -0.0529659580386482
12.75 -0.0414191675658878
13 -0.0338702692961804
13.25 -0.0302796667360318
13.5 -0.0305934313126949
13.75 -0.0347354643094496
14 -0.0425656270914728
14.25 -0.0429783914266633
14.5 -0.0421685752304192
14.75 -0.043711265275571
15 -0.0479851095363582
15.25 -0.0548651781989292
15.5 -0.0641381692255585
15.75 -0.0754573716605926
16 -0.0791823485784915
16.25 -0.150416418802503
16.5 -0.150514905105775
16.75 -0.0541398928919829
17 -0.0500663412739889
17.25 -0.0487527616354508
17.5 -0.0501070445426274
17.75 -0.0528341767031907
18 -0.0531204553770341
18.25 -0.0514352191910898
18.5 -0.0509686966751135
18.75 -0.0526601064579604
19 -0.0563983147301061
19.25 -0.0708425771021673
19.5 -0.0754750453190944
19.75 -0.116517698592088
20 -0.150514171286558
};
\addplot [semithick, c3,densely dotted]
table {%
0.1 -14.6132282145491
0.2 -14.7060790276532
0.3 -14.439986686344
0.4 -14.6091025850598
0.5 -14.5413701388734
0.6 -14.4882894779408
0.7 -14.5008169135492
0.8 -14.5958763236144
0.9 -14.4264709407009
1 -14.1707621044267
1.1 -14.0814769199378
1.2 -13.7890181903104
1.3 -13.393370008537
1.4 -12.9642508586468
1.5 -12.5517092616481
1.6 -12.1634590940191
1.7 -11.7971024760348
1.8 -11.4525189273122
1.9 -11.1263677265078
2 -10.8171686577469
2.1 -10.5233525337513
2.2 -10.2434063744091
2.3 -9.97620717508218
2.4 -9.72063923941615
2.5 -9.47577634574831
2.6 -9.24078869958801
2.7 -9.01494100136555
2.8 -8.79757642677352
2.9 -8.58811052595393
3 -8.38601727032896
3.1 -8.19082187980759
3.2 -8.00209584397576
3.3 -7.81944941220033
3.4 -7.64252812831004
3.5 -7.47100840024716
3.6 -7.30459392452551
3.7 -7.1430128508521
3.8 -6.98601506333062
3.9 -6.83336996961877
4 -6.68486454905576
4.1 -6.54030158561004
4.2 -6.39949813790797
4.3 -6.26228422482406
4.4 -6.12850160505637
4.5 -5.99800274353127
4.6 -5.8706498635864
4.7 -5.74631410624953
4.8 -5.62487478051355
4.9 -5.50621869487069
5 -5.39023954676295
5 -5.39023954676295
5.25 -5.11136160861462
5.5 -4.84719521536012
5.75 -4.59651913813342
6 -4.35826933734815
6.25 -4.13151335452001
6.5 -3.91542973321983
6.75 -3.70929132987513
7 -3.51245166495285
7.25 -3.32433367423724
7.5 -3.14442037210072
7.75 -2.97224705057488
8 -2.80739472194902
8.25 -2.64948457538876
8.5 -2.49817326631375
8.75 -2.35314889393429
9 -2.21412755098305
9.25 -2.08085035191495
9.5 -1.95308086339129
9.75 -1.83060287473428
10 -1.71321845710464
10.25 -1.60074626902837
10.5 -1.4930200730587
10.75 -1.38988743415841
11 -1.29120857511408
11.25 -1.19685536815479
11.5 -1.10671044511139
11.75 -1.02066641104665
12 -0.938625148412029
12.25 -0.860497200518091
12.5 -0.786201224500948
12.75 -0.715663505068056
13 -0.648817521140199
13.25 -0.585603558087194
13.5 -0.525968358584607
13.75 -0.46986480518506
14 -0.417251627474083
14.25 -0.368093126120475
14.5 -0.322358905168171
14.75 -0.0619087037896777
15 -0.141270193246499
15.25 -0.0970012977240838
15.5 -0.150514987975657
15.75 -0.14432365321666
16 -0.118756623072049
16.25 -0.0965228889382488
16.5 -0.0776210630220034
16.75 -0.0620501512100387
17 -0.0498080478444988
17.25 -0.0408895562667373
17.5 -0.0352837847600222
17.75 -0.0329707265860647
18 -0.0339167849887229
18.25 -0.0380689522742221
18.5 -0.0431980805864973
18.75 -0.0415369604309477
19 -0.0413201134531207
19.25 -0.0436259151291075
19.5 -0.0484805467670702
19.75 -0.0558265371643836
20 -0.0654568690605047
};
\addplot [semithick, c4,dashed]
table {%
0.1 -14.6428177443205
0.2 -14.7060790276532
0.3 -14.3606532253891
0.4 -14.6091025850598
0.5 -14.5413701388734
0.6 -14.3471678461672
0.7 -14.2684790864161
0.8 -14.556649761519
0.9 -14.2042249645525
1 -14.4687199747471
1.1 -14.344111004132
1.2 -14.3162152996754
1.3 -14.3253908681432
1.4 -14.4911614157162
1.5 -14.4999287845363
1.6 -14.4762654805884
1.7 -14.3710580861357
1.8 -14.3405480653939
1.9 -14.4063085510039
2 -14.4632280763567
2.1 -14.4238635306474
2.2 -14.3372519301265
2.3 -14.2177550659845
2.4 -14.004864199993
2.5 -13.8681971160571
2.6 -13.5834236290124
2.7 -13.3231163680733
2.8 -13.0600949318266
2.9 -12.7920063405001
3 -12.5334045752032
3.1 -12.2794270560391
3.2 -12.0362404608683
3.3 -11.8000521625749
3.4 -11.5703411199776
3.5 -11.3478694495421
3.6 -11.1316030643268
3.7 -10.921470540256
3.8 -10.7170799004082
3.9 -10.5182802562591
4 -10.3246960414563
4.1 -10.1361148268145
4.2 -9.95227136634341
4.3 -9.77298571203342
4.4 -9.5980421248004
4.5 -9.42725048757266
4.6 -9.26043056050298
4.7 -9.09741922259846
4.8 -8.93805923656218
4.9 -8.78220474023576
5 -8.6297211883498
5 -8.6297211883498
5.25 -8.26242669946711
5.5 -7.91358823378854
5.75 -7.58163131844347
6 -7.26518268768467
6.25 -6.96303786278284
6.5 -6.67413485540213
6.75 -6.39753254953742
7 -6.13239319638627
7.25 -5.87796788954336
7.5 -5.63358458406329
7.75 -5.39863802887239
8 -5.17258131801435
8.25 -4.9549187434909
8.5 -4.74519971338027
8.75 -4.54301355522113
9 -4.34798505393612
9.25 -4.1597706013538
9.5 -3.97805486238074
9.75 -3.80254787554141
10 -3.63298252270597
10.25 -3.46911231350237
10.5 -3.31070943870955
10.75 -3.15756305500349
11 -3.00947776922024
11.25 -2.86627229538973
11.5 -2.72777826193467
11.75 -2.59383914979521
12 -2.46430934512964
12.25 -2.33905329261418
12.5 -2.21794473732109
12.75 -2.10086604485709
13 -1.98770759085882
13.25 -1.8783672121243
13.5 -1.77274971269412
13.75 -1.67076641905534
14 -1.57233477938715
14.25 -1.47737800239643
14.5 -1.38582473183776
14.75 -1.29760875327329
15 -1.21266873002146
15.25 -1.13094796558019
15.5 -1.05239419008608
15.75 -0.976959368602341
16 -0.904599529207735
16.25 -0.835274608993186
16.5 -0.768948316158505
16.75 -0.705588006435389
17 -0.645164572040513
17.25 -0.587652341273703
17.5 -0.533028986709902
17.75 -0.481275439672452
18 -0.432375808297086
18.25 -0.386317295970504
18.5 -0.343090116215341
18.75 -0.302687399142449
19 -0.150514981826002
19.25 -0.150514677523846
19.5 -0.144117289407415
19.75 -0.169279077459961
20 -0.142988583700012
};
\addplot [semithick, c5,densely dashed]
table {%
0.1 -14.6428177443205
0.2 -14.7060790276532
0.3 -14.3606532253891
0.4 -14.6091025850598
0.5 -14.5413701388734
0.6 -14.3471678461672
0.7 -14.5008169135492
0.8 -14.5958763236144
0.9 -14.4264709407009
1 -14.5071549418184
1.1 -14.3183208815034
1.2 -14.454589770191
1.3 -14.3253908681432
1.4 -14.4911614157162
1.5 -14.3187560755316
1.6 -14.4762654805884
1.7 -14.3710580861357
1.8 -14.4612869931631
1.9 -14.4063085510039
2 -14.4632280763567
2.1 -14.3728726570178
2.2 -14.2022693996738
2.3 -14.3198843803899
2.4 -14.4140962467356
2.5 -14.3239583357948
2.6 -14.2566427638203
2.7 -14.4494190840838
2.8 -14.1958433670214
2.9 -14.4976828439991
3 -14.4293749677236
3.1 -14.238661726422
3.2 -14.31843684799
3.3 -14.4186027656394
3.4 -14.1682084078762
3.5 -14.1766459643192
3.6 -14.3491630875336
3.7 -14.1078774502773
3.8 -14.2916011626515
3.9 -14.2494331822437
4 -14.0562196932782
4.1 -13.8693232378618
4.2 -13.7504464292936
4.3 -13.5898808105331
4.4 -13.4175323211033
4.5 -13.2195970253315
4.6 -13.0266727429467
4.7 -12.8317851266862
4.8 -12.6358598535848
4.9 -12.4444419809295
5 -12.258229717953
5 -12.2580771918302
5.25 -11.8045280348156
5.5 -11.3725307569938
5.75 -10.9608802812743
6 -10.5678093730096
6.25 -10.1918428142685
6.5 -9.83177283829872
6.75 -9.4864378797235
7 -9.15480921500625
7.25 -8.83596937370537
7.5 -8.52910123807817
7.75 -8.23345832988962
8 -7.94837068040926
8.25 -7.67322827483534
8.5 -7.40747690276279
8.75 -7.15061085079027
9 -6.90216748377836
9.25 -6.6617226414779
9.5 -6.428886520408
9.75 -6.20330010491297
10 -5.9846321074833
10.25 -5.77257620696505
10.5 -5.56684867976825
10.75 -5.36718628869606
11 -5.17334441873728
11.25 -4.98509542193617
11.5 -4.80222714660386
11.75 -4.62454162486537
12 -4.45185389955838
12.25 -4.28399097266655
12.5 -4.12079086182563
12.75 -3.96210175058164
13 -3.8077812225536
13.25 -3.65769556979965
13.5 -3.51171916683212
13.75 -3.36973390355788
14 -3.23162867048714
14.25 -3.09729889094982
14.5 -2.96664609539084
14.75 -2.83957753364099
15 -2.71600582131248
15.25 -2.59584861708814
15.5 -2.47902832799969
15.75 -2.36547184010088
16 -2.25511027226211
16.25 -2.14787875104913
16.5 -2.0437162048867
16.75 -1.94256517588948
17 -1.84437164792633
17.25 -1.74908488963461
17.5 -1.65665731123302
17.75 -1.56704433410123
18 -1.48020427219656
18.25 -1.39609822447261
18.5 -1.31468997753655
18.75 -1.23594591785143
19 -1.1598349528454
19.25 -1.08632844033018
19.5 -1.01540012566437
19.75 -0.947026086115371
20 -0.881184681877768
};
\addplot [semithick, c6, dashdotted]
table {%
0.1 -14.6428177443205
0.2 -14.7060790276532
0.3 -14.3606532253891
0.4 -14.6091025850598
0.5 -14.6173242462285
0.6 -14.4882894779408
0.7 -14.5008169135492
0.8 -14.556649761519
0.9 -14.4264709407009
1 -14.4687199747471
1.1 -14.344111004132
1.2 -14.3162152996754
1.3 -14.3253908681432
1.4 -14.4911614157162
1.5 -14.3187560755316
1.6 -14.4762654805884
1.7 -14.3710580861357
1.8 -14.4612869931631
1.9 -14.389061275134
2 -14.3858237055472
2.1 -14.3728726570178
2.2 -14.2022693996738
2.3 -14.3198843803899
2.4 -14.4140962467356
2.5 -14.3239583357948
2.6 -14.4890516217157
2.7 -14.3380599706427
2.8 -14.1958433670214
2.9 -14.4976828439991
3 -14.4485932128614
3.1 -14.238661726422
3.2 -14.31843684799
3.3 -14.2845413016212
3.4 -14.0696704184656
3.5 -14.1766459643192
3.6 -14.3779491739492
3.7 -14.2043171687917
3.8 -14.3629548483496
3.9 -14.3796724218331
4 -14.3336920545434
4.1 -14.3069060714109
4.2 -14.0240915518124
4.3 -14.3128260877602
4.4 -14.3026762050971
4.5 -14.2713954392986
4.6 -14.3438826512698
4.7 -14.1057027455321
4.8 -14.3746558466447
4.9 -14.2110736379336
5 -14.1967025065831
5 -14.3240475927287
5.25 -14.2959744612517
5.5 -14.2746720290211
5.75 -14.0240915518124
6 -13.859472084101
6.25 -13.6675650887139
6.5 -13.2722855334777
6.75 -12.8757430602775
7 -12.4944844144743
7.25 -12.1175748674043
7.5 -11.7501637257563
7.75 -11.395180978226
8 -11.0523175952268
8.25 -10.7209027277859
8.5 -10.4003658771915
8.75 -10.0901598478046
9 -9.78971453733848
9.25 -9.49848412547852
9.5 -9.21603254564076
9.75 -8.94192603860657
10 -8.67575338730077
10.25 -8.41717261676628
10.5 -8.16584295721379
10.75 -7.92145209261457
11 -7.68370774500527
11.25 -7.45233837185967
11.5 -7.22709027284788
11.75 -7.00772662685122
12 -6.79402569047102
12.25 -6.58577980441811
12.5 -6.38279422488456
12.75 -6.18488601504007
13 -5.9918832506233
13.25 -5.80362413462885
13.5 -5.61995626704658
13.75 -5.44073596737581
14 -5.26582765570009
14.25 -5.09510329414625
14.5 -4.92844187436145
14.75 -4.76572895077669
15 -4.60685621324645
15.25 -4.45172109585311
15.5 -4.30022641681957
15.75 -4.1522800502573
16 -4.0077946224471
16.25 -3.86668723327516
16.5 -3.72887919961902
16.75 -3.59429581858357
17 -3.46286614929688
17.25 -3.33452281136512
17.5 -3.20920179878568
17.75 -3.08684230796666
18 -2.9673865787365
18.25 -2.85077974736867
18.5 -2.7369697106479
18.75 -2.62590700018527
19 -2.51754466623301
19.25 -2.41183817031368
19.5 -2.30874528606759
19.75 -2.20822600775688
20 -2.11024246593524
};
\addplot [semithick, c7, densely dashdotted]
table {%
0.1 -14.6132282145491
0.2 -14.7060790276532
0.3 -14.3606532253891
0.4 -14.6091025850598
0.5 -14.5413701388734
0.6 -14.4882894779408
0.7 -14.5008169135492
0.8 -14.556649761519
0.9 -14.4264709407009
1 -14.4687199747471
1.1 -14.344111004132
1.2 -14.454589770191
1.3 -14.2195315936155
1.4 -14.4911614157162
1.5 -14.3187560755316
1.6 -14.4762654805884
1.7 -14.5187303485316
1.8 -14.4612869931631
1.9 -14.4063085510039
2 -14.4632280763567
2.1 -14.3728726570178
2.2 -14.4604627482322
2.3 -14.3198843803899
2.4 -14.4140962467356
2.5 -14.3239583357948
2.6 -14.2566427638203
2.7 -14.3380599706427
2.8 -14.1958433670214
2.9 -14.4976828439991
3 -14.4293749677236
3.1 -14.2035811627774
3.2 -14.31843684799
3.3 -14.4186027656394
3.4 -14.1682084078762
3.5 -14.1766459643192
3.6 -14.3779491739492
3.7 -14.2043171687917
3.8 -14.3629548483496
3.9 -14.3796724218331
4 -14.3681424213149
4.1 -14.3046115174278
4.2 -14.154525509722
4.3 -14.3128260877602
4.4 -14.3026762050971
4.5 -14.2713954392986
4.6 -14.3438826512698
4.7 -14.3669426093129
4.8 -14.2340369057513
4.9 -14.2110736379336
5 -14.1967025065831
5 -14.1967025065831
5.25 -14.2959744612517
5.5 -14.1395157266568
5.75 -14.2417411635097
6 -14.3797877927245
6.25 -14.2241408488127
6.5 -14.3382743351326
6.75 -14.3725249191999
7 -14.2263877730539
7.25 -14.2973862114603
7.5 -14.2496232085254
7.75 -14.2903935968738
8 -14.202014781031
8.25 -14.1822602798917
8.5 -14.1521394012771
8.75 -14.2244227972284
9 -14.1901314200522
9.25 -13.9438659047992
9.5 -13.7581411772599
9.75 -13.5117066951645
10 -13.2071529468623
10.25 -12.8888830920129
10.5 -12.5692479804614
10.75 -12.2690657176469
11 -11.9680651680881
11.25 -11.6749993918909
11.5 -11.3885340059743
11.75 -11.1089039760804
12 -10.835860381873
12.25 -10.5692608214855
12.5 -10.3089344339641
12.75 -10.0545873813666
13 -9.80601674738179
13.25 -9.56303171966923
13.5 -9.32543871747429
13.75 -9.09304913795329
14 -8.86569897428835
14.25 -8.64323724103892
14.5 -8.42550406065145
14.75 -8.21235988775065
15 -8.00366668603197
15.25 -7.79929529599227
15.5 -7.59912265034785
15.75 -7.40303161495358
16 -7.2109107837795
16.25 -7.02265418297814
16.5 -6.83816094066578
16.75 -6.65733483159654
17 -6.4800841449588
17.25 -6.30632138607122
17.5 -6.1359630339294
17.75 -5.96892934819544
18 -5.80514412847697
18.25 -5.64453455827181
18.5 -5.48703100523152
18.75 -5.33256687202665
19 -5.18107843632685
19.25 -5.03250470958181
19.5 -4.8867873060192
19.75 -4.74387031538769
20 -4.60370018820649
};

\addlegendentry{$n=5$}
\addlegendentry{$n=9$}
\addlegendentry{$n=13$}
\addlegendentry{$n=17$}
\addlegendentry{$n=21$}
\addlegendentry{$n=25$}
\addlegendentry{$n=31$}
\end{axis}

\end{tikzpicture}

%% file: epsivsepsi.tex
\begin{tikzpicture}

\begin{axis}[
tick align=outside,
tick pos=left,
unbounded coords=jump,
x grid style={darkgray176},
xlabel={Coefficient error \(\displaystyle -\log_{10}\left(\epsilon_{coeff}^{-1}\right)\)},
xmin=-11.0967361580687, xmax=-1.56896768244198,
xtick style={color=black},
y grid style={darkgray176},
ylabel={Error in QSP \(\displaystyle -\log_{10}\left(\epsilon_{QSP}^{-1}\right)\)},
ymin=-10.9462101608534, ymax=-1.41845320839018,
ytick style={color=black}
]
\addplot [semithick, c1, mark=o, mark size=0.5]
table {%
-10.663655772813 -10.5131302993778
-8.85789472866251 -8.70737946547059
-7.80202586484185 -7.65151084418403
-7.05334377699551 -6.90282877796809
-5.99951244176484 -5.84899744370021
-5.59959802801996 -5.44908303001197
-5.25368517834791 -5.10317018052217
-4.94908171338536 -4.79856671552794
-4.43162405161 -4.28110905376825
-4.20802529182446 -4.05751029398562
-4.00285986340195 -3.85234486556879
-5.86608958930383 -5.71557459171442
-5.40869581217498 -5.25818081432433
-5.20166541800579 -5.05115042018827
-5.00689854328671 -4.85638354541831
-4.82309268497851 -4.67257768712651
-4.48411815935483 -4.33360316151837
-4.32720018709605 -4.17668518926219
-4.17769117653182 -4.02717617869711
-4.03497984453182 -3.8844648467012
-5.48182803977205 -5.33131304206714
-5.32276541211913 -5.17225041426416
-5.16985072189568 -5.01933572402159
-5.02266591116367 -4.87215091337674
-4.74402107060014 -4.59350607274334
-4.61191642817794 -4.46140143034508
-4.4842437656212 -4.33372876778687
-4.36075021151687 -4.21023521369787
-4.12539675021077 -3.9748817523732
-4.01313179581253 -3.86261679798244
-5.44820203614568 -5.29768703821695
-5.31889487644179 -5.16837987868596
-5.07088235910601 -4.92036736127995
-4.95185150728908 -4.80133650942506
-4.83593986577684 -4.68542486793546
-4.72301199461771 -4.57249699681882
-4.50561087600591 -4.35509587816883
-4.4009094912752 -4.25039449343423
-4.29873424033246 -4.14821924248232
-4.19898831955883 -4.04847332172776
-4.10158083812312 -3.95106584029506
-5.26187660654167 -5.11136160861462
-4.99771021322272 -4.84719521536012
-4.74703413597701 -4.59651913813342
-4.50878433520429 -4.35826933734815
-4.28202835236102 -4.13151335452001
-4.06594473105005 -3.9154297332119
-5.14743368289477 -4.99691868498471
-4.91613575625706 -4.76562075841954
-4.69454017605209 -4.54402517821663
-4.48205187723327 -4.33153687938591
-4.27813529565779 -4.12762029779862
-4.08230686127571 -3.93179186344042
-5.10543374142992 -4.9549187434995
-4.89571471120007 -4.74519971338027
-4.69352855307467 -4.54301355522113
-4.49850005175574 -4.34798505393612
-4.31028559918012 -4.1597706013538
-4.12856986022274 -3.97805486238991
-5.10657480688137 -4.95605980915432
-4.91219540273511 -4.76168040494386
-4.72395916220892 -4.57344416436343
-4.54160995214566 -4.39109495429938
-4.36491034426616 -4.21439534643526
-4.1936399237284 -4.04312492590884
-4.02759378918041 -3.87707879133802
-4.95274214445894 -4.8022271466161
-4.77505662274293 -4.62454162494861
-4.60236889736588 -4.45185389955838
-4.43450597048679 -4.28399097266655
-4.27130585966445 -4.12079086182563
-4.11261674842218 -3.962101750575
-5.0098829920164 -4.8593679940754
-4.84048027668864 -4.68996527880976
-4.67542000880588 -4.52490501097056
-4.5145710237308 -4.3640560259103
-4.35780969408627 -4.20729469623864
-4.20501941465568 -4.05450441681516
-4.05609013289193 -3.9055751350609
-4.91624394853877 -4.76572895077669
-4.75737121115423 -4.60685621324645
-4.60223609365386 -4.45172109581382
-4.45074141465414 -4.30022641681957
-4.30279504808885 -4.1522800502573
-4.15830962026149 -4.00779462245359
-4.01720223110011 -3.86668723327516
-4.84596703906496 -4.6954520411612
-4.69558391477328 -4.54506891693898
-4.54848365483306 -4.39796865701256
-4.40459179824706 -4.25407680039867
-4.26383744279765 -4.11332244496552
-4.12615305633271 -3.97563805847339
-4.93964828686162 -4.78913328891843
-4.79328426185854 -4.64276926399336
-4.64991069145403 -4.49939569364341
-4.50946625508546 -4.35895125720641
-4.37189236839271 -4.22137737055921
-4.23713305115197 -4.0866180533345
-4.10513480424022 -3.95461980640372
-4.89438531310476 -4.74387031538769
-4.75421518609789 -4.60370018820649
-4.99371089923524 -4.84319590140706
-3.79298604433538 -3.64247104650219
-3.09428545904104 -2.94377046120927
-2.60247317578908 -2.451958177957
-2.22506650773899 -2.07455150990701
-3.63761291911349 -3.48709792128036
-3.24703832226931 -3.09652332443728
-2.91193188642591 -2.76141688859393
-2.61961441677135 -2.46909941893952
-2.36143481905678 -2.21091982122477
-3.58182427901443 -3.43130928118345
-3.29536789454299 -3.14485289671109
-3.03455106360758 -2.88403606577562
-2.79579439179727 -2.64527939396516
-2.57626426684969 -2.42574926901766
-2.37368037922729 -2.22316538139518
-2.18618171800146 -2.03566672016945
-2.01223094822007 -1.86171595038807
-3.01053583219918 -2.86002083436752
-2.81057202266061 -2.6600570248291
-2.6227356201129 -2.47222062228086
-2.44602895178479 -2.29551395395299
-2.2795903203419 -2.12907532250985
-2.12267105104809 -1.97215605321606
-3.05152099700384 -2.90100599917143
-2.87295041205059 -2.72243541421825
-2.7032072212676 -2.55269222343541
-2.54174286668011 -2.39122786884799
-2.38806864100141 -2.23755364316949
-2.24174770415588 -2.09123270632385
-2.10238841764672 -1.95187341981468
-2.95790971978117 -2.80739472194902
-2.79999957322089 -2.64948457538884
-2.64868826414554 -2.49817326631371
-2.50366389176624 -2.35314889393421
-2.36464254881506 -2.21412755098304
-2.23136534974688 -2.08085035191487
-2.10359586122322 -1.95308086339128
-2.90573142080799 -2.75521642297606
-2.76171642030352 -2.61120142247144
-2.62297999623628 -2.4724649984043
-2.4893239425279 -2.33880894469587
-2.36056535966448 -2.21005036183246
-2.2365353198204 -2.08602032198851
-2.11707768636777 -1.96656268853571
-2.00204806769775 -1.85153306986578
-2.74435414762744 -2.59383914979521
-2.61482434296167 -2.46430934512964
-2.48956829044644 -2.33905329261413
-2.36845973515305 -2.21794473732123
-2.25138104268905 -2.10086604485711
-2.1382225886908 -1.98770759085886
-2.02888220995635 -1.87836721212436
-2.74064611703844 -2.59013111920627
-2.61819201675249 -2.46767701892015
-2.49943484043635 -2.34891984260413
-2.38428303394628 -2.23376803611416
-2.27265076169541 -2.12213576386348
-2.16445756308906 -2.01394256525703
-2.05962804057119 -1.90911304273924
-2.7463636149198 -2.59584861708814
-2.62954332583166 -2.47902832799963
-2.51598683793308 -2.36547184010111
-2.40562527009399 -2.25511027226201
-2.29839374888099 -2.14787875104895
-2.19423120271869 -2.04371620488669
-2.0930801737215 -1.94256517588948
-2.75886779844223 -2.60835280061012
-2.64665564810053 -2.49614065026853
-2.53736987363084 -2.38685487579898
-2.43095715261277 -2.28044215478039
-2.32736709476908 -2.17685209693699
-2.22655211262165 -2.07603711478967
-2.12846730273368 -1.97795230490154
-2.03307033679167 -1.88255533895962
-2.66805966406468 -2.51754466623301
-2.56235316814566 -2.41183817031374
-2.45926028389979 -2.30874528606777
-2.35874100558887 -2.20822600775684
-2.2607574637673 -2.11024246593518
};
\end{axis}

\end{tikzpicture}

%% file: depthvstau.tex
\begin{tikzpicture}

\begin{axis}[
legend style={at={(axis cs:0,65)},anchor=north west},
tick align=outside,
tick pos=left,
x grid style={darkgray176},
xlabel={Simulation time $\tau$},
xmin=-0.895, xmax=20.995,
xtick style={color=black},
y grid style={darkgray176},
ylabel={Growth of QSP Circuit},
ymin=0, ymax=66,
ytick style={color=black},
mark size=1,
]
\addplot [semithick, c4, mark=*]
table {%
0.1 5
0.2 5
0.3 5
0.4 5
0.5 5
0.6 5
0.7 5
0.8 5
0.9 5
1 5
1.1 5
1.2 5
1.3 5
1.4 7
1.5 7
1.6 7
1.7 7
1.8 7
1.9 7
2 7
2.1 7
2.2 7
2.3 7
2.4 7
2.5 9
2.6 9
2.7 9
2.8 9
2.9 9
3 9
3.1 9
3.2 9
3.3 9
3.4 9
3.5 9
3.6 9
3.7 9
3.8 11
3.9 11
4 11
4.1 11
4.2 11
4.3 11
4.4 11
4.5 11
4.6 11
4.7 11
4.8 11
4.9 11
5 11
5 11
5.25 13
5.5 13
5.75 13
6 13
6.25 13
6.5 13
6.75 15
7 15
7.25 15
7.5 15
7.75 15
8 15
8.25 17
8.5 17
8.75 17
9 17
9.25 17
9.5 17
9.75 19
10 19
10.25 19
10.5 19
10.75 19
11 19
11.25 19
11.5 21
11.75 21
12 21
12.25 21
12.5 21
12.75 21
13 23
13.25 23
13.5 23
13.75 23
14 23
14.25 23
14.5 23
14.75 25
15 25
15.25 25
15.5 25
15.75 25
16 25
16.25 25
16.5 27
16.75 27
17 27
17.25 27
17.5 27
17.75 27
18 29
18.25 29
18.5 29
18.75 29
19 29
19.25 29
19.5 29
19.75 31
20 31
};
\addplot [semithick, c1, solid]
table {%
0.1 11
0.2 11
0.3 11
0.4 11
0.5 11
0.6 11
0.7 11
0.8 11
0.9 11
1 11
1.1 11
1.2 11
1.3 11
1.4 15
1.5 15
1.6 15
1.7 15
1.8 15
1.9 15
2 15
2.1 15
2.2 15
2.3 15
2.4 15
2.5 19
2.6 19
2.7 19
2.8 19
2.9 19
3 19
3.1 19
3.2 19
3.3 19
3.4 19
3.5 19
3.6 19
3.7 19
3.8 23
3.9 23
4 23
4.1 23
4.2 23
4.3 23
4.4 23
4.5 23
4.6 23
4.7 23
4.8 23
4.9 23
5 23
5 23
5.25 27
5.5 27
5.75 27
6 27
6.25 27
6.5 27
6.75 31
7 31
7.25 31
7.5 31
7.75 31
8 31
8.25 35
8.5 35
8.75 35
9 35
9.25 35
9.5 35
9.75 39
10 39
10.25 39
10.5 39
10.75 39
11 39
11.25 39
11.5 43
11.75 43
12 43
12.25 43
12.5 43
12.75 43
13 47
13.25 47
13.5 47
13.75 47
14 47
14.25 47
14.5 47
14.75 51
15 51
15.25 51
15.5 51
15.75 51
16 51
16.25 51
16.5 55
16.75 55
17 55
17.25 55
17.5 55
17.75 55
18 59
18.25 59
18.5 59
18.75 59
19 59
19.25 59
19.5 59
19.75 63
20 63
};
\addplot [semithick, c5, mark=star]
table {%
0.1 5
0.2 5
0.3 5
0.4 5
0.5 5
0.6 5
0.7 5
0.8 5
0.9 5
1 5
1.1 5
1.2 5
1.3 5
1.4 5
1.5 5
1.6 5
1.7 5
1.8 5
1.9 5
2 5
2.1 5
2.2 5
2.3 5
2.4 5
2.5 5
2.6 5
2.7 5
2.8 5
2.9 5
3 7
3.1 7
3.2 7
3.3 7
3.4 7
3.5 7
3.6 7
3.7 7
3.8 7
3.9 7
4 7
4.1 7
4.2 7
4.3 7
4.4 7
4.5 7
4.6 9
4.7 9
4.8 9
4.9 9
5 9
5 9
5.25 9
5.5 9
5.75 9
6 9
6.25 11
6.5 11
6.75 11
7 11
7.25 11
7.5 11
7.75 11
8 13
8.25 13
8.5 13
8.75 13
9 13
9.25 13
9.5 13
9.75 15
10 15
10.25 15
10.5 15
10.75 15
11 15
11.25 15
11.5 15
11.75 17
12 17
12.25 17
12.5 17
12.75 17
13 17
13.25 17
13.5 19
13.75 19
14 19
14.25 19
14.5 19
14.75 19
15 19
15.25 21
15.5 21
15.75 21
16 21
16.25 21
16.5 21
16.75 21
17 23
17.25 23
17.5 23
17.75 23
18 23
18.25 23
18.5 23
18.75 23
19 25
19.25 25
19.5 25
19.75 25
20 25
};
\addplot [semithick, c2, dashed]
table {%
0.1 11
0.2 11
0.3 11
0.4 11
0.5 11
0.6 11
0.7 11
0.8 11
0.9 11
1 11
1.1 11
1.2 11
1.3 11
1.4 11
1.5 11
1.6 11
1.7 11
1.8 11
1.9 11
2 11
2.1 11
2.2 11
2.3 11
2.4 11
2.5 11
2.6 11
2.7 11
2.8 11
2.9 11
3 15
3.1 15
3.2 15
3.3 15
3.4 15
3.5 15
3.6 15
3.7 15
3.8 15
3.9 15
4 15
4.1 15
4.2 15
4.3 15
4.4 15
4.5 15
4.6 19
4.7 19
4.8 19
4.9 19
5 19
5 19
5.25 19
5.5 19
5.75 19
6 19
6.25 23
6.5 23
6.75 23
7 23
7.25 23
7.5 23
7.75 23
8 27
8.25 27
8.5 27
8.75 27
9 27
9.25 27
9.5 27
9.75 31
10 31
10.25 31
10.5 31
10.75 31
11 31
11.25 31
11.5 31
11.75 35
12 35
12.25 35
12.5 35
12.75 35
13 35
13.25 35
13.5 39
13.75 39
14 39
14.25 39
14.5 39
14.75 39
15 39
15.25 43
15.5 43
15.75 43
16 43
16.25 43
16.5 43
16.75 43
17 47
17.25 47
17.5 47
17.75 47
18 47
18.25 47
18.5 47
18.75 47
19 51
19.25 51
19.5 51
19.75 51
20 51
};
\addlegendentry{degree $n$, $\epsilon_{QSP}=10^{-4}$}
\addlegendentry{depth $D$, $\epsilon_{QSP}=10^{-4}$}
\addlegendentry{degree $n$, $\epsilon_{QSP}=10^{-2}$}
\addlegendentry{depth $D$, $\epsilon_{QSP}=10^{-2}$}
\end{axis}

\end{tikzpicture}

%% file: noiseappendix.tex
\subsection{Validity of ZNE for QSP}
The validity of the ZNE method for a given problem relies on the expansion of the noisy expectation value $\braket{O}_{\lambda}$ as a function of noise parameter $\lambda$. If the expectation is infinitely differentiable with respect to $\lambda$, then one can obtain the expectation value expansion with a truncated Taylor expansion around $\lambda=0$. Generally, one cannot assume that the arbitrary noisy expectation value is differentiable with respect to the noise parameter. However, $\langle O \rangle _\lambda$ is infinitely differentiable w.r.t. $\lambda$ when the noise is a convex combination of Pauli channels, except for some non-analytic cases. 

%% file: results-math.tex
\subsection{Lower Bounds on Mitigation Cost}\label{sec:appendixbounds}

There have been several works discussing the sampling cost of performing error mitigation techniques, including \cite{FI_pap, stat_pap, Eisert}.

A general lower bound for the sampling cost of any mitigation technique under the noise model we consider can be found in \cite{Eisert}, and it is exponential in circuit depth. While there are proofs of the existence of circuits scaling exponentially in the volume of the circuit, we do not expect our instances to adhere to this worst-case scenario (rapidly mixing circuits). Therefore, the general bound for local depolarizing noise, which is the focus of this work, can be reformulated from \cite{Eisert}:
given an error mitigation algorithm to approximate the expectation value within $\epsilon<0.5$ with high probability $1-\delta$, there exists an observable $A$ and input state $\rho$ such that the number of noisy copies needed is lower bounded by
\begin{equation}
    M_e = \Omega(p^{-2D}),
\end{equation}
where $D$ is circuit depth and $p$ is noise parameter.

\begin{figure}[h]
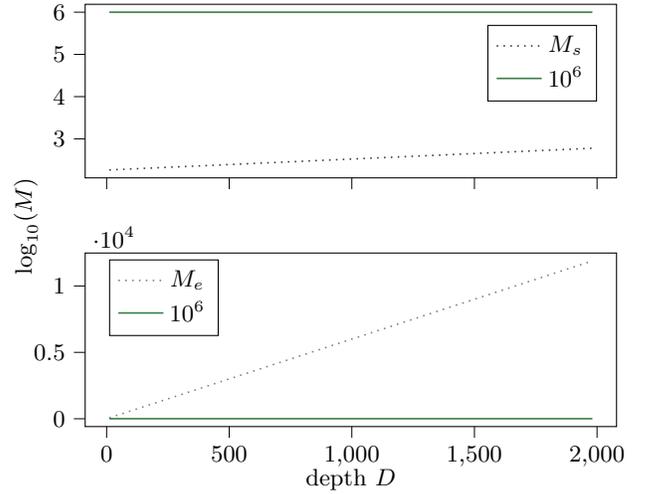

    \centering
    \include{shotsvstau}
    \caption{Sufficient number of shots $M$ for different shot determination strategies.  Above: $M_s$ computed with \cref{eq:statboundqsp} vs with a fixed $10^{6}$ shots. Below: $M_e$ computed from \cite{Eisert} vs with a fixed $10^{6}$ shots. In both cases, we assume the precision is $\mathcal{O}\left(10^{-2}\right)$ and $p_{QSP}=1$.}
    \label{fig:shotsvstau}
\end{figure}

$M_e$ gives an unattainable number of samples required for mitigation to be successful up to $\epsilon = 0.5$ with high probability, even for the smallest of our test cases, as can be seen in \ref{fig:shotsvstau}. Therefore, we choose to check a bound with looser restrictions on allowed error, $M_s$, which provides an executable number of samples required for the mitigation protocol to work up to error $\epsilon_{max} = b_{max}+\epsilon$ with failure probability $\delta = \mathcal{O}(\epsilon)$; this can be found in \cite{stat_pap}.

We are going to tailor the bound for $M_s$ from \cite{stat_pap} for any QSP protocol. The parameter specifying how far we allow the bias to be erroneous, $\epsilon$, is set to the same order as the noisy error coming from QSP, $\epsilon_{QSP}$. This ensures that the full algorithm error, including the QSP error and the mitigation error, remains bound. The second parameter fixed is the probability of failure of QEM, $\delta$. This is fixed to be $(1-p_{QSP})$ where $p_{QSP}$ is the probability of success of QSP approximation after post-selection given in \cref{eq:qspsuccprob}. \footnote{We are assuming access to $M_s$ required samples after samples have been disregarded during post-selection. If post-selection were included in the protocol explicitly, one would need to double the samples to account for the ones disregarded in post-selection.} Finally, the depth of the circuit, or the number of layers, is fixed to be $d_{QSP}$, the depth of the QSP circuit used from \cref{eq:qspdepth}, \cref{eq:qspdepthhs}. The number of samples needed for the QSP circuit is finally given by
\begin{equation}
    M_s \geq \frac{\log(\frac{2}{(1-p_{QSP})})}{(1-\lambda)^{d_{QSP}} 4\ln(2)n(R+1) \epsilon_{QSP}^2}.\label{eq:statboundqsp}
\end{equation}


In \cref{fig:shotsvstau}, we compute $M_s$ for our QEM protocol with precision $10^{-2}$, depth $D$, and success probability $\mathcal{O}(1)$

$M_s$ is not considered loose, as it reaches the value required, i.e. $\epsilon_{max}$, with the number of shots provided. However, as it requires the estimate to be accurate up to $\epsilon_{max}$, it leads to a large error allowed when the maximum bias of the estimator is large\footnote{In our case, it was set to 1, due to the trivial estimator being able to reach the following bias for our chosen observable}. 

A lower bound $M_s$ lies between $\mathcal{O}(10^2)$ and $\mathcal{O}(10^3)$ for $\epsilon = \mathcal{O}(10^{-2})$ and $\mathcal{O}(10^5)$ and $\mathcal{O}(10^8)$ for $\epsilon = \mathcal{O}(10^{-4})$. In our case, this leads to an artificially low estimate in comparison to other bounds.

While we do observe in numerical simulations that even using the bound $M_s$ found in \cite{stat_pap}, the recovery of the expectation value within $\epsilon$ is possible in certain very short-time scale regimes, we see that the bound does not provide sufficient number of samples required for QEM to perform up to precision of the algorithm $\epsilon$, and only performs up to $\epsilon_{\max}$.

\begin{figure}
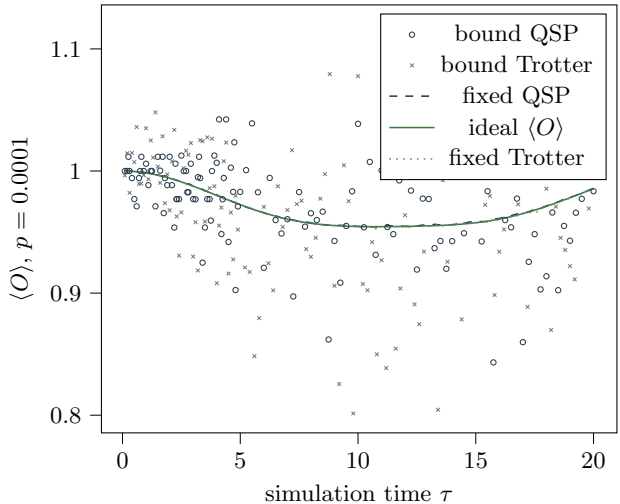

    \centering
    \include{fixedboundcompq40.1}
    \caption{Bias vs simulation time, for both a fixed and bound number of shots, for $4$-qubit QSP simulations with $p=0.0001$, scaling steps of $[1, 2, 3]$. We select the best-performing mitigation for each plot. Already for the very low noise, $M_s$ undersamples, and we cannot recover expectation values. }
    \label{fig:fixedboundcomp}
\end{figure}

In \cref{fig:fixedboundcomp} we see that, while the QEM protocol does not always succeed within the error set by algorithmic precision, fixing the number of shots allows much better recovery of error than the $M_s$ bound version. Therefore, we have set a fixed number of shots to be performed in our simulations. The number of shots is equal to $5\cdot10^{6}$. Thus, while under-sampling according to $M_e$, it still allows us to reach the desired estimation result in some cases.

\subsection{Costs of Trotter vs QSP}
The sampling cost required for ZNE for the Trotter-based Hamiltonian simulation can be derived similarly to the QSP case, as all bounds are determined by noise level and depth. While depth for the same time $\tau$ depends on the method used, this does not alter the underlying derivations on sampling bounds. {The depth of our trotter simulations is always longer than the depth of the QSP simulations due to our artificially generous assumption that the QSP oracle has depth one. Therefore, like in the QSP case, the bound from \cite{Eisert} is orders of magnitude above a feasible number of shots.}

The bound $M_s$ derived in \cite{stat_pap} can also be easily adapted to Trotter simulation. The simplest scaling of an algorithmic error in Trotterisation is dependent on the order of Trotterisation used as well as the time step size $dt$. In our numerical calculation, the first-order Trotter expansion is considered to obtain timescales comparable to QSP. The estimated algorithmic error scales as $\mathcal{O}(dt^2)$ \cite{Trotter}. Hence, the accuracy we want to achieve is set to $dt^2$ while calculating the number of shots $M_s$. The depth of the circuit is calculated during numeric simulations on Pennylane after circuit optimization.

However, as in the case with QSP, the bound $M_s$ calculates a number of shots wished to reach $b_{max} + \epsilon$, hence leaving a wide gap for error. So instead, we use the same fixed number of shots for the Trotter and QSP methods, $M=5\cdot10^{6}$.

\subsection{Feasible QEM regimes}\label{sec:qemfeasibleregimes}
In \cref{sec:appendixbounds}, we discussed the sampling cost involved in performing error mitigation and the provable infeasibility of QEM for large circuits due to exponential costs. In this section, we will discuss the wider infeasibility of ZNE by using physical intuition about the system in question.

ZNE relies on the fact that increasing the noise on the circuit will lead to expectation values changing as a function of the noise parameter. However, this is not the case once the system of interest reaches a steady state. While the thermalization of the arbitrary system is an open question, certain noise models do reach a steady state as the noise parameter is increased. 
Either adding more noise to the circuit or increasing its length accumulates error, resulting in the simulation evolving towards the fully mixed state in our chosen noise model. 
If this steady state is reached, the bounds discussed in the previous section become trivial.

This trend toward a steady state in our noise model sets a limit on how far mitigation can go, given a certain noise strength, even in the case of infinite resources.

\begin{figure}
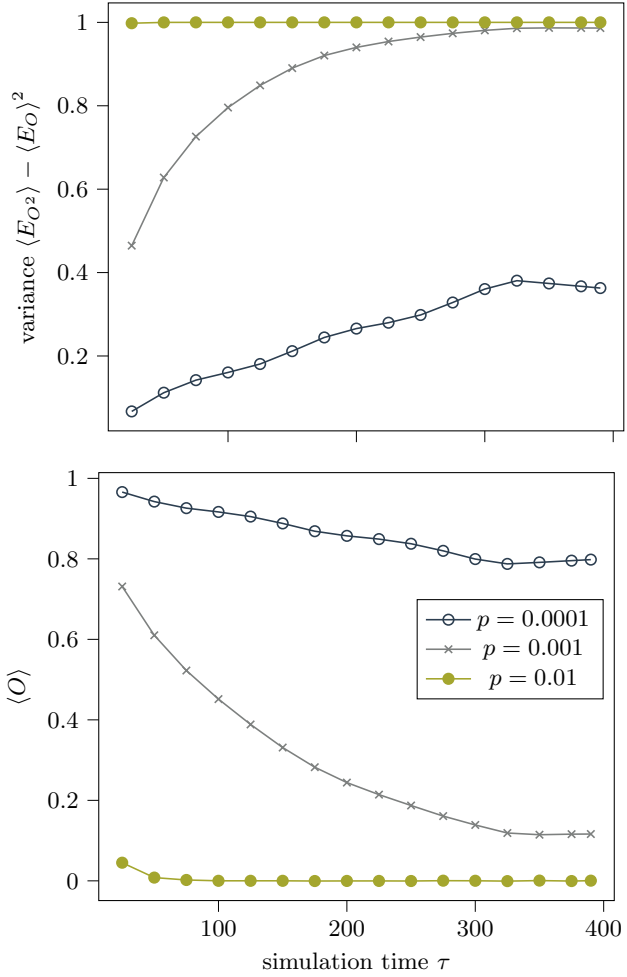

    \centering
    \begin{subfigure}[b]{0.47\textwidth}
      \include{varvslongtime}
    \end{subfigure}
    \begin{subfigure}[b]{0.47\textwidth}
       \include{expvslongtime}
    \end{subfigure}
    \caption{Limiting scenarios for our $4$-qubit QEM-QSP protocol. Above: variance vs increasing simulation time (and thereby depth). Below: noisy expectation values vs increasing simulation times. Note that the decay is steeper for increasing $p$}
    \label{fig:longtimeexpvsnoise}
\end{figure}

In addition to the decay to a steady state we discussed in the main text with increasing noise, we consider a trend to a steady state arising from increasing circuit length. We extend simulation times up to $300$ for the $4$-qubit QSP simulations, and \cref{fig:longtimeexpvsnoise} shows how the variance increases and decays as circuits get longer. For large noise parameters, namely $0.01$, the time scales chosen are beyond reach for ZNE, as we can see variance and expectation values have decayed to $1$ and $0$ respectively; that is, the expected maximally mixed state was reached. From the plots, it is also visible that the noise level $0.001$ is headed for the same stable values but on a longer timescale; therefore, a longer timescale would be needed to see the expectation value decay to $0$ as expected. The smallest noise parameter $0.0001$ does not exhibit the exponential decay to $1$ in variance and $0$ in expectation value in our chosen time-scale. For this noise level, even longer times would be required to see the trend to a steady, fully mixed state. 

\subsection{Extrapolation Techniques}
\label{sec:appendixextrapolation}
Within the second step of ZNE, the extrapolation of noisy data, we will consider three functional fits.

Richardson extrapolation is a popular choice in existing QEM literature due to its convergence behavior \cite{temmeerrormitigation2017}. Richardson's estimation for the noiseless expectation value is given by 
\begin{equation}
    E_O  = \sum ^n _{k=1} \beta_k\langle O\rangle _{c_k\lambda},
\end{equation}
where $\alpha$ are scaling factors obeying following order $1=c_0<c_1<c_2<...<c_n$. The coefficients $\beta$ are calculated as follows
\begin{equation}
    \beta_k = \prod_{i\neq k} \frac{c_i \lambda}{c_k \lambda-c_i\lambda},
\end{equation}
which are derived from conditions $\sum _{k=0}^n\beta_k = 1$ and $\sum^n_{l=0} \beta_l c^k_l = 0$.

This holds for scaling factors that are equally spaced. However, using higher-degree polynomial extrapolation should be avoided, as it may lead to instability in the method.

We also consider a linear fit,
\begin{equation}
    E_O = b-a\langle O\rangle _{c_k\lambda}.
\end{equation}
We consider a linear fit because, for shallow circuits or small noise values, the linear fit might be sufficient for the ZNE to perform well.

Finally, we consider an exponential fit,
\begin{equation}
    E_O = b+e^{-a*\langle O\rangle _{c_k\lambda}}.
\end{equation}
As the noise accumulates in the circuit with every layer, the exponential fit might be most accurate in describing how the expectation value depends on the noise parameter $\lambda$

The variance of the final estimate is computed for each of the extrapolation methods using the variance collected for each scaled value during computation. As we consider variances of each scaled expectation value to be independent variables, the variance of the linear and Richardson fits are computed using the variance of a weighted sum of independent variables, \begin{equation}
    \text{Var}(aX+bY)=a^2\text{Var}(X)+b^2\text{Var}(Y)
\end{equation}. The variance of exponential extrapolation is computed using the delta method, i.e., based on a second-order Taylor expansion.

We consider six different methods of extrapolation (three functional fits and two sets of scaling parameters) for each of our cases, and the one with the smallest MSE is chosen, see \cref{sec:numericextrapolationresults}

\subsection{Numeric Comparison of Extrapolation Techniques}\label{sec:numericextrapolationresults}
\begin{figure*}[h]
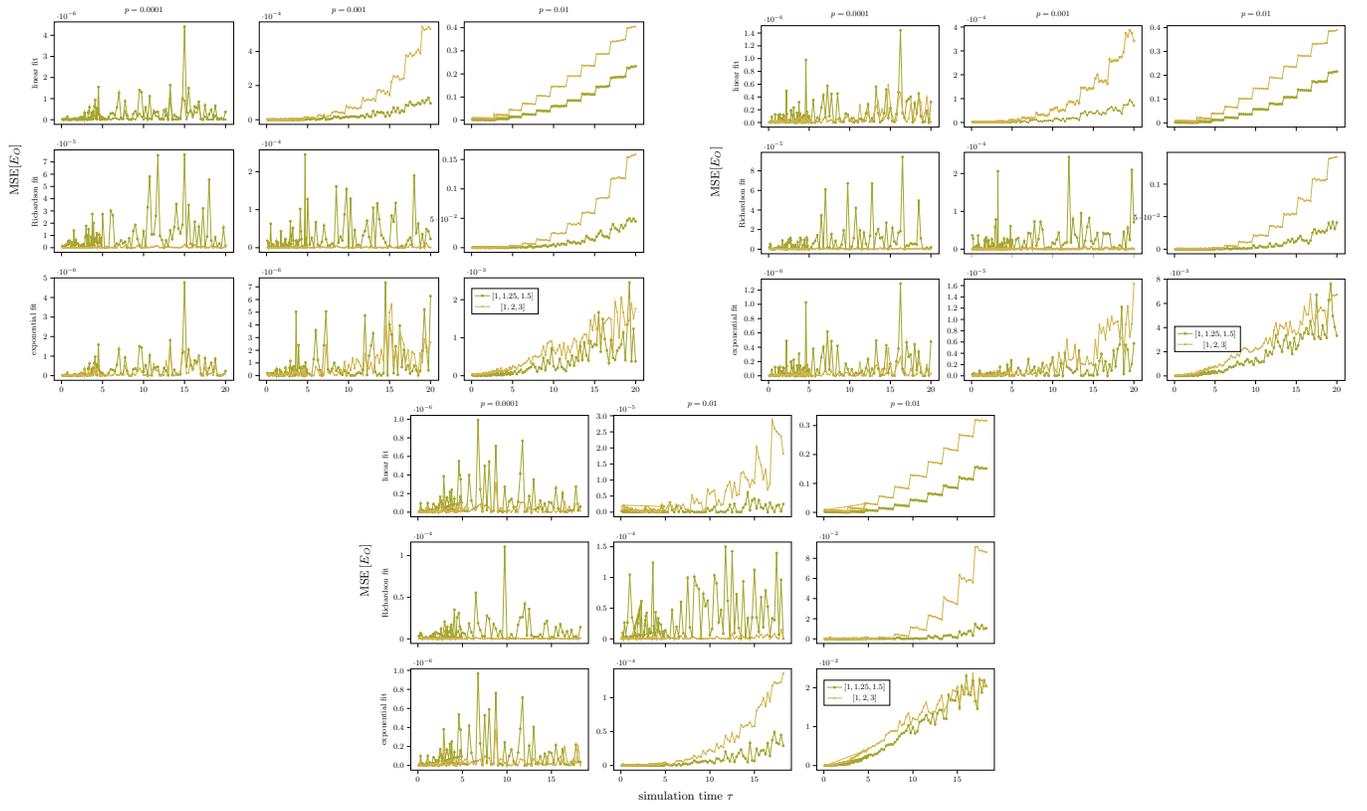

    \centering
     \begin{subfigure}[b]{0.48\textwidth}
         \centering
         \resizebox{\linewidth}{!}{
         \include{msescalingcompqsp4}
         }
     \end{subfigure}
     \hfill
      \begin{subfigure}[b]{0.48\textwidth}
         \centering
        \resizebox{\linewidth}{!}{
         \include{msescalingcompqsp6}
         }
     \end{subfigure}
     \hfill
     \begin{subfigure}[b]{0.48\textwidth}
         \centering
          \resizebox{\linewidth}{!}{
          \include{msescalingcompqsp8}
         }
     \end{subfigure}
    \caption{Scaling plots for QEM-QSP. For $4$-qubit (above left), $6$-qubit (middle), and $8$-qubit (above right) Ising models, we show the mean-squared error $\text{MSE}[E_A]$ against increasing simulation time $\tau$. For the $4, 6$-qubit Hamiltonians we consider $\tau\in[0.1, 20.0]$. For the $8$-qubit case we consider $\tau\in[0.1, 18.2]$.}
    \label{fig:scalingplots}
\end{figure*}

In this section, we compare the performance of ZNE with three types of extrapolation functions: linear, Richardson, and exponential scaling. For each extrapolation technique, we use three noise scaling steps, either $[1, 1.25, 1.5]$ or $[1, 2, 3]$. We plot the mean-squared error of extrapolated expectation values $E_O$ at simulation time steps of $0.1$ up to $\tau=5$. For longer times ($\tau\geq 5$), the QSP simulations go up to $\tau=20$ in $0.25$ time-steps. For Trotter, the longer time steps are up to $\tau=20$ in $0.2$ steps. We show the $4, 6, 8$-qubit and precision $\mathcal{O}(10^-2)$ cases in \cref{fig:scalingplots}. In \cref{tab:allscalingrefs}, we reproduce the best fix and scaling for each of the instances considered in this work. We can see that for $P>10^{-4}$, the smaller noise scaling increments achieve a better fit, and when it succeeds, the exponential fit is usually optimal.

\begin{table}[h]
    \centering
     \resizebox{\linewidth}{!}{
    \begin{tabular}{ c | c | c }\hline \hline
    Case & Extrapolation method & Scaling\\ \hline
     QSP $4$ qubits; $\mathcal{O}\left(10^{-4}\right)$; $p=10^{-4}$ &  exponential & $[1, 2, 3]$ \\
    QSP $4$ qubits; $\mathcal{O}\left(10^{-2}\right)$; $p=10^{-4}$ &  exponential & $[1, 2, 3]$ \\
   QSP $4$ qubits; $\mathcal{O}\left(10^{-2}\right)$; $p=10^{-3}$ &  exponential & $[1, 2, 3]$ \\
   QSP $4$ qubits; $\mathcal{O}\left(10^{-2}\right)$; $p=10^{-2}$ &  exponential & $[1, 1.25, 1.5]$ \\
   QSP $6$ qubits; $\mathcal{O}\left(10^{-2}\right)$; $p=10^{-4}$ &  exponential & $[1, 2, 3]$ \\
   QSP $6$ qubits; $\mathcal{O}\left(10^{-4}\right)$; $p=10^{-4}$ &  exponential & $[1, 2, 3]$ \\
   QSP $6$ qubits; $\mathcal{O}\left(10^{-2}\right)$; $p=10^{-3}$ &  exponential & $[1, 1.25, 1.5]$ \\
   QSP $6$ qubits; $\mathcal{O}\left(10^{-2}\right)$; $p=10^{-2}$ &  exponential & $[1, 1.25, 1.5]$ \\
   QSP $8$ qubits; $\mathcal{O}\left(10^{-2}\right)$; $p=10^{-4}$ &  exponential/linear & $[1, 2, 3]$ \\
   QSP $8$ qubits; $\mathcal{O}\left(10^{-2}\right)$; $p=10^{-3}$ &  exponential & $[1, 1.25, 1.5]$ \\
   QSP $8$ qubits; $\mathcal{O}\left(10^{-2}\right)$; $p=10^{-2}$ &  Richardson & $[1, 2, 3]$ \\
    Trotter $4$ qubits; $\mathcal{O}\left(10^{-2}\right)$; $p=10^{-4}$ &  exponential & $[1, 2, 3]$ \\
   Trotter $4$ qubits; $\mathcal{O}\left(10^{-2}\right)$; $p=10^{-3}$ &  exponential & $[1, 2, 3]$ \\
   Trotter $4$ qubits; $\mathcal{O}\left(10^{-2}\right)$; $p=10^{-2}$ &  linear/richardson & $[1, 1.25, 1.5]$ \\
    Trotter $6$ qubits; $\mathcal{O}\left(10^{-2}\right)$; $p=10^{-4}$ &  exponential & $[1, 2, 3]$ \\
   Trotter $6$ qubits; $\mathcal{O}\left(10^{-2}\right)$; $p=10^{-3}$ &  exponential & $[1, 2, 3]$ \\
   Trotter $6$ qubits; $\mathcal{O}\left(10^{-2}\right)$; $p=10^{-2}$ &  Richardson/linear & $[1, 1.25, 1.5]$ \\
    Trotter $8$ qubits; $\mathcal{O}\left(10^{-2}\right)$; $p=10^{-4}$ &  exponential & $[1, 2, 3]$ \\
   Trotter $8$ qubits; $\mathcal{O}\left(10^{-2}\right)$; $p=10^{-3}$ &  exponential & $[1, 1.25, 1.5]$ \\
   Trotter $8$ qubits; $\mathcal{O}\left(10^{-2}\right)$; $p=10^{-2}$ &  Richardson/linear & $[1, 1.25, 1.5]$ \\\hline\hline
    \end{tabular}
    }
    \caption{Best scaling parameters and extrapolation fits for all simulations considered in this work, with a fixed number of shots.}
    \label{tab:allscalingrefs}
\end{table}

\begin{figure*}[h]
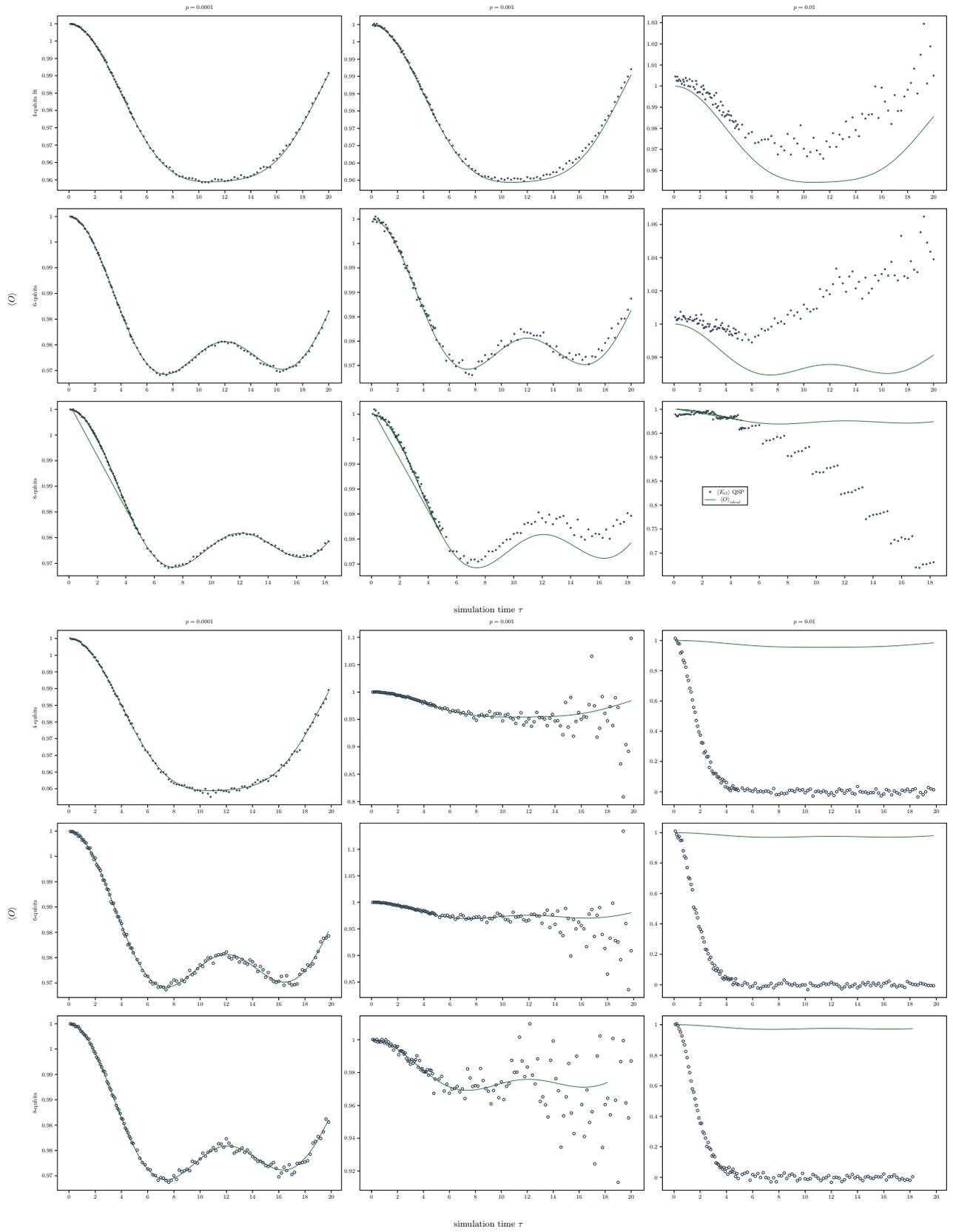

    \centering
    \begin{subfigure}[b]{0.95\textwidth}
         \centering 
          \resizebox{\linewidth}{!}{
         \include{idealmitcompplotallq}
         }
     \end{subfigure}
     \hfill
     \begin{subfigure}[b]{0.95\textwidth}
         \centering
          \resizebox{\linewidth}{!}{
         \include{idealmitcompplotallqtrotter}
         }
     \end{subfigure}
    \caption{Best performing mitigation for all qubits ($4, 6, 8$) and noise levels ($10^{-4}, 10^{-3}, 10^{-2}$). The behavior of the mitigated expectation values already demonstrates that the amount of noise is more influential than the number of qubits for both the Trotter (below) and QSP (above) QEM protocols.}
    \label{fig:allqexpvstau}
\end{figure*}

While the error mitigation protocols do worse as the number of qubits increases, increasing the number of qubits is far less impactful than increasing the amount of noise or depth of the circuit, as is shown in \cref{fig:allqexpvstau}. For this reason, we focus on the $N=4$ case within the main text.

%% file: shotsvstau.tex
\begin{tikzpicture}

\begin{groupplot}[group style={group size=1 by 2}]
\nextgroupplot[
legend style={at={(axis cs:2000, 5.7)},anchor=north east},
tick align=outside,
tick pos=left,
x grid style={darkgray176},
xmin=-87.5, xmax=2079.5,
xtick style={color=black},
y grid style={darkgray176},
ymin=2.07557364421695, ymax=6.18687744551348,
ytick style={color=black},
 width =\linewidth, 
 height = 0.45*\linewidth,
 xticklabel=\empty,
]
\addplot [semithick, c1, dotted]
table {%
11 2.26245108973043
21 2.26717172840301
31 2.26951294421792
41 2.2718416065365
51 2.27415784926368
61 2.27646180417324
71 2.27875360095283
81 2.28103336724773
91 2.28330122870355
101 2.28780172993023
111 2.29003461136252
121 2.29225607135648
131 2.29446622616159
141 2.29666519026153
151 2.30102999566398
161 2.30319605742049
171 2.30535136944662
181 2.30749603791321
191 2.3096301674259
201 2.31386722036915
211 2.31597034545692
221 2.31806333496276
231 2.32014628611105
241 2.32428245529769
251 2.32633586092875
261 2.32837960343874
271 2.33243845991561
281 2.33445375115093
291 2.33645973384853
301 2.3384564936046
311 2.34242268082221
321 2.34439227368511
331 2.34635297445064
341 2.35024801833416
351 2.35218251811136
361 2.3541084391474
371 2.35793484700045
381 2.35983548233989
391 2.36361197989214
401 2.3654879848909
411 2.36735592102602
421 2.37106786227174
431 2.37291200297011
441 2.37657695705651
451 2.37839790094814
461 2.38021124171161
471 2.38381536598043
481 2.38560627359831
491 2.38916608436453
521 2.39619934709574
541 2.40140054078154
561 2.40654018043396
581 2.41161970596323
601 2.41830129131975
621 2.42324587393681
641 2.42813479402879
661 2.43296929087441
681 2.43775056282039
701 2.44404479591808
721 2.44870631990508
741 2.45484486000851
761 2.45939248775923
781 2.46389298898591
801 2.46982201597816
821 2.47567118832443
841 2.48000694295715
861 2.48572142648158
881 2.49136169383427
901 2.49554433754645
921 2.50105926221775
941 2.50650503240487
961 2.51188336097887
981 2.51719589794997
1001 2.52244423350632
1021 2.52762990087134
1041 2.5327543789925
1061 2.53781909507327
1081 2.54282542695918
1101 2.54777470538782
1121 2.55388302664387
1141 2.55870857053317
1161 2.56348108539441
1181 2.56937390961505
1201 2.57403126772772
1221 2.57978359661681
1241 2.58433122436753
1261 2.58994960132571
1281 2.59549622182557
1301 2.59988307207369
1321 2.60530504614111
1341 2.61066016308988
1361 2.6159500516564
1381 2.62117628177504
1401 2.62634036737504
1421 2.63144376901317
1441 2.63648789635337
1461 2.64246452024212
1481 2.64738297011462
1501 2.65224634100332
1521 2.65801139665711
1541 2.66275783168157
1561 2.66838591669
1581 2.6730209071289
1601 2.67851837904011
1621 2.68394713075151
1641 2.68930885912362
1661 2.69460519893357
1681 2.69983772586725
1701 2.70500795933334
1721 2.71011736511182
1741 2.71516735784846
1761 2.72015930340596
1781 2.72591163229505
1801 2.73078227566639
1821 2.73559889969818
1841 2.74115159885178
1861 2.74663419893758
1881 2.75127910398334
1901 2.75663610824585
1921 2.76192783842053
1941 2.76715586608218
1961 2.77232170672292
1981 2.77742682238931
};
\addplot [semithick, c3]
table {%
11 6
21 6
31 6
41 6
51 6
61 6
71 6
81 6
91 6
101 6
111 6
121 6
131 6
141 6
151 6
161 6
171 6
181 6
191 6
201 6
211 6
221 6
231 6
241 6
251 6
261 6
271 6
281 6
291 6
301 6
311 6
321 6
331 6
341 6
351 6
361 6
371 6
381 6
391 6
401 6
411 6
421 6
431 6
441 6
451 6
461 6
471 6
481 6
491 6
521 6
541 6
561 6
581 6
601 6
621 6
641 6
661 6
681 6
701 6
721 6
741 6
761 6
781 6
801 6
821 6
841 6
861 6
881 6
901 6
921 6
941 6
961 6
981 6
1001 6
1021 6
1041 6
1061 6
1081 6
1101 6
1121 6
1141 6
1161 6
1181 6
1201 6
1221 6
1241 6
1261 6
1281 6
1301 6
1321 6
1341 6
1361 6
1381 6
1401 6
1421 6
1441 6
1461 6
1481 6
1501 6
1521 6
1541 6
1561 6
1581 6
1601 6
1621 6
1641 6
1661 6
1681 6
1701 6
1721 6
1741 6
1761 6
1781 6
1801 6
1821 6
1841 6
1861 6
1881 6
1901 6
1921 6
1941 6
1961 6
1981 6
};

\addlegendentry{$M_s$}
\addlegendentry{$10^{6}$}

\nextgroupplot[
legend style={at={(axis cs:10,12000)},anchor=north west},
tick align=outside,
tick pos=left,
x grid style={darkgray176},
xmin=-87.5, xmax=2079.5,
xtick style={color=black},
y grid style={darkgray176},
ymin=-588, ymax=12480,
ytick style={color=black},
 width =\linewidth, 
 height = 0.45*\linewidth
]
\addplot [semithick, c2, dotted]
table {%
11 66
21 126
31 186
41 246
51 306
61 366
71 426
81 486
91 546
101 606
111 666
121 726
131 786
141 846
151 906
161 966
171 1026
181 1086
191 1146
201 1206
211 1266
221 1326
231 1386
241 1446
251 1506
261 1566
271 1626
281 1686
291 1746
301 1806
311 1866
321 1926
331 1986
341 2046
351 2106
361 2166
371 2226
381 2286
391 2346
401 2406
411 2466
421 2526
431 2586
441 2646
451 2706
461 2766
471 2826
481 2886
491 2946
521 3126
541 3246
561 3366
581 3486
601 3606
621 3726
641 3846
661 3966
681 4086
701 4206
721 4326
741 4446
761 4566
781 4686
801 4806
821 4926
841 5046
861 5166
881 5286
901 5406
921 5526
941 5646
961 5766
981 5886
1001 6006
1021 6126
1041 6246
1061 6366
1081 6486
1101 6606
1121 6726
1141 6846
1161 6966
1181 7086
1201 7206
1221 7326
1241 7446
1261 7566
1281 7686
1301 7806
1321 7926
1341 8046
1361 8166
1381 8286
1401 8406
1421 8526
1441 8646
1461 8766
1481 8886
1501 9006
1521 9126
1541 9246
1561 9366
1581 9486
1601 9606
1621 9726
1641 9846
1661 9966
1681 10086
1701 10206
1721 10326
1741 10446
1761 10566
1781 10686
1801 10806
1821 10926
1841 11046
1861 11166
1881 11286
1901 11406
1921 11526
1941 11646
1961 11766
1981 11886
};
\addplot [semithick, c3]
table {%
11 6
21 6
31 6
41 6
51 6
61 6
71 6
81 6
91 6
101 6
111 6
121 6
131 6
141 6
151 6
161 6
171 6
181 6
191 6
201 6
211 6
221 6
231 6
241 6
251 6
261 6
271 6
281 6
291 6
301 6
311 6
321 6
331 6
341 6
351 6
361 6
371 6
381 6
391 6
401 6
411 6
421 6
431 6
441 6
451 6
461 6
471 6
481 6
491 6
521 6
541 6
561 6
581 6
601 6
621 6
641 6
661 6
681 6
701 6
721 6
741 6
761 6
781 6
801 6
821 6
841 6
861 6
881 6
901 6
921 6
941 6
961 6
981 6
1001 6
1021 6
1041 6
1061 6
1081 6
1101 6
1121 6
1141 6
1161 6
1181 6
1201 6
1221 6
1241 6
1261 6
1281 6
1301 6
1321 6
1341 6
1361 6
1381 6
1401 6
1421 6
1441 6
1461 6
1481 6
1501 6
1521 6
1541 6
1561 6
1581 6
1601 6
1621 6
1641 6
1661 6
1681 6
1701 6
1721 6
1741 6
1761 6
1781 6
1801 6
1821 6
1841 6
1861 6
1881 6
1901 6
1921 6
1941 6
1961 6
1981 6
};

\addlegendentry{$M_e$}
\addlegendentry{$10^{6}$}

\end{groupplot}

\draw ({$(current bounding box.south west)!0.0!(current bounding box.south east)$}|-{$(current bounding box.south west)!0.4!(current bounding box.north west)$}) node[
  scale=1.0,
  anchor=west,
  text=black,
  rotate=90.0
]{$\log_{10}(M)$};
\draw ({$(current bounding box.south west)!0.55!(current bounding box.south east)$}|-{$(current bounding box.south west)!-0.05!(current bounding box.north west)$}) node[
  scale=1.0,
  anchor=south,
  text=black,
  rotate=0.0
]{depth $D$};
\end{tikzpicture}

%% file: fixedboundcompq40.1.tex
\begin{tikzpicture}

\begin{axis}[
tick align=outside,
tick pos=left,
x grid style={darkgray176},
xlabel={simulation time $\tau$},
xmin=-0.895, xmax=20.995,
xtick style={color=black},
y grid style={darkgray176},
ylabel={$\braket{O}$, $p=0.0001$},
ymin=0.785521089215934, ymax=1.13604277213384,
ytick style={color=black},
mark size=1
]
\addplot [draw=c1, fill=c1, mark=o, only marks]
table{%
x  y
0.1 0.999999995494715
0.2 0.999999995494715
0.25 1.01167017977762
0.3 0.999999995494715
0.4 0.994203635449189
0.5 0.976987805948561
0.6 0.971091068487431
0.7 0.994203635449189
0.75 0.999999995494715
0.8 1.01167017977762
0.9 0.999999995494715
1 1.00597687720774
1.1 0.988409289633162
1.2 0.999999995494715
1.25 0.999999995494715
1.4 0.971091068487431
1.5 1.01167017977762
1.6 1.00028607635675
1.7 1.01167017977762
1.75 0.965495729124125
1.8 0.994203635449189
1.9 0.988405797007
2 1.01167017977762
2.1 0.988409289633162
2.2 0.95370113423577
2.25 1.00597687720774
2.3 0.976987805948561
2.3 0.976987805948561
2.4 0.976987805948561
2.5 1.01262565129154
2.6 0.999974102501299
2.7 0.999999995494715
2.75 0.982608697500758
2.8 0.982608697500758
2.9 1.00597687720774
3 0.976987805948561
3.1 0.976809914281059
3.2 0.994949412688519
3.25 1.01167017977762
3.3 0.994203635449189
3.4 0.924953547165927
3.5 0.95370113423577
3.6 0.976987805948561
3.7 0.976987805948561
3.75 0.95942031253443
3.8 0.999974102501299
3.9 1.01262565129154
4 1.00675178456337
4.1 1.04225074319562
4.2 0.94833548618291
4.25 0.976809914281059
4.3 0.994280229338014
4.4 1.04225074319562
4.5 0.941896721437294
4.6 1.00274931434017
4.7 0.98290001019489
4.75 1.02349378826661
4.8 0.902473062290167
4.9 0.971011640826207
5 0.98290001019489
5.25 1.00071571613042
5.5 1.03903073319878
5.75 0.982582342559991
6 0.920733246607058
6.25 0.994353644158942
6.5 0.959770114943072
6.75 0.948822151844981
7 0.96053119672054
7.25 0.897259118657519
7.5 0.982732738279605
7.75 0.954398024466262
8 0.965408851645662
8.25 0.959848873053887
8.5 0.966722372696755
8.75 0.86196051251304
9 0.942708104344479
9.25 0.908541168199623
9.5 0.955014807802035
9.75 0.983368402054103
10 1.03865481268847
10.25 0.953891601049176
10.5 1.00744975560405
10.75 0.931317851397654
11 1.0005168152342
11.25 0.95409957487502
11.5 0.948277507253641
11.75 0.992299608561651
12 1.03937968827188
12.25 0.983941236611447
12.5 0.919298819500217
12.75 0.977406570749186
13 0.977082675569553
13.25 0.936781616871707
13.5 0.942708104344479
13.75 0.919928868701207
14 0.942538267389427
14.25 1.04381350102994
14.5 0.949019390119042
14.75 1.0249610967965
15 1.01936187166721
15.25 0.942291135007294
15.5 0.983474706152546
15.75 0.843229529769823
16 0.999895546184618
16.25 0.959770112198836
16.5 0.953891601049176
16.75 0.977406570749186
17 0.85979224703136
17.25 0.925597681927449
17.5 0.948275854865055
17.75 0.903028376001165
18 0.913787779004357
18.25 0.966053339718256
18.5 0.902379749756563
18.75 0.955014807802035
19 0.942892285007738
19.25 0.965811963461198
19.5 0.977206891743577
19.75 1.01239166052051
20 0.98342303036739
};
\addplot [draw=c2, fill=c2, mark=x, only marks]
table{%
x  y
0.1 0.996356106746213
0.2 1.01453198788985
0.3 0.982107987050133
0.4 1.01475963383904
0.5 1.00736242443749
0.6 1.0359835934016
0.7 0.98946417938094
0.8 0.989747633474245
0.9 0.996564311710719
1 1.03503809765511
1.1 0.993270414455942
1.2 1.02502917206633
1.3 1.01100533876437
1.4 1.04798876144051
1.5 1.00364213949681
1.6 0.99028134156343
1.7 1.02839838881553
1.8 0.973625915472808
1.9 0.997191935736799
2 0.967871421351035
2.1 1.00788815829113
2.2 1.03371733808288
2.3 0.962802577753722
2.4 0.930721674265052
2.5 0.966086761472214
2.6 1.00230393201011
2.7 1.01664952322312
2.8 1.01333151529011
2.9 1.02296898211978
3 0.929944016429459
3.1 0.958055785685724
3.2 0.918544489714649
3.3 1.02523725786792
3.4 1.02443288391779
3.5 0.95858592350384
3.6 1.02754735776937
3.7 0.986048524109915
3.8 1.02650039456368
3.9 0.987540928340576
4 0.908043915915563
4.1 0.955317256454541
4.2 0.970020870665351
4.3 1.00798499461221
4.4 1.02331703602409
4.5 0.916004355558275
4.6 0.966031200904942
4.7 0.927539367401502
4.8 0.979834729127589
4.9 0.95200117401388
5.2 0.920991494600727
5.4 0.917314372408012
5.6 0.848367059787962
5.8 0.87949835569087
6 0.970644694866464
6.2 0.924407947620105
6.4 0.902208070233132
6.6 0.98750086150682
6.8 0.954141417831607
7 0.968126812662634
7.2 1.01704047030581
7.4 0.972886730621302
7.6 0.975694624683811
7.8 0.936034825381174
8 0.929737893670503
8.2 0.945321317352098
8.4 0.977363280881345
8.6 0.997462255350857
8.8 1.07938470754497
9 0.906045725553306
9.2 0.825561956886789
9.4 1.00469973421311
9.6 0.975322584501901
9.8 0.80145389298493
10 1.07759757843214
10.2 0.90740392934572
10.4 0.902331362628235
10.6 0.941761143328016
10.8 0.849848613692385
11 0.926945894579256
11.2 0.838604575552857
11.4 1.02668896161946
11.6 0.854557714596887
11.8 0.903777990204309
12 0.963759171039505
12.2 1.0019491594502
12.4 0.890794031453138
12.6 0.874495654346076
12.8 0.934008241821432
13 1.07091604053852
13.2 1.03125048672092
13.4 0.804445148669027
13.6 0.992422461391589
13.8 0.925705885001453
14 1.04162720469675
14.2 1.00401360878636
14.4 0.878411002701412
14.6 1.12010996836485
14.8 1.02437952140907
15 1.00474722162805
15.2 0.994285475900434
15.4 0.972310868428641
15.6 0.979636818531301
15.8 0.898440300362597
16 0.933939736630401
16.2 0.952111131095319
16.4 1.02993834477728
16.6 1.01123514499747
16.8 0.972721079911138
17 1.01922058817233
17.2 0.888735491070722
17.4 1.00343133665774
17.6 1.0516478126235
17.8 1.09692912517923
18 0.985508751777322
18.2 0.869726616186737
18.4 0.970822747295559
18.6 0.945270113713465
18.8 0.935409538228814
19 0.922128135056976
19.2 0.911193403746127
19.4 1.04240901662533
19.6 1.03048399544669
19.8 0.969321342898918
};
\addplot [semithick, c1, dashed]
table {%
0.1 0.999924022470059
0.2 0.9999265354598
0.25 0.999911779705014
0.3 0.999865781226853
0.4 0.999854598251404
0.5 0.999545374483015
0.6 0.999394180852329
0.7 0.999335423582628
0.75 0.999136376813519
0.8 0.999113249037908
0.9 0.998859194084344
1 0.998381365518054
1.1 0.998137834000742
1.2 0.997679260242739
1.25 0.997719849835131
1.4 0.996947742072037
1.5 0.996731949196309
1.6 0.996255986521746
1.7 0.995652303378198
1.75 0.995434882455833
1.8 0.995198230676519
1.9 0.994705858528655
2 0.994197869530671
2.1 0.993501935368055
2.2 0.993148508174522
2.25 0.99260303245699
2.3 0.992399255048476
2.3 0.992399255048476
2.4 0.99198842573019
2.5 0.991173998059529
2.6 0.990629015363984
2.7 0.99011390164071
2.75 0.989391718617101
2.8 0.989516054734701
2.9 0.988884412157464
3 0.987516546218509
3.1 0.98683778928989
3.2 0.986074650379576
3.25 0.98593296311583
3.3 0.985714996808797
3.4 0.984971781442856
3.5 0.983972156102084
3.6 0.983268052351912
3.7 0.982503792044402
3.75 0.981822833820323
3.8 0.981702211658601
3.9 0.981331901722868
4 0.980490501646336
4.1 0.979472154292166
4.2 0.978790396816198
4.25 0.978651152880428
4.3 0.978046007817905
4.4 0.977414560657856
4.5 0.976911114834839
4.6 0.975631710974177
4.7 0.974568488222116
4.75 0.97425789208788
4.8 0.974115819016961
4.9 0.973027827840941
5 0.972511822925215
5.25 0.970403245857965
5.5 0.969061934998307
5.75 0.967534543685753
6 0.965799037327236
6.25 0.964137312684246
6.5 0.963327472649565
6.75 0.961567640113805
7 0.96079698162849
7.25 0.959838041340581
7.5 0.958904050223959
7.75 0.958591083519592
8 0.957513539475852
8.25 0.956573185986441
8.5 0.955991042673364
8.75 0.955955160819951
9 0.955510417631882
9.25 0.955503237776611
9.5 0.955419700621846
9.75 0.955128529164552
10 0.954696280357008
10.25 0.954271801995136
10.5 0.954332390273906
10.75 0.954267996169302
11 0.954667501213132
11.25 0.95520664450238
11.5 0.955009260392798
11.75 0.955052149643205
12 0.954616561675725
12.25 0.954762593370389
12.5 0.95495836591798
12.75 0.95576323392039
13 0.955381620631046
13.25 0.95634097860188
13.5 0.955792533079931
13.75 0.955688821100123
14 0.956076549913744
14.25 0.956549591256112
14.5 0.957142076215365
14.75 0.958279374716224
15 0.958748932089414
15.25 0.958593172028664
15.5 0.958563340683626
15.75 0.960604508892263
16 0.961138456751951
16.25 0.962384188322429
16.5 0.963412363997328
16.75 0.964838794813757
17 0.965144396666414
17.25 0.966713727120818
17.5 0.968551531923784
17.75 0.969525245814461
18 0.971383084228807
18.25 0.973098952619186
18.5 0.97466658983429
18.75 0.97700369839992
19 0.978425594422306
19.25 0.979989178123493
19.5 0.981813808494213
19.75 0.983917026260331
20 0.985794277971866
};
\addplot [semithick, c3]
table {%
0.1 0.999984322253531
0.2 0.999937316048458
0.25 0.999902087993421
0.3 0.999859062423781
0.4 0.999749696232385
0.5 0.999609405823361
0.6 0.999438432598445
0.7 0.999237070443749
0.75 0.99912509954066
0.8 0.999005665038244
0.9 0.998744613040839
1 0.998454361158164
1.1 0.998135405095548
1.2 0.997788288393915
1.25 0.997604352428722
1.4 0.99701197866293
1.5 0.99658409989175
1.6 0.996130685926211
1.7 0.995652498276588
1.75 0.995404362911638
1.8 0.995150337106013
1.9 0.994625039370295
2 0.994077476875803
2.1 0.993508554260528
2.2 0.992919206903649
2.25 0.992617174129473
2.3 0.992310398769103
2.3 0.992310398769103
2.4 0.991683120188853
2.5 0.991038385591691
2.6 0.99037723118357
2.7 0.989700712585584
2.75 0.989357026316278
2.8 0.989009902435811
2.9 0.98830588796137
3 0.987589768527063
3.1 0.986862653167096
3.2 0.986125658106365
3.25 0.985753805863718
3.3 0.985379904277848
3.4 0.984626514842655
3.5 0.983866612719245
3.6 0.983101318128335
3.7 0.982331746159953
3.75 0.981945702629522
3.8 0.981559004369023
3.9 0.9807841904058
4 0.98000838968738
4.1 0.979232673116355
4.2 0.978458094852616
4.25 0.978071556897716
4.3 0.977685690144076
4.4 0.976916473222
4.5 0.976151435266367
4.6 0.975391542446561
4.7 0.974637734042424
4.75 0.974263397225317
4.8 0.973890920650495
4.9 0.973151982480018
5 0.972421767743045
5.25 0.97063965615509
5.5 0.96892881138133
5.75 0.967299502778876
6 0.965760428420085
6.25 0.964318652091784
6.5 0.962979569793162
6.75 0.961746905975175
7 0.960622739223592
7.25 0.959607556551009
7.5 0.958700334940377
7.75 0.957898648283958
8 0.957198797396691
8.25 0.956595960360986
8.5 0.956084360089292
8.75 0.95565744567886
9 0.955308083886447
9.25 0.955028756874138
9.5 0.954811762275025
9.75 0.954649411601136
10 0.954534223066697
10.25 0.954459105026463
10.5 0.954417526429264
10.75 0.954403670956914
11 0.954412571853028
11.25 0.954440224838352
11.5 0.954483676950941
11.75 0.954541089632261
12 0.954611774894168
12.25 0.954696203936569
12.5 0.954795988130704
12.75 0.954913832827626
13 0.955053464984911
13.25 0.955219536116504
13.5 0.955417502551073
13.75 0.955653485424212
14 0.955934113221046
14.25 0.956266350021176
14.5 0.956657312871512
14.75 0.957114081919665
15 0.957643507078036
15.25 0.958252015054501
15.5 0.958945420579272
15.75 0.959728745579925
16 0.960606049909829
16.25 0.961580277022944
16.5 0.962653117714371
16.75 0.963824894717154
17 0.965094470567682
17.25 0.966459180732235
17.5 0.967914793533262
17.75 0.969455497934343
18 0.97107391974594
18.25 0.972761166308634
18.5 0.974506899205412
18.75 0.976299434058091
19 0.978125865983579
19.25 0.979972218831163
19.5 0.981823615899892
19.75 0.983664469452122
20 0.985478686001584
};
\addplot [semithick, c2, dotted]
table {%
0.1 1.00004273558581
0.2 0.999814421633299
0.3 0.999793386423507
0.4 0.99980927257291
0.5 0.999621458551089
0.6 0.999537777765303
0.7 0.999269859461762
0.8 0.999166113168338
0.9 0.998898485754368
1 0.9982380803605
1.1 0.997939185610932
1.2 0.997752637938528
1.3 0.997453019831099
1.4 0.997057836298118
1.5 0.996430685555453
1.6 0.996086201113156
1.7 0.995587524063577
1.8 0.995228627289901
1.9 0.99435947779336
2 0.994438578885452
2.1 0.993425878513494
2.2 0.99315459574174
2.3 0.9922403482222
2.4 0.991738556838991
2.5 0.991322958471822
2.6 0.990153991108062
2.7 0.98965482976823
2.8 0.9889365035641
2.9 0.98844808354424
3 0.987295458379659
3.1 0.986880139123544
3.2 0.986106803386428
3.3 0.985239663095875
3.4 0.984296522107508
3.5 0.983396493340983
3.6 0.982833603544777
3.7 0.982658010398332
3.8 0.981769239226172
3.9 0.980420693324012
4 0.980047220126914
4.1 0.979636953289062
4.2 0.978440389333702
4.3 0.977129253711187
4.4 0.977050422154811
4.5 0.976520440651408
4.6 0.975449572331949
4.7 0.974488846004635
4.8 0.974600519349106
4.9 0.973231783240233
5.2 0.971405111896078
5.4 0.969776530411138
5.6 0.96783379572022
5.8 0.966546241809953
6 0.965986928857969
6.2 0.964955214947986
6.4 0.963406066479205
6.6 0.962630501944407
6.8 0.960983033052032
7 0.959945212153019
7.2 0.959926612566008
7.4 0.959279987013426
7.6 0.957821961277929
7.8 0.957076932208192
8 0.956620006664987
8.2 0.956307934674164
8.4 0.956203494023377
8.6 0.955998553340842
8.8 0.956073265138447
9 0.955436915478839
9.2 0.955350212827988
9.4 0.955271061437004
9.6 0.954438383244055
9.8 0.954552002122562
10 0.954348743754094
10.2 0.953772598897841
10.4 0.954965005033504
10.6 0.953525399715829
10.8 0.952561737603247
11 0.954488320643465
11.2 0.953823625619815
11.4 0.954667093604239
11.6 0.954539163232741
11.8 0.954291338307883
12 0.954263404428174
12.2 0.953981526605268
12.4 0.955121140915562
12.6 0.955143513065586
12.8 0.954496663291347
13 0.954504657108951
13.2 0.955670736168954
13.4 0.955721529308193
13.6 0.955153160547476
13.8 0.955577435024528
14 0.95633809823467
14.2 0.95691921914914
14.4 0.957649625570683
14.6 0.957123154439261
14.8 0.957538505282965
15 0.957825631507807
15.2 0.957294384252731
15.4 0.9588900684306
15.6 0.959040687682585
15.8 0.958762354170819
16 0.960438005764873
16.2 0.962137080924514
16.4 0.961906377991976
16.6 0.962711168697195
16.8 0.963598167513563
17 0.96532604760529
17.2 0.966241963713086
17.4 0.966054441803917
17.6 0.96657298805083
17.8 0.969339005965222
18 0.97162471905426
18.2 0.97272200384982
18.4 0.973715337622528
18.6 0.974872946830473
18.8 0.977330170641752
19 0.976948182819261
19.2 0.979446356576863
19.4 0.980632037333613
19.6 0.981900332422848
19.8 0.984572540137959
};
\addlegendentry{bound QSP}
\addlegendentry{bound Trotter}
\addlegendentry{fixed QSP}
\addlegendentry{ideal $\braket{O}$}
\addlegendentry{fixed Trotter}
\end{axis}

\end{tikzpicture}

%% file: varvslongtime.tex
\begin{tikzpicture}

\begin{axis}[
tick align=outside,
tick pos=left,
x grid style={darkgray176},
xmin=6.75, xmax=408.25,
xtick style={color=black},
y grid style={darkgray176},
ymin=0.0204245998277766, ymax=1.04664644762725,
ytick style={color=black},
ylabel={variance $\braket{E_{O^2}}-\braket{E_O}^2$},
xticklabel=\empty,
]
\addplot [semithick, c1, mark=o]
table {%
25 0.0670710474550253
50 0.111931366965978
75 0.142049993302464
100 0.160507764289502
125 0.180739192249023
150 0.211578069897155
175 0.244425734279185
200 0.265699072372067
225 0.279784530686969
250 0.298328460826228
275 0.328228705603742
300 0.360431635959844
325 0.380552068113657
350 0.37390765265161
375 0.367052208613834
390 0.362853402060898
};
\addplot [semithick, c2, mark=x]
table {%
25 0.464585750674881
50 0.627955978783905
75 0.726121912356039
100 0.795745017787886
125 0.848891134820119
150 0.890273183392092
175 0.920496947967541
200 0.940005909852242
225 0.954086083068403
250 0.964731216770623
275 0.973698118273647
300 0.980816109166292
325 0.98588279231561
350 0.986738091137946
375 0.98651649653127
390 0.986419005847059
};
\addplot [semithick, c4, mark=*]
table {%
25 0.997977606728085
50 0.999939983914074
75 0.999997456228487
100 0.999999891479995
125 0.999999994861175
150 0.999999999730008
175 0.9999999999858
200 0.999999999999169
225 0.999999999999956
250 0.999999999999997
275 1
300 1
325 1
350 1
375 1
390 1
};
\end{axis}

\end{tikzpicture}

%% file: expvslongtime.tex
\begin{tikzpicture}

\begin{axis}[
legend style={at={(axis cs:400, 0.7)},anchor=north east},
tick align=outside,
tick pos=left,
x grid style={darkgray176},
xmin=6.75, xmax=408.25,
xtick style={color=black},
y grid style={darkgray176},
ymin=-0.04899768, ymax=1.01423768,
ytick style={color=black},
ylabel={$\braket{O}$},
xlabel={simulation time $\tau$}
]
\addplot [semithick, c1, mark=o]
table {%
25 0.9659088
50 0.9422552
75 0.925992
100 0.9165632
125 0.9050324
150 0.8880132
175 0.8686516
200 0.857156
225 0.8490832
250 0.8375368
275 0.8198944
300 0.7994916
325 0.7874332
350 0.7912612
375 0.7955484
390 0.7981132
};
\addplot [semithick, c2, mark=x]
table {%
25 0.7313196
50 0.6101416
75 0.5225376
100 0.4515584
125 0.3885776
150 0.3312084
175 0.2823532
200 0.2441532
225 0.2142168
250 0.1872052
275 0.1609184
300 0.1388436
325 0.1188512
350 0.1145992
375 0.1160232
390 0.1163168
};
\addplot [semithick, c4, mark=*]
table {%
25 0.0450604
50 0.0080836
75 0.0021588
100 0.0001092
125 0.000102
150 9.36e-05
175 -0.000424
200 -0.0002176
225 -0.000194
250 -0.0004036
275 0.0004172
300 0.0001828
325 -0.0006688
350 0.0006416
375 -0.0005156
390 0.0004568
};
\addlegendentry{$p=0.0001$}
\addlegendentry{$p=0.001$}
\addlegendentry{$p=0.01$}
\end{axis}

\end{tikzpicture}

%% file: idealmitcompplotallqtrotter.tex
\begin{tikzpicture}

\begin{groupplot}[group style={group size=3 by 3}]
\nextgroupplot[
mark size=1,
tick align=outside,
tick pos=left,
title={$p=0.0001$},
x grid style={darkgray176},
xmin=-0.885000000000001, xmax=20.785,
xtick style={color=black},
y grid style={darkgray176},
ymin=0.950187687704119, ymax=1.00241678548494,
ytick style={color=black},
ylabel={$4$-qubits},
width =\linewidth, 
height = 0.65*\linewidth
]
\addplot [draw=c1, fill=c1, mark=o, only marks]
table{%
x  y
0.1 1.00004273558581
0.2 0.999814421633299
0.3 0.999793386423507
0.4 0.99980927257291
0.5 0.999621458551089
0.6 0.999537777765303
0.7 0.999269859461762
0.8 0.999166113168338
0.9 0.998898485754368
1 0.9982380803605
1.1 0.997939185610932
1.2 0.997752637938528
1.3 0.997453019831099
1.4 0.997057836298118
1.5 0.996430685555453
1.6 0.996086201113156
1.7 0.995587524063577
1.8 0.995228627289901
1.9 0.99435947779336
2 0.994438578885452
2.1 0.993425878513494
2.2 0.99315459574174
2.3 0.9922403482222
2.4 0.991738556838991
2.5 0.991322958471822
2.6 0.990153991108062
2.7 0.98965482976823
2.8 0.9889365035641
2.9 0.98844808354424
3 0.987295458379659
3.1 0.986880139123544
3.2 0.986106803386428
3.3 0.985239663095875
3.4 0.984296522107508
3.5 0.983396493340983
3.6 0.982833603544777
3.7 0.982658010398332
3.8 0.981769239226172
3.9 0.980420693324012
4 0.980047220126914
4.1 0.979636953289062
4.2 0.978440389333702
4.3 0.977129253711187
4.4 0.977050422154811
4.5 0.976520440651408
4.6 0.975449572331949
4.7 0.974488846004635
4.8 0.974600519349106
4.9 0.973231783240233
5.2 0.971405111896078
5.4 0.969776530411138
5.6 0.96783379572022
5.8 0.966546241809953
6 0.965986928857969
6.2 0.964955214947986
6.4 0.963406066479205
6.6 0.962630501944407
6.8 0.960983033052032
7 0.959945212153019
7.2 0.959926612566008
7.4 0.959279987013426
7.6 0.957821961277929
7.8 0.957076932208192
8 0.956620006664987
8.2 0.956307934674164
8.4 0.956203494023377
8.6 0.955998553340842
8.8 0.956073265138447
9 0.955436915478839
9.2 0.955350212827988
9.4 0.955271061437004
9.6 0.954438383244055
9.8 0.954552002122562
10 0.954348743754094
10.2 0.953772598897841
10.4 0.954965005033504
10.6 0.953525399715829
10.8 0.952561737603247
11 0.954488320643465
11.2 0.953823625619815
11.4 0.954667093604239
11.6 0.954539163232741
11.8 0.954291338307883
12 0.954263404428174
12.2 0.953981526605268
12.4 0.955121140915562
12.6 0.955143513065586
12.8 0.954496663291347
13 0.954504657108951
13.2 0.955670736168954
13.4 0.955721529308193
13.6 0.955153160547476
13.8 0.955577435024528
14 0.95633809823467
14.2 0.95691921914914
14.4 0.957649625570683
14.6 0.957123154439261
14.8 0.957538505282965
15 0.957825631507807
15.2 0.957294384252731
15.4 0.9588900684306
15.6 0.959040687682585
15.8 0.958762354170819
16 0.960438005764873
16.2 0.962137080924514
16.4 0.961906377991976
16.6 0.962711168697195
16.8 0.963598167513563
17 0.96532604760529
17.2 0.966241963713086
17.4 0.966054441803917
17.6 0.96657298805083
17.8 0.969339005965222
18 0.97162471905426
18.2 0.97272200384982
18.4 0.973715337622528
18.6 0.974872946830473
18.8 0.977330170641752
19 0.976948182819261
19.2 0.979446356576863
19.4 0.980632037333613
19.6 0.981900332422848
19.8 0.984572540137959
};
\addplot [semithick, c3]
table {%
0.1 0.999984320102678
0.2 0.999937307449658
0.3 0.999859043095476
0.4 0.999749661918819
0.5 0.999609352304798
0.6 0.9994383557013
0.7 0.999236966050588
0.8 0.999005529097635
0.9 0.998744441577026
1 0.99845415028054
1.1 0.998135151007896
1.2 0.997787987403395
1.3 0.99741324968153
1.4 0.997011573244923
1.5 0.996583637198211
1.6 0.996130162761802
1.7 0.995651911589677
1.8 0.99514968399566
1.9 0.994624317092837
2 0.99407668285102
2.1 0.993507686077368
2.2 0.992918262325496
2.3 0.992309375738581
2.4 0.991682016832153
2.5 0.991037200222399
2.6 0.990375962305996
2.7 0.989699358897582
2.8 0.98900846283108
2.9 0.988304361531228
3 0.987588154561699
3.1 0.986860951156281
3.2 0.98612386773964
3.3 0.98537802544418
3.4 0.984624547629552
3.5 0.983864557411349
3.6 0.983099175205489
3.7 0.982329516294752
3.8 0.981556688423849
3.9 0.980781789429364
4 0.980005904910758
4.1 0.979230105948557
4.2 0.978455446875682
4.3 0.977682963107717
4.4 0.976913669037799
4.5 0.976148556001564
4.6 0.975388590317433
4.7 0.974634711407276
4.8 0.97388783000229
4.9 0.973148826438666
5.2 0.970987397109026
5.4 0.969600465848411
5.6 0.96826324896746
5.8 0.966980609363759
6 0.965756753708355
6.2 0.964595214665008
6.4 0.963498840880886
6.6 0.962469794788419
6.8 0.961509558167136
7 0.960618945323675
7.2 0.959798123659158
7.4 0.959046641306591
7.6 0.958363461438052
7.8 0.95774700276327
8 0.95719518566863
8.2 0.956705483379663
8.4 0.956274977471574
8.6 0.955900417001974
8.8 0.955578280498487
9 0.955304840001777
9.2 0.955076226342281
9.4 0.954888494816744
9.6 0.954737690428853
9.8 0.95461991186675
10 0.954531373409004
10.2 0.954468463979411
10.4 0.954427802609503
10.6 0.954406289615278
10.8 0.954401152850956
11 0.95440998846663
11.2 0.95443079566786
11.4 0.954462005052472
11.6 0.954502500182241
11.8 0.954551632133556
12 0.954609226860607
12.2 0.954675585295796
12.4 0.954751476204028
12.6 0.954838121898869
12.8 0.954937177018298
13 0.955050700644747
13.2 0.955181122137236
13.4 0.955331201121688
13.6 0.955503982157924
13.8 0.955702744667704
14 0.955930948766484
14.2 0.956192177691863
14.4 0.956490077563316
14.6 0.956828295240304
14.8 0.957210415068972
15 0.957639895321126
15.2 0.958120005132929
15.4 0.958653762744875
15.6 0.959243875829151
15.8 0.959892684665825
16 0.960602108895725
16.2 0.96137359853586
16.4 0.962208089893405
16.6 0.963105966957151
16.8 0.964067028781767
17 0.965090463310899
17.2 0.966174828010924
17.4 0.967318037608904
17.6 0.968517359146984
17.8 0.969769414481878
18 0.971070190273289
18.2 0.972415055419997
18.4 0.973798785817768
18.6 0.975215596230274
18.8 0.976659178983568
19 0.978122749117346
19.2 0.979599095552863
19.4 0.981080637768961
19.6 0.982559487414655
19.8 0.984027514230016
};

\nextgroupplot[
width =\linewidth, 
height = 0.65*\linewidth,
tick align=outside,
tick pos=left,
title={$p=0.001$},
x grid style={darkgray176},
xmin=-0.885000000000001, xmax=20.785,
xtick style={color=black},
y grid style={darkgray176},
ymin=0.794006360776428, ymax=1.11257495812451,
ytick style={color=black}
]
\addplot [draw=c1, fill=c1, mark=o, only marks]
table{%
x  y
0.1 0.999920604709652
0.2 1.00019958804987
0.3 1.00046068016486
0.4 1.00016128104782
0.5 1.00049873878133
0.6 0.999783958084749
0.7 0.999388832995195
0.8 0.999071791330344
0.9 0.998988056265268
1 0.998732751459638
1.1 0.998596780514075
1.2 0.997323272233515
1.3 0.998050244515554
1.4 0.996982756987634
1.5 0.997331777345219
1.6 0.996888441180036
1.7 0.996669960955048
1.8 0.995350957861403
1.9 0.994028660583104
2 0.994251032769894
2.1 0.993844950627215
2.2 0.993811601703029
2.3 0.992145112460823
2.4 0.991935427496949
2.5 0.991648628382856
2.6 0.990509423934426
2.7 0.991120919402927
2.8 0.99154315625529
2.9 0.989649123885936
3 0.989041769914982
3.1 0.98806068536154
3.2 0.987307676517498
3.3 0.986472494565697
3.4 0.986048308872847
3.5 0.984841001623649
3.6 0.984404429442294
3.7 0.983371105755594
3.8 0.982806553278271
3.9 0.979927536359773
4 0.982854117149365
4.1 0.981222031791367
4.2 0.979350298047618
4.3 0.978678372710702
4.4 0.976405877979324
4.5 0.979017015171729
4.6 0.975724960906214
4.7 0.976539839368594
4.8 0.97541432495305
4.9 0.972980863691509
5.2 0.971101291682574
5.4 0.969279408891896
5.6 0.971228047206695
5.8 0.968955594547036
6 0.963723659137131
6.2 0.966289318430633
6.4 0.964769322717908
6.6 0.965819354178229
6.8 0.961243469699415
7 0.962601108958498
7.2 0.958055269671894
7.4 0.963775757511647
7.6 0.960244858408883
7.8 0.953074361593756
8 0.948424124812799
8.2 0.956671286460129
8.4 0.958961301878112
8.6 0.956276132892681
8.8 0.957601040571353
9 0.964443897804209
9.2 0.954452306227882
9.4 0.960023740338382
9.6 0.960854119793363
9.8 0.960077019862541
10 0.947472919727129
10.2 0.957868274927717
10.4 0.959610693798749
10.6 0.95320346047487
10.8 0.950423834231231
11 0.942570385616561
11.2 0.953259748398342
11.4 0.962137195139982
11.6 0.939833593647116
11.8 0.945365900487872
12 0.951322300641677
12.2 0.93737772741333
12.4 0.945682288584205
12.6 0.962908648096355
12.8 0.953572018808845
13 0.953483332580111
13.2 0.950833626716968
13.4 0.938553659891183
13.6 0.952189442365051
13.8 0.959092021020704
14 0.947166317745177
14.2 0.94777795263133
14.4 0.938406589211413
14.6 0.921965041261686
14.8 0.981087757277307
15 0.935943514311435
15.2 0.990331808714475
15.4 0.919345062640182
15.6 0.946372241030105
15.8 0.962641214955458
16 0.950371600229735
16.2 0.955368986529503
16.4 0.954094108573785
16.6 0.977048702788969
16.8 1.06543127805524
17 0.974935140451634
17.2 0.917678881683298
17.4 0.933919665487682
17.6 0.991743098665896
17.8 0.961067868577
18 0.947296312785057
18.2 0.973329595242358
18.4 0.938943472963783
18.6 0.989558619926999
18.8 0.971901717291474
19 0.868722506445955
19.2 0.808486751564977
19.4 0.903923229265623
19.6 0.89163968490037
19.8 1.09809456733596
};
\addplot [semithick, c3]
table {%
0.1 0.999984320102678
0.2 0.999937307449658
0.3 0.999859043095476
0.4 0.999749661918819
0.5 0.999609352304798
0.6 0.9994383557013
0.7 0.999236966050588
0.8 0.999005529097635
0.9 0.998744441577026
1 0.99845415028054
1.1 0.998135151007896
1.2 0.997787987403395
1.3 0.99741324968153
1.4 0.997011573244923
1.5 0.996583637198211
1.6 0.996130162761802
1.7 0.995651911589677
1.8 0.99514968399566
1.9 0.994624317092837
2 0.99407668285102
2.1 0.993507686077368
2.2 0.992918262325496
2.3 0.992309375738581
2.4 0.991682016832153
2.5 0.991037200222399
2.6 0.990375962305996
2.7 0.989699358897582
2.8 0.98900846283108
2.9 0.988304361531228
3 0.987588154561699
3.1 0.986860951156281
3.2 0.98612386773964
3.3 0.98537802544418
3.4 0.984624547629552
3.5 0.983864557411349
3.6 0.983099175205489
3.7 0.982329516294752
3.8 0.981556688423849
3.9 0.980781789429364
4 0.980005904910758
4.1 0.979230105948557
4.2 0.978455446875682
4.3 0.977682963107717
4.4 0.976913669037799
4.5 0.976148556001564
4.6 0.975388590317433
4.7 0.974634711407276
4.8 0.97388783000229
4.9 0.973148826438666
5.2 0.970987397109026
5.4 0.969600465848411
5.6 0.96826324896746
5.8 0.966980609363759
6 0.965756753708355
6.2 0.964595214665008
6.4 0.963498840880886
6.6 0.962469794788419
6.8 0.961509558167136
7 0.960618945323675
7.2 0.959798123659158
7.4 0.959046641306591
7.6 0.958363461438052
7.8 0.95774700276327
8 0.95719518566863
8.2 0.956705483379663
8.4 0.956274977471574
8.6 0.955900417001974
8.8 0.955578280498487
9 0.955304840001777
9.2 0.955076226342281
9.4 0.954888494816744
9.6 0.954737690428853
9.8 0.95461991186675
10 0.954531373409004
10.2 0.954468463979411
10.4 0.954427802609503
10.6 0.954406289615278
10.8 0.954401152850956
11 0.95440998846663
11.2 0.95443079566786
11.4 0.954462005052472
11.6 0.954502500182241
11.8 0.954551632133556
12 0.954609226860607
12.2 0.954675585295796
12.4 0.954751476204028
12.6 0.954838121898869
12.8 0.954937177018298
13 0.955050700644747
13.2 0.955181122137236
13.4 0.955331201121688
13.6 0.955503982157924
13.8 0.955702744667704
14 0.955930948766484
14.2 0.956192177691863
14.4 0.956490077563316
14.6 0.956828295240304
14.8 0.957210415068972
15 0.957639895321126
15.2 0.958120005132929
15.4 0.958653762744875
15.6 0.959243875829151
15.8 0.959892684665825
16 0.960602108895725
16.2 0.96137359853586
16.4 0.962208089893405
16.6 0.963105966957151
16.8 0.964067028781767
17 0.965090463310899
17.2 0.966174828010924
17.4 0.967318037608904
17.6 0.968517359146984
17.8 0.969769414481878
18 0.971070190273289
18.2 0.972415055419997
18.4 0.973798785817768
18.6 0.975215596230274
18.8 0.976659178983568
19 0.978122749117346
19.2 0.979599095552863
19.4 0.981080637768961
19.6 0.982559487414655
19.8 0.984027514230016
};

\nextgroupplot[
width =\linewidth, 
height = 0.65*\linewidth,
tick align=outside,
tick pos=left,
title={$p=0.01$},
x grid style={darkgray176},
xmin=-0.885000000000001, xmax=20.785,
xtick style={color=black},
y grid style={darkgray176},
ymin=-0.0856695, ymax=1.0659403,
ytick style={color=black}
]
\addplot [draw=c1, fill=c1, mark=o, only marks]
table{%
x  y
0.1 1.0135944
0.2 0.999681200000001
0.3 0.980602
0.4 0.97717
0.5 0.9164668
0.6 0.923576400000001
0.7 0.8684048
0.8 0.8524164
0.9 0.823062
1 0.7644128
1.1 0.7359432
1.2 0.6841368
1.3 0.6592736
1.4 0.6070732
1.5 0.5575548
1.6 0.52594
1.7 0.4708144
1.8 0.433806
1.9 0.3964248
2 0.3742664
2.1 0.3245452
2.2 0.3224016
2.3 0.2572704
2.4 0.266116
2.5 0.2304884
2.6 0.2327424
2.7 0.1605248
2.8 0.196136
2.9 0.1373524
3 0.119316
3.1 0.1219788
3.2 0.1202416
3.3 0.0986108
3.4 0.0985568
3.5 0.085934
3.6 0.059456
3.7 0.0822648
3.8 0.0736876
3.9 0.0326948
4 0.0615656
4.1 0.0269388
4.2 0.0190292
4.3 0.0080568
4.4 0.0266432
4.5 0.0386728
4.6 0.0191652
4.7 0.0164868
4.8 0.020058
4.9 0.0086044
5.2 0.0250908
5.4 0.00945
5.6 0.0196492
5.8 0.008814
6 0.0035072
6.2 0.0177832
6.4 -0.000777600000000002
6.6 0.0053288
6.8 -0.0012664
7 0.0060712
7.2 0.00241
7.4 -0.007026
7.6 0.021648
7.8 -0.0188808
8 -0.0183524
8.2 -0.0081124
8.4 -0.0129612
8.6 -0.0087292
8.8 0.0098672
9 0.009958
9.2 0.0011756
9.4 0.0190856
9.6 0.0100912
9.8 0.013886
10 -0.0061548
10.2 0.0086844
10.4 0.0177604
10.6 -0.0281128
10.8 0.0010624
11 0.0050252
11.2 -0.003238
11.4 0.0067872
11.6 -0.0015892
11.8 -0.0199524
12 -0.0065704
12.2 -0.0240956
12.4 0.000292400000000001
12.6 0.0185844
12.8 0.0097204
13 -0.0093552
13.2 0.008364
13.4 0.0128856
13.6 -0.0084124
13.8 -0.015002
14 0.0350616
14.2 0.0020452
14.4 -0.0026408
14.6 -0.0013972
14.8 -0.0096696
15 -0.0055624
15.2 -0.005106
15.4 0.0182256
15.6 -0.0077372
15.8 -0.0265276
16 -0.0138856
16.2 0.0155432
16.4 0.006702
16.6 -0.02025
16.8 0.0057116
17 -0.0061552
17.2 -0.0147628
17.4 0.010606
17.6 0.0050412
17.8 0.0170704
18 0.0010824
18.2 0.013888
18.4 0.020094
18.6 -0.0333236
18.8 -0.0040384
19 -0.0189328
19.2 -0.000247600000000001
19.4 0.027898
19.6 0.019648
19.8 0.0136288
};
\addplot [semithick, c3]
table {%
0.1 0.999984320102678
0.2 0.999937307449658
0.3 0.999859043095476
0.4 0.999749661918819
0.5 0.999609352304798
0.6 0.9994383557013
0.7 0.999236966050588
0.8 0.999005529097635
0.9 0.998744441577026
1 0.99845415028054
1.1 0.998135151007896
1.2 0.997787987403395
1.3 0.99741324968153
1.4 0.997011573244923
1.5 0.996583637198211
1.6 0.996130162761802
1.7 0.995651911589677
1.8 0.99514968399566
1.9 0.994624317092837
2 0.99407668285102
2.1 0.993507686077368
2.2 0.992918262325496
2.3 0.992309375738581
2.4 0.991682016832153
2.5 0.991037200222399
2.6 0.990375962305996
2.7 0.989699358897582
2.8 0.98900846283108
2.9 0.988304361531228
3 0.987588154561699
3.1 0.986860951156281
3.2 0.98612386773964
3.3 0.98537802544418
3.4 0.984624547629552
3.5 0.983864557411349
3.6 0.983099175205489
3.7 0.982329516294752
3.8 0.981556688423849
3.9 0.980781789429364
4 0.980005904910758
4.1 0.979230105948557
4.2 0.978455446875682
4.3 0.977682963107717
4.4 0.976913669037799
4.5 0.976148556001564
4.6 0.975388590317433
4.7 0.974634711407276
4.8 0.97388783000229
4.9 0.973148826438666
5.2 0.970987397109026
5.4 0.969600465848411
5.6 0.96826324896746
5.8 0.966980609363759
6 0.965756753708355
6.2 0.964595214665008
6.4 0.963498840880886
6.6 0.962469794788419
6.8 0.961509558167136
7 0.960618945323675
7.2 0.959798123659158
7.4 0.959046641306591
7.6 0.958363461438052
7.8 0.95774700276327
8 0.95719518566863
8.2 0.956705483379663
8.4 0.956274977471574
8.6 0.955900417001974
8.8 0.955578280498487
9 0.955304840001777
9.2 0.955076226342281
9.4 0.954888494816744
9.6 0.954737690428853
9.8 0.95461991186675
10 0.954531373409004
10.2 0.954468463979411
10.4 0.954427802609503
10.6 0.954406289615278
10.8 0.954401152850956
11 0.95440998846663
11.2 0.95443079566786
11.4 0.954462005052472
11.6 0.954502500182241
11.8 0.954551632133556
12 0.954609226860607
12.2 0.954675585295796
12.4 0.954751476204028
12.6 0.954838121898869
12.8 0.954937177018298
13 0.955050700644747
13.2 0.955181122137236
13.4 0.955331201121688
13.6 0.955503982157924
13.8 0.955702744667704
14 0.955930948766484
14.2 0.956192177691863
14.4 0.956490077563316
14.6 0.956828295240304
14.8 0.957210415068972
15 0.957639895321126
15.2 0.958120005132929
15.4 0.958653762744875
15.6 0.959243875829151
15.8 0.959892684665825
16 0.960602108895725
16.2 0.96137359853586
16.4 0.962208089893405
16.6 0.963105966957151
16.8 0.964067028781767
17 0.965090463310899
17.2 0.966174828010924
17.4 0.967318037608904
17.6 0.968517359146984
17.8 0.969769414481878
18 0.971070190273289
18.2 0.972415055419997
18.4 0.973798785817768
18.6 0.975215596230274
18.8 0.976659178983568
19 0.978122749117346
19.2 0.979599095552863
19.4 0.981080637768961
19.6 0.982559487414655
19.8 0.984027514230016
};

\nextgroupplot[
width =\linewidth, 
height = 0.65*\linewidth,
tick align=outside,
tick pos=left,
ylabel={$6$-qubits},
x grid style={darkgray176},
xmin=-0.885000000000001, xmax=20.785,
xtick style={color=black},
y grid style={darkgray176},
ymin=0.967139291397986, ymax=1.00154836908903,
ytick style={color=black}
]
\addplot [draw=c1, fill=c1, mark=o, only marks]
table{%
x  y
0.1 0.999933519111768
0.2 0.999865557246038
0.3 0.999884756939229
0.4 0.999647970095817
0.5 0.999493339171725
0.6 0.999553764808069
0.7 0.998972550410286
0.8 0.999105882752058
0.9 0.998813626654731
1 0.998199601361061
1.1 0.998269557254408
1.2 0.99783534131868
1.3 0.997633603559627
1.4 0.996707027653243
1.5 0.996597881360915
1.6 0.996656234665378
1.7 0.995640016033648
1.8 0.995108637590011
1.9 0.994377774633812
2 0.994668616077234
2.1 0.993357678405353
2.2 0.992971674604782
2.3 0.992603549725808
2.4 0.99222798109375
2.5 0.991241576831118
2.6 0.990306709106558
2.7 0.990256233245692
2.8 0.989498663837269
2.9 0.988532409529617
3 0.988415834475084
3.1 0.98745074487837
3.2 0.986025344053827
3.3 0.985689603171832
3.4 0.985749228216856
3.5 0.984437954456846
3.6 0.983995793808154
3.7 0.983133894389218
3.8 0.983022217048948
3.9 0.982252436103267
4 0.981751144149264
4.1 0.98053826711733
4.2 0.979738143327529
4.3 0.979281473433388
4.4 0.979566953354591
4.5 0.977548848711093
4.6 0.977730941203897
4.7 0.97691335366357
4.8 0.976890353376943
4.9 0.975978643199749
5.2 0.974623637590823
5.4 0.973899698062626
5.6 0.972465042590144
5.8 0.971517994707575
6 0.971326068214242
6.2 0.970264496079581
6.4 0.969811222595692
6.6 0.969971625275375
6.8 0.969178549422575
7 0.969267507306328
7.2 0.969177207117199
7.4 0.968703340383942
7.6 0.96933363777815
7.8 0.970031028820228
8 0.970620465496067
8.2 0.97037610566876
8.4 0.96981119521889
8.6 0.970560377567306
8.8 0.970295018094528
9 0.971087862829267
9.2 0.972493000002315
9.4 0.97165298495692
9.6 0.97199660379137
9.8 0.97284255245411
10 0.973592883809749
10.2 0.973273223910247
10.4 0.974375086299437
10.6 0.974961592004134
10.8 0.975018490229603
11 0.973917338101811
11.2 0.975397377453571
11.4 0.975596386166345
11.6 0.975611970251255
11.8 0.975760938319983
12 0.976134099126876
12.2 0.975086851223396
12.4 0.974762121782027
12.6 0.975272229031822
12.8 0.974987674106329
13 0.974154977640472
13.2 0.97467333403093
13.4 0.973993488578421
13.6 0.97369079468164
13.8 0.974565045188216
14 0.974210044144957
14.2 0.973497401658553
14.4 0.972284135340668
14.6 0.973277715129754
14.8 0.972605773641241
15 0.971156127107636
15.2 0.972084836452201
15.4 0.970824610396422
15.6 0.97032897518154
15.8 0.970125446170412
16 0.970352508886505
16.2 0.971310709087766
16.4 0.969579268367378
16.6 0.971451654030211
16.8 0.970022553028943
17 0.969610120585218
17.2 0.969691590561307
17.4 0.969807676233546
17.6 0.971723317033178
17.8 0.972619901101059
18 0.972452186828612
18.2 0.973588274960409
18.4 0.973345257160303
18.6 0.973779236478079
18.8 0.975022664855894
19 0.976232467327016
19.2 0.978151654255463
19.4 0.978844626579177
19.6 0.978894649076699
19.8 0.979282999736198
};
\addplot [semithick, c3]
table {%
0.1 0.999984320103076
0.2 0.999937315438487
0.3 0.999859090996366
0.4 0.999749821453131
0.5 0.999609750711651
0.6 0.999439191259224
0.7 0.999238523345126
0.8 0.99900819397994
0.9 0.998748715759405
1 0.998460665515978
1.1 0.998144682801773
1.2 0.997801468207023
1.3 0.997431781518629
1.4 0.997036439723843
1.5 0.99661631486452
1.6 0.996172331747815
1.7 0.99570546551959
1.8 0.99521673910718
1.9 0.99470722053854
2 0.994178020145133
2.1 0.993630287656259
2.2 0.993065209192826
2.3 0.992484004168859
2.4 0.991887922109315
2.5 0.991278239392984
2.6 0.99065625592953
2.7 0.990023291779866
2.8 0.989380683729249
2.9 0.988729781822644
3 0.988071945871989
3.1 0.987408541945103
3.2 0.986740938846037
3.3 0.986070504596701
3.4 0.985398602929594
3.5 0.984726589801464
3.6 0.984055809937661
3.7 0.983387593416845
3.8 0.98272325230565
3.9 0.982064077352706
4 0.9814113347513
4.1 0.980766262979749
4.2 0.980130069728317
4.3 0.97950392892128
4.4 0.97888897784247
4.5 0.978286314372284
4.6 0.977696994343882
4.7 0.977122029025882
4.8 0.976562382738529
4.9 0.976018970609911
5.2 0.974494509593795
5.4 0.973573757025971
5.6 0.972735285299742
5.8 0.971983014054916
6 0.971319868822285
6.2 0.970747765295913
6.4 0.970267605215136
6.6 0.969879283830399
6.8 0.969581708788715
7 0.969372830138749
7.2 0.969249681024175
7.4 0.969208428508717
7.6 0.969244433858799
7.8 0.969352321501379
8 0.969526055776785
8.2 0.969759024520131
8.4 0.970044128431337
8.6 0.970373875133532
8.8 0.970740476773256
9 0.971135949983831
9.2 0.971552217015614
9.4 0.971981206833631
9.6 0.97241495499408
9.8 0.972845701136032
10 0.973265982962844
10.2 0.973668725638681
10.4 0.974047325588307
10.6 0.974395727762113
10.8 0.974708495512227
11 0.974980872318392
11.2 0.975208834703017
11.4 0.975389135782218
11.6 0.975519339012558
11.8 0.975597841810205
12 0.975623888839248
12.2 0.975597574887394
12.4 0.975519837369126
12.6 0.975392438617042
12.8 0.975217938240494
13 0.974999655945133
13.2 0.974741625316647
13.4 0.974448539175176
13.6 0.974125687202825
13.8 0.973778886633896
14 0.973414406875035
14.2 0.973038888989503
14.4 0.972659261035235
14.6 0.972282650289758
14.8 0.971916293425556
15 0.971567445717006
15.2 0.971243290364018
15.4 0.970950849008144
15.6 0.970696894494311
15.8 0.970487866895608
16 0.970329793770529
16.2 0.97022821556219
16.4 0.970188116978336
16.6 0.970213865110391
16.8 0.970309154960624
17 0.970476962949999
17.2 0.97071950887681
17.4 0.971038226689363
17.6 0.971433744326195
17.8 0.97190587276619
18 0.972453604320036
18.2 0.973075120085197
18.4 0.973767806380427
18.6 0.97452827987417
18.8 0.975352421025099
19 0.97623541536384
19.2 0.97717180206327
19.4 0.978155529171781
19.6 0.979180014819873
19.8 0.980238213656284
};

\nextgroupplot[
width =\linewidth, 
height = 0.65*\linewidth,
tick align=outside,
tick pos=left,
x grid style={darkgray176},
xmin=-0.885000000000001, xmax=20.785,
xtick style={color=black},
y grid style={darkgray176},
ymin=0.820605188202058, ymax=1.1491293264341,
ytick style={color=black}
]
\addplot [draw=c1, fill=c1, mark=o, only marks]
table{%
x  y
0.1 0.999799363214202
0.2 1.0001338466218
0.3 1.00003741309705
0.4 0.999695008024883
0.5 1.00012723252529
0.6 0.999617388267092
0.7 0.999127544311912
0.8 0.998814177256781
0.9 0.999308323298622
1 0.998941599833129
1.1 0.998122950839296
1.2 0.998152072195904
1.3 0.99695419692813
1.4 0.997377074565696
1.5 0.996652739770779
1.6 0.995410001858429
1.7 0.995382465108667
1.8 0.995200019379063
1.9 0.993872456734431
2 0.994987649047695
2.1 0.993565094710202
2.2 0.992933551269661
2.3 0.991235521568498
2.4 0.991756628371549
2.5 0.992326518430812
2.6 0.990880442441544
2.7 0.991388422255743
2.8 0.98997046496909
2.9 0.989350117097804
3 0.989563886042246
3.1 0.988083607844415
3.2 0.98777481927492
3.3 0.986180950724755
3.4 0.985283035507732
3.5 0.98678773421393
3.6 0.985190448630898
3.7 0.985342528638491
3.8 0.982364908321861
3.9 0.984185118941008
4 0.984004158005447
4.1 0.983234229149655
4.2 0.982134224053916
4.3 0.981173070407545
4.4 0.978649176433475
4.5 0.97856391170714
4.6 0.980558739552746
4.7 0.979905467640546
4.8 0.978233945541242
4.9 0.975656901987808
5.2 0.971763012524963
5.4 0.975488750536676
5.6 0.97455230760453
5.8 0.975681196050727
6 0.97213915954915
6.2 0.974279274407397
6.4 0.968169477147962
6.6 0.97094208664202
6.8 0.977014451702294
7 0.974482159441004
7.2 0.968294989368837
7.4 0.970144759655962
7.6 0.96855940642142
7.8 0.971172037353459
8 0.972396734862075
8.2 0.972502919430081
8.4 0.967470647737198
8.6 0.972356927306738
8.8 0.962177351646686
9 0.96972394050382
9.2 0.969048490409719
9.4 0.973002353515556
9.6 0.979674859277333
9.8 0.9764217133491
10 0.973446604011526
10.2 0.967646779663817
10.4 0.966645734869351
10.6 0.963093873160596
10.8 0.981388200127775
11 0.971992688190418
11.2 0.97269920396589
11.4 0.973779186085872
11.6 0.967075978483974
11.8 0.964644829935361
12 0.96646378407647
12.2 0.977217943144568
12.4 0.977509364055212
12.6 0.974502643841796
12.8 0.974104105423104
13 0.960607465788274
13.2 0.978270380019737
13.4 0.985283198513448
13.6 0.952592724335809
13.8 0.976703587774579
14 0.95841288127702
14.2 0.983631787465947
14.4 0.9432690011942
14.6 0.937328602059321
14.8 0.98852174233836
15 0.969319006662145
15.2 0.898844986655575
15.4 0.967642267223758
15.6 0.949988435736228
15.8 0.96251781513698
16 0.956344697308444
16.2 0.950163462210378
16.4 0.916520920854326
16.6 0.977721330511944
16.8 0.98631699854261
17 0.935676994347833
17.2 0.975606692492611
17.4 0.990099417151571
17.6 0.939558988793392
17.8 0.913269719443275
18 0.864728405951947
18.2 0.932387063621137
18.4 0.998267734649914
18.6 0.927835145830859
18.8 0.92522122006762
19 0.891933446832272
19.2 1.13419641105992
19.4 0.960071662976401
19.6 0.835538103576242
19.8 0.908653633946348
};
\addplot [semithick, c3]
table {%
0.1 0.999984320103076
0.2 0.999937315438487
0.3 0.999859090996366
0.4 0.999749821453131
0.5 0.999609750711651
0.6 0.999439191259224
0.7 0.999238523345126
0.8 0.99900819397994
0.9 0.998748715759405
1 0.998460665515978
1.1 0.998144682801773
1.2 0.997801468207023
1.3 0.997431781518629
1.4 0.997036439723843
1.5 0.99661631486452
1.6 0.996172331747815
1.7 0.99570546551959
1.8 0.99521673910718
1.9 0.99470722053854
2 0.994178020145133
2.1 0.993630287656259
2.2 0.993065209192826
2.3 0.992484004168859
2.4 0.991887922109315
2.5 0.991278239392984
2.6 0.99065625592953
2.7 0.990023291779866
2.8 0.989380683729249
2.9 0.988729781822644
3 0.988071945871989
3.1 0.987408541945103
3.2 0.986740938846037
3.3 0.986070504596701
3.4 0.985398602929594
3.5 0.984726589801464
3.6 0.984055809937661
3.7 0.983387593416845
3.8 0.98272325230565
3.9 0.982064077352706
4 0.9814113347513
4.1 0.980766262979749
4.2 0.980130069728317
4.3 0.97950392892128
4.4 0.97888897784247
4.5 0.978286314372284
4.6 0.977696994343882
4.7 0.977122029025882
4.8 0.976562382738529
4.9 0.976018970609911
5.2 0.974494509593795
5.4 0.973573757025971
5.6 0.972735285299742
5.8 0.971983014054916
6 0.971319868822285
6.2 0.970747765295913
6.4 0.970267605215136
6.6 0.969879283830399
6.8 0.969581708788715
7 0.969372830138749
7.2 0.969249681024175
7.4 0.969208428508717
7.6 0.969244433858799
7.8 0.969352321501379
8 0.969526055776785
8.2 0.969759024520131
8.4 0.970044128431337
8.6 0.970373875133532
8.8 0.970740476773256
9 0.971135949983831
9.2 0.971552217015614
9.4 0.971981206833631
9.6 0.97241495499408
9.8 0.972845701136032
10 0.973265982962844
10.2 0.973668725638681
10.4 0.974047325588307
10.6 0.974395727762113
10.8 0.974708495512227
11 0.974980872318392
11.2 0.975208834703017
11.4 0.975389135782218
11.6 0.975519339012558
11.8 0.975597841810205
12 0.975623888839248
12.2 0.975597574887394
12.4 0.975519837369126
12.6 0.975392438617042
12.8 0.975217938240494
13 0.974999655945133
13.2 0.974741625316647
13.4 0.974448539175176
13.6 0.974125687202825
13.8 0.973778886633896
14 0.973414406875035
14.2 0.973038888989503
14.4 0.972659261035235
14.6 0.972282650289758
14.8 0.971916293425556
15 0.971567445717006
15.2 0.971243290364018
15.4 0.970950849008144
15.6 0.970696894494311
15.8 0.970487866895608
16 0.970329793770529
16.2 0.97022821556219
16.4 0.970188116978336
16.6 0.970213865110391
16.8 0.970309154960624
17 0.970476962949999
17.2 0.97071950887681
17.4 0.971038226689363
17.6 0.971433744326195
17.8 0.97190587276619
18 0.972453604320036
18.2 0.973075120085197
18.4 0.973767806380427
18.6 0.97452827987417
18.8 0.975352421025099
19 0.97623541536384
19.2 0.97717180206327
19.4 0.978155529171781
19.6 0.979180014819873
19.8 0.980238213656284
};

\nextgroupplot[
width =\linewidth, 
height = 0.65*\linewidth,
tick align=outside,
tick pos=left,
x grid style={darkgray176},
xmin=-0.885000000000001, xmax=20.785,
xtick style={color=black},
y grid style={darkgray176},
ymin=-0.0843843800000002, ymax=1.06221118,
ytick style={color=black}
]
\addplot [draw=c1, fill=c1, mark=o, only marks]
table{%
x  y
0.1 1.0100932
0.2 0.986496000000002
0.3 0.966326000000001
0.4 0.974581599999999
0.5 0.948293200000002
0.6 0.949648000000001
0.7 0.8807356
0.8 0.844233199999999
0.9 0.8327176
1 0.7700552
1.1 0.7041984
1.2 0.6974108
1.3 0.6592364
1.4 0.624253599999999
1.5 0.510044
1.6 0.4961128
1.7 0.4719516
1.8 0.4392852
1.9 0.3815904
2 0.3649224
2.1 0.3483388
2.2 0.309018
2.3 0.2812764
2.4 0.2328396
2.5 0.2288764
2.6 0.2032136
2.7 0.1671112
2.8 0.1803596
2.9 0.1420012
3 0.1044684
3.1 0.1278424
3.2 0.0927784
3.3 0.1060768
3.4 0.0874504
3.5 0.0691036
3.6 0.0895684
3.7 0.0438236
3.8 0.0365532
3.9 0.0540392
4 0.0385756
4.1 0.0438632
4.2 0.0324092
4.3 0.0366856
4.4 0.0388772
4.5 0.000931199999999997
4.6 0.0133504
4.7 0.039054
4.8 0.0045508
4.9 0.000207600000000001
5.2 -0.0045568
5.4 0.013964
5.6 0.0052172
5.8 -0.0157132
6 -0.0089628
6.2 0.00949
6.4 -0.0208892
6.6 -0.0243636
6.8 -0.011312
7 -0.0085776
7.2 -0.0088296
7.4 0.0086792
7.6 -0.0211652
7.8 -0.0031136
8 0.0074188
8.2 0.0077808
8.4 0.0312992
8.6 0.0126608
8.8 0.0030104
9 -0.0026
9.2 0.0062496
9.4 -0.006942
9.6 -0.0073548
9.8 -0.0009404
10 -0.0016492
10.2 -0.0322664
10.4 0.0105964
10.6 -0.0043116
10.8 -0.0084768
11 -0.0098844
11.2 -0.0015364
11.4 0.0095848
11.6 0.0031672
11.8 0.0037984
12 0.0155176
12.2 0.0044536
12.4 0.0025296
12.6 -0.0107744
12.8 -0.0267996
13 -0.0144252
13.2 0.005882
13.4 -0.0144136
13.6 -0.008532
13.8 0.0041148
14 -0.0204816
14.2 -0.0080136
14.4 0.012572
14.6 -0.000102799999999999
14.8 0.015086
15 -0.0061608
15.2 0.0115948
15.4 -0.0028316
15.6 0.0089924
15.8 -0.0113544
16 0.0088744
16.2 0.013322
16.4 0.0230488
16.6 -0.0065276
16.8 -0.0004476
17 -0.0154268
17.2 -0.0053996
17.4 -0.0079352
17.6 -0.0183516
17.8 0.0211072
18 0.0098484
18.2 0.0027952
18.4 0.0011356
18.6 -0.00177
18.8 0.0099664
19 0.0052432
19.2 -0.000516
19.4 -0.0049268
19.6 -0.004862
19.8 -0.0067916
};
\addplot [semithick, c3]
table {%
0.1 0.999984320103076
0.2 0.999937315438487
0.3 0.999859090996366
0.4 0.999749821453131
0.5 0.999609750711651
0.6 0.999439191259224
0.7 0.999238523345126
0.8 0.99900819397994
0.9 0.998748715759405
1 0.998460665515978
1.1 0.998144682801773
1.2 0.997801468207023
1.3 0.997431781518629
1.4 0.997036439723843
1.5 0.99661631486452
1.6 0.996172331747815
1.7 0.99570546551959
1.8 0.99521673910718
1.9 0.99470722053854
2 0.994178020145133
2.1 0.993630287656259
2.2 0.993065209192826
2.3 0.992484004168859
2.4 0.991887922109315
2.5 0.991278239392984
2.6 0.99065625592953
2.7 0.990023291779866
2.8 0.989380683729249
2.9 0.988729781822644
3 0.988071945871989
3.1 0.987408541945103
3.2 0.986740938846037
3.3 0.986070504596701
3.4 0.985398602929594
3.5 0.984726589801464
3.6 0.984055809937661
3.7 0.983387593416845
3.8 0.98272325230565
3.9 0.982064077352706
4 0.9814113347513
4.1 0.980766262979749
4.2 0.980130069728317
4.3 0.97950392892128
4.4 0.97888897784247
4.5 0.978286314372284
4.6 0.977696994343882
4.7 0.977122029025882
4.8 0.976562382738529
4.9 0.976018970609911
5.2 0.974494509593795
5.4 0.973573757025971
5.6 0.972735285299742
5.8 0.971983014054916
6 0.971319868822285
6.2 0.970747765295913
6.4 0.970267605215136
6.6 0.969879283830399
6.8 0.969581708788715
7 0.969372830138749
7.2 0.969249681024175
7.4 0.969208428508717
7.6 0.969244433858799
7.8 0.969352321501379
8 0.969526055776785
8.2 0.969759024520131
8.4 0.970044128431337
8.6 0.970373875133532
8.8 0.970740476773256
9 0.971135949983831
9.2 0.971552217015614
9.4 0.971981206833631
9.6 0.97241495499408
9.8 0.972845701136032
10 0.973265982962844
10.2 0.973668725638681
10.4 0.974047325588307
10.6 0.974395727762113
10.8 0.974708495512227
11 0.974980872318392
11.2 0.975208834703017
11.4 0.975389135782218
11.6 0.975519339012558
11.8 0.975597841810205
12 0.975623888839248
12.2 0.975597574887394
12.4 0.975519837369126
12.6 0.975392438617042
12.8 0.975217938240494
13 0.974999655945133
13.2 0.974741625316647
13.4 0.974448539175176
13.6 0.974125687202825
13.8 0.973778886633896
14 0.973414406875035
14.2 0.973038888989503
14.4 0.972659261035235
14.6 0.972282650289758
14.8 0.971916293425556
15 0.971567445717006
15.2 0.971243290364018
15.4 0.970950849008144
15.6 0.970696894494311
15.8 0.970487866895608
16 0.970329793770529
16.2 0.97022821556219
16.4 0.970188116978336
16.6 0.970213865110391
16.8 0.970309154960624
17 0.970476962949999
17.2 0.97071950887681
17.4 0.971038226689363
17.6 0.971433744326195
17.8 0.97190587276619
18 0.972453604320036
18.2 0.973075120085197
18.4 0.973767806380427
18.6 0.97452827987417
18.8 0.975352421025099
19 0.97623541536384
19.2 0.97717180206327
19.4 0.978155529171781
19.6 0.979180014819873
19.8 0.980238213656284
};

\nextgroupplot[
width =\linewidth, 
height = 0.65*\linewidth,
tick align=outside,
tick pos=left,
ylabel={$8$-qubits},
x grid style={darkgray176},
xmin=-0.885000000000001, xmax=20.785,
xtick style={color=black},
y grid style={darkgray176},
ymin=0.967151310440494, ymax=1.00154779675409,
ytick style={color=black}
]
\addplot [draw=c1, fill=c1, mark=o, only marks]
table{%
x  y
0.1 0.999965087864546
0.2 0.999938472804196
0.3 0.999815750046322
0.4 0.999831465333365
0.5 0.999442243023397
0.6 0.999452796678686
0.7 0.999431970164732
0.8 0.99903850210069
0.9 0.998532257309732
1 0.998378065720129
1.1 0.998019239500998
1.2 0.99775815289307
1.3 0.997547190595131
1.4 0.997080325492973
1.5 0.996981481127734
1.6 0.996147539569741
1.7 0.995824056225454
1.8 0.995254114378185
1.9 0.994592438667518
2 0.994189546796835
2.1 0.993414019036146
2.2 0.993088850420025
2.3 0.99261533544711
2.4 0.991965996534135
2.5 0.991429354807697
2.6 0.990700231170261
2.7 0.990577721033196
2.8 0.989956798894
2.9 0.988565600472308
3 0.988122002550209
3.1 0.987339105252991
3.2 0.98693611109702
3.3 0.985951807703771
3.4 0.985663384151316
3.5 0.985298000109455
3.6 0.983921502920348
3.7 0.98331509714366
3.8 0.982980186240333
3.9 0.982310153735164
4 0.981308616040896
4.1 0.981179730701665
4.2 0.980321536879315
4.3 0.979413181581095
4.4 0.979044429936845
4.5 0.978484712818665
4.6 0.977801785526355
4.7 0.977359998522417
4.8 0.976509507976681
4.9 0.976412582240822
5.2 0.975090214755872
5.4 0.973448946379546
5.6 0.972601401760749
5.8 0.972183569974781
6 0.97135408238023
6.2 0.97116997795427
6.4 0.970202587287959
6.6 0.970024640456273
6.8 0.969128738127807
7 0.96965280135809
7.2 0.969508854412285
7.4 0.96909516038361
7.6 0.968714787091112
7.8 0.969102389656449
8 0.970012014887738
8.2 0.969306354085241
8.4 0.971131920280235
8.6 0.970820850432692
8.8 0.970996804937258
9 0.971284515975554
9.2 0.971001315406712
9.4 0.972854521962688
9.6 0.97267758618204
9.8 0.972500505457718
10 0.973342872212674
10.2 0.974459621112674
10.4 0.973875935882907
10.6 0.974657035698131
10.8 0.974953007107993
11 0.975558444158672
11.2 0.975010766969411
11.4 0.976314805149429
11.6 0.976278638314504
11.8 0.975644996378694
12 0.977316775371521
12.2 0.976442049941512
12.4 0.97583103514158
12.6 0.974494046615359
12.8 0.974855077433961
13 0.975040077677285
13.2 0.975504927746652
13.4 0.974614178523059
13.6 0.97510111941805
13.8 0.974474431884432
14 0.973819851325075
14.2 0.973336852805995
14.4 0.972834747822622
14.6 0.972888110771869
14.8 0.972581067057784
15 0.972022651647001
15.2 0.971845204786934
15.4 0.972227086837526
15.6 0.972634420079357
15.8 0.971483714271619
16 0.969802931406262
16.2 0.97055707985805
16.4 0.971683932969814
16.6 0.971026505409966
16.8 0.972181113042308
17 0.970263936883518
17.2 0.970675707098323
17.4 0.972443918426316
17.6 0.972487013404301
17.8 0.972626856446906
18 0.972717100349431
18.2 0.972973075715614
18.4 0.974262865875459
18.6 0.976437524055681
18.8 0.976783414518277
19 0.977752326869258
19.2 0.977353781234294
19.4 0.9786774825992
19.6 0.981197065718218
19.8 0.980588269573699
};
\addplot [semithick, c3]
table {%
0.1 0.999984320103473
0.2 0.999937315439281
0.3 0.999859090997558
0.4 0.999749821454722
0.5 0.999609750713641
0.6 0.999439191261625
0.7 0.999238523347968
0.8 0.999008193983297
0.9 0.998748715763447
1 0.998460665521047
1.1 0.998144682808517
1.2 0.997801468216591
1.3 0.997431781532966
1.4 0.997036439746098
1.5 0.99661631489962
1.6 0.996172331803226
1.7 0.995705465606342
1.8 0.995216739241186
1.9 0.99470722074228
2 0.994178020449779
2.1 0.993630288104328
2.2 0.993065209841435
2.3 0.992484005093724
2.4 0.991887923409542
2.5 0.991278241196836
2.6 0.990656258401264
2.7 0.990023295127801
2.8 0.98938068821524
2.9 0.988729787773094
3 0.988071953690614
3.1 0.987408552127615
3.2 0.986740951996979
3.3 0.986070521448611
3.4 0.98539862436476
3.5 0.984726616876487
3.6 0.984055843911066
3.7 0.983387635780014
3.8 0.982723304817308
3.9 0.982064142077253
4 0.981411414101235
4.1 0.980766359762474
4.2 0.980130187197606
4.3 0.979504070833663
4.4 0.978889148518803
4.5 0.9782865187648
4.6 0.977697238108954
4.7 0.977122318602772
4.8 0.976562725434341
4.9 0.976019374690983
5.2 0.974495158532728
5.4 0.9735746328512
5.6 0.972736453451686
5.8 0.971984554932471
6 0.971321880273392
6.2 0.970750365338678
6.4 0.97027093497902
6.6 0.969883510700707
6.8 0.969587029730794
7 0.969379475171267
7.2 0.969257916803466
7.4 0.969218561978989
7.6 0.969256815915807
7.8 0.969367350610332
8 0.969544181478917
8.2 0.969780750756762
8.4 0.970070016609432
8.6 0.970404546853031
8.8 0.970776616133885
9 0.971178305387839
9.2 0.971601602383075
9.4 0.97203850214859
9.6 0.972481106103205
9.8 0.972921718726236
10 0.973352940650868
10.2 0.973767757113702
10.4 0.974159620758244
10.6 0.974522527865277
10.8 0.974851087168291
11 0.97514058050605
11.2 0.975387014666074
11.4 0.975587163880983
11.6 0.97573860255298
11.8 0.975839727899167
12 0.975889772330283
12.2 0.97588880549681
12.4 0.975837726057687
12.6 0.975738243346859
12.8 0.97559284923019
13 0.975404780558692
13.2 0.975177972732115
13.4 0.974917004988777
13.6 0.974627038131469
13.8 0.97431374548481
14 0.973983237954932
14.2 0.973641984127443
14.4 0.973296726393068
14.6 0.9729543941317
14.8 0.972622015014265
15 0.972306625497372
15.2 0.972015181588017
15.4 0.971754470944588
15.6 0.971531027356245
15.8 0.971351048605775
16 0.971220318671694
16.2 0.971144135164522
16.4 0.971127242820516
16.6 0.971173773794891
16.8 0.97128719540674
17 0.971470265890879
17.2 0.971724998608966
17.4 0.972052635065095
17.6 0.972453626961039
17.8 0.972927627414935
18 0.973473491356129
18.2 0.974089284999431
18.4 0.974772304195733
18.6 0.975519101354039
18.8 0.976325520533911
19 0.977186740217769
19.2 0.978097323191021
19.4 0.979051272884884
19.6 0.980042095473023
19.8 0.981062866959086
};

\nextgroupplot[
width =\linewidth, 
height = 0.65*\linewidth,
tick align=outside,
tick pos=left,
x grid style={darkgray176},
xmin=-0.894999999999999, xmax=20.995,
xtick style={color=black},
y grid style={darkgray176},
ymin=0.908353749916597, ymax=1.01450009143205,
ytick style={color=black}
]
\addplot [draw=c1, fill=c1, mark=o, only marks]
table{%
x  y
0.1 1.00011402643519
0.2 0.999697316084261
0.3 0.999115976049858
0.4 1.00013874935181
0.5 0.998605334988965
0.6 0.999075484679482
0.7 0.9985097285253
0.8 1.00014289773748
0.9 0.998973827776299
1 0.99811761076149
1.1 0.99997025589486
1.2 0.997532910226437
1.3 0.99862979130424
1.4 0.996164971874238
1.5 0.997292136245755
1.6 0.996172944554579
1.7 0.993675104806826
1.8 0.996208384060124
1.9 0.995468384245022
2 0.995941134078743
2.1 0.990827044068051
2.2 0.993071349949545
2.3 0.993254484146934
2.4 0.99291171346503
2.5 0.991227994835766
2.6 0.991919029982138
2.7 0.988578525780601
2.8 0.987119259914978
2.9 0.987011390923236
3 0.98504929014223
3.1 0.987911549518036
3.2 0.985725653252595
3.3 0.986989640250386
3.4 0.987909394104942
3.5 0.990242864024517
3.6 0.98331060634909
3.7 0.98733103517796
3.8 0.981726132056424
3.9 0.982433728594261
4 0.980654102772627
4.1 0.980364178572037
4.2 0.980502807836555
4.3 0.981705182683556
4.4 0.978400083878064
4.5 0.981574826084706
4.6 0.981275244585845
4.7 0.979272841755816
4.8 0.967356594561257
4.9 0.97606383772901
5.2 0.976022812084995
5.4 0.977714746362528
5.6 0.96879440897131
5.8 0.972947780602295
6 0.967447886449306
6.2 0.969913963962334
6.4 0.970342386274559
6.6 0.97156735068189
6.8 0.968271896620273
7 0.970122110784474
7.2 0.964168861275232
7.4 0.982250798527237
7.6 0.976754933453685
7.8 0.971517091170633
8 0.97209857064739
8.2 0.971811341572117
8.4 0.982523327451364
8.6 0.975172041260433
8.8 0.970273348310984
9 0.968627180973447
9.2 0.960948076767031
9.4 0.969099119183849
9.6 0.97249428188023
9.8 0.96475413187281
10 0.974394941441469
10.2 0.963595920257041
10.4 0.971638570759023
10.6 0.973294244322143
10.8 0.987449771097253
11 0.980119545742292
11.2 0.980293178658902
11.4 1.00153629260405
11.6 0.987105946239412
11.8 0.977570397579093
12 0.983356397908205
12.2 1.00967525772681
12.4 0.97841954248112
12.6 0.973096633029125
12.8 0.988375870069399
13 0.96246898743032
13.2 0.965286124076004
13.4 0.960323948774429
13.6 0.952989348862589
13.8 0.987487798499098
14 0.999215922650212
14.2 0.976318672988648
14.4 0.968990308982267
14.6 0.934597851657325
14.8 0.953694657175674
15 0.965271440886401
15.2 0.985509148987085
15.4 0.955357415483856
15.6 0.942857876006587
15.8 0.989966532013482
16.2 0.960492573646277
16.4 0.941082120071758
16.6 0.969405429678129
16.8 0.949644070500239
17 0.956189259949208
17.2 0.924428311389638
17.4 0.990060096008149
17.6 1.0023078404636
17.8 0.934363526556467
18 0.96052661004285
18.2 0.964391033359176
18.4 0.954266270630012
18.6 1.00072415252076
18.8 0.963147226107668
19 0.913178583621845
19.2 0.986681858742802
19.4 0.999451674033691
19.6 0.961588718946669
19.8 0.952377670974784
20 0.987096156630219
};
\addplot [semithick, c3]
table {%
0.1 0.999984320103473
0.2 0.999937315439281
0.3 0.999859090997558
0.4 0.999749821454722
0.5 0.999609750713641
0.6 0.999439191261625
0.7 0.999238523347968
0.8 0.999008193983297
0.9 0.998748715763447
1 0.998460665521047
1.1 0.998144682808517
1.2 0.997801468216591
1.3 0.997431781532966
1.4 0.997036439746098
1.5 0.99661631489962
1.6 0.996172331803226
1.7 0.995705465606342
1.8 0.995216739241186
1.9 0.99470722074228
2 0.994178020449779
2.1 0.993630288104328
2.2 0.993065209841435
2.3 0.992484005093724
2.4 0.991887923409542
2.5 0.991278241196836
2.6 0.990656258401264
2.7 0.990023295127801
2.8 0.98938068821524
2.9 0.988729787773094
3 0.988071953690614
3.1 0.987408552127615
3.2 0.986740951996979
3.3 0.986070521448611
3.4 0.98539862436476
3.5 0.984726616876487
3.6 0.984055843911066
3.7 0.983387635780014
3.8 0.982723304817308
3.9 0.982064142077253
4 0.981411414101235
4.1 0.980766359762474
4.2 0.980130187197606
4.3 0.979504070833663
4.4 0.978889148518803
4.5 0.9782865187648
4.6 0.977697238108954
4.7 0.977122318602772
4.8 0.976562725434341
4.9 0.976019374690983
5.2 0.974495158532728
5.4 0.9735746328512
5.6 0.972736453451686
5.8 0.971984554932471
6 0.971321880273392
6.2 0.970750365338678
6.4 0.97027093497902
6.6 0.969883510700707
6.8 0.969587029730794
7 0.969379475171267
7.2 0.969257916803466
7.4 0.969218561978989
7.6 0.969256815915807
7.8 0.969367350610332
8 0.969544181478917
8.2 0.969780750756762
8.4 0.970070016609432
8.6 0.970404546853031
8.8 0.970776616133885
9 0.971178305387839
9.2 0.971601602383075
9.4 0.97203850214859
9.6 0.972481106103205
9.8 0.972921718726236
10 0.973352940650868
10.2 0.973767757113702
10.4 0.974159620758244
10.6 0.974522527865277
10.8 0.974851087168291
11 0.97514058050605
11.2 0.975387014666074
11.4 0.975587163880983
11.6 0.97573860255298
11.8 0.975839727899167
12 0.975889772330283
12.2 0.97588880549681
12.4 0.975837726057687
12.6 0.975738243346859
12.8 0.97559284923019
13 0.975404780558692
13.2 0.975177972732115
13.4 0.974917004988777
13.6 0.974627038131469
13.8 0.97431374548481
14 0.973983237954932
14.2 0.973641984127443
14.4 0.973296726393068
14.6 0.9729543941317
14.8 0.972622015014265
15 0.972306625497372
15.2 0.972015181588017
15.4 0.971754470944588
15.6 0.971531027356245
15.8 0.971351048605775
16.2 0.971144135164522
16.4 0.971127242820516
16.6 0.971173773794891
16.8 0.97128719540674
17 0.971470265890879
17.2 0.971724998608966
17.4 0.972052635065095
17.6 0.972453626961039
17.8 0.972927627414935
18 0.973473491356129
18.2 0.974089284999431
};

\nextgroupplot[
width =\linewidth, 
height = 0.65*\linewidth,
tick align=outside,
tick pos=left,
x grid style={darkgray176},
xmin=-0.885000000000001, xmax=20.785,
xtick style={color=black},
y grid style={darkgray176},
ymin=-0.08468386, ymax=1.05762346,
ytick style={color=black}
]
\addplot [draw=c1, fill=c1, mark=o, only marks]
table{%
x  y
0.1 1.0020524
0.2 1.0057004
0.3 0.9929484
0.4 0.970924000000001
0.5 0.951776
0.6 0.9262028
0.7 0.892046000000001
0.8 0.8678948
0.9 0.824936
1 0.784182400000001
1.1 0.7246304
1.2 0.6862204
1.3 0.6307128
1.4 0.5796392
1.5 0.5587904
1.6 0.5139924
1.7 0.4716688
1.8 0.437588
1.9 0.3888744
2 0.3535856
2.1 0.3540764
2.2 0.29808
2.3 0.2870532
2.4 0.257334
2.5 0.2014852
2.6 0.2286304
2.7 0.1918436
2.8 0.182114
2.9 0.1390336
3 0.146992
3.1 0.1348952
3.2 0.0948768
3.3 0.090526
3.4 0.0834288
3.5 0.0653988
3.6 0.0681424
3.7 0.0540104
3.8 0.0637372
3.9 0.03826
4 0.0579852
4.1 0.0204768
4.2 0.0651592
4.3 0.03349
4.4 0.0368416
4.5 0.0538184
4.6 0.0311388
4.7 0.00544
4.8 0.0247928
4.9 -0.0079468
5.2 0.019232
5.4 0.0265656
5.6 0.0330716
5.8 -0.018428
6 0.0183964
6.2 0.006662
6.4 0.025298
6.6 -0.0007608
6.8 0.030498
7 -0.0010444
7.2 -0.0063988
7.4 -0.029814
7.6 -0.0030664
7.8 -0.0294076
8 0.0111832
8.2 0.0087792
8.4 0.027982
8.6 -0.0074484
8.8 0.0126672
9 0.02164
9.2 0.0086384
9.4 -0.0217096
9.6 -0.010332
9.8 0.0257932
10 -0.0100232
10.2 -0.0136724
10.4 -0.0085656
10.6 0.018772
10.8 0.0078952
11 -0.0056616
11.2 -0.0212888
11.4 -0.0070056
11.6 -0.0017228
11.8 0.0113904
12 -0.0327608
12.2 -0.0016604
12.4 2.36000000000004e-05
12.6 -0.0053208
12.8 -0.0277152
13 -0.0149168
13.2 0.01169
13.4 0.000152
13.6 0.0195032
13.8 -0.0034888
14 0.0063004
14.2 -0.0079516
14.4 -0.0209244
14.6 0.001648
14.8 0.0062752
15 -0.0047364
15.2 -0.0072948
15.4 0.0059204
15.6 0.0029056
15.8 0.002798
16 0.0217656
16.2 -0.0178252
16.4 0.003268
16.6 0.004358
16.8 0.0157256
17 -0.000450000000000001
17.2 -0.0146544
17.4 -0.0074268
17.6 -0.0181384
17.8 -0.0137088
18 -0.0189732
18.2 0.0032588
};
\addplot [semithick, c3]
table {%
0.1 0.999984320103473
0.2 0.999937315439281
0.3 0.999859090997558
0.4 0.999749821454722
0.5 0.999609750713641
0.6 0.999439191261625
0.7 0.999238523347968
0.8 0.999008193983297
0.9 0.998748715763447
1 0.998460665521047
1.1 0.998144682808517
1.2 0.997801468216591
1.3 0.997431781532966
1.4 0.997036439746098
1.5 0.99661631489962
1.6 0.996172331803226
1.7 0.995705465606342
1.8 0.995216739241186
1.9 0.99470722074228
2 0.994178020449779
2.1 0.993630288104328
2.2 0.993065209841435
2.3 0.992484005093724
2.4 0.991887923409542
2.5 0.991278241196836
2.6 0.990656258401264
2.7 0.990023295127801
2.8 0.98938068821524
2.9 0.988729787773094
3 0.988071953690614
3.1 0.987408552127615
3.2 0.986740951996979
3.3 0.986070521448611
3.4 0.98539862436476
3.5 0.984726616876487
3.6 0.984055843911066
3.7 0.983387635780014
3.8 0.982723304817308
3.9 0.982064142077253
4 0.981411414101235
4.1 0.980766359762474
4.2 0.980130187197606
4.3 0.979504070833663
4.4 0.978889148518803
4.5 0.9782865187648
4.6 0.977697238108954
4.7 0.977122318602772
4.8 0.976562725434341
4.9 0.976019374690983
5.2 0.974495158532728
5.4 0.9735746328512
5.6 0.972736453451686
5.8 0.971984554932471
6 0.971321880273392
6.2 0.970750365338678
6.4 0.97027093497902
6.6 0.969883510700707
6.8 0.969587029730794
7 0.969379475171267
7.2 0.969257916803466
7.4 0.969218561978989
7.6 0.969256815915807
7.8 0.969367350610332
8 0.969544181478917
8.2 0.969780750756762
8.4 0.970070016609432
8.6 0.970404546853031
8.8 0.970776616133885
9 0.971178305387839
9.2 0.971601602383075
9.4 0.97203850214859
9.6 0.972481106103205
9.8 0.972921718726236
10 0.973352940650868
10.2 0.973767757113702
10.4 0.974159620758244
10.6 0.974522527865277
10.8 0.974851087168291
11 0.97514058050605
11.2 0.975387014666074
11.4 0.975587163880983
11.6 0.97573860255298
11.8 0.975839727899167
12 0.975889772330283
12.2 0.97588880549681
12.4 0.975837726057687
12.6 0.975738243346859
12.8 0.97559284923019
13 0.975404780558692
13.2 0.975177972732115
13.4 0.974917004988777
13.6 0.974627038131469
13.8 0.97431374548481
14 0.973983237954932
14.2 0.973641984127443
14.4 0.973296726393068
14.6 0.9729543941317
14.8 0.972622015014265
15 0.972306625497372
15.2 0.972015181588017
15.4 0.971754470944588
15.6 0.971531027356245
15.8 0.971351048605775
16 0.971220318671694
16.2 0.971144135164522
16.4 0.971127242820516
16.6 0.971173773794891
16.8 0.97128719540674
17 0.971470265890879
17.2 0.971724998608966
17.4 0.972052635065095
17.6 0.972453626961039
17.8 0.972927627414935
18 0.973473491356129
18.2 0.974089284999431
};
\end{groupplot}

\draw ({$(current bounding box.south west)!0.5!(current bounding box.south east)$}|-{$(current bounding box.south west)!-0.05!(current bounding box.north west)$}) node[
  scale=1.5,
  anchor=south,
  text=black,
  rotate=0.0
]{simulation time $\tau$};
\draw ({$(current bounding box.south west)!-0.02!(current bounding box.south east)$}|-{$(current bounding box.south west)!0.5!(current bounding box.north west)$}) node[
  scale=1.5,
  anchor=west,
  text=black,
  rotate=90.0
]{$\braket{O}$};
\end{tikzpicture}

%% file: main.bbl
\begin{thebibliography}{36}%
\makeatletter
\providecommand \@ifxundefined [1]{%
 \@ifx{#1\undefined}
}%
\providecommand \@ifnum [1]{%
 \ifnum #1\expandafter \@firstoftwo
 \else \expandafter \@secondoftwo
 \fi
}%
\providecommand \@ifx [1]{%
 \ifx #1\expandafter \@firstoftwo
 \else \expandafter \@secondoftwo
 \fi
}%
\providecommand \natexlab [1]{#1}%
\providecommand \enquote  [1]{``#1''}%
\providecommand \bibnamefont  [1]{#1}%
\providecommand \bibfnamefont [1]{#1}%
\providecommand \citenamefont [1]{#1}%
\providecommand \href@noop [0]{\@secondoftwo}%
\providecommand \href [0]{\begingroup \@sanitize@url \@href}%
\providecommand \@href[1]{\@@startlink{#1}\@@href}%
\providecommand \@@href[1]{\endgroup#1\@@endlink}%
\providecommand \@sanitize@url [0]{\catcode `\\12\catcode `\$12\catcode
  `\&12\catcode `\#12\catcode `\^12\catcode `\_12\catcode `\%12\relax}%
\providecommand \@@startlink[1]{}%
\providecommand \@@endlink[0]{}%
\providecommand \url  [0]{\begingroup\@sanitize@url \@url }%
\providecommand \@url [1]{\endgroup\@href {#1}{\urlprefix }}%
\providecommand \urlprefix  [0]{URL }%
\providecommand \Eprint [0]{\href }%
\providecommand \doibase [0]{https://doi.org/}%
\providecommand \selectlanguage [0]{\@gobble}%
\providecommand \bibinfo  [0]{\@secondoftwo}%
\providecommand \bibfield  [0]{\@secondoftwo}%
\providecommand \translation [1]{[#1]}%
\providecommand \BibitemOpen [0]{}%
\providecommand \bibitemStop [0]{}%
\providecommand \bibitemNoStop [0]{.\EOS\space}%
\providecommand \EOS [0]{\spacefactor3000\relax}%
\providecommand \BibitemShut  [1]{\csname bibitem#1\endcsname}%
\let\auto@bib@innerbib\@empty
\bibitem [{\citenamefont {Tsubouchi}\ \emph {et~al.}(2023)\citenamefont
  {Tsubouchi}, \citenamefont {Sagawa},\ and\ \citenamefont
  {Yoshioka}}]{FI_pap}%
  \BibitemOpen
  \bibfield  {author} {\bibinfo {author} {\bibfnamefont {K.}~\bibnamefont
  {Tsubouchi}}, \bibinfo {author} {\bibfnamefont {T.}~\bibnamefont {Sagawa}},\
  and\ \bibinfo {author} {\bibfnamefont {N.}~\bibnamefont {Yoshioka}},\
  }\bibfield  {title} {\bibinfo {title} {Universal cost bound of quantum error
  mitigation based on quantum estimation theory},\ }\bibfield  {journal}
  {\bibinfo  {journal} {Physical Review Letters}\ }\textbf {\bibinfo {volume}
  {131}},\ \href {https://doi.org/10.1103/physrevlett.131.210601}
  {10.1103/physrevlett.131.210601} (\bibinfo {year} {2023})\BibitemShut
  {NoStop}%
\bibitem [{\citenamefont {Takagi}\ \emph {et~al.}(2022)\citenamefont {Takagi},
  \citenamefont {Endo}, \citenamefont {Minagawa},\ and\ \citenamefont
  {Gu}}]{stat_pap}%
  \BibitemOpen
  \bibfield  {author} {\bibinfo {author} {\bibfnamefont {R.}~\bibnamefont
  {Takagi}}, \bibinfo {author} {\bibfnamefont {S.}~\bibnamefont {Endo}},
  \bibinfo {author} {\bibfnamefont {S.}~\bibnamefont {Minagawa}},\ and\
  \bibinfo {author} {\bibfnamefont {M.}~\bibnamefont {Gu}},\ }\bibfield
  {title} {\bibinfo {title} {Fundamental limits of quantum error mitigation},\
  }\bibfield  {journal} {\bibinfo  {journal} {npj Quantum Information}\
  }\textbf {\bibinfo {volume} {8}},\ \href
  {https://doi.org/10.1038/s41534-022-00618-z} {10.1038/s41534-022-00618-z}
  (\bibinfo {year} {2022})\BibitemShut {NoStop}%
\bibitem [{\citenamefont {Quek}\ \emph {et~al.}(2024)\citenamefont {Quek},
  \citenamefont {Stilck~França}, \citenamefont {Khatri}, \citenamefont
  {Meyer},\ and\ \citenamefont {Eisert}}]{Eisert}%
  \BibitemOpen
  \bibfield  {author} {\bibinfo {author} {\bibfnamefont {Y.}~\bibnamefont
  {Quek}}, \bibinfo {author} {\bibfnamefont {D.}~\bibnamefont
  {Stilck~França}}, \bibinfo {author} {\bibfnamefont {S.}~\bibnamefont
  {Khatri}}, \bibinfo {author} {\bibfnamefont {J.~J.}\ \bibnamefont {Meyer}},\
  and\ \bibinfo {author} {\bibfnamefont {J.}~\bibnamefont {Eisert}},\
  }\bibfield  {title} {\bibinfo {title} {Exponentially tighter bounds on
  limitations of quantum error mitigation},\ }\href
  {https://doi.org/10.1038/s41567-024-02536-7} {\bibfield  {journal} {\bibinfo
  {journal} {Nature Physics}\ }\textbf {\bibinfo {volume} {20}},\ \bibinfo
  {pages} {1648–1658} (\bibinfo {year} {2024})}\BibitemShut {NoStop}%
\bibitem [{\citenamefont {Aharonov}\ \emph {et~al.}(2025)\citenamefont
  {Aharonov}, \citenamefont {Alberton}, \citenamefont {Arad}, \citenamefont
  {Atia}, \citenamefont {Bairey}, \citenamefont {Brakerski}, \citenamefont
  {Cohen}, \citenamefont {Golan}, \citenamefont {Gurwich}, \citenamefont
  {Kenneth}, \citenamefont {Leviatan}, \citenamefont {Lindner}, \citenamefont
  {Melcer}, \citenamefont {Meyer}, \citenamefont {Schul},\ and\ \citenamefont
  {Shutman}}]{newpaper}%
  \BibitemOpen
  \bibfield  {author} {\bibinfo {author} {\bibfnamefont {D.}~\bibnamefont
  {Aharonov}}, \bibinfo {author} {\bibfnamefont {O.}~\bibnamefont {Alberton}},
  \bibinfo {author} {\bibfnamefont {I.}~\bibnamefont {Arad}}, \bibinfo {author}
  {\bibfnamefont {Y.}~\bibnamefont {Atia}}, \bibinfo {author} {\bibfnamefont
  {E.}~\bibnamefont {Bairey}}, \bibinfo {author} {\bibfnamefont
  {Z.}~\bibnamefont {Brakerski}}, \bibinfo {author} {\bibfnamefont
  {I.}~\bibnamefont {Cohen}}, \bibinfo {author} {\bibfnamefont
  {O.}~\bibnamefont {Golan}}, \bibinfo {author} {\bibfnamefont
  {I.}~\bibnamefont {Gurwich}}, \bibinfo {author} {\bibfnamefont
  {O.}~\bibnamefont {Kenneth}}, \bibinfo {author} {\bibfnamefont
  {E.}~\bibnamefont {Leviatan}}, \bibinfo {author} {\bibfnamefont {N.~H.}\
  \bibnamefont {Lindner}}, \bibinfo {author} {\bibfnamefont {R.~A.}\
  \bibnamefont {Melcer}}, \bibinfo {author} {\bibfnamefont {A.}~\bibnamefont
  {Meyer}}, \bibinfo {author} {\bibfnamefont {G.}~\bibnamefont {Schul}},\ and\
  \bibinfo {author} {\bibfnamefont {M.}~\bibnamefont {Shutman}},\ }\href
  {https://arxiv.org/abs/2503.17243} {\bibinfo {title} {On the importance of
  error mitigation for quantum computation}} (\bibinfo {year} {2025}),\ \Eprint
  {https://arxiv.org/abs/2503.17243} {arXiv:2503.17243} \BibitemShut {NoStop}%
\bibitem [{\citenamefont {Low}\ and\ \citenamefont
  {Chuang}(2017)}]{lowoptimal2017}%
  \BibitemOpen
  \bibfield  {author} {\bibinfo {author} {\bibfnamefont {G.~H.}\ \bibnamefont
  {Low}}\ and\ \bibinfo {author} {\bibfnamefont {I.~L.}\ \bibnamefont
  {Chuang}},\ }\bibfield  {title} {\bibinfo {title} {Optimal hamiltonian
  simulation by quantum signal processing},\ }\href
  {https://doi.org/10.1103/PhysRevLett.118.010501} {\bibfield  {journal}
  {\bibinfo  {journal} {Phys. Rev. Lett.}\ }\textbf {\bibinfo {volume} {118}},\
  \bibinfo {pages} {010501} (\bibinfo {year} {2017})}\BibitemShut {NoStop}%
\bibitem [{\citenamefont {Gily\'{e}n}\ \emph {et~al.}(2019)\citenamefont
  {Gily\'{e}n}, \citenamefont {Su}, \citenamefont {Low},\ and\ \citenamefont
  {Wiebe}}]{gilyen_quantum_2019}%
  \BibitemOpen
  \bibfield  {author} {\bibinfo {author} {\bibfnamefont {A.}~\bibnamefont
  {Gily\'{e}n}}, \bibinfo {author} {\bibfnamefont {Y.}~\bibnamefont {Su}},
  \bibinfo {author} {\bibfnamefont {G.~H.}\ \bibnamefont {Low}},\ and\ \bibinfo
  {author} {\bibfnamefont {N.}~\bibnamefont {Wiebe}},\ }\bibfield  {title}
  {\bibinfo {title} {Quantum singular value transformation and beyond:
  Exponential improvements for quantum matrix arithmetics},\ }in\ \href
  {https://doi.org/10.1145/3313276.3316366} {\emph {\bibinfo {booktitle}
  {Proceedings of the 51st Annual ACM SIGACT Symposium on Theory of
  Computing}}},\ \bibinfo {series and number} {STOC 2019}\ (\bibinfo
  {publisher} {Association for Computing Machinery},\ \bibinfo {address} {New
  York, NY, USA},\ \bibinfo {year} {2019})\ p.\ \bibinfo {pages}
  {193–204}\BibitemShut {NoStop}%
\bibitem [{\citenamefont {Magano}\ and\ \citenamefont
  {Mur\ifmmode~\mbox{\c{c}}\else \c{c}\fi{}a}(2022)}]{magano2022}%
  \BibitemOpen
  \bibfield  {author} {\bibinfo {author} {\bibfnamefont {D.}~\bibnamefont
  {Magano}}\ and\ \bibinfo {author} {\bibfnamefont {M.}~\bibnamefont
  {Mur\ifmmode~\mbox{\c{c}}\else \c{c}\fi{}a}},\ }\bibfield  {title} {\bibinfo
  {title} {Simplifying a classical-quantum algorithm interpolation with quantum
  singular value transformations},\ }\href
  {https://doi.org/10.1103/PhysRevA.106.062419} {\bibfield  {journal} {\bibinfo
   {journal} {Phys. Rev. A}\ }\textbf {\bibinfo {volume} {106}},\ \bibinfo
  {pages} {062419} (\bibinfo {year} {2022})}\BibitemShut {NoStop}%
\bibitem [{\citenamefont {Dong}\ \emph {et~al.}(2022)\citenamefont {Dong},
  \citenamefont {Whaley},\ and\ \citenamefont {Lin}}]{Dong_2022}%
  \BibitemOpen
  \bibfield  {author} {\bibinfo {author} {\bibfnamefont {Y.}~\bibnamefont
  {Dong}}, \bibinfo {author} {\bibfnamefont {K.~B.}\ \bibnamefont {Whaley}},\
  and\ \bibinfo {author} {\bibfnamefont {L.}~\bibnamefont {Lin}},\ }\bibfield
  {title} {\bibinfo {title} {A quantum hamiltonian simulation benchmark},\
  }\bibfield  {journal} {\bibinfo  {journal} {npj Quantum Information}\
  }\textbf {\bibinfo {volume} {8}},\ \href
  {https://doi.org/10.1038/s41534-022-00636-x} {10.1038/s41534-022-00636-x}
  (\bibinfo {year} {2022})\BibitemShut {NoStop}%
\bibitem [{\citenamefont {Dong}\ and\ \citenamefont
  {Lin}(2021)}]{Dong2020RandomCB}%
  \BibitemOpen
  \bibfield  {author} {\bibinfo {author} {\bibfnamefont {Y.}~\bibnamefont
  {Dong}}\ and\ \bibinfo {author} {\bibfnamefont {L.}~\bibnamefont {Lin}},\
  }\bibfield  {title} {\bibinfo {title} {Random circuit block-encoded matrix
  and a proposal of quantum linpack benchmark},\ }\href
  {https://doi.org/10.1103/PhysRevA.103.062412} {\bibfield  {journal} {\bibinfo
   {journal} {Phys. Rev. A}\ }\textbf {\bibinfo {volume} {103}},\ \bibinfo
  {pages} {062412} (\bibinfo {year} {2021})}\BibitemShut {NoStop}%
\bibitem [{\citenamefont {Tan}\ \emph {et~al.}(2023)\citenamefont {Tan},
  \citenamefont {Liu}, \citenamefont {Tran},\ and\ \citenamefont
  {Chuang}}]{tan2023errorcorrectionquantumalgorithms}%
  \BibitemOpen
  \bibfield  {author} {\bibinfo {author} {\bibfnamefont {A.~K.}\ \bibnamefont
  {Tan}}, \bibinfo {author} {\bibfnamefont {Y.}~\bibnamefont {Liu}}, \bibinfo
  {author} {\bibfnamefont {M.~C.}\ \bibnamefont {Tran}},\ and\ \bibinfo
  {author} {\bibfnamefont {I.~L.}\ \bibnamefont {Chuang}},\ }\href
  {https://arxiv.org/abs/2301.08542} {\bibinfo {title} {Error correction of
  quantum algorithms: Arbitrarily accurate recovery of noisy quantum signal
  processing}} (\bibinfo {year} {2023}),\ \Eprint
  {https://arxiv.org/abs/2301.08542} {arXiv:2301.08542} \BibitemShut {NoStop}%
\bibitem [{\citenamefont {Kikuchi}\ \emph {et~al.}(2023)\citenamefont
  {Kikuchi}, \citenamefont {Mc~Keever}, \citenamefont {Coopmans}, \citenamefont
  {Lubasch},\ and\ \citenamefont {Benedetti}}]{Kikuchi_2023}%
  \BibitemOpen
  \bibfield  {author} {\bibinfo {author} {\bibfnamefont {Y.}~\bibnamefont
  {Kikuchi}}, \bibinfo {author} {\bibfnamefont {C.}~\bibnamefont {Mc~Keever}},
  \bibinfo {author} {\bibfnamefont {L.}~\bibnamefont {Coopmans}}, \bibinfo
  {author} {\bibfnamefont {M.}~\bibnamefont {Lubasch}},\ and\ \bibinfo {author}
  {\bibfnamefont {M.}~\bibnamefont {Benedetti}},\ }\bibfield  {title} {\bibinfo
  {title} {Realization of quantum signal processing on a noisy quantum
  computer},\ }\bibfield  {journal} {\bibinfo  {journal} {npj Quantum
  Information}\ }\textbf {\bibinfo {volume} {9}},\ \href
  {https://doi.org/10.1038/s41534-023-00762-0} {10.1038/s41534-023-00762-0}
  (\bibinfo {year} {2023})\BibitemShut {NoStop}%
\bibitem [{\citenamefont {Trotter}(1959)}]{trotter1959}%
  \BibitemOpen
  \bibfield  {author} {\bibinfo {author} {\bibfnamefont {H.~F.}\ \bibnamefont
  {Trotter}},\ }\bibfield  {title} {\bibinfo {title} {On the product of
  semi-groups of operators},\ }\href {https://doi.org/10.2307/2033649}
  {\bibfield  {journal} {\bibinfo  {journal} {Proceedings of the American
  Mathematical Society}\ }\textbf {\bibinfo {volume} {10}},\ \bibinfo {pages}
  {545} (\bibinfo {year} {1959})}\BibitemShut {NoStop}%
\bibitem [{\citenamefont {Suzuki}(1976)}]{suzuki1976}%
  \BibitemOpen
  \bibfield  {author} {\bibinfo {author} {\bibfnamefont {M.}~\bibnamefont
  {Suzuki}},\ }\bibfield  {title} {\bibinfo {title} {Generalized trotter’s
  formula and systematic approximants of exponential operators and inner
  derivations with applications to many-body problems},\ }\href
  {https://doi.org/10.1007/BF01609348} {\bibfield  {journal} {\bibinfo
  {journal} {Communications in Mathematical Physics}\ }\textbf {\bibinfo
  {volume} {51}},\ \bibinfo {pages} {183} (\bibinfo {year} {1976})}\BibitemShut
  {NoStop}%
\bibitem [{\citenamefont {Suzuki}(1985)}]{suzuki1985}%
  \BibitemOpen
  \bibfield  {author} {\bibinfo {author} {\bibfnamefont {M.}~\bibnamefont
  {Suzuki}},\ }\bibfield  {title} {\bibinfo {title} {Decomposition formulas of
  exponential operators and lie exponentials with some applications to quantum
  mechanics and statistical physics},\ }\href
  {https://doi.org/10.1063/1.526596} {\bibfield  {journal} {\bibinfo  {journal}
  {Journal of Mathematical Physics}\ }\textbf {\bibinfo {volume} {26}},\
  \bibinfo {pages} {601} (\bibinfo {year} {1985})}\BibitemShut {NoStop}%
\bibitem [{\citenamefont {Wahl}\ \emph {et~al.}(2023)\citenamefont {Wahl},
  \citenamefont {Mari}, \citenamefont {Shammah}, \citenamefont {Zeng},\ and\
  \citenamefont {Ravi}}]{ZNE+QEC}%
  \BibitemOpen
  \bibfield  {author} {\bibinfo {author} {\bibfnamefont {M.~A.}\ \bibnamefont
  {Wahl}}, \bibinfo {author} {\bibfnamefont {A.}~\bibnamefont {Mari}}, \bibinfo
  {author} {\bibfnamefont {N.}~\bibnamefont {Shammah}}, \bibinfo {author}
  {\bibfnamefont {W.~J.}\ \bibnamefont {Zeng}},\ and\ \bibinfo {author}
  {\bibfnamefont {G.~S.}\ \bibnamefont {Ravi}},\ }\bibfield  {title} {\bibinfo
  {title} {Zero noise extrapolation on logical qubits by scaling the error
  correction code distance},\ }in\ \href
  {https://doi.org/10.1109/qce57702.2023.00103} {\emph {\bibinfo {booktitle}
  {2023 IEEE International Conference on Quantum Computing and Engineering
  (QCE)}}}\ (\bibinfo  {publisher} {IEEE},\ \bibinfo {year} {2023})\ p.\
  \bibinfo {pages} {888–897}\BibitemShut {NoStop}%
\bibitem [{\citenamefont {Piveteau}\ \emph {et~al.}(2021)\citenamefont
  {Piveteau}, \citenamefont {Sutter}, \citenamefont {Bravyi}, \citenamefont
  {Gambetta},\ and\ \citenamefont {Temme}}]{QEM+encoded}%
  \BibitemOpen
  \bibfield  {author} {\bibinfo {author} {\bibfnamefont {C.}~\bibnamefont
  {Piveteau}}, \bibinfo {author} {\bibfnamefont {D.}~\bibnamefont {Sutter}},
  \bibinfo {author} {\bibfnamefont {S.}~\bibnamefont {Bravyi}}, \bibinfo
  {author} {\bibfnamefont {J.~M.}\ \bibnamefont {Gambetta}},\ and\ \bibinfo
  {author} {\bibfnamefont {K.}~\bibnamefont {Temme}},\ }\bibfield  {title}
  {\bibinfo {title} {Error mitigation for universal gates on encoded qubits},\
  }\bibfield  {journal} {\bibinfo  {journal} {Physical Review Letters}\
  }\textbf {\bibinfo {volume} {127}},\ \href
  {https://doi.org/10.1103/physrevlett.127.200505}
  {10.1103/physrevlett.127.200505} (\bibinfo {year} {2021})\BibitemShut
  {NoStop}%
\bibitem [{\citenamefont {Low}\ \emph {et~al.}(2016)\citenamefont {Low},
  \citenamefont {Yoder},\ and\ \citenamefont {Chuang}}]{low_method_2016}%
  \BibitemOpen
  \bibfield  {author} {\bibinfo {author} {\bibfnamefont {G.~H.}\ \bibnamefont
  {Low}}, \bibinfo {author} {\bibfnamefont {T.~J.}\ \bibnamefont {Yoder}},\
  and\ \bibinfo {author} {\bibfnamefont {I.~L.}\ \bibnamefont {Chuang}},\
  }\bibfield  {title} {\bibinfo {title} {Methodology of resonant equiangular
  composite quantum gates},\ }\href {https://doi.org/10.1103/PhysRevX.6.041067}
  {\bibfield  {journal} {\bibinfo  {journal} {Phys. Rev. X}\ }\textbf {\bibinfo
  {volume} {6}},\ \bibinfo {pages} {041067} (\bibinfo {year}
  {2016})}\BibitemShut {NoStop}%
\bibitem [{\citenamefont {Haah}(2019)}]{Haah2019product}%
  \BibitemOpen
  \bibfield  {author} {\bibinfo {author} {\bibfnamefont {J.}~\bibnamefont
  {Haah}},\ }\bibfield  {title} {\bibinfo {title} {Product {D}ecomposition of
  {P}eriodic {F}unctions in {Q}uantum {S}ignal {P}rocessing},\ }\href
  {https://doi.org/10.22331/q-2019-10-07-190} {\bibfield  {journal} {\bibinfo
  {journal} {{Quantum}}\ }\textbf {\bibinfo {volume} {3}},\ \bibinfo {pages}
  {190} (\bibinfo {year} {2019})}\BibitemShut {NoStop}%
\bibitem [{\citenamefont {Berry}\ \emph {et~al.}(2024)\citenamefont {Berry},
  \citenamefont {Motlagh}, \citenamefont {Pantaleoni},\ and\ \citenamefont
  {Wiebe}}]{berrydoubling}%
  \BibitemOpen
  \bibfield  {author} {\bibinfo {author} {\bibfnamefont {D.~W.}\ \bibnamefont
  {Berry}}, \bibinfo {author} {\bibfnamefont {D.}~\bibnamefont {Motlagh}},
  \bibinfo {author} {\bibfnamefont {G.}~\bibnamefont {Pantaleoni}},\ and\
  \bibinfo {author} {\bibfnamefont {N.}~\bibnamefont {Wiebe}},\ }\bibfield
  {title} {\bibinfo {title} {Doubling the efficiency of hamiltonian simulation
  via generalized quantum signal processing},\ }\href
  {https://doi.org/10.1103/PhysRevA.110.012612} {\bibfield  {journal} {\bibinfo
   {journal} {Phys. Rev. A}\ }\textbf {\bibinfo {volume} {110}},\ \bibinfo
  {pages} {012612} (\bibinfo {year} {2024})}\BibitemShut {NoStop}%
\bibitem [{\citenamefont
  {Skelton}(2025)}]{skelton2025hitchhikersguideqsppreprocessing}%
  \BibitemOpen
  \bibfield  {author} {\bibinfo {author} {\bibfnamefont {S.~E.}\ \bibnamefont
  {Skelton}},\ }\bibfield  {title} {\bibinfo {title} {The hitchhiker's guide to
  qsp pre-processing},\ }\Eprint {https://arxiv.org/abs/2501.05977}
  {arXiv:2501.05977}  (\bibinfo {year} {2025})\BibitemShut {NoStop}%
\bibitem [{\citenamefont {Temme}\ \emph {et~al.}(2017)\citenamefont {Temme},
  \citenamefont {Bravyi},\ and\ \citenamefont
  {Gambetta}}]{temmeerrormitigation2017}%
  \BibitemOpen
  \bibfield  {author} {\bibinfo {author} {\bibfnamefont {K.}~\bibnamefont
  {Temme}}, \bibinfo {author} {\bibfnamefont {S.}~\bibnamefont {Bravyi}},\ and\
  \bibinfo {author} {\bibfnamefont {J.~M.}\ \bibnamefont {Gambetta}},\
  }\bibfield  {title} {\bibinfo {title} {Error mitigation for short-depth
  quantum circuits},\ }\href {https://doi.org/10.1103/PhysRevLett.119.180509}
  {\bibfield  {journal} {\bibinfo  {journal} {Phys. Rev. Lett.}\ }\textbf
  {\bibinfo {volume} {119}},\ \bibinfo {pages} {180509} (\bibinfo {year}
  {2017})}\BibitemShut {NoStop}%
\bibitem [{\citenamefont {Giurgica-Tiron}\ \emph {et~al.}(2020)\citenamefont
  {Giurgica-Tiron}, \citenamefont {Hindy}, \citenamefont {LaRose},
  \citenamefont {Mari},\ and\ \citenamefont {Zeng}}]{DQEM}%
  \BibitemOpen
  \bibfield  {author} {\bibinfo {author} {\bibfnamefont {T.}~\bibnamefont
  {Giurgica-Tiron}}, \bibinfo {author} {\bibfnamefont {Y.}~\bibnamefont
  {Hindy}}, \bibinfo {author} {\bibfnamefont {R.}~\bibnamefont {LaRose}},
  \bibinfo {author} {\bibfnamefont {A.}~\bibnamefont {Mari}},\ and\ \bibinfo
  {author} {\bibfnamefont {W.~J.}\ \bibnamefont {Zeng}},\ }\bibfield  {title}
  {\bibinfo {title} {Digital zero noise extrapolation for quantum error
  mitigation},\ }in\ \href {https://doi.org/10.1109/qce49297.2020.00045} {\emph
  {\bibinfo {booktitle} {2020 IEEE International Conference on Quantum
  Computing and Engineering (QCE)}}}\ (\bibinfo  {publisher} {IEEE},\ \bibinfo
  {year} {2020})\ p.\ \bibinfo {pages} {306–316}\BibitemShut {NoStop}%
\bibitem [{\citenamefont {Majumdar}\ \emph {et~al.}(2023)\citenamefont
  {Majumdar}, \citenamefont {Rivero}, \citenamefont {Metz}, \citenamefont
  {Hasan},\ and\ \citenamefont {Wang}}]{ZNEbestpractices}%
  \BibitemOpen
  \bibfield  {author} {\bibinfo {author} {\bibfnamefont {R.}~\bibnamefont
  {Majumdar}}, \bibinfo {author} {\bibfnamefont {P.}~\bibnamefont {Rivero}},
  \bibinfo {author} {\bibfnamefont {F.}~\bibnamefont {Metz}}, \bibinfo {author}
  {\bibfnamefont {A.}~\bibnamefont {Hasan}},\ and\ \bibinfo {author}
  {\bibfnamefont {D.~S.}\ \bibnamefont {Wang}},\ }\bibfield  {title} {\bibinfo
  {title} {{ Best Practices for Quantum Error Mitigation with Digital
  Zero-Noise Extrapolation }},\ }in\ \href
  {https://doi.org/10.1109/QCE57702.2023.00102} {\emph {\bibinfo {booktitle}
  {2023 IEEE International Conference on Quantum Computing and Engineering
  (QCE)}}}\ (\bibinfo  {publisher} {IEEE Computer Society},\ \bibinfo {address}
  {Los Alamitos, CA, USA},\ \bibinfo {year} {2023})\ pp.\ \bibinfo {pages}
  {881--887}\BibitemShut {NoStop}%
\bibitem [{\citenamefont {Javanmard}\ \emph {et~al.}(2022)\citenamefont
  {Javanmard}, \citenamefont {Liaubaite}, \citenamefont {Osborne},\ and\
  \citenamefont {Santos}}]{javanmard2022quantumsimulationdynamicalphase}%
  \BibitemOpen
  \bibfield  {author} {\bibinfo {author} {\bibfnamefont {Y.}~\bibnamefont
  {Javanmard}}, \bibinfo {author} {\bibfnamefont {U.}~\bibnamefont
  {Liaubaite}}, \bibinfo {author} {\bibfnamefont {T.~J.}\ \bibnamefont
  {Osborne}},\ and\ \bibinfo {author} {\bibfnamefont {L.}~\bibnamefont
  {Santos}},\ }\bibfield  {title} {\bibinfo {title} {Quantum simulation of
  dynamical phase transitions in noisy quantum devices},\ }\Eprint
  {https://arxiv.org/abs/2211.08318} {arXiv:2211.08318}  (\bibinfo {year}
  {2022})\BibitemShut {NoStop}%
\bibitem [{\citenamefont {Low}\ and\ \citenamefont
  {Chuang}(2019)}]{low_hamiltonian_2019}%
  \BibitemOpen
  \bibfield  {author} {\bibinfo {author} {\bibfnamefont {G.~H.}\ \bibnamefont
  {Low}}\ and\ \bibinfo {author} {\bibfnamefont {I.~L.}\ \bibnamefont
  {Chuang}},\ }\bibfield  {title} {\bibinfo {title} {Hamiltonian {Simulation}
  by {Qubitization}},\ }\href {https://doi.org/10.22331/q-2019-07-12-163}
  {\bibfield  {journal} {\bibinfo  {journal} {Quantum}\ }\textbf {\bibinfo
  {volume} {3}},\ \bibinfo {pages} {163} (\bibinfo {year} {2019})}\BibitemShut
  {NoStop}%
\bibitem [{Note1()}]{Note1}%
  \BibitemOpen
  \bibinfo {note} {Preparing the exponentiation of $\arccos (H)$ into an oracle
  with the same dimension as $H$ is not a standard routine; one generally
  embeds it within a larger unitary known as a block-encoding. For example, one
  can use quantum walks to generate a block-encoding. However, the
  block-encoding is prepared, it will have eigenvalues $ie^{\pm i\arccos
  \lambda }$ and then one can act on the appropriate eigensubspaces of the
  Hilbert space of the block-encoding.}\BibitemShut {Stop}%
\bibitem [{\citenamefont {Zhang}\ and\ \citenamefont
  {Yuan}(2024)}]{Zhang_2024}%
  \BibitemOpen
  \bibfield  {author} {\bibinfo {author} {\bibfnamefont {X.-M.}\ \bibnamefont
  {Zhang}}\ and\ \bibinfo {author} {\bibfnamefont {X.}~\bibnamefont {Yuan}},\
  }\bibfield  {title} {\bibinfo {title} {Circuit complexity of quantum access
  models for encoding classical data},\ }\bibfield  {journal} {\bibinfo
  {journal} {npj Quantum Information}\ }\textbf {\bibinfo {volume} {10}},\
  \href {https://doi.org/10.1038/s41534-024-00835-8}
  {10.1038/s41534-024-00835-8} (\bibinfo {year} {2024})\BibitemShut {NoStop}%
\bibitem [{\citenamefont {Childs}\ \emph {et~al.}(2021)\citenamefont {Childs},
  \citenamefont {Su}, \citenamefont {Tran}, \citenamefont {Wiebe},\ and\
  \citenamefont {Zhu}}]{Trotter}%
  \BibitemOpen
  \bibfield  {author} {\bibinfo {author} {\bibfnamefont {A.~M.}\ \bibnamefont
  {Childs}}, \bibinfo {author} {\bibfnamefont {Y.}~\bibnamefont {Su}}, \bibinfo
  {author} {\bibfnamefont {M.~C.}\ \bibnamefont {Tran}}, \bibinfo {author}
  {\bibfnamefont {N.}~\bibnamefont {Wiebe}},\ and\ \bibinfo {author}
  {\bibfnamefont {S.}~\bibnamefont {Zhu}},\ }\bibfield  {title} {\bibinfo
  {title} {Theory of trotter error with commutator scaling},\ }\bibfield
  {journal} {\bibinfo  {journal} {Physical Review X}\ }\textbf {\bibinfo
  {volume} {11}},\ \href {https://doi.org/10.1103/physrevx.11.011020}
  {10.1103/physrevx.11.011020} (\bibinfo {year} {2021})\BibitemShut {NoStop}%
\bibitem [{\citenamefont {Lötstedt}\ and\ \citenamefont
  {Yamanouchi}(2024)}]{LOTSTEDT2024140975}%
  \BibitemOpen
  \bibfield  {author} {\bibinfo {author} {\bibfnamefont {E.}~\bibnamefont
  {Lötstedt}}\ and\ \bibinfo {author} {\bibfnamefont {K.}~\bibnamefont
  {Yamanouchi}},\ }\bibfield  {title} {\bibinfo {title} {Comparison of current
  quantum devices for quantum computing of heisenberg spin chain dynamics},\
  }\href {https://doi.org/https://doi.org/10.1016/j.cplett.2023.140975}
  {\bibfield  {journal} {\bibinfo  {journal} {Chemical Physics Letters}\
  }\textbf {\bibinfo {volume} {836}},\ \bibinfo {pages} {140975} (\bibinfo
  {year} {2024})}\BibitemShut {NoStop}%
\bibitem [{\citenamefont {IBMQuantum}(2025)}]{ibmQuantum}%
  \BibitemOpen
  \bibfield  {author} {\bibinfo {author} {\bibnamefont {IBMQuantum}},\ }\href
  {https://quantum.ibm.com/services/resources} {\bibinfo {title} {Quantum
  processing units}} (\bibinfo {year} {2025}),\ \bibinfo {note} {[Accessed
  14-04-2025]}\BibitemShut {NoStop}%
\bibitem [{\citenamefont {Quantinuum}(2024)}]{quantiniummh1datasheet}%
  \BibitemOpen
  \bibfield  {author} {\bibinfo {author} {\bibnamefont {Quantinuum}},\ }\href
  {https://docs.quantinuum.com/systems/data_sheets/Quantinuum%20H1%20Product%20Data%20Sheet.pdf}
  {\bibinfo {title} {Quantinuum system model h1}} (\bibinfo {year} {2024}),\
  \bibinfo {note} {[Accessed 14-04-2025]}\BibitemShut {NoStop}%
\bibitem [{Note2()}]{Note2}%
  \BibitemOpen
  \bibinfo {note} {That is, a block encoding with subnormalization $1$, $0$
  ancilla qubits, and precision $\epsilon _{BE}=0$}\BibitemShut {NoStop}%
\bibitem [{\citenamefont
  {Skelton}(2024)}]{skelton2024harmlessmethodsqspprocessinglaurent}%
  \BibitemOpen
  \bibfield  {author} {\bibinfo {author} {\bibfnamefont {S.~E.}\ \bibnamefont
  {Skelton}},\ }\bibfield  {title} {\bibinfo {title} {Mostly harmless methods
  for qsp-processing with laurent polynomials},\ }in\ \href
  {https://doi.org/10.1109/QCE60285.2024.00027} {\emph {\bibinfo {booktitle}
  {{2024 International Conference on Quantum Computing and Engineering}}}}\
  (\bibinfo {year} {2024})\ pp.\ \bibinfo {pages} {150--160}\BibitemShut
  {NoStop}%
\bibitem [{Note3()}]{Note3}%
  \BibitemOpen
  \bibinfo {note} {Note that each of these could be decomposed into a more
  standard form of control-zero or control-one gate with a constant overhead of
  single-qubit gates.}\BibitemShut {Stop}%
\bibitem [{Note4()}]{Note4}%
  \BibitemOpen
  \bibinfo {note} {We are assuming access to $M_s$ required samples after
  samples have been disregarded during post-selection. If post-selection were
  included in the protocol explicitly, one would need to double the samples to
  account for the ones disregarded in post-selection.}\BibitemShut {Stop}%
\bibitem [{Note5()}]{Note5}%
  \BibitemOpen
  \bibinfo {note} {In our case, it was set to 1, due to the trivial estimator
  being able to reach the following bias for our chosen observable}\BibitemShut
  {NoStop}%
\end{thebibliography}%
